\newcommand{\shortversion}{}

\documentclass[10pt]{article}
\usepackage[margin=.8in,left=.8in]{geometry}

\newcommand{\ourrowcolor}[1]
{}

\newcommand{\vspaceabove}{$\phantom{\mathclap{\vert^{\vert^{\vert^{\vert}}}}}$}

\newcommand{\vspacebelow}{$\phantom{\mathclap{\vert_{\vert_{\vert_{\vert}}}}}$}

\usepackage{amsthm}
\usepackage{amssymb}
\usepackage{amsmath}
\usepackage{mathtools}
\usepackage{adjustbox}
\usepackage{xcolor}
\usepackage{stmaryrd}    
\usepackage[all]{xy}
\usepackage{rotating}
\usepackage{xfrac}
\usepackage{enumitem}\setlist{nosep}

\usepackage{hhline}

\usepackage{tikz}
\usetikzlibrary{
  backgrounds,
  decorations.pathmorphing,
  decorations.markings,
  decorations.text
}
\usetikzlibrary{cd}

\usepackage{mathrsfs} 
\DeclareMathAlphabet{\mathpzc}{OT1}{pzc}{m}{it} 

\usepackage{hyperref}
\hypersetup{
  colorlinks = true,
  allcolors=.
}

\let\PLAINthebibliography\thebibliography
\renewcommand\thebibliography[1]{
  \PLAINthebibliography{#1}
  \setlength{\parskip}{0.5pt}
  \setlength{\itemsep}{0.5pt plus .3ex}
}

\usepackage[titletoc]{appendix}

\usepackage[safe]{tipa} 

\definecolor{nyulight}{RGB}{107, 33, 158}

\definecolor{darkblue}{rgb}{0.05,0.25,0.65}
\definecolor{greenii}{RGB}{20,140,10}
\definecolor{darkgreen}{rgb}{0.00,0.85,0.1}
\definecolor{lightgray}{rgb}{0.9,0.9,0.9}
\definecolor{orangeii}{RGB}{200,100,5}
\definecolor{darkyellow}{rgb}{.91,.91,0}

\usepackage{multirow}

\usepackage{enumerate} 

\newcommand{\defneq}{\equiv}

\newcommand{\closed}{\mathrm{clsd}}

\newcommand{\HilbertSpace}[1]{\mathcal{#1}}

\newcommand{\Differential}{\mathrm{d}}
\newcommand{\differential}{\Differential}

\newcommand{\shape}{
  \raisebox{1pt}{\rm\normalfont\textesh}
}

\newcommand{\mapsup}{\rotatebox[origin=c]{90}{$\mapsto$}}

\usepackage[new]{old-arrows}   

\newdir{> }{{}*!/10pt/@{>}}

\usepackage{amssymb}

\DeclareRobustCommand{\rchi}{{\mathpalette\irchi\relax}}
\newcommand{\irchi}[2]{\raisebox{\depth}{$#1\chi$}} 

\makeatletter
\newif\if@sup
\newtoks\@sups
\def\append@sup#1{\edef\act{\noexpand\@sups={\the\@sups #1}}\act}%
\def\reset@sup{\@supfalse\@sups={}}%
\def\mk@scripts#1#2{\if #2/ \if@sup ^{\the\@sups}\fi \else%
  \ifx #1_ \if@sup ^{\the\@sups}\reset@sup \fi {}_{#2}%
  \else \append@sup#2 \@suptrue \fi%
  \expandafter\mk@scripts\fi}
\def\tensor#1#2{\reset@sup#1\mk@scripts#2_/}
\def\multiscripts#1#2#3{\reset@sup{}\mk@scripts#1_/#2%
  \reset@sup\mk@scripts#3_/}
\makeatother

\makeatletter
\newbox\slashbox \setbox\slashbox=\hbox{$/$}
\def\itex@pslash#1{\setbox\@tempboxa=\hbox{$#1$}
  \@tempdima=0.5\wd\slashbox \advance\@tempdima 0.5\wd\@tempboxa
  \copy\slashbox \kern-\@tempdima \box\@tempboxa}
\def\slash{\protect\itex@pslash}
\makeatother

\def\clap#1{\hbox to 0pt{\hss#1\hss}}
\def\mathllap{\mathpalette\mathllapinternal}
\def\mathrlap{\mathpalette\mathrlapinternal}
\def\mathclap{\mathpalette\mathclapinternal}
\def\mathllapinternal#1#2{\llap{$\mathsurround=0pt#1{#2}$}}
\def\mathrlapinternal#1#2{\rlap{$\mathsurround=0pt#1{#2}$}}
\def\mathclapinternal#1#2{\clap{$\mathsurround=0pt#1{#2}$}}

\let\oldroot\root
\def\root#1#2{\oldroot #1 \of{#2}}
\renewcommand{\sqrt}[2][]{\oldroot #1 \of{#2}}

\DeclareSymbolFont{symbolsC}{U}{txsyc}{m}{n}
\SetSymbolFont{symbolsC}{bold}{U}{txsyc}{bx}{n}
\DeclareFontSubstitution{U}{txsyc}{m}{n}

\DeclareSymbolFont{stmry}{U}{stmry}{m}{n}
\SetSymbolFont{stmry}{bold}{U}{stmry}{b}{n}

\DeclareFontFamily{OMX}{MnSymbolE}{}
\DeclareSymbolFont{mnomx}{OMX}{MnSymbolE}{m}{n}
\SetSymbolFont{mnomx}{bold}{OMX}{MnSymbolE}{b}{n}
\DeclareFontShape{OMX}{MnSymbolE}{m}{n}{
    <-6>  MnSymbolE5
   <6-7>  MnSymbolE6
   <7-8>  MnSymbolE7
   <8-9>  MnSymbolE8
   <9-10> MnSymbolE9
  <10-12> MnSymbolE10
  <12->   MnSymbolE12}{}

\usepackage{cleveref}

\crefformat{section}{\S#2#1#3} 
\crefformat{subsection}{\S#2#1#3}
\crefformat{subsubsection}{\S#2#1#3}

\newtheorem{theorem}{Theorem}[section]

\newtheorem{proposition}[theorem]{Proposition}

\theoremstyle{definition}
\newtheorem{definition}[theorem]{Definition}

\newtheorem{example}[theorem]{Example}

\newtheorem{remark}[theorem]{Remark}

\usepackage{amsfonts}
\usepackage{colortbl}

\renewcommand{\emph}{\textit}

\newcommand{\plus}{{\sqcup\{\infty\}}}

\begin{document}

\setlength{\abovedisplayskip}{3pt}
\setlength{\belowdisplayskip}{3pt}
\setlength{\abovedisplayshortskip}{-8pt}
\setlength{\belowdisplayshortskip}{3pt}

\title{Flux Quantization}

\author{
  Hisham Sati${}^{\ast \dagger}$
  \;\;
  and
  \;\;
  Urs Schreiber${}^{\ast}$
}

\maketitle

\thispagestyle{empty}

\begin{abstract}
Flux- and charge-quantization laws for higher gauge fields of Maxwell type — e.g. the common electromagnetic field (the ``A-field'') but also the B-, RR-,
and C-fields considered in string/M-theory —  specify non-perturbative completions of these fields by encoding their solitonic behaviour and hence
by specifying the discrete  charges carried by the individual branes (the higher-dimensional monopoles or solitons) that source the field fluxes.

This article surveys the general (rational-)homotopy theoretic understanding 
of flux- and charge-quantization via the Chern-Dold character map generalized to the non-linear (self-sourcing) Bianchi identities that appear
in higher-dimensional supergravity theories, notably for B-\&RR-fields in $D=10$, for the C-field in $D=11$ supergravity, and for the B-field on fivebrane worldvolumes.
\end{abstract}

\vspace{.1cm}

\begin{center}
\begin{minipage}{11.5cm}
\tableofcontents
\end{minipage}
\end{center}

\medskip

\vfill

\hrule
\vspace{5pt}

{
\footnotesize
\noindent
\def\arraystretch{1}
\tabcolsep=0pt
\begin{tabular}{ll}
${}^*$\,
&
Mathematics, Division of Science; and
\\
&
Center for Quantum and Topological Systems,
\\
&
NYUAD Research Institute,
\\
&
New York University Abu Dhabi, UAE.
\end{tabular}
\hfill
\adjustbox{raise=-15pt}{
\includegraphics[width=3cm]{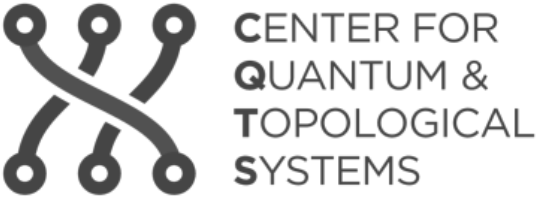}
}

\vspace{1mm}
\noindent ${}^\dagger$The Courant Institute for Mathematical Sciences, NYU, NY

\vspace{.2cm}

\noindent
The authors acknowledge the support by {\it Tamkeen} under the
{\it NYU Abu Dhabi Research Institute grant} {\tt CG008}.
}

\newpage

\noindent
{\bf History} {\small (see more references below).}
{\small

\vspace{.1cm}
\noindent
In 1852 Faraday observes magnetic field flux lines emanating from magnetic poles \cite{Faraday1852}, cf. \cref{ElectromagneticFluxAndItsBranes} below.

\vspace{.1cm}

\noindent
In 1931 Dirac invokes quantum mechanics to argue that, if there were unpaired such (mono-)poles, then the total flux emanating from them — and thus the magnetic charge carried by them — had to come in integer multiples of a unit quantum \cite{Dirac1931}, cf. Ex. \ref{FluxQuantizationLawsForOrdinaryelectromagnetism} below.

\vspace{.1cm}

\noindent
In 1957 Abrikosov essentially notices that the same electromagnetic flux-\&charge-quantization mechanism makes vortex strings in type II superconductors carry units of localized magnetic flux \cite{Abrikosov1957}, cf. p. \pageref{AbrikosovVortices} below.

\vspace{.1cm}

\noindent
In 1985 Alvarez understands such solitonic magnetic fields as 2-cocycles in (differential) ordinary cohomology \cite{Alvarez85}.

\vspace{.1cm}

\noindent In 1988 Gawedzki observes that the B-field flux felt by a string, hence the NS5-brane charge, must similarly be quantized as a 3-cocycle in Deligne cohomology \cite{Gawedzki88} \cite[\S 6]{FreedWitten99}, cf. Ex. \ref{HigherCircleGaugePotentials} below.

\vspace{.1cm}

\noindent
In the 1990s, string theorists hypothesize that the flux of RR-fields and hence the charge of D-branes is analogously quantized in a generalized cohomology theory called
``topological K-theory'' \cite{MinasianMoore97} \cite{Witten98}; or more generally in ``twisted'' such K-theory \cite{BouwknegtMathai01}, cf. \cref{RRFieldFluxQuantization} below;

\vspace{.05cm}

\noindent
and that the flux of the C-field and hence the charge of M-branes is quantized in a ``shifted half-integral'' cohomology theory \cite{Witten97} whose proper mathematical home motivates \cite{HopkinsSinger05} but for a long time remains somewhat mysterious, cf. \eqref{ProofOfShiftedFluxQuantizationOfCField} below.

\vspace{.1cm}

\noindent
In the 2020s  \cite{FSS20-H} develop a systematic understanding
of flux quantization of any higher gauge theory (of Maxwell-type, Def. \ref{HigherMaxwellEquations} below) in generalized non-abelian cohomology theory, using tools from dg-algebraic rational homotopy theory related to the ``FDA''-method in the supergravity literature \cite{FSS23Char}, cf. \cref{FluxQuantizationLaws} below.

This gives transparent re-derivation of previous flux quantization laws and allows to discuss C-field flux- and M-brane charge-quantization, cf. \cref{CFieldFluxQuantization} below.
}

\medskip

\section{Overview}

\vspace{-.2cm}
{\bf (\cref{FluxDensitiesAndBraneCharges})}
A {\it higher gauge theory} (for review in this volume see \cite[\S 2]{Alfonsi24}\cite{BFJKNRSW24}, but beware Rem. \ref{LInfinityOfGaugePotentialsVsFLuxDensities} below) of Maxwell-type (Def. \ref{HigherMaxwellEquations} below) is a (quantum) field theory analogous to vacuum-electromagnetism (on curved spacetimes), but with the analog of the electromagnetic flux density $F_2$ (which ordinarily is a differential 2-form on 3+1 dimensional spacetime $X^4$) allowed to be a system of differential forms $\vec F = \big\{F^{(i)}\}_{i \in I}$ of any degree $\mathrm{deg}_i \geq 1$ on a $D$-dimensional spacetime $X^D$ of any dimension $D = d + 1 \geq 2$, and satisfying a higher analog of Maxwell's equations \eqref{DualitySymmetricHigherMaxwellEquations}. Such higher gauge theories famously appear as the gauge field-sector in higher-dimensional supergravity (e.g. \cite{CastellaniDAuriaFre91}
\cite{Tanii96}
\cite{deWitLouis99}
\cite{Sezgin21}) and hence in super-string/M-theory (e.g. \cite{Duff99} \cite{BLT13}), which motivates their deeper investigation.

Like for ordinary Maxwell theory, where one may think of singularities or stable bumps in the electromagnetic flux density $F_2$ as being sourced by charges carried by (hypothetical) Dirac monopoles or by (observed) Abrikosov vortex strings, respectively (cf. \cref{ElectromagneticFluxAndItsBranes}), so one may think of singularities or stable bumps in these higher flux densities as sourced by singular branes or solitonic branes, respectively (cf. \cref{SingularVersusSolitonicBranes}), for suitably higher dimensional (mem-)branes carrying suitable higher charges.

\medskip

\noindent
{\bf (\cref{FluxQuantizationLaws})}
But for such singular/solitonic branes to be “elementary” objects of individually discernible nature, their total charges, and hence the total fluxes which they source, should take discrete (``quantized'') values (as indeed observed for Abrikosov vortex strings). This is what {\it flux quantization} is about.

Traditionally one declares the full higher gauge field to be given by gauge potentials $\widehat A$ whose curvature is the flux density $\vec F$ and which is globally subjected to a topological condition that implies the flux and charge quantization. While this is classical for electromagnetism (and Yang-Mills theory), it is not transparent from this point of view how to identify the structure of gauge potentials for general higher gauge theories.

More systematically, one may understand flux quantization as the specification of a generalized (non-abelian) cohomology theory for which the charges are required to be cocycles,
and of which the total fluxes are then the differential-geometric (Chern-Dold-){\it characters}.
From this streamlined point of view the higher gauge potential, and hence the full field content of the higher gauge theory, arises as the homotopy/gauge theoretic witness of the matching of total fluxes with the character of the charges, making the full higher gauge fields be cocycles in a corresponding generalized (nonabelian) {\it differential} cohomology theory.

\medskip

\noindent
{\bf (\cref{ExamplesInStringTheory})} Examples of flux quantization beyond Dirac charge quantization in electromagnetism play a key role in string/M-theory:

\begin{itemize}
\item the seminal ``Hypothesis K'' postulating D-brane charge quantization in K-theory,

\item the more recent ``Hypothesis H'' postulating M-brane charge quantization in unstable twisted Cohomotopy.
\end{itemize}

\smallskip

These hypotheses are not unrelated: Under “double dimensional Kaluza-Klein reduction” along a circle fiber, the C-field fluxes on $X^{10+1}$ are to give rise to most of the B\&RR-field-fluxes in $X^{9+1}$ (as part of the duality between M-theory and type IIA string theory, which is either conjectural or defining, depending on attitude towards the definition of the elusive M-theory). Therefore Hypothesis H may be understood as providing a non-perturbative/M-theoretic lift of Hypothesis K to the extent that it reduces to the latter under double dimensional reduction, and as a non-perturbative/M-theoretic correction to the extent that it does not quite reduce to Hypothesis K.

\medskip

\noindent
{\bf Perspective.}
This highlights that a {\it choice} of flux quantization is (depending on perspective of how that higher gauge theory is ultimately defined):
(i) a {\it hypothesis} about
or else
(ii) a {\it specification} of
the {\it non-perturbative completion} of the given higher gauge theory, which is generally an issue that deserves (more) attention.

Traditionally, flux quantization laws have been postulated sporadically and in ad-hoc fashion, in order to patch up ``anomalous'' theories: Since the ancient past it has been common to define physical theories by stationary action principles embodied by Lagrangian densities, from which perturbative BRST complexes ares extracted, whose quantization (e.g. \cite{HenneauxTeitelboim92}) is generally afflicted with problems (“anomalies”) some of which are dealt with by ad-hoc flux quantization: For example the original Dirac charge quantization was postulated to cure an anomaly in the quantum theory of an electron propagating in the background field of a magnetic monopole (a “0-brane”), while the enigmatic shifted C-field flux quantization similarly serves to cure an anomaly in the quantum theory of the M2-brane propagating in the background field of an M5-brane (\cite{Witten97}, \S2.2).

More systematically,
the available choices of flux-quantization laws $\mathcal{A}$ are algebro-topologically determined by the form of the higher Gauss law on any Cauchy surface, and any such choice, given by a compatible non-abelian cohomology-theory, determines the non-perturbative phase space stack of flux-quantized gauge fields. This process makes no reference to Lagrangian densities and applies seamlessly to field theories that do not even have a natural Lagrangian description, such as self-dual higher gauge theories.

Typically there is an ``evident'' choice of flux quantization and this is the choice tacitly made in the literature, where considered at all. But it is important to notice that there are other admissible choices, embodying hypotheses about (or definitions of) non-evident nonperturbative completions of the given higher gauge theory.

\medskip

\noindent
{\bf The logic of flux quantization.}
The table on p. \pageref{LogicOfFluxQuantization} shows in outline the logic of algebro-topological flux quantization as reviewed here; on the left in generality and on the right for our {\bf running examples}:

\begin{itemize}[leftmargin=.5cm]
\item[1.]    traditional Dirac charge quantization of the electromagnetic field (experimentally well-supported);

\item[2.]   traditional D-brane charge quantization in twisted topological K-theory (``Hypothesis K'');

\item[3.] more recent M-brane charge quantization in unstable twisted Cohomotopy (``Hypothesis H'').
\end{itemize}

\medskip

\noindent
{\bf The role of $L_\infty$-algebras.}
As the table on p. \pageref{LogicOfFluxQuantization} indicates, the algebro-topological nature of flux\&charge quantization is higher Lie-theoretic (explained in \cref{FluxQuantizationLaws}), by matching two $L_\infty$-algebras associated with a given higher gauge theory of Maxwell type (\cref{FluxDensitiesAndBraneCharges}):

\smallskip

\noindent
{\bf (i) Bianchi-Gau{\ss} $L_\infty$-algebras.}
The {\it higher Bianchi identities} on duality-symmetric higher fluxes (Def. \ref{HigherMaxwellEquations} below) and hence their {\it higher Gauss law} (Prop. \ref{SolutionSpaceViaGaussLaw} below) are equivalent to the condition that the flux densities jointly constitute a closed $L_\infty$-algebra valued differential form with coefficients in a characteristic $L_\infty$-algebra $\mathfrak{a}$ (Prop. \ref{FluxSolutionsAsLInfinityValuedForms} below):
\vspace{-.4cm}
$$
  \hspace{-20pt}
    \def\arraystretch{1}
    \begin{array}{l}
    {
      \scalebox{.7}{
        \color{darkblue}
        \bf
        \begin{tabular}{c}
          Space of flux densities
          \\
          on spacetime, solving
          \\
          the equations of motion
        \end{tabular}
      }
    }
    \mathrm{SolSpace}(X^D)
    \;\;\defneq\;\;
    \left\{
      \overset{
        \mathclap{
          \raisebox{3pt}{
            \scalebox{.7}{
              \color{gray}
              \rm
              electromagnetic flux densities on spacetime
            }
          }
        }
      }{
      \vec F \,\defneq\,
      \Big(
        F^{(i)}
        \,\in\,
        \Omega^{\mathrm{deg}_i}_{\mathrm{dR}}\big(
          X^D
        \big)
       \Big)_{i \in I}
       }
       \;\;\Bigg\vert
  \def\arraystretch{1.6}
  \begin{array}{l}
    \overset{
      \mathclap{
        \raisebox{3pt}{
          \scalebox{.7}{
            \color{gray}
            \rm
            Bianchi identities
          }
        }
      }
    }{
    \differential
    \, \vec F \,=\,
    \vec P\big( \vec F \big)
    }
    \\
    \underset{
      \mathclap{
        \raisebox{-1pt}{
          \scalebox{.7}{
            \color{gray}
            \rm
            self-duality
          }
        }
      }
    }{
    \star \, \vec F
    \,=\,
    \vec \mu\big( \vec F \big)
    }
  \end{array}
  \!\!  \right\}
  {
    \scalebox{.8}{
      \color{darkblue}
      \bf
      covariant form
    }
  }
    \\[25pt]
    \;\;
     \underset{
       \iota^\ast
     }{
       \simeq
     }
   \;
    \left\{
    \overset{
      \mathclap{
        \raisebox{3pt}{
          \scalebox{.7}{
            \color{gray}
            \rm
            magnetic flux densities on Cauchy surface
          }
        }
      }
    }{
      \vec B \,\defneq\,
      \Big(
        B^{(i)}
        \,\in\,
        \Omega^{\mathrm{deg}_i}_{\mathrm{dR}}\big(
          \,X^d\,
        \big)
       \Big)_{i \in I}
       }
       \;\;\bigg\vert
       \begin{array}{l}
    \overset{
      \mathclap{
        \raisebox{3pt}{
          \scalebox{.7}{
            \color{gray}
            \rm
            Gau{\ss} law
          }
        }
      }
    }{
         \differential
         \, \vec B \,=\,
         \vec P\big( \vec B \big)
    }
    \\
    {}
       \end{array}
  \!\!\!  \right\}
  {
    \scalebox{.8}{
      \color{darkblue}
      \bf
      canonical form
    }
  }
  \;\;\simeq\;\;
    \Omega^1_{\mathrm{dR}}\big(
      X^d
      ;\,
      \mathfrak{a}
    \big)_\closed
    \;\;
    \mathrlap{
      \scalebox{.7}{
        \bf
        \color{purple}
        \begin{tabular}{l}
        space of
        closed (flat)
        \\
        $\mathfrak{a}$-valued
        \\
        differential forms
        \end{tabular}
      }
    }
  \end{array}
$$

\smallskip

\noindent
{\bf (ii) Whitehead $L_\infty$-algebras.} The classifying space $\mathcal{A}$ of any charge quantization law is rationally characterized by its rational Whitehead $L_\infty$-algebra $\mathfrak{l}\mathcal{A}$ (the ``Koszul-dual'' of $\mathcal{A}$: that $L_\infty$-algebra whose Chevalley-Eilenberg algebra $\mathrm{CE}(\mathfrak{l}\mathcal{A})$ is the Sullivan model of $\mathcal{A}$) and the nonabelian Chern-Dold character map extracts from $\mathcal{A}$-cohomology its image in $\mathfrak{l}\mathcal{A}$-valued nonabelian de Rham cohomology \eqref{NonAbDeRhamCohomologyIsTargteOfCharacter}:
$$
\hspace{-2cm}
\adjustbox{raise=-17pt}{
  \begin{tikzcd}[
    column sep=20pt,
    row sep=6pt
  ]
    H^1(X;\Omega\mathcal{A})
    \ar[
      rr,
      "{
        \scalebox{.7}{
          \color{greenii}
          \bf
          rationalization
        }
      }"
    ]
    \ar[
      d,
      equals
    ]
    \ar[
      rrrrrr,
      rounded corners,
      to path={
           ([yshift=00pt]\tikztostart.north)
        -- ([yshift=8pt]\tikztostart.north)
        -- node{
          \scalebox{.7}{
            \colorbox{white}{
              \color{orangeii}
              \bf
              character map on $\mathcal{A}$-cohomology
            }
          }
        }
           ([yshift=8pt]\tikztotarget.north)
        -- ([yshift=00pt]\tikztotarget.north)
      }
    ]
    &&
    H^1\big(
      X
      ;\,
      L^{\mathbb{Q}}
      \Omega
      \mathcal{A}
    \big)
    \ar[
      rr,
      "{
        \scalebox{.7}{
          \color{greenii}
          \bf
          extension
        }
      }",
      "{
        \scalebox{.7}{
          \color{greenii}
          \bf
          of scalars
        }
      }"{swap}
    ]
    \ar[
      d,
      equals
    ]
    &&
    H^1\big(
      X
      ;\,
      L^{\mathbb{R}}
      \Omega
      \mathcal{A}
    \big)
    \ar[
      d,
      equals
    ]
    \ar[
      rr,
      "{
        \scalebox{.7}{
          \color{greenii}
          \bf
          nonabelian
        }
      }",
      "{
        \scalebox{.7}{
          \color{greenii}
          \bf
          de Rham theorem
        }
      }"{swap}
    ]
    &&
    H^1_{\mathrm{dR}}\big(
      X
      ;\,
      \mathfrak{l}\mathcal{A}
    \big)
    \ar[
      d,
      equals
    ]
    \\
    \pi_0
    \mathrm{Map}\big(
      X
      ,\,
      \mathcal{A}
    \big)
    \ar[
      rr,
      "{
        (\eta_{\mathcal{A}}^{\mathbb{Q}})_\ast
      }"
    ]
    &&
    \pi_0
    \mathrm{Map}\big(
      X
      ,\,
      L^{\mathbb{Q}}\mathcal{A}
    \big)
    \ar[
      rr,
      "{
        (\eta_{L^{\mathbb{Q}}\mathcal{A}}^{\mathrm{ext}})_\ast
      }"
    ]
    &&
    \pi_0
    \mathrm{Map}\big(
      X
      ,\,
      L^{\mathbb{R}}\mathcal{A}
    \big)
    \ar[
      rr,
      "\sim",
      "{
        \scalebox{.7}{
          \color{gray}
          \def\arraystretch{.9}
          \begin{tabular}{c}
            fundamental theorem
            \\
            of dg-algebraic RHT
          \end{tabular}
        }
      }"{swap, yshift=-5pt}
    ]
    &&
    \mathrm{Hom}_{\mathrm{dgAlg}}\big(
      \mathrm{CE}(\mathfrak{l}\mathcal{A})
      ,\,
      \Omega^\bullet_{\mathrm{dR}}(X)
    \big)_{\!/\mathrm{cncrd}}
  \end{tikzcd}
  }
  \hspace{-2cm}
$$

\medskip

\noindent
\hspace{-.4cm}
\adjustbox{}{
\small\label{LogicOfFluxQuantization}
\def\tabcolsep{2pt}
\def\arraystretch{1}
\begin{tabular}{|c|c||c|c|c|c|}
  \hline
  \multicolumn{2}{|c||}{}
  &&&&
  \\[-3pt]
  \multicolumn{2}{|c||}{
    {
      \bf
      \def\arraystretch{.9}
      \begin{tabular}{c}
        Logic of
        \\
        flux quantization
      \end{tabular}
    }
  }
  &
  \def\arraystretch{.9}
  \begin{tabular}{c}
    \bf
    Higher gauge theory
    \\
    of Maxwell-type
  \end{tabular}
  &
  \def\arraystretch{.9}
  \begin{tabular}{c}
    {\bf A-field}
    \\
    in $D = 4$
  \end{tabular}
  &
  \def\arraystretch{.9}
  \begin{tabular}{c}
    {\bf B\&RR-field}
    \\
    in $D = 10$
  \end{tabular}
  &
  \def\arraystretch{.9}
  \begin{tabular}{c}
    {\bf C-field}
    \\
    in $D = 11$
  \end{tabular}
  \\[+10pt]
  \hline
  \hline
  \ourrowcolor{lightgray}
  &&&&&
  \\[-8pt]
  \ourrowcolor{lightgray}
  \cellcolor{white}
  &
  \def\arraystretch{.9}
  \begin{tabular}{c}
    Flux
    \\
    densities
  \end{tabular}
  &
  $
  \begin{array}{l}
    \vec F \,\defneq\,
    \\
    \Big(
      F^{(i)} \in
      \Omega^{\mathrm{deg}_i}_{\mathrm{dR}}
      \big(
        X^D
      \big)
    \Big)_{i \in I}
  \end{array}
  $
  &
  $
    \def\arraystretch{1}
    \def\arraycolsep{3pt}
    \begin{array}{ll}
      F_2
      &\scalebox{.8}{\color{gray} magnetic}
      \\
      G_2
      &\scalebox{.8}{\color{gray} electric}
    \end{array}
  $
  &
  $
    \def\arraystretch{1}
    \begin{array}{rl}
      H_3
      &\scalebox{.8}{\color{gray} NS5}
      \\
      H_7
      &\scalebox{.8}{\color{gray} F1}
      \\
      F_{2\bullet}
      &\scalebox{.8}{\color{gray} $D_{8-2\bullet}$ flux densities on}
      \\[-7pt]
      {}
    \end{array}
  $
  &
  $
    \def\arraystretch{1}
    \begin{array}{rl}
      G_4
      &\scalebox{.8}{\color{gray} M5}
      \\
      G_7
      &\scalebox{.8}{\color{gray} M2}
      \\[-7pt]
      {}
    \end{array}
  $
  \\[+14pt]
  \cline{2-6}
  &&&&&
  \\[-8pt]
  \cref{FluxDensitiesAndBraneCharges}
  &
  \def\arraystretch{.9}
  \begin{tabular}{c}
    Self-
    \\
    duality
  \end{tabular}
  &
  $
    \def\arraystretch{1.2}
    \def\arraycolsep{2pt}
    \begin{array}{rcl}
      \star \, \vec F
      \,=\,
      \vec \mu\big(
        \vec F
      \big)
    \end{array}
  $
  &
  $
    \def\arraystretch{1.2}
    \def\arraycolsep{2pt}
    \begin{array}{rcl}
      \star \, F_2
      &=&
      G_2
    \end{array}
  $
  &
  $
    \def\arraystretch{1.2}
    \def\arraycolsep{2pt}
    \begin{array}{rcl}
      \star H_3 &=& H_7
      \\
      \star F_{2\bullet}
      &=&
      F_{10-2\bullet}
    \end{array}
  $
  &
  $
    \def\arraystretch{1.2}
    \def\arraycolsep{2pt}
    \begin{array}{rcl}
      \star \, G_4 &=& G_7
    \end{array}
  $
  \\[+10pt]
  \cline{2-6}
  &&&&&
  \\[-8pt]
  \ourrowcolor{lightgray}
  &
  \def\arraystretch{.9}
  \begin{tabular}{c}
    Bianchi
    \\
    identities
  \end{tabular}
  &
  $
    \def\arraystretch{1.6}
    \begin{array}{l}
      \differential
      \,
      \vec F
      \,=\,
      \vec P\big(
        \vec F
      \big)
    \end{array}
  $
  &
  $
    \def\arraystretch{1.2}
    \def\arraycolsep{2pt}
    \begin{array}{rcl}
      \differential \, F_2
      &=&
      0
      \\
      \differential\, G_2
      &=&
      0
    \end{array}
  $
  &
  $
    \def\arraystretch{1.2}
    \def\arraycolsep{2pt}
    \begin{array}{rcl}
      \differential H_3 &=& 0
      \\
      \differential H_7 &=& 0
      \\
      \differential F_{2\bullet}
      &=&
      H_3 \wedge F_{2\bullet-2}
      \\[-10pt]
      {}
    \end{array}
  $
  &
  $
    \def\arraystretch{1.2}
    \def\arraycolsep{2pt}
    \begin{array}{rcl}
      \differential G_4
      &=&
      0
      \\
      \differential G_7
      &=&
      -\tfrac{1}{2} G_4 \wedge G_4
    \end{array}
  $
  \\[+10pt]
  \hline
  \hline
  &&&&&
  \\[-8pt]
  &
  \begin{tabular}{c}
    CE-algebra of
    \\
    characteristic
    \\
    $L_\infty$-algebra
  \end{tabular}
  &
  $
    \def\arraystretch{1.4}
    \def\arraycolsep{2pt}
    \begin{array}{l}
    \mathrm{CE}(\mathfrak{a})
    \defneq
    \\
    \mathbb{R}\big[
      \vec b
    \big]
    \big/
    \Big(
    \differential \vec b
    \,=\,
    \vec P\big(
      \vec b
    \big)
    \Big)
    \end{array}
  $
  &
  $
    \def\arraystretch{1.2}
    \def\arraycolsep{2pt}
    \begin{array}{rcl}
      \differential f_2
      &=& 0
      \\
      \differential g_2
      &=& 0
    \end{array}
  $
  &
  $
    \def\arraystretch{1.2}
    \def\arraycolsep{2pt}
    \begin{array}{rcl}
      \differential h_3
      &=& 0
      \\
      \differential h_7
      &=& 0
      \\
      \differential f_{2\bullet}
      &=&
      h_3 \wedge f_{2\bullet-2}
      \\[-10pt]
      {}
    \end{array}
  $
  &
  $
    \def\arraystretch{1.2}
    \def\arraycolsep{2pt}
    \begin{array}{rcl}
      \differential g_4
      &=&
      0
      \\
      \differential g_7
      &=&
      -\tfrac{1}{2}
      g_4
      \wedge
      g_4
    \end{array}
  $
  \\[+10pt]
  \cline{2-6}
  &&&&&
  \\[-10pt]
 \ourrowcolor{lightgray}   \cref{FluxQuantizationLaws}
  &
 \def\arraystretch{.9}
  \begin{tabular}{c}
    Solution space
    \\
    of fluxes on
    \\
    $X^D = \mathbb{R}^{0,1} \times X^d$
  \end{tabular}
  &
  $
  \def\arraystretch{1.4}
  \begin{array}{l}
   \overset{
     \mathclap{
       \scalebox{.7}{
         \color{gray}
         Gau{\ss} law =
         $\mathfrak{a}$-closedness
       }
     }
   }{
      \mathbf{\Omega}_{\mathrm{dR}}\big(
        X^d
        ;\,
        \mathfrak{a}
      \big)_{\closed}
    }
    \,\defneq\,
    \\
    \mathrm{Hom}\big(
      \mathrm{CE}(\mathfrak{a})
      ,\,
      \Omega^\bullet_{\mathrm{dR}}(X^d)
    \big)
  \end{array}
  $
  &
  $
    \def\arraystretch{1.4}
    \begin{array}{r}
    \Omega^2_{\mathrm{dR}}(X^d)_{\mathrm{clsd}}
    \\
    \times
    \underset{
      \mathclap{
        \scalebox{.7}{
          \color{gray}
          can. momenta
        }
      }
    }{
    \, \Omega^2_{\mathrm{dR}}(X^d)_{\mathrm{clsd}}
    }
    \\[-12pt]
    {}
    \end{array}
  $
  &
  \scalebox{.9}{
  \begin{tabular}{c}
    3-twisted
    \\
    de Rham
    \\
    cocycles
  \end{tabular}
  }
  &
  \scalebox{.9}{
  \begin{tabular}{c}
    ``4-twisted''
    \\
    de Rham
    \\
    cocycles
  \end{tabular}
  }
  \\[+12pt]
  \cline{2-6}
  &&&&&
  \\[-5pt]
  &
  \def\arraystretch{.9}
  \begin{tabular}{c}
    Characteristic
    \\
    $L_\infty$-algebra
  \end{tabular}
  &
  $\mathfrak{a}$
  &
  $
   b \mathfrak{u}(1)
   \oplus
   b \mathfrak{u}(1)
  $
  &
  $
  \begin{array}{c}
    \big[
      b_2, v_{2\bullet-1}
    \big]
    =
    v_{2\bullet+1}
  \end{array}
  $
  &
  $
  \underset{
    \mathclap{
      \scalebox{.7}{
        \color{gray}
        M-theory gauge algebra
      }
    }
  }{
  \big[
    v_3,\, v_3
  \big]
  =
  v_6
  }
  $
  \\[+10pt]
  \cline{2-6}
  &&&&&
  \\[-4pt]
  \ourrowcolor{lightgray}  &
 \def\arraystretch{.9}
   \begin{tabular}{c}
    as rational White-
    \\
    head $L_\infty$-algebra
  \end{tabular}
  &
  $
    \mathfrak{a}
    \,\simeq\,
    \mathfrak{l}\mathcal{A}
  $
  &
  $
    \mathfrak{l}
    \big(
      B^2 \mathbb{Z}
      \times
      B^2 \mathbb{Z}
    \big)
  $
  &
  $
    \mathfrak{l}\Big(
      \big(
      \mathrm{KU}_0
      \!\sslash\!
      B^2 \mathbb{Z}
      \big)
      \times
      B^7 \mathbb{Z}
    \Big)
  $
  &
  $
  \mathfrak{l}\big(
    S^4
  \big)
  $
  \\[+8pt]
  \hline
  \hline
  &&&&&
  \\[-6pt]
  &
  \def\arraystretch{.9}
  \begin{tabular}{c}
    \\[-6pt]
    Evident choice of
    \\
    classifying space
  \end{tabular}
  &
  $\mathcal{A}$
  &
  $
    \underset{
      \mathclap{
        \raisebox{-3pt}{
        \scalebox{.7}{
          \color{purple}
          Dirac's hypothesis
        }
        }
      }
    }{
    B^2 \mathbb{Z} \times B^2 \mathbb{Z}
    }
  $
  &
  $
    \underset{
      \mathclap{
        \raisebox{-0pt}{
        \scalebox{.7}{
          \color{purple}
          Hypothesis K
        }
        }
      }
    }{
    \big(
    \mathrm{KU}_0
    \!\sslash\!
    B^2 \mathbb{Z}
    \big)
    \times
    B^7 \mathbb{Z}
    }
  $
  &
  $
    \underset{
      \mathclap{
        \raisebox{-4pt}{
        \scalebox{.7}{
          \color{purple}
          Hypothesis H
        }
        }
      }
    }{
    S^4
    }
  $
  \\[+12pt]
  \cline{2-6}
  &&&&&
  \\[-4pt]
 \ourrowcolor{lightgray}   \cref{ExamplesInStringTheory}
  &
 \def\arraystretch{.9}
  \begin{tabular}{c}
    Corresponding
    \\
    cohomology theory
  \end{tabular}
  &
  \def\arraystretch{.9}
  \begin{tabular}{c}
    generalized
    \\
    cohomology
  \end{tabular}
  &
  \def\arraystretch{.9}
  \begin{tabular}{c}
    ordinary
    \\
    cohomology
  \end{tabular}
  &
  \def\arraystretch{.9}
  \begin{tabular}{c}
    twisted
    \\
    K-theory
  \end{tabular}
  &
  \def\arraystretch{.9}
  \begin{tabular}{c}
    unstable
    \\
    CoHomotopy
  \end{tabular}
  \\[+10pt]
  \cline{2-6}
  &&&&&
  \\[-6pt]
  &
  \def\arraystretch{.9}
  \begin{tabular}{c}
    Flux-quantized
    \\
    phase space
  \end{tabular}
  &
  \adjustbox{raise=3pt}{
  $
    \phantom{\vert^{\vert^{\vert^{\vert}}}}
    \mathbf{\Omega}_{\mathrm{dR}}\big(
      X^d; \mathfrak{a}
    \big)_{\closed}
    \underset{
      \mathclap{
        \;\;\;\;
        \raisebox{-3pt}{\scalebox{.6}{$
        L^{\mathbb{R}}\mathcal{A}(X^d)
        $}}
      }
    }{\times}
    \mathcal{A}(X^d)
  $
  }
  &
  \def\arraystretch{.9}
  \begin{tabular}{c}
    differential
    \\
    cohomology
  \end{tabular}
  &
  \def\arraystretch{.9}
  \begin{tabular}{c}
    differential
    \\
    twisted K-theory
  \end{tabular}
  &
  \def\arraystretch{.9}
  \begin{tabular}{c}
    differential
    \\
    CoHomotopy
  \end{tabular}
  \\[+10pt]
  \hline
\end{tabular}
}

\medskip

\noindent
\def\tabcolsep{0pt}
\hspace{-.25cm}
\begin{tabular}{p{6.7cm}l}
The admissible flux quantization laws for a higher gauge theory with Bianchi-Gau{\ss} $L_\infty$-algebra $\mathfrak{a}$ are hence those classified by spaces $\mathcal{A}$ with Whitehead $L_\infty$-algebra $\mathfrak{l}\mathcal{A} \simeq \mathfrak{a}$. Given such a choice, then quantizing a flux density $\vec B$ globally is to lift its $\mathfrak{a}$-valued de Rham-class to a class in $\mathcal{A}$-valued nonabelian cohomology.
&
\adjustbox{raise=-1.2cm}{
$
  \begin{tikzcd}[
    column sep=25pt,
    row sep=15pt
  ]
    &[-10pt]
    &&
      H^1\big(
        X^d
        ;\,
        \Omega\mathcal{A}
      \big)
    \mathrlap{
      \raisebox{3pt}{
      \scalebox{.7}{
       \color{darkblue}
       \bf
        \def\arraystretch{.9}
        \begin{tabular}{c}
          choice of
          \\
          \scalebox{1}{$\mathcal{A}$}-cohomology
          \\
          \color{purple}
          with
          $\mathfrak{l}\mathcal{A} \,\simeq\,
          \mathfrak{a}$
        \end{tabular}
      }
      }
    }
    \ar[
      dd,
      "{
        \underset{
          \mathclap{
            \scalebox{.7}{
              \color{greenii}
              \bf
              sourced flux
            }
          }
        }{
        \mathrm{ch}_{\mathcal{A}}(X^d)
        }
      }"{description},
    ]
    \\
    \\
    \ast
    \ar[
      rr,
      "{
        \vec B
      }"{description},
      "{
        \scalebox{.7}{
          \color{darkblue}
          \bf
          \def\arraystretch{.9}
          \begin{tabular}{c}
            flux densities on Cauchy surface
            \\
            satisfying their higher Gau{\ss} law
          \end{tabular}
        }
      }"{swap, yshift=-6pt},
    ]
    \ar[
      uurrr,
      dashed,
      "{
        \scalebox{.7}{
          \color{orangeii}
          \bf
          charge quantum in
          \scalebox{1.2}{$\mathcal{A}$}-cohomology
        }
      }"{sloped},
      "{ c }"{description}
    ]
    &
    &
    \Omega^1_{\mathrm{dR}}(
      X^d
      ;\,
      \mathfrak{a}
    )_{\closed}
    \ar[
      r,
      "{
        \scalebox{.7}{
          \color{greenii}
          \bf
          \def\arraystretch{.9}
          \begin{tabular}{c}
            total flux
          \end{tabular}
        }
      }"{pos=.4}
    ]
    &
    \underset{
      \mathclap{
        \raisebox{-4pt}{
          \scalebox{.7}{
            \color{darkblue}
            \bf
            \def\arraystretch{.9}
            \begin{tabular}{c}
              $\mathfrak{a}$-valued
              \\
              de Rham cohomology
            \end{tabular}
          }
        }
      }
    }{
    H^1_{\mathrm{dR}}(
      X^d
      ;\,
      \mathfrak{a}
    )
    }
  \end{tikzcd}
$
}
\end{tabular}

\noindent
\hspace{-.2cm}
\def\tabcolsep{0pt}
\begin{tabular}{p{10.2cm}l}
Locally, a {\it flux-quantized higher gauge field} is (i) a flux density $\vec B$ being a cocycle in $\mathfrak{a}$-de Rham cohomology,
(ii) a charge $\rchi$ being a cocycle in $\mathcal{A}$-cohomology and (iii) a {\it gauge potential} $\widehat A$ being a coboundary
between their joint images (thus exhibiting the above identification of their cohomology classes).
This makes the flux-quantized higher gauge fields be cocycles in {\it differential} $\mathcal{A}$-cohomology.
&
\hspace{-5pt}
\adjustbox{raise=-.9cm}{
$
  \begin{tikzcd}[
    column sep=23pt,
    row sep=15pt
  ]
    \widehat{X}{}^d
    \ar[
      rr,
      dashed,
      "{
        \scalebox{.7}{
          \color{greenii}
          \bf
          charges
        }
      }",
      "{ \rchi }"{
        swap,
        name=s
      }
    ]
    \ar[
      dd,
      dashed,
      "{
        \scalebox{.7}{
          \color{greenii}
          \bf
          \def\arraystretch{.9}
          \begin{tabular}{c}
            flux
            \\
            densities
          \end{tabular}
        }
      }"{swap, xshift=2pt},
      "{
        \vec B
      }"{
        name=t
      }
    ]
    &&
    \mathcal{A}
    \ar[
      dd,
      "{
        \mathbf{ch}_{\mathcal{A}}
      }"{swap},
      "{
        \scalebox{.7}{
          \color{greenii}
          \bf
          \def\arraystretch{.9}
          \begin{tabular}{c}
            differential
            \\
            character
          \end{tabular}
        }
      }"
    ]
    \\
    \\
    \Omega^1_{\mathrm{dR}}\big(
      -;
      \mathfrak{a}
    \big)_{\closed}
    \ar[
      rr,
      "{
        \eta^{\,\scalebox{.62}{$\shape$}}
      }",
      "{
        \scalebox{.7}{
          \color{greenii}
          \bf
          \def\arraystretch{.9}
          \begin{tabular}{c}
            shape
            unit
          \end{tabular}
        }
      }"{swap}
    ]
    &&
    \shape
    \;
    \Omega^1_{\mathrm{dR}}\big(
      -;
      \mathfrak{a}
    \big)_{\closed}
    \ar[
      from=s,
      to=t,
      Rightarrow,
      dashed,
      shorten=7pt,
      "{
        \widehat{A}
      }"{swap},
      "{
        \scalebox{.7}{
          \color{orangeii}
          \bf
          \def\arraystretch{.9}
          \begin{tabular}{c}
            gauge
            \\
            potentials
          \end{tabular}
        }
      }"{xshift=-9pt, yshift=+3pt}
    ]
  \end{tikzcd}
$
}
\end{tabular}
\begin{remark}[$L_\infty$-algebra of gauge potentials vs $L_\infty$-algebra of flux densities]
\label{LInfinityOfGaugePotentialsVsFLuxDensities}
This means that the $L_\infty$-algebras of concern here are {\it not} the coefficients of the gauge potentials as familar from Yang-Mills theory, but serve as coefficients for their flux densities (field strengths). Even for higher $\mathrm{U}(1)$-gauge theories (Ex. \ref{HigherCircleGaugePotentials}) these differ by a degree shift. For RR-fields flux-quantized in K-theory it is mathematically a ``coincidence'' that $\mathfrak{su}(n)$-valued gauge potentials present differential K-theory classes; while for the C-field it remains unclear whether $\mathfrak{e}_8$-valued gauge potentials play an analogous role (cf. \cref{GreenSchwarzMechanicsm} and footnote \footref{RoleOfE8InDFM} below). In general,  there is no reason to expect that flux-quantized higher gauge potentials are $L_\infty$-algebra valued at all.

\end{remark}

\section{Flux Densities and Brane Charges}
\label{FluxDensitiesAndBraneCharges}

\subsection{Electromagnetic flux and its ``branes''}
\label{ElectromagneticFluxAndItsBranes}

Faraday observed ``lines of force'' -- now called flux of the magnetic field -- concentrating towards the poles of rod magnets. In modern differential-geometric formulation, the density of these flux lines through any given surface-element is encoded in a differential 2-form $F_2$:

\vspace{-.3cm}
\noindent
\def\tabcolsep{0pt}
\begin{tabular}{ll}
\hspace{-.3cm}
\begin{tikzpicture}
\node at (0,0) {
\includegraphics[width=8cm]{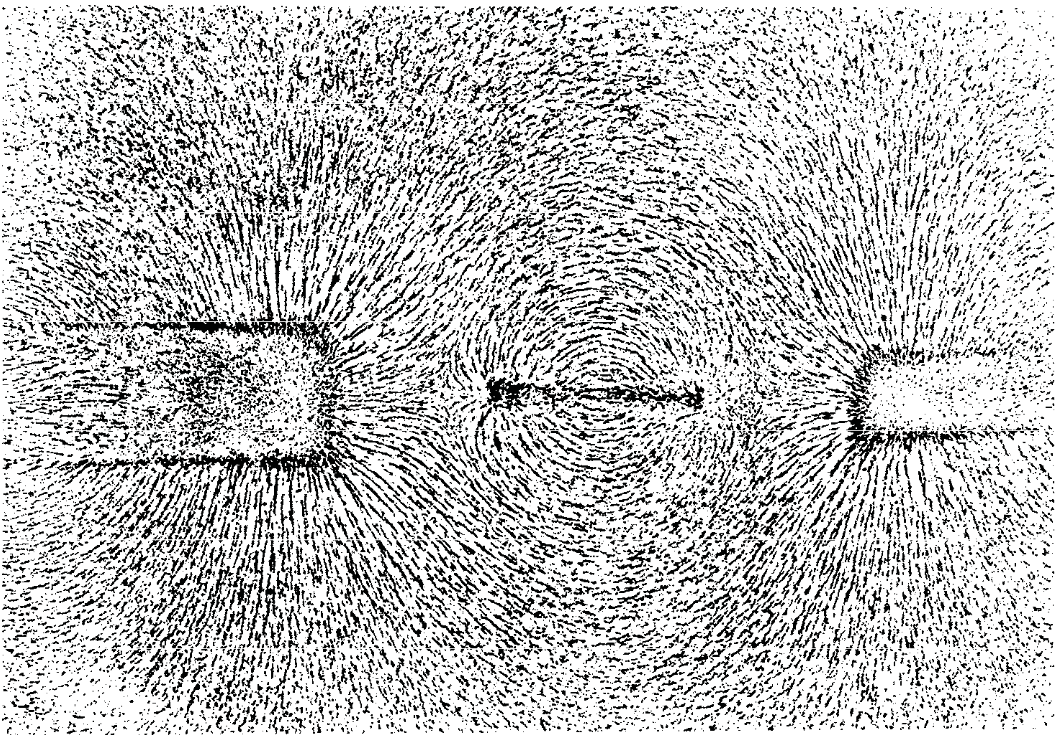}
};

\begin{scope}[shift={(-1.8,-.2)}]
\draw[
  blue,
  draw opacity=.12,
  line width=7
]
  (180+25:1.1) arc (180+25:360+180-30:1.1);
\end{scope}
\end{tikzpicture}
&
\begin{tikzpicture}

  \draw (0,0) node {
    \includegraphics[width=8.8cm]{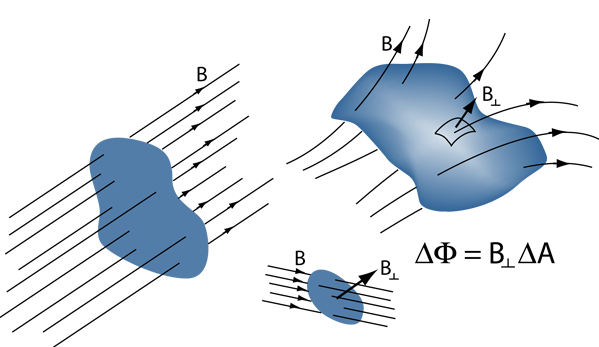}
  };

\node[rotate=-52]
  at (-.15,-.15) {
    \scalebox{.7}{
      \color{darkblue}
      \bf
      magnetic flux lines
    }
  };

\draw[
  draw=white,
  fill=white
]
  (1.7,-.9) rectangle (3.8,-1.5);

\node[
 rotate=-43
]
  at (2,.4) {
   \scalebox{.8}{
     $\Delta \, x^{{}_1}$
   }
  };

\node[
 rotate=+40
]
  at (2.06,.88) {
   \scalebox{.8}{
     $\Delta \, x^{{}_2}$
   }
  };

\node[
  rotate=-40
]
  at (.7,-2.6) {
    \scalebox{.9}{$
      \vec \Delta x^{{}_1}
      \!\!
      \wedge
      \vec \Delta x^{{}_2}
    $}
  };

\node
  at (3.3,-1.6) {
    \scalebox{1}{
      $
        \def\arraystretch{1.2}
        \begin{array}{l}
        \overset{
          \mathclap{
            \scalebox{.7}{
              \def\arraystretch{.9}
              \color{darkblue}
              \bf
              \begin{tabular}{c}
                magnetic flux
                \\
                through surface element
              \end{tabular}
            }
          }
        }{
        F_2\big(\vec \Delta x^{_1}, \vec \Delta x^{_2}\big)
        }
        \\
        \,=\,
        B_{\perp}
          \cdot
        \Delta x^{_1}
          \cdot
        \Delta x^{_2    }
        \end{array}
      $
    }
  };

\end{tikzpicture}
\\
\begin{minipage}{8cm}
From Faraday’s {\it Diary of experimental investigation}, vol VI, entry from 11th Dec. 1851, as reproduced in \cite{Martin09}; the colored arc is our addition, for ease of comparison with the schematics on the right.
\end{minipage}
&
\hspace{.3cm}
\begin{minipage}{9cm}
The density and orientation of magnetic field flux lines are encoded in a differential 2-form whose integral over a given surface is proportional to the total magnetic flux through that surface.
(Graphics adapted from \cite{HyperphysicsMagneticFlux}.)\end{minipage}
\end{tabular}

\medskip

\noindent
\begin{tabular}{p{10cm}l}
More in detail, with respect to any foliation $X^4 \simeq \mathbb{R} \times X^3$ of a globally hyperbolic spacetime $X^4$ by spacelike Cauchy surfaces $X^3$, the spatial component of $F_2$ is the magnetic flux density $B$, while the Hodge dual (with respect to $X^4$) of the temporal component is the electric flux density $E$.
& \qquad
\adjustbox{raise=-.6cm}{ \small
\def\arraystretch{1}
\def\tabcolsep{4pt}
\begin{tabular}{|l|l|}
\hline
\multicolumn{2}{|l|}{
{\bf Electromagnetic flux density.}
}
\\
\hline
$
  \mathclap{\phantom{\vert^{\vert^{\vert}}}}
  X^{4}
$
&
spacetime 4-fold
\\
\hline
$
  \mathclap{\phantom{\vert^{\vert}}}
  F_2
  \;\in\; \Omega^2_{\mathrm{dR}}(X^{4})
$
&
Faraday tensor
\\
\hhline{~-}
  $\;= \;
  \star
  \big(
  E_{i j} \,
  \Differential x^{{}_i}
    \wedge
  \Differential x^{{}_j}
  \big)
  $
  &
  electric flux density
  \\
\hhline{~-}
 $
  \;
  \phantom{=}
  +
  \;
  B_{i j}
    \,
  \Differential x^{{}_i}
    \wedge
  \Differential x^{{}_j}
 $
 &
 magnetic flux density
 \\
 \hline
\end{tabular}
}
\end{tabular}

\smallskip
\noindent
\hspace{-1.5mm}
\def\tabcolsep{0pt}
\begin{tabular}{p{5.3cm}l}
Imagining, with Dirac, that Faraday’s rod magnet could be made {\it infinitely} long and thin, any one of its poles would look like an isolated mono-pole with flux concentrating towards it from all directions.

\medskip
At the point of the idealized monopole itself, the flux density $B$ per unit volume would diverge -- a ``singularity'' much in the sense of black holes, which therefore is not to be regarded as part of space(-time): The spacetime domain on which to discuss the fluxes sourced by a magnetic monopole is (more on all this below in \cref{SingularVersusSolitonicBranes}) not Minkowski spacetime $\mathbb{R}^{3,1}$ itself, but its complement around the worldline $\mathbb{R}^{0,1}$ of the would-be monopole.
&
\hspace{-.5cm}
\adjustbox{raise=-4.2cm}{
\begin{minipage}{13.6cm}
\begin{equation}
  \label{DiracMonopole}
  \adjustbox{raise=-5cm}{
  \includegraphics[width=11cm]{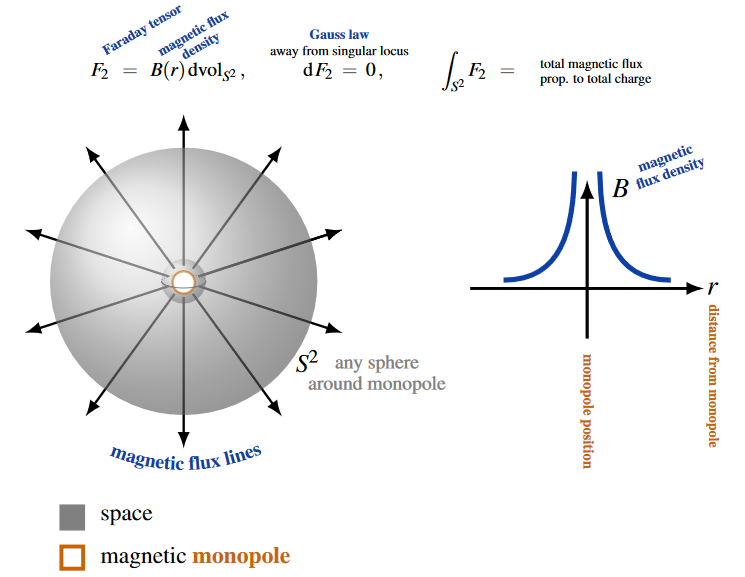}
  }
\end{equation}
\end{minipage}
}
\end{tabular}

\bigskip
$$
  \mathbb{R}^{3,1}
    \setminus
  \mathbb{R}^{0,1}
  \;\;
  \underset{\mathrm{homeo}}{\simeq}
  \;\;
  \mathbb{R}^{0,1}
  \times
  \big(
    \mathbb{R}^3 \setminus \{0\}
  \big)
  \;\;
  \underset{\mathrm{homeo}}{\simeq}
  \;\;
  \mathbb{R}^{0,1}
  \times
  R^1_{> 0}
  \times S^2
  \;\;
  \underset{\mathrm{hmtp}}{\simeq}
  \;\;
  S^2
  \,.
$$

As such, magnetic monopoles are the {\bf singular 0-branes} of electromagnetism (cf. \cref{SingularVersusSolitonicBranes}) -- in theory: Whether magnetic monopoles exist in nature remains open; they have not been seen in experiment, but there are decent theoretical arguments that they should exist if the standard model symmetry is a broken ``grand unified'' symmetry.

\noindent
\hspace{-.1cm}
\def\tabcolsep{0pt}
\begin{tabular}{p{5cm}l}
\label{AbrikosovVortices}
However, in the EM-field there are also {\bf solitonic 1-branes} which are  experimentally well-established as the {\it Abrikosov vortices} formed in type II superconductors within a transverse magnetic field
\cite{Abrikosov1957} \cite{LoudonMidgley09} \cite[\S 6.5]{Timm20}.  These may be regarded as {\it strings} approximated by a Nambu-Goto action \cite{NielsenOlesen73} \cite{BeekmanZaanen11}.

& \qquad
\adjustbox{raise=-5.5cm}{
\includegraphics[width=.67\textwidth]{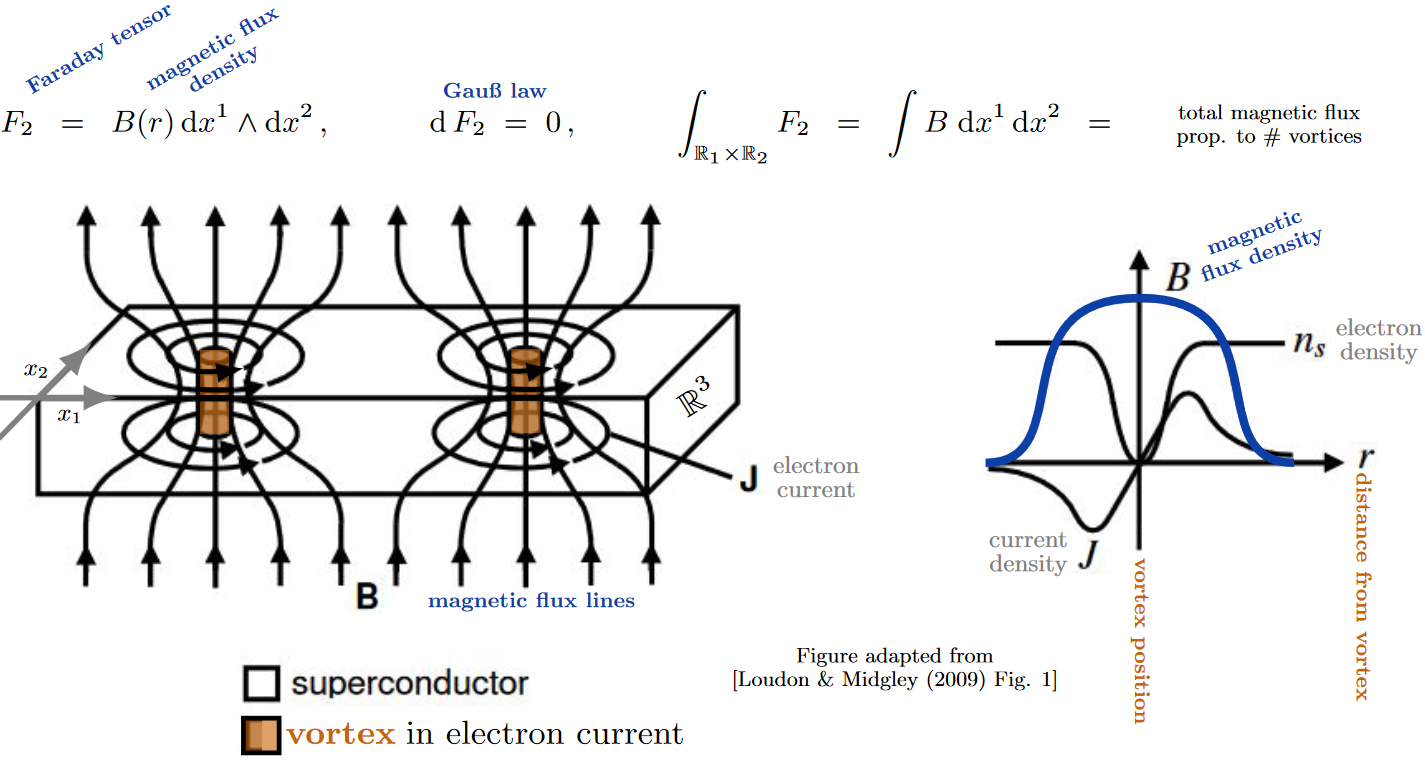}
}
\end{tabular}

\vspace{-.2cm}

\noindent
\hspace{-.1cm}
\def\tabcolsep{0pt}
\begin{tabular}{p{5cm}l}
In this case, the ``sphere'' through which the total magnetic flux density is measured is nominally the $(x_1, x_2)$-plane filled by the superconducting material; but since far away from any vortex the magnetic flux has to vanish, this plane appears to the fluxes via its one-point compactification with the ``point at infinity'' adjoined.

\medskip
These vortex strings are {\it solitons} in that the flux density is everywhere finite, and yet the ``bumps'' in the flux density are topologically stable. Much like a bump in a rug cannot be flattened as long as the boundary of the rug is fixed in place, so the requirement that flux densities ``vanish at infinity'' keeps the vortex strings in place -- or at least this is the case once we take account of Dirac flux quantization below
(the corresponding classifying maps for which are previewed on the right).
&
\hspace{14pt}
\adjustbox{raise=-11.5cm, margin=0pt, fbox}{
\includegraphics[width=.64\textwidth]{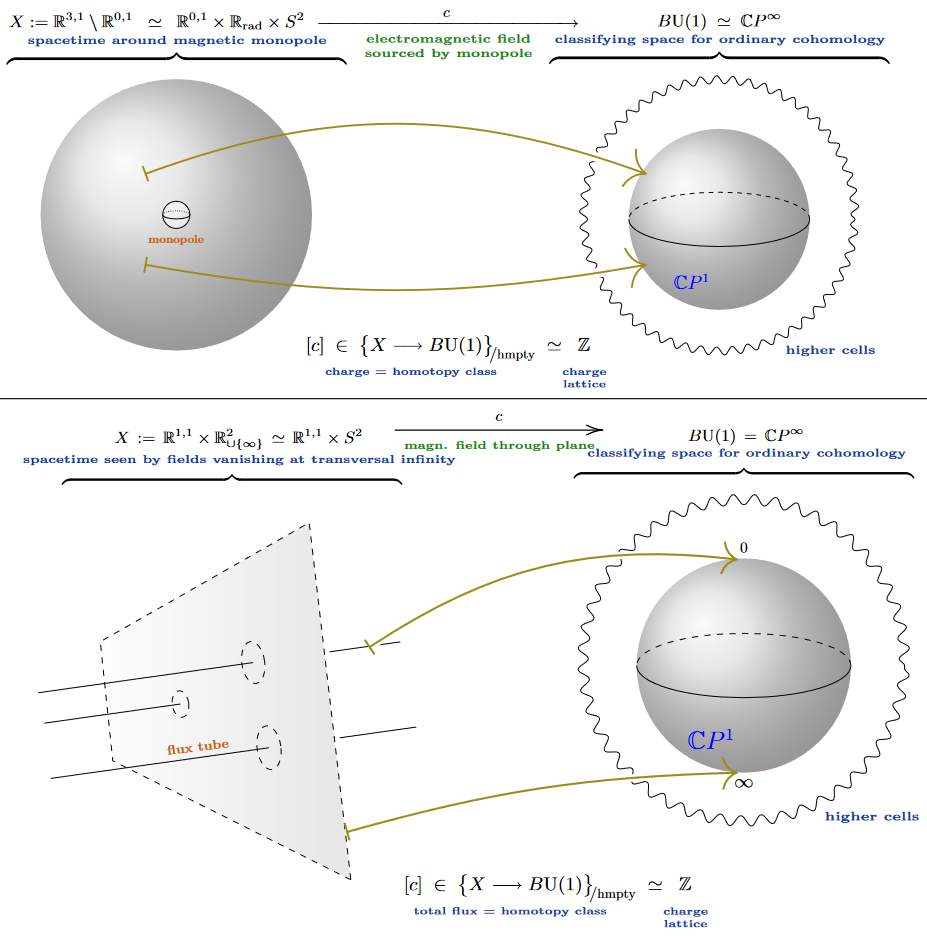}
}
\end{tabular}

\subsection{Singular versus solitonic branes}
\label{SingularVersusSolitonicBranes}

Generally, imprinted on flux densities may be two kinds of branes, here to be called:

\begin{itemize}[leftmargin=.8cm]
  \item[{\bf (i)}] {\it singular branes} (black branes), reflected in {\it diverging flux density} at {\it singular loci} that are to be removed from spacetime,

  {\footnotesize Beware that in supergravity these are also called ``elementary branes'' \cite{DuffLu1994}, in reference to how black holes carry the same quantum numbers as elementary particles -- but here we rather not conflate these two aspects.}

  \item[{\bf (ii)}] {\it solitonic branes}, reflected in {\it finite flux density} which is localized in that it {\it vanishes at infinity}, transversally.

  {\footnotesize The terminology ``solitonic brane'' was introduced by \cite{DuffKhuriLu92}\cite{DuffKhuriLu95} and Duff \& Lu 1993, 1994 to mean stable but non-singular brane-like solutions to (supergravity/flux) equations of motion (``solitons'').}
\end{itemize}

\medskip

This general distinction between singular branes and solitonic branes is important for the correct identification of the implications of choices of flux quantization laws on the corresponding brane charges.

\paragraph{Spacetime domains for brane fluxes.} More formally, one may encode these two cases by slightly adjusting the nature of the {\it spacetime domain} on which fluxes are actually defined \cite[\S 2.1]{SS23MF}:

\noindent
\hspace{-.2cm}
\def\tabcolsep{0pt}
\begin{tabular}{p{5.2cm}l}
$\bullet$ fluxes sourced by singular branes of dimension $p+1$ inside spacetime $X^{d+1}$ are actually defined on the complement $X^{d+1}\setminus Q^{p+1}$ of their singular worldvolume,

$\bullet$ fluxes sourced by solitonic branes of codimension $d - p$ are

&
\qquad
\adjustbox{raise=-1.2cm}{\small
\def\arraystretch{2}\def\tabcolsep{5pt}
\begin{tabular}{|c|lc|}
  \hline
  {\bf Type of brane}
  &
  {\bf Spacetime domain of flux density}
  &
  \\
  \hline
  \hline
 {\bf Singular brane}
  &
  \def\arraystretch{.9}
  \begin{tabular}{l}
    complement of singular
    worldvolume
    \\
    locus $Q^{p+1}$ inside spacetime $X^{d+1}$
  $\mathclap{\phantom{\vert_{\vert_{\vert}}}}$
  \end{tabular}
  &
  $X^{d+1} \setminus Q^{p+1}$
  \\
  \hline
  {\bf Solitonic brane}
  &
  \def\arraystretch{.9}
  \begin{tabular}{l}
    Transverse space $Y^{d-p}$
    to worldvolume
    \\
    equipped with a
    ``point at infinity''
   $\mathclap{\phantom{\vert_{\vert_{\vert}}}}$  \end{tabular}
  &
  $
    \big(
      Y^{d-p}
      ,\,
      \infty_{{}_Y}
    \big)
  $
  \\
  \hline
\end{tabular}
}
\end{tabular}

\vspace{-.5cm}
\noindent
 actually defined on their transverse space $T^{d-p}$ equipped with a ``point at infinity'' on which they are required to vanish.

The condition of flux densities vanishing at infinity on some space is naturally formalized by considering the larger category of pointed topological spaces $(X, x \in X)$ (we discuss further below how to properly speak of differential geometric smoothness in this context) and regarding their given “base point” as being the “point at infinity”, whence we shall write $(X, \infty_{{}_X})$ for the generic pointed space. Then a function “vanishing at infinity” on $(X, \infty_{{}_X})$  is a function on $X$ that literally vanishes at $\infty_{{}_X}$.

For example:

\begin{itemize}
\item The result of {\it adjoining} to $\mathbb{R}^n$ its "point at infinity" (this is called its {\it one-point compactification}, here to be denoted $\mathbb{R}^n_{\cup \{\infty\}}$) is homeomorphic to the $n$-sphere with any basepoint:

   $$
     \mathbb{R}^n_{\cup \{\infty\}}
       \,\underset{homeo}{\simeq}\,
     S^n
     \,.
   $$

\item On the other hand, to consider unconstrained functions on some $X$ in this context, we may regard all the points of $X$ as being at finite distance by declaring that the ``point at infinity'' is disjoint from $X$, hence by considering the disjoint union (denoted "$\sqcup$" as opposed to "$\cup$"):

   $$
     X_{\sqcup \{\infty\}}
     \;\coloneqq\;
     X \sqcup \{ \infty \}
     \,.
   $$
\end{itemize}

Given two such pointed spaces, their {\it smash product} "$\wedge$" is their Cartesian product with all points that are at infinity in either factor identified with a single new point at infinity:

$$
  \big(
    X,\, \infty_X
  \big)
  \wedge
  \big(
    Y,\, \infty_Y
  \big)
  \;\;
  \coloneqq
  \;\;
  \frac{
    X \times Y
  }{
    X \times \{\infty_Y\}
    \,\cup\,
    \{\infty_X\} \times Y
  }
$$

\begin{example}[{\bf Flat branes}]
\label{FlatBranes}
In the case of flat branes -- i.e. with Cartesian worldvolumes inside Minkowski spacetime -- both these spacetime domains are homotopy equivalent to spheres, but of different dimensions:

\noindent
\hspace{-.2cm}
\def\tabcolsep{0pt}
\begin{tabular}{p{5cm}l}
{\bf (i)} The spacetime domain for flat singular branes is homotopy-equivalent to the unit sphere in the transverse space, hence the sphere around the singular brane locus.
&
\hspace{2pt}
\adjustbox{raise=-1.8cm}{
  \includegraphics[width=12.4cm]{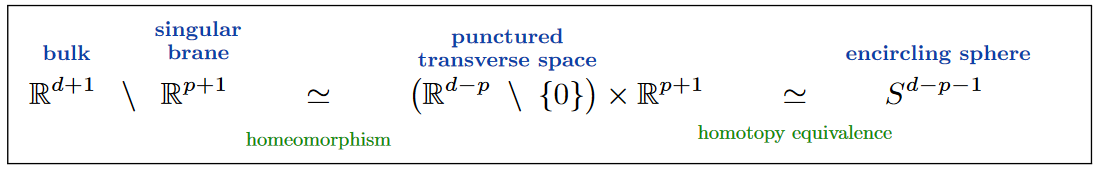}
}
\\
{\bf (ii)} The spacetime domain for flat solitonic branes is homotopy equivalent to the sphere which is the one-point compactification of the transverse space (its stereographic projection).
&
\hspace{2pt}
\adjustbox{raise=-2.3cm}{
  \includegraphics[width=12.4cm]{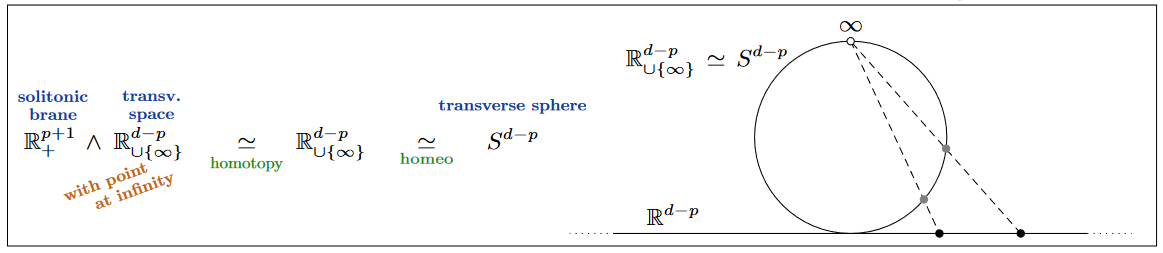}
}
\end{tabular}
\end{example}

\begin{example}[{\bf Flat branes of electromagnetism}]
Specifying Ex. \ref{FlatBranes} to the case of ordinary electromagnetic flux (\cref{ElectromagneticFluxAndItsBranes}) it follows from this general reasoning that a flux density 2-form $F_2$ in $D = 3+1$ may reflect the presence of
\begin{itemize}
\item singular 0-branes with spacetime domain
   $
     \mathbb{R}^{3,1} \setminus \mathbb{R}^{0,1}
       \,\underset{\mathrm{homeo}}{\simeq}\,
     \mathbb{R}^{0,1} \times \mathbb{R}_{> 0} \times S^2
       \,\underset{\mathrm{hmtp}}{\simeq}\,
     S^2
   $

\item solitonic 1-branes with spacetime domain
  $
     \mathbb{R}^{1,1}_+
       \wedge
     \mathbb{R}^{2}_{\cup \{\infty\}}
     \;\underset{\mathrm{homeo}}{\simeq}\;
     \mathbb{R}^{1,1}_+
       \wedge
     S^2
     \;\underset{\mathrm{hmtp}}{\simeq}\;
     S^2
   $
\end{itemize}
which are just the magnetic monopoles (hypothetical) and Abrikosov vortex strings (observed) from \cref{ElectromagneticFluxAndItsBranes}.
\end{example}

\newpage
\begin{example}[{\bf Near-horizon geometries of singular branes}] $\;$

\noindent
\hspace{-.1cm}
\begin{tabular}{p{5.9cm}l}
The idea of regarding singular branes from the complement of their singular locus in spacetime is familiar from the AdS/CFT correspondence:

The near-horizon geometry of any $> 1/4$ BPS black brane are products of an anti de Sitter spacetime with a (free discrete quotient of) a sphere around the singularity \cite{AFHS99}. On a causal chart of AdS spacetime, this is homeomorphic to the flat brane complements.
& \;\,
\adjustbox{raise=-5cm}{
\includegraphics[width=11.3cm]{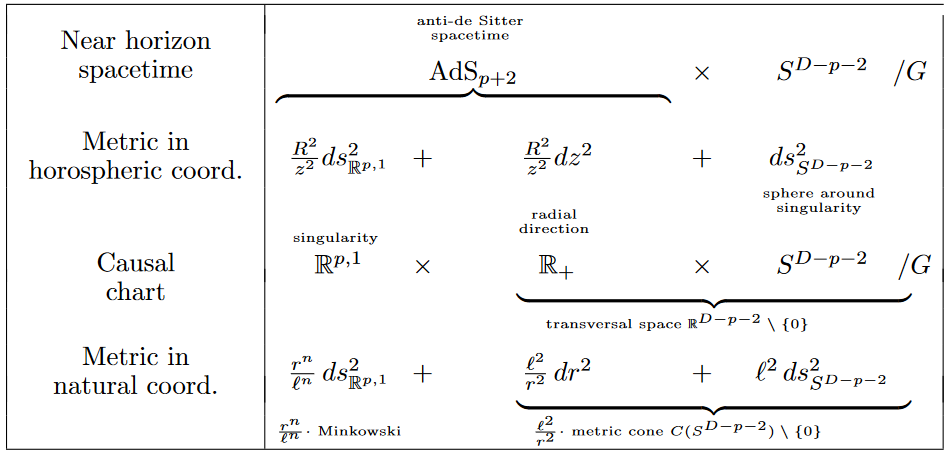}
}
\end{tabular}

\end{example}

In general, the spacetime domain on which to measure flux densities may be a mix of these purely singular and purely solitonic situations, in which case the notions of singular and of solitonic branes blend into each other:

\begin{example}[{\bf solitonic branes in KK-compactifications}]
In mild generalization of Ex. \ref{FlatBranes}, consider the case that spacetime is a trivial circle-fiber bundle over a Minkowski spacetime
$
  X^D \,\defneq\,
  \mathbb{R}^{1,d-1}
  \times
  S^1
  \,.
$
In this case, the flux sourced by solitonic branes of codimension $n$
as seen on the base spacetime $\mathbb{R}^{1, d-1}$ is measured on the smash product space
\vspace{2mm}
$$
  \big(
    S^1 \times \mathbb{R}^n
  \big)_{\cup \{\infty\}}
  \;\;\underset{\mathrm{homeo}}{\simeq}\;\;
  S^1_{\sqcup \{\infty\}}
  \wedge
  \mathbb{R}^n_{\cup \{\infty\}}
  \;\;\underset{\mathrm{homeo}}{\simeq}\;\;
  S^1_{\sqcup \{\infty\}}
  \wedge
  S^n
  \,.
$$
This was first understood by
\cite[\S 2.2 \& \S 2.3]{BGH99}.

\end{example}

\subsection{Higher fluxes and their brane sources}
\label{HigherFluxesAndTheirBraneSources}

On this backdrop of ordinary electromagnetic flux (\cref{ElectromagneticFluxAndItsBranes}) and of the general rule for measuring flux sourced by singular branes or solitonic branes (\cref{SingularVersusSolitonicBranes}) it clearly makes sense to consider physical theories of higher gauge fields whose precise nature remains to be discussed, but whose flux densities are reflected in higher-degree differential forms
\begin{equation}
  \label{FluxDensities}
    F^{(i)}
    \;\in\;
    \Omega_{\mathrm{dR}}^{\mathrm{deg}_i}(X^D)
  \,,
\end{equation}
these possibly being of different field species to be labeled by a finite index set $I \in \mathrm{FinSet}$ and jointly to be denoted as follows:
\begin{equation}
  \label{TheHigherFluxDensities}
  \vec F
  \;\equiv\;
  \Big\{
    F^{(i)}
    \,\in\,
    \Omega_{\mathrm{dR}}^{\mathrm{deg}_i}\big(
      X^D
    \big)
  \Big\}
  \,.
\end{equation}

\noindent
\hspace{-.2cm}
\def\tabcolsep{0pt}
\begin{tabular}{p{7cm}l}
Such higher flux densities appear in higher dimensional supergravity, namely as “superpartners” of the gravitino field that cannot be accounted for by the graviton itself. In particular, in $D=10$ supergravity and $D=11$ supergravity these higher flux densities are known under the (now) fairly standard symbols shown on the right, along with the standard name of the corresponding singular branes (the “higher-dimensional monopoles”), e.g. \cite[\S18.5]{BLT13}.
&
\hspace{-.6cm}
\adjustbox{raise=-2.2cm}{
\begin{minipage}{11cm}
\begin{equation}
\label{TableOfFluxSpeciesAndTheirSources}
\adjustbox{}{\small
\def\tabcolsep{4pt}
\def\arraystretch{1.4}
\begin{tabular}{|c||c|c|c|}
  \hline
  &
  {\bf Field} & {\bf Flux} & {\bf Singular source}
  \\
  \hline
  \hline
  $D\!=\!4$ Maxwell theory
  &
  A-field
    &
  $F_2$
  &
  \def\arraystretch{.9}
  \begin{tabular}{c}
    monopole
    \\
    0-branes
  \end{tabular}
  \\
  \hline
  \multirow{3}{*}{
  $D\!=\!10$ supergravity
  }
  &
  \multirow{2}{*}{
    B-field
  }
  &
  $H_3$
  &
  NS5-brane
  \\
  \cline{3-4}
  &
  &
  $H_7$
  &
  F$1$-branes
  \\
  \cline{2-4}
  &
  RR-field
  &
  $F_{8-p}$
  &
  D$p$-branes
  \\
  \hline
  \multirow{2}{*}{
  $D\!=\!11$ supergravity
  }
  &
  \multirow{2}{*}{
  C-field
  }
  &
  $G_4$ & M5-branes
  \\
  \cline{3-4}
  &&
  $G_7$ & M2-branes
  \\
  \hline
\end{tabular}
}
\end{equation}
\end{minipage}
}
\end{tabular}

\medskip

The above flux densities in 11d and 10d are closely related:

\begin{example}[{\bf Double dimensional reduction of fluxes form 11d to 10d}]

Consider the case of C-field flux densities $G_4$ and $G_7$ on an 11-dimensional spacetime $Y^{11}$ which is the total space of a circle-principal bundle and denote by

\begin{itemize}
\item $\theta \in \Omega^1_{\mathrm{dR}}(Y^{11})$ any fiberwise Maurer-Cartan form along the fibers (i.e. an Ehresmann connection form),

\item $F_2 \in \Omega^2_{\mathrm{dR}}(X^{10})$ the corresponding (first) Chern class-characteristic form, i.e.,
the curvature form whose pullback to $Y^{11}$ is $\differential \theta = p^\ast F_2$.
\end{itemize}

\noindent
\hspace{-.2cm}
\def\tabcolsep{0pt}
\begin{tabular}{p{7.5cm}l}
Assuming that all flux densities are $S^1$-invariant (hence focusing on their 0th KK-modes) they decompose into a basic component (a differential form on $X^{10}$, pulled back along the projection $p$) and the wedge product of a basic differential form with the Maurer-Cartan form $\theta$ on the $S^1$-fibers:
&
\adjustbox{raise=-.9cm}{
\begin{minipage}{9.2cm}
\begin{equation}
\label{DDReductionOfCFieldFlux}
\adjustbox{raise=-1cm}{
\includegraphics[width=7.6cm]{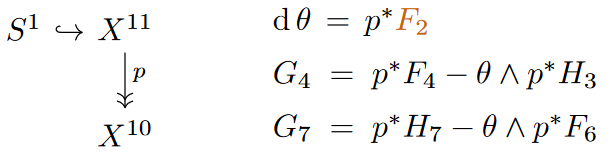}
}
\end{equation}
\end{minipage}
}
\end{tabular}

\vspace{1mm}
This is the process of ``double dimensional reduction'' \cite{DHIS87} \cite[\S 2.2]{BMSS19} -- called this way since both spacetime dimension is reduced by Kaluza-Klein reduction on a fiber space, but also the degrees of densities of fluxes “through the fiber space” are decreased – known as part of the duality between M-theory and type IIA string theory: The new component flux densities $H_3$ and $H_7$ are interpreted as those of the B-field and the component flux densities $F_4$ and $F_6$ (and $F_2$) as those of the RR-field in type IIA supergravity.

Here the flux density $F_2$, which in 10d is understood as witnessing singular D6-brane sources, is a gravitational flux from the 11d point of view: If $X^{10} \,\defneq\, \mathbb{R}^{6,1} \times \mathbb{R}_{>0} \times S^2$ is the spacetime domain around a flat singular D6-brane, then the total space of the circle-principal bundle $Y^{11}$ (a multiple of the complex Hopf fibration) is known as the corresponding “KK-monopole” spacetime.
\end{example}

This transmutation, under Kaluza-Klein compactification, of parts of the gravitational field in higher dimensions into gauge fields in lower dimensions is a major subtlety in choosing flux quantization laws: Since these laws apply to higher gauge fields but not directly to the field of gravity, there may appear new possibilities for flux quantization after KK-reduction to lower dimensions which do not come from flux quantization in higher dimensions.

\subsection{Equations of motion of higher flux}
\label{EquationsOfMotionOfHigherFlux}

As we now turn to the equations of motion for flux densities (the analogs of Maxwell's equations for electromagnetic flux), the key move towards identifying possible flux quantization laws (in \cref{FluxQuantizationLaws}) is to arrange these equations, equivalently, as:

\begin{itemize}
\item[{\bf (i)}] a purely cohomological system of differential equations known as higher {\it Bianchi identities},

\item[{\bf (ii)}] a purely geometric system of linear equations expressing a Hodge self-duality,
\end{itemize}
the point being that the first item is entirely “algebro-topological” (homotopy-theoretic), while dependency on geometry, namely on the spacetime metric (the field of gravity) is all isolated in the second item.

It turns out \cite{SS23FQ} that from such duality-symmetric laws of flux, the canonical phase space of the higher gauge theory, including the flux-quantization structure, may be obtained straightforwardly, without going through the traditional and thorny route of BRST-BV analysis based on a stationary action principle given by a Lagrangian density.
\vspace{-.1cm}
$$
  \begin{tikzcd}[
    column sep=10pt,
    row sep=-22pt
  ]
    &&
    \fbox{
      \scalebox{.9}{
        \hspace{-12pt}
        \def\arraystretch{1}
        \begin{tabular}{c}
          \bf
          Bianchi identities
          \\
          (cohomological)
        \end{tabular}
        \hspace{-11pt}
      }
    }
    \ar[
      rr,
      rounded corners,
      to path={
           ([yshift=0pt, xshift=15pt]\tikztostart.north)
        -- ([yshift=10pt, xshift=15pt]\tikztostart.north)
        -- node[yshift=6pt]{
          \scalebox{.7}{
            \color{greenii}
            \bf
            enhance flux densities to
            differential cohomology
          }
        }
          node[yshift=-6pt] {
            \scalebox{.7}{
              \cref{FluxQuantizationLaws}
            }
          }
           ([yshift=10pt]\tikztotarget.north)
        -- ([yshift=00pt]\tikztotarget.north)
      }
    ]
    \ar[
      r,
      shorten=-1.5pt
    ]
    \ar[
      ddr,
      rounded corners,
      to path={
           ([xshift=-0pt]\tikztostart.east)
        -- ([xshift=5pt]\tikztostart.east)
        -- ([xshift=-13pt]\tikztotarget.west)
        -- ([xshift=0pt]\tikztotarget.west)
      }
    ]
    &
    \fbox{
      \scalebox{.9}{
        \hspace{-12pt}
        \def\arraystretch{1}
        \begin{tabular}{c}
          \bf
          Gau{\ss} law constraint
          \\
          on Cauchy data
        \end{tabular}
        \hspace{-10pt}
      }
    }
    \ar[
      r,
      shorten=-1pt
    ]
    &
    \fbox{
      \scalebox{.9}{
        \hspace{-12pt}
        \def\arraystretch{1}
        \begin{tabular}{c}
          \bf
          Canonical Phase space
          \\
          with flux quantization
        \end{tabular}
        \hspace{-10pt}
      }
    }
    \\
    \fbox{
      \scalebox{.9}{
        \hspace{-12pt}
        \def\arraystretch{1}
        \begin{tabular}{c}
          \bf
          Equations of motion
          \\
          for higher flux densities
          \\
          in background gravity
        \end{tabular}
        \hspace{-10pt}
      }
    }
    \ar[
      urr,
      rounded corners,
      to path={
           ([yshift=0pt]\tikztostart.north)
        -- ([yshift=23.5pt]\tikztostart.north)
        -- node[yshift=6pt]{
          \scalebox{.7}{
            \color{greenii}
            \bf
            duality-symmetric formulation
          }
        }
        node[yshift=-6pt]{
          \scalebox{.7}{
            \cref{EquationsOfMotionOfHigherFlux}
          }
        }
           ([yshift=10pt, xshift=-15pt]\tikztotarget.north)
        -- ([yshift=00pt, xshift=-15pt]\tikztotarget.north)
      }
    ]
    \ar[
      urr,
      <->,
      rounded corners,
      to path={
        ([xshift=-0pt]\tikztostart.east)
        --
        ([xshift=10pt]\tikztostart.east)
        --
        ([xshift=-8pt]\tikztotarget.west)
        --
        ([xshift=-0pt]\tikztotarget.west)
      }
    ]
    \ar[
      drr,
      <->,
      rounded corners,
      to path={
        ([xshift=-0pt]\tikztostart.east)
        --
        ([xshift=10pt]\tikztostart.east)
        --
        ([xshift=-19pt]\tikztotarget.west)
        --
        ([xshift=-0pt]\tikztotarget.west)
      }
    ]
    \\
    &&
    \fbox{
      \scalebox{.9}{
        \hspace{-12pt}
        \def\arraystretch{1}
        \begin{tabular}{c}
          \bf
          Self-Duality
          \\
          (geometric)
        \end{tabular}
        \hspace{-10pt}
      }
    }
    \ar[
      r
    ]
    &
    \fbox{
      \scalebox{.9}{
        \hspace{-12pt}
        \def\arraystretch{1}
        \begin{tabular}{c}
          \bf
          Time evolution
          \\
          of Cauchy data
        \end{tabular}
        \hspace{-10pt}
      }
    }
    \ar[
      r,
      gray
    ]
    &
    \fcolorbox{gray}{white}{
      \scalebox{.9}{
        \color{gray}
        \hspace{-12pt}
        \def\arraystretch{1}
        \begin{tabular}{c}
          \bf
          Hamiltonian
          \\
          on phase space
        \end{tabular}
        \hspace{-10pt}
      }
    }
  \end{tikzcd}
$$

This move of isolating “pre-metric flux equations” supplemented by a “constitutive” duality constraint has a curious status in the literature. On the one hand, it is elementary and immediate as an equivalent re-formulation of the usual form of (higher) Maxwell-type equations of motion, and as such has been highlighted a century ago \cite[\S 80]{Cartan24} and again more recently \cite{HehlObhukov03} (historical survey in \cite{HIO16}). While the broader community does not seem to have taken much note of “premetric electromagnetism” as such, one may notice that just the same perspective is evidently what in supergravity and string theory is called “duality-symmetric” \cite{BandosBerkovitsSorokin98} or “democratic” \cite{MkrtchyanValach23} formulations of fluxes in supergravity (see Ex. \ref{MotionOfUnboundedRRFieldFluxes} and Ex. \ref{MotionOfCFieldFlux} below).

\begin{definition}[{\bf Higher Maxwell-type equations}]
\label{HigherMaxwellEquations}

A system of {\it higher Maxwell-type equations} on a tuple \eqref{FluxDensities} of flux densities on a spacetime $X^D$ is

\begin{itemize}
  \item a system of polynomial  ($\vec P$) first-order exterior-differential equations $\differential \vec F \,=\, \vec P\big(\vec F\big)$ (the higher {\it Bianchi identities}, crucially admitting polynomial “self-sourcing” of fluxes);
  \item subject to a system of linear ($\vec \mu$) Hodge-self-duality relation $\star \vec F \,=\, \vec \mu\big(\vec F\,\big)$  (the “{\it constitutive relations}”):
\end{itemize}

\begin{equation}
  \label{DualitySymmetricHigherMaxwellEquations}
  \adjustbox{raise=-1cm}{
    \includegraphics[width=15.2cm]{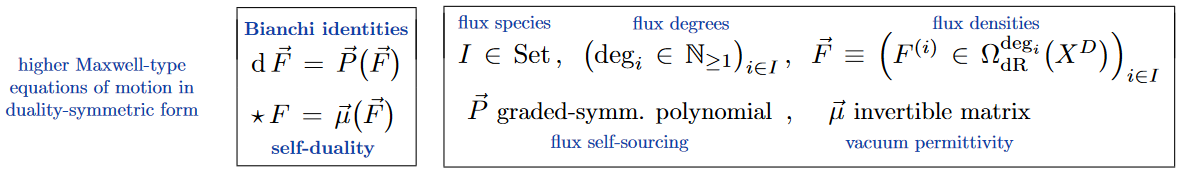}
  }
\end{equation}
Concretely:
\begin{itemize}
\item $\vec P$ is an $I$-tuple of graded-symmetric polynomials with rational coefficients in $I$ variables of degrees $\vec {\mathrm{deg}}$,

\item $\vec \mu$ is a linear endomorphism on the vector space spanned by these variables.
\end{itemize}

\end{definition}

\begin{remark}
  The equations in Def. \ref{HigherMaxwellEquations} imply that $\vec P$ and $\vec \mu$ respect degrees in a certain evident way. Moreover, the following property of the Hodge star operator on Lorentzian manifolds (e.g. \cite[\S 14]{Frankel97}) implies further constraints on the available higher Maxwell-type equations:
\begin{equation}
\label{HodgeSquareOnLorentzianSpacetim}
  \star
  \,
  \star
  \,
  F_{\mathrm{deg}}
  \;\;
  =
  \;\;
  -(-1)^{\mathrm{deg}(D-\mathrm{deg})}
  \,
  F_{\mathrm{deg}}
  \,,
  \;\;\;\;\;\;\;\;
  \text{for}
  \;
  F_{deg}
  \,\in\,
  \Omega^{\mathrm{deg}}_{\mathrm{dR}}(X^D)
\end{equation}
This controls notably the existence of genuinely self-dual higher gauge theories, see Ex. \ref{MotionOfSelfDualHigherGaugeField} below.
\end{remark}

\begin{remark}
Not all higher gauge theories are of the higher Maxwell-form (Def. \ref{HigherMaxwellEquations}): For instance,
higher Chern-Simons type theories are different.
\end{remark}

\begin{example}[{\bf Motion of the ordinary electromagnetic fluxes}]
\label{MotionOfOrdinaryElectromagneticFluxes}
$\,$

\noindent
\hspace{-.1cm}
\begin{tabular}{p{9cm}l}
 The classical Maxwell equations expressed in terms of differential forms are as shown on the left (e.g. \cite[\S 3.5 \& \S7.2b]{Frankel97}), with their ``premetric'' form shown on the right.

 Here the differential 3-form $J_3$ embodies the density of an electric current carrying an electric field and inducing a magnetic field. &
\adjustbox{raise=-1cm}{
\begin{minipage}{8.5cm}
\begin{equation}
  \label{PremetricFormulationOfElectromagnetism}
  \begin{tikzcd}[row sep=-10pt]
    &&
    \adjustbox{scale=.9, fbox}{$
      \def\arraystretch{.9}
      \begin{array}{c}
        \mathrm{d} \, F_2
        \\
        \mathrm{d} \, G_2
      \end{array}
      \begin{array}{c}
        =
        \\
        =
      \end{array}
      \begin{array}{c}
        0
        \\
        J_3
      \end{array}
    $}
    \\[-8pt]
    \adjustbox{scale=.9, fbox}{$
      \def\arraystretch{.9}
      \begin{array}{r}
        \mathrm{d} \, F_2
        \\
        \mathrm{d} \star F_2
      \end{array}
      \begin{array}{c}
        =
        \\
        =
      \end{array}
      \begin{array}{l}
        0
        \\
        J_3
      \end{array}
    $}
    \ar[
      urr,
      <->,
      rounded corners,
      to path={
        ([xshift=-0pt]\tikztostart.east)
        --
        ([xshift=30pt]\tikztostart.east)
        --
        ([xshift=-13.5pt]\tikztotarget.west)
        --
        ([xshift=-0pt]\tikztotarget.west)
      }
    ]
    \ar[
      drr,
      <->,
      rounded corners,
      to path={
        ([xshift=-0pt]\tikztostart.east)
        --
        ([xshift=30pt]\tikztostart.east)
        --
        ([xshift=-22pt]\tikztotarget.west)
        --
        ([xshift=-0pt]\tikztotarget.west)
      }
    ]
    \\
    &&
    \adjustbox{scale=.9, fbox}{$
      \def\arraystretch{.9}
        G_2 \;=\; \star \,F_2
    $}
  \end{tikzcd}
\end{equation}
\end{minipage}
}
\\
This kind of {\it external} or {\it background} source term, where the source is not given by (a polynomial in) the flux densities themselves, does not fit into the Definition \ref{HigherMaxwellEquations} and will be disregarded for the purpose of the present discussion, meaning that we focus on the special case of Maxwell’s equations “in vacuum”.
&
\adjustbox{raise=-1cm}{
\begin{minipage}{8.5cm}
\begin{equation}
  \label{PremetricFormulationOfVacuumElectromagnetism}
  \begin{tikzcd}[row sep=-10pt]
    &&
    \adjustbox{scale=.9, fbox}{$
      \def\arraystretch{.9}
      \begin{array}{c}
        \mathrm{d} \, F_2
        \\
        \mathrm{d} \, G_2
      \end{array}
      \begin{array}{c}
        =
        \\
        =
      \end{array}
      \begin{array}{c}
        0
        \\
        0
      \end{array}
    $}
    \\[-8pt]
    \adjustbox{scale=.9, fbox}{$
      \def\arraystretch{.9}
      \begin{array}{r}
        \mathrm{d} \, F_2
        \\
        \mathrm{d} \star F_2
      \end{array}
      \begin{array}{c}
        =
        \\
        =
      \end{array}
      \begin{array}{l}
        0
        \\
        0
      \end{array}
    $}
    \ar[
      urr,
      <->,
      rounded corners,
      to path={
        ([xshift=-0pt]\tikztostart.east)
        --
        ([xshift=30pt]\tikztostart.east)
        --
        ([xshift=-13.5pt]\tikztotarget.west)
        --
        ([xshift=-0pt]\tikztotarget.west)
      }
    ]
    \ar[
      drr,
      <->,
      rounded corners,
      to path={
        ([xshift=-0pt]\tikztostart.east)
        --
        ([xshift=30pt]\tikztostart.east)
        --
        ([xshift=-22pt]\tikztotarget.west)
        --
        ([xshift=-0pt]\tikztotarget.west)
      }
    ]
    \\
    &&
    \adjustbox{scale=.9, fbox}{$
      \def\arraystretch{.9}
        G_2 \;=\; \star \,F_2
    $}
  \end{tikzcd}
\end{equation}
\end{minipage}
}
\end{tabular}
\end{example}

It is clear that, mathematically at least, Ex. \ref{MotionOfOrdinaryElectromagneticFluxes}, makes sense more generally for flux densities of any degree. In particular:

\begin{example}[{\bf Motion of unbounded RR-field fluxes}]
\label{MotionOfUnboundedRRFieldFluxes}
The equations of motion of the RR-field fluxes in $D=10$ supergravity in the case of vanishing B-field-fluxes are often taken to be as follows (e.g. \cite{MkrtchyanValach23}):
\begin{equation}  \label{PremetricFormulationOfPlainRRFieldTheory}
  \begin{tikzcd}[
    row sep=-13pt,
    column sep=27pt
  ]
    &&
    \adjustbox{scale=.9, fbox}{$
      \def\arraystretch{1}
      \begin{array}{l}
        \mathrm{d}\, F_{2\bullet + \sigma}
        \;=\;
        0
        \;\;
        \adjustbox{scale=.8}{
          $\forall\, 2\!\bullet + \sigma$
        }
        \hspace{-8pt}
      \end{array}
    $}
    \\[-8pt]
    \adjustbox{scale=.9, fbox}{$
      \begin{array}{c}
      \def\arraystretch{1}
      \begin{array}{r}
        \mathrm{d} \, F_{2\bullet + \sigma}
        \\
        \mathrm{d} \star F_{2\bullet + \sigma}
      \end{array}
      \begin{array}{c}
        =
        \\
        =
      \end{array}
      \begin{array}{l}
        0
        \\
        0
      \end{array}
      \\[+3pt]
      \adjustbox{scale=.8}{
        $
        \forall \,
        2 \bullet  + \sigma
          \;\leq\;
        5
        $
      }
      \\
      \star F_5 \,=\, F_5
      \;
      \mbox{if $\sigma = 1$}
      \end{array}
    $}
    \ar[
      urr,
      <->,
      rounded corners,
      to path={
        ([xshift=-0pt]\tikztostart.east)
        --
        ([xshift=30pt]\tikztostart.east)
        --
        ([xshift=-15pt]\tikztotarget.west)
        --
        ([xshift=-0pt]\tikztotarget.west)
      }
    ]
    \ar[
      drr,
      <->,
      rounded corners,
      to path={
        ([xshift=-0pt]\tikztostart.east)
        --
        ([xshift=30pt]\tikztostart.east)
        --
        ([xshift=-13pt]\tikztotarget.west)
        --
        ([xshift=-0pt]\tikztotarget.west)
      }
    ]
    \\[-3pt]
    &&
    \adjustbox{scale=.9, fbox}{$
      \;\;
      \def\arraystretch{.9}
        F_{{}_{(10-2\bullet-\sigma)}}
          \;=\;
        \star \,F_{2\bullet+\sigma}
      \;\;
    $}
  \end{tikzcd}
  \hspace{.5cm}
  \adjustbox{raise=-8pt}{
  $
  \def\arraystretch{1}
  \begin{array}{l}
    \bullet \in \mathbb{N}
    \\[+8pt]
    \sigma
      =
    \left\{
    \def\arraystretch{1.2}
    \begin{array}{ll}
      0 & \mbox{for type IIA}
      \\
      1 & \mbox{for type IIB}
    \end{array}
    \right.
  \end{array}
  $
  }
\end{equation}
and, more generally, those with non-vanishing B-field as follows:
\begin{equation}
  \label{MotionOfUnboundedBRRFields}
  \begin{tikzcd}[
    row sep=-20pt,
    column sep=20pt
  ]
    &&
    \adjustbox{fbox}{
      $
        \hspace{-3pt}
        \begin{array}{l}
          \differential
          \, F_{2\bullet + \sigma}
          \,=\,
          H_3
            \wedge
          F_{2\bullet+\sigma-2}
        \end{array}
        \;
        \begin{array}{l}
          \differential
          \,
          H_3 \,=\, 0
          \\
          \differential
          \,
          H_7 \,=\, \cdots
        \end{array}
        \hspace{-3pt}
      $
    }
    \\
    \adjustbox{fbox}{$
      \def\arraystretch{1.5}
      \begin{array}{l}
        \phantom{\star}\;\;
        \differential
        \,
        F_{2 \bullet + \sigma}
        \,=\,
        H_3
          \wedge
        F_{2 \bullet + \sigma -2}
        \\
        \differential
        \star
        F_{2 \bullet + \sigma}
        \,=\,
        H_3
          \wedge
        \star F_{D - 2 \bullet - \sigma + 2}
      \end{array}
      \def\arraystretch{1.5}
      \begin{array}{l}
        \differential
        \,
        H_3 \,=\, 0
        \\
        \differential
        \,
        \star H_3 \,=\, \cdots
      \end{array}
    $}
    \ar[
      urr,
      <->,
      rounded corners,
      to path={
        ([xshift=-0pt]\tikztostart.east)
        --
        ([xshift=20pt]\tikztostart.east)
        --
        ([xshift=-12pt]\tikztotarget.west)
        --
        ([xshift=-0pt]\tikztotarget.west)
      }
    ]
    \ar[
      drr,
      <->,
      rounded corners,
      to path={
        ([xshift=-0pt]\tikztostart.east)
        --
        ([xshift=20pt]\tikztostart.east)
        --
        ([xshift=-11pt]\tikztotarget.west)
        --
        ([xshift=-0pt]\tikztotarget.west)
      }
    ]
    \\
    &&
    \adjustbox{fbox}{
      \hspace{+3pt}
      $
        F_{D = 2\bullet - \sigma}
        \;=\;
        \star \, F_{2\bullet + \sigma}
        \hspace{.5cm}
        \begin{array}{c}
          H_7
          \,=\,
          \star
          \,
          H_3
        \end{array}
      $
      \hspace{+3pt}
    }
  \end{tikzcd}
\end{equation}

Beware, while these equations are now often stated in this form, and while this is the form that motivates the traditional {\it Hypothesis K} (\cref{RRFieldFluxQuantization}), it is at least subtle to see them in entirety as actually arising from ordinary $D=10$ supergravity (namely from KK-compactification of $D=11$ supergravity, in the case $\sigma = 0$), since in that context:

\noindent
$\bullet$ The fluxes $F_0$ and $F_{1-}$ are not actually present: They are from {\it massive} type IIA, which has its own subtleties.

\noindent
$\bullet$ The flux $H_7$ has a non-linear Bianchi ($\differential H_7 = - F_4 \wedge  F_4 +  F_2 \wedge F_6$)
which does not fit the pattern (cf. Ex. \ref{MotionOfBRRFieldFluxes}).
\end{example}

Notice that in type IIB, \eqref{PremetricFormulationOfPlainRRFieldTheory} describes a flux density ($F_5$) which is Hodge dual (not just to any other flux in the tuple but) to itself, $F_5 = \star F_5$. Generally we have:

\begin{example}[{\bf Motion of self-dual higher gauge field fluxes}]
\label{MotionOfSelfDualHigherGaugeField}
$\,$

\noindent
\hspace{-.2cm}
\def\tabcolsep{0pt}
\begin{tabular}{p{7cm}l}
Since Def. \ref{HigherMaxwellEquations} regards {\it every} higher gauge theory (of Maxwell-type) as being “self-dual” in a sense, the equations of motion of flux densities of actual self-dual higher gauge fields --- in the strict sense that one and the same flux density form is required to be Hodge dual to itself --- are readily an example of Def. \ref{HigherMaxwellEquations}:
&
\hspace{-5pt}
\adjustbox{raise=-1.2cm}{
\begin{minipage}{10.5cm}
  \begin{equation}
  \label{EquationsOfMotionOfSelfDualField}
  \begin{tikzcd}[
    column sep=24pt,
    row sep=-20pt
  ]
    &&
    \adjustbox{fbox}{$
      \;\;
      \differential
      \;
      F_{{}_{D/2}}
      \;=\;
      0
      \;\;
    $}
    \\
    \adjustbox{fbox}{
      \begin{tabular}{c}
        equations of motion of
        \\
        self-dual higher
        gauge field
        \\
        in $D = 4k+2$
      \end{tabular}
    }
    \ar[
      urr,
      <->,
      rounded corners,
      to path={
        ([xshift=-0pt]\tikztostart.east)
        --
        ([xshift=25pt]\tikztostart.east)
        --
        ([xshift=-15pt]\tikztotarget.west)
        --
        ([xshift=-0pt]\tikztotarget.west)
      }
    ]
    \ar[
      drr,
      <->,
      rounded corners,
      to path={
        ([xshift=-0pt]\tikztostart.east)
        --
        ([xshift=25pt]\tikztostart.east)
        --
        ([xshift=-15pt]\tikztotarget.west)
        --
        ([xshift=-0pt]\tikztotarget.west)
      }
    ]
    \\
    &&
    \adjustbox{fbox}{$
      F_{{}_{D/2}}
      \;=\;
      \star
      \,
      F_{{}_{D/2}}
    $}
  \end{tikzcd}
  \end{equation}
 \end{minipage}
}
\end{tabular}

Due to the properties of the square of the Hodge operator \eqref{HodgeSquareOnLorentzianSpacetim}, this has non-trivial solutions iff the degree of the flux is odd, $\mathrm{deg}=2k+1$, and hence iff spacetime dimension is $D=4k+2$, $k\in \mathbb{N}$.
\end{example}

\begin{example}[{\bf Motion of C-field fluxes}]
\label{MotionOfCFieldFlux} $\,$

\noindent
\hspace{-.1cm}
\begin{tabular}{p{7cm}l}
The equations of motion of the C-field in $D=11$ supergravity (originally the “3-index A-field” due to \cite{CremmerJuliaScherk79}
(cf. \cite[p. 32]{MiemiecSchnakenburg06})
are traditionally as shown on the left here, with their equivalent “duality-symmetric” reformulation \cite{BandosBerkovitsSorokin98} shown on the right, cf. \cite{GSS-11dSuperFlux}.
&
\hspace{-.3cm}
\adjustbox{raise=-1cm}{
\begin{minipage}{10.7cm}
\begin{equation}
  \label{PremetricFormulationOfCField}
  \begin{tikzcd}[row sep=-9pt]
    &&
    \adjustbox{scale=.9, fbox}{$
      \def\arraystretch{.9}
      \begin{array}{c}
        \mathrm{d} \, G_4
        \\
        \mathrm{d} \, G_7
      \end{array}
      \begin{array}{c}
        =
        \\
        =
      \end{array}
      \begin{array}{l}
        0
        \\
        \mathllap{-}\tfrac{1}{2}
        G_4 \wedge G_4
      \end{array}
    $}
    \\[-8pt]
    \adjustbox{scale=.9, fbox}{$
      \def\arraystretch{.9}
      \begin{array}{r}
        \mathrm{d} \, G_4
        \\
        \mathrm{d} \star G_4
      \end{array}
      \begin{array}{c}
        =
        \\
        =
      \end{array}
      \begin{array}{l}
        0
        \\
        \mathllap{-}\tfrac{1}{2}
        G_4 \wedge G_4
      \end{array}
    $}
    \ar[
      urr,
      <->,
      rounded corners,
      to path={
        ([xshift=-0pt]\tikztostart.east)
        --
        ([xshift=30pt]\tikztostart.east)
        --
        ([xshift=-13pt]\tikztotarget.west)
        --
        ([xshift=-0pt]\tikztotarget.west)
      }
    ]
    \ar[
      drr,
      <->,
      rounded corners,
      to path={
        ([xshift=-0pt]\tikztostart.east)
        --
        ([xshift=30pt]\tikztostart.east)
        --
        ([xshift=-32pt]\tikztotarget.west)
        --
        ([xshift=-0pt]\tikztotarget.west)
      }
    ]
    \\
    &&
    \adjustbox{scale=.9, fbox}{$
      \def\arraystretch{.9}
        G_7 \;=\; \star \,G_4
    $}
  \end{tikzcd}
\end{equation}
\end{minipage}
}
\end{tabular}
\end{example}

\begin{example}[{\bf Motion of type IIA B\&RR-field fluxes}]
\label{MotionOfBRRFieldFluxes}
   \cite[\S 4]{MathaiSati04}\cite[\S 3]{FSS17-Sph}

\noindent
\hspace{-.2cm}
\def\tabcolsep{0pt}
\begin{tabular}{p{2.8cm}l}
Under double dimensional reduction \eqref{DDReductionOfCFieldFlux} of the
C-field flux along a circle-bundle, the equations of motion \eqref{PremetricFormulationOfCField}
  are equivalently expressed in terms of its B\&RR-field-components as shown.
 &
\hspace{.1cm}
\adjustbox{raise=-2.2cm}{
\begin{minipage}{15.7cm}
  \adjustbox{raise=-2.4cm}{
    \includegraphics[width=14.4cm]{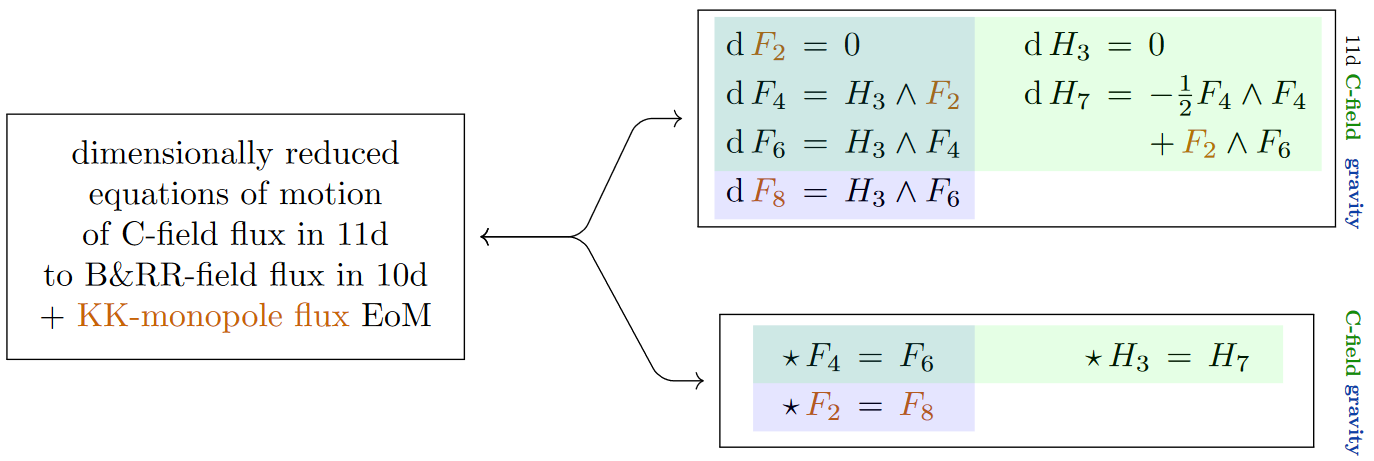}
  }
\end{minipage}
}
\end{tabular}

These are the equations of motion of the flux densities of type IIA supergravity in their duality-symmetric formulation \cite[\S 3]{CJLP98}.

Several terms above deserve special attention, either for how they appear or for how they do {\it not} appear:

\begin{itemize}
  \item The Hodge dual $F_8 = \star F_2$ is {\it not} part of the C-field in 11d, but is part of the gravitational field (the Hodge star encodes the 10d metric and $F_2$ is an aspect of the 11d fiber geometry). The expression for $\differential H_8$ arises as part of the gravitational field equations \cite[(3.4)]{CJLP98}.

\item The presence in type IIA supergravity of the non-linear Bianchi identity for $H_7$, albeit readily verified and “well-known” at least since \cite[(3.4)]{CJLP98}, is not as widely appreciated as the pattern \eqref{MotionOfUnboundedBRRFields} from Ex. \ref{MotionOfUnboundedRRFieldFluxes} – which it {\it breaks} (cf. also Rem. \ref{ShortcomingOfHigherCircleFluxQuantization} below).

\item No flux densities $F_0$ nor $F_{10}$ appear in 10d from KK-reduction on a circle, nor are they part of type IIA supergravity. These fluxes are instead part of massive type IIA supergravity whose relation to $D=11$ supergravity/M-theory remains less understood.
\end{itemize}
\end{example}

\subsection{Solution space of the flux equations}
\label{SolutionSpaceOfTheFluxEquations}

{\bf The phase space}. Abstractly, the phase space of any field theory is nothing but the space of all those field histories that satisfy the given equations of motion (the “on-shell” field histories). Phrased this way, this is sometimes called the {\it covariant phase space} \cite{CrnkovicWitten87}, to emphasize that no choice of foliation of spacetime by Cauchy surfaces has been or needs to be made.

\label{PhaseSpaceIdea}
The more traditionally familiar {\it canonical phase space} (physics jargon which, beware, is somewhat incompatible with the mathematician’s “canonical”) is instead a parameterization of the covariant phase space by initial value

\noindent
\hspace{-.2cm}
\def\tabcolsep{0pt}
\begin{tabular}{p{4.5cm}l}
data on a choice of Cauchy surface.
This choice breaks the “manifest covariance” of the covariant phase space.

Nevertheless, if a Cauchy surface exists at all (hence on globally hyperbolic spacetimes), then both these phase spaces are equivalent, by definition, the equivalence being the map that generates from initial value data the essentially unique on-shell field history that evolves from it (disregarding gauge transformations for the moment).
& \qquad
\adjustbox{raise=-7cm}{
  \includegraphics[width=11cm]{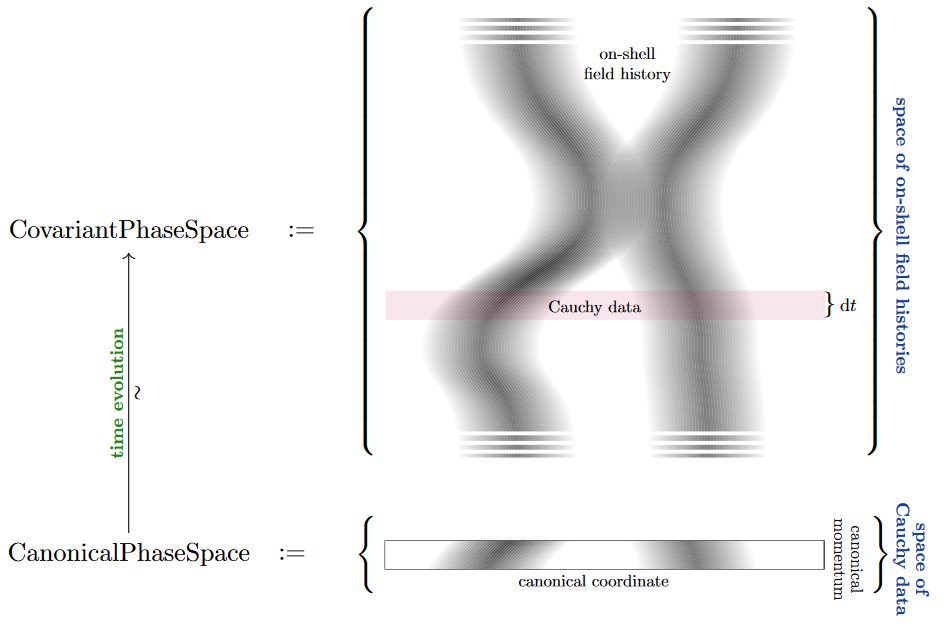}
}
\end{tabular}

\vspace{-.3cm}

\paragraph{Solution space of on-shell flux densities.} At this point in the discussion the full gauge field content is not yet determined --- this will only be implied by a choice of flux quantization below in \cref{PhaseSpacesAsDifferentialNonabelianCohomology} --- so far we are only considering the flux densities of the would-be gauge fields. To remember this, we shall call the space of flux densities solving their equations of motion (Def. \ref{HigherMaxwellEquations}) the {\it solution space}; and we are after its incarnation as a canonical solution space of initial value data on a Cauchy surface. But this goes a long way, since the higher Maxwell-type equations of motion constrain exclusively the flux densities: Once the flux-quantization the canonical phase will simply consist of all flux-quantized gauge potentials compatible with the flux densities in the canonical solution space.

\begin{proposition}[{\cite{SS23FQ}}]
 \label{SolutionSpaceViaGaussLaw}
 On a globally hyperbolic spacetime $X^D \,\simeq\, \mathbb{R}^{0,1} \times X^d$, the solution space given higher Maxwell-equations of motion (Def. \ref{HigherMaxwellEquations}) is isomorphic to the solution of (just) the duality-symmetric Bianchi identities restricted (i.e.: pulled back to) to any Cauchy surface $\iota \,\colon\, X^d \hookrightarrow X^D$, there to be called the {\it higher Gau{ss} law}:
 \begin{equation}
    \label{GaussLawFromBianchiIdentities}
    \def\arraystretch{1}
    \begin{array}{rcl}
    \mathllap{
      \scalebox{.7}{
        \color{darkblue}
        \bf
        \begin{tabular}{c}
          Space of flux densities
          \\
          on spacetime, solving
          \\
          the equations of motion
        \end{tabular}
      }
    }
    \mathrm{SolSpace}
    &\defneq&
    \Bigg\{\;
      \overset{
        \mathclap{
          \raisebox{3pt}{
            \scalebox{.7}{
              \color{darkblue}
              \rm
              electromagnetic flux densities on spacetime
            }
          }
        }
      }{
      \vec F \,\defneq\,
      \Big(
        F^{(i)}
        \,\in\,
        \Omega^{\mathrm{deg}_i}_{\mathrm{dR}}\big(
          X^D
        \big)
       \Big)_{i \in I}
       }
       \;\; \Bigg\vert\,
  \def\arraystretch{1.3}
  \begin{array}{l}
    \overset{
      \mathclap{
        \raisebox{3pt}{
          \scalebox{.7}{
            \color{darkblue}
            \rm
            Bianchi identities
          }
        }
      }
    }{
    \differential
    \, \vec F \,=\,
    \vec P\big( \vec F \big)
    }
    \\
    \underset{
      \mathclap{
        \raisebox{-1pt}{
          \scalebox{.7}{
            \color{darkblue}
            \rm
            self-duality
          }
        }
      }
    }{
    \star \, F
    \,=\,
    \vec \mu\big( \vec F \big)
    }
  \end{array}
   \!\! \Bigg\}
  \mathrlap{
    \scalebox{.8}{
      \color{darkblue}
      \bf
      covariant form
    }
  }
    \\[30pt]
   &
     \underset{
       \iota^\ast
     }{
       \simeq
     }
    &
    \Bigg\{\;
    \overset{
      \mathclap{
        \raisebox{3pt}{
          \scalebox{.7}{
            \color{darkblue}
            \rm
            magnetic flux densities on Cauchy surface
          }
        }
      }
    }{
      \vec B \,\defneq\,
      \Big(
        B^{(i)}
        \,\in\,
        \Omega^{\mathrm{deg}_i}_{\mathrm{dR}}\big(
          \,X^d\,
        \big)
       \Big)_{i \in I}
       }
       \;\; \bigg\vert\,
       \begin{array}{l}
    \overset{
      \mathclap{
        \raisebox{3pt}{
          \scalebox{.7}{
            \color{darkblue}
            \rm
            Gau{\ss} law
          }
        }
      }
    }{
         \differential
         \, \vec B \,=\,
         \vec P\big( \vec B \big)
    }
    \\
    {}
       \end{array}
  \!\!  \Bigg\}
  \mathrlap{
    \scalebox{.8}{
      \color{darkblue}
      \bf
      canonical form
    }
  }
    \end{array}
  \end{equation}
\end{proposition}

\begin{example}[{\bf Solution- and phase-space of ordinary electromagnetism}]
\label{SolutionSpaceOfOrdinaryElectromagnetism}

In the case of ordinary vacuum electromagnetism, Prop. \ref{SolutionSpaceViaGaussLaw} applied to the ordinary Maxwell equations (Ex. \ref{MotionOfOrdinaryElectromagneticFluxes}) says that the initial value data on a Cauchy surface $X^3$ is given by {\it independently} specifying magnetic and electric flux densities
$
  B,\, E \,\in\, \Omega^2_{\mathrm{dR}}(X^3)
$
subject only to the ordinary Gau{\ss} laws
$
  \differential\, B \,=\, 0
  \,,\;\;
  \differential\, E \,=\, 0
  \,.
$
Indeed, the actual phase space of electromagnetism is well-known (e.g. \cite[\S 5]{BlaschkeGieres21}) to have as

(i) canonical coordinate the gauge potential $\widehat{A}$,
\;
(ii) canonical momentum the electric flux density $E$

Thereby $B \defneq \mathrm{curv}\big(\widehat A\big)$ is indeed independent from $E$ (and satisfies its Gau{\ss} law definitionally, while the Gau{\ss} law on $E$ is a phase space constraint).
\end{example}
\begin{remark}[\bf Canonical coordinates/momenta from duality-symmetry]
Notice how, thereby, the traditional split of initial value data into canonical coordinates and canonical momenta (whose definition requires assumption and variation of a Lagrangian density) is preempted here, under Prop. \ref{SolutionSpaceViaGaussLaw}, already by the pregeometric/duality-symmetric formulation of Maxwell’s equations (Ex. \ref{HigherMaxwellEquations}), in the sense that the spacetime archetypes of the canonical coordinates and momenta on a Cauchy surface (the former seen under the differential) are just the ordinary flux density $F_2$ (since $B = \iota^\ast F_2$) and its “duality partner” $G_2$ (since $E = \iota^\ast G_2$).
\end{remark}

\begin{remark}[{\bf Gravity “decouples” on canonical phase space}]
The inverse isomorphism \eqref{GaussLawFromBianchiIdentities} is given by time evolution of initial value data. Notice that the pseudo-Riemannian metric on $X^D$ --- the background field of gravity --- enters only in determining the nature of this isomorphism $\iota^\ast$ (the time evolution away from the Cauchy surface), but does not affect the nature of the initial value data (of the canonical phase space) as such.
\end{remark}


\ifdefined\shortversion
\else

\subsection{Qualitative solutions: Brane intersections}
\label{QualitativeSolutions}

Prop. \ref{SolutionSpaceViaGaussLaw} implies that on globally hyperbolic spacetimes the structure of on-shell flux densities in supergravity may be analyzed already by solving the (non-linear) Gau{\ss} law \eqref{GaussLawFromBianchiIdentities} for duality-symmetric fluxes on any Cauchy surface and ignoring the coupling to gravity there (assuming only that there exists at least one gravitational field configuration which solves its Einstein equations with source terms of this form). Since the same Gau{\ss} law also governs the admissible flux quantization laws below in \cref{FluxQuantizationLaws} we showcase a couple of qualitative solutions to highlight just how much non-trivial (brane-)physics is encoded in these equations.

\begin{example}[{\bf D6/D8-intersections and the Hanany-Witten effect}]
\cite{HananyWitten97} conjectured that D$p$-branes stretching between NS5 and D$p+2$ -- cf. \eqref{TableOfFluxSpeciesAndTheirSources} --  are “created” as the D$p+2$-branes are “dragged over” the NS5, intuitively like a pole will cause a spike in a rubber sheet that is pulled over its tip. It was suggested by \cite[\S2]{Marolf01} that this {\it Hanany-Witten effect} should be understandable entirely from analysis of the flux Bianchi identities \eqref{GaussLawFromBianchiIdentities}.

For the case of NS5/D6/D8-brane intersections (e.g. \cite[\S 2.4]{HananyZaffaroni98})\cite[p. 60]{BergshoeffLozanoOrtin98} this may be seen as follows (the other cases work analogously):

Here, by the flux equations \eqref{MotionOfUnboundedBRRFields}
\begin{equation}
  \label{D8D8FluxEquations}
 \begin{array}{l}
   \differential F_0 \,=\, 0
   \\
   \differential F_2 \,=\, H_3 \wedge F_0
  \end{array}
\end{equation}
the flux density $F_0$ of D8-branes (the “Romans mass”) is a locally constant function that vanishes in the vacuum and jumps by $N$ units across the singular locus of N D8 -branes (cf. e.g. \cite[p. 40]{Fazzi17}). But this means that:

(1.) When the NS5 -brane is located in the vacuum where $F_0 = 0$, then its sourcing of F 2-flux is “switched off” by the vanishing $F_0$-factor in \eqref{D8D8FluxEquations}, hence if $F_2$ vanishes at infinity then the PDE demands it vanishes everywhere, reflecting the absence of D6 -branes.

(2.) When the NS5 -brane is located on the other side of the D8-branes, where $F_0 = N$, then the equation \eqref{D8D8FluxEquations} shows that $F_2$-flux/D6-number density which vanishes far away will increase along the axis orthogonal axis $x^9$ to the D8-branes in proportionality to the $\differential x^9$-component of the flux $H_3$, and hence pronouncedly so as one crosses the NS5-brane locus.
\begin{center}
 \includegraphics[width=.72\textwidth]{FluxesOfD6BetweenNS5AndD8}
\end{center}
\end{example}

\begin{example}[{\bf M2$\perp$M5-brane intersections on ``M-strings'')}]

Consider the singular loci of two parallel flat M5-branes at a distance $2d > 0$
$$
  \begin{tikzcd}[sep=0pt]
    \mathbb{R}^{1,5}_{{}_{(i)}}
    \ar[rr, hook]
    &&
    \mathbb{R}^{1,D}
    \\
    (t,\vec x)
      &\mapsto&
    \big(t,\vec x, (-1)^i d, \vec 0\big)
  \end{tikzcd}
$$
each reflected by unit 4-flux through their surrounding 4-spheres
$$
    G_4^{{}^{(i)}}
    \;\coloneqq\;
    \mathrm{dvol}_{S^4}
    \,\in\,
    \Omega^4_{\mathrm{dR}}(S^4)
    \xhookrightarrow{ \quad \mathrm{pr}_{S^4}^\ast \quad}
    \Omega^4_{\mathrm{dR}}\Big(
      \mathbb{R}^{1,5}_{{}_{(i)}}
      \times
      \mathbb{R}_{\plus}
      \times
      S^4
    \Big)
    \;\simeq\;
    \Omega^4_{\mathrm{dR}}\Big(
      \mathbb{R}^{1,10}
      \setminus
      \mathbb{R}^{1,5}_{{}_{(i)}}
    \Big)
  \,.
$$
With the total 4-flux thus
$
  G_4
  \;\coloneqq\;
  G_4^{(1)}
  +
  G_4^{(2)}
  \,,
$
the C-field Bianchi identity (Ex. \ref{MotionOfCFieldFlux}) says that the flux $G_7$ sourced by M2-branes obeys the equation
\[
  \label{M2FluxSourcedSourceByTwoM5Branes}
  \begin{array}{rcl}
    \mathrm{d}
    G_7
    &=&
    -
    \tfrac{1}{2} G_4 \wedge G_4
    \\
    &=&
    - G_4^{(1)} \wedge G_4^{(2)}
    \,,
  \end{array}
\]
where the wedge product of the two translated $S^4$-volume forms acts as an effective "electric potential" source term.

For the purpose of illustration we consider a qualitative sketch of  the 2-dimensional analog of this situation, where the 4-forms are replaced by 2-forms:

\noindent
\hspace{-.3cm}
\def\tabcolsep{5pt}
\begin{tabular}{p{7cm}l}
Effective dipole of quadratic brane flux. The figure means to indicate the nature of the differential 2-form which is the wedge product of two copies of the pullback of dvol S 1 to around either of the punctures (the brane loci) in the 2-punctured plane. Here:

    the strength of the circular lines indicates the absolute value of the flux density sourced by the respective 5-brane,

    the arrows indicate the orientation of the flux density of either 5-brane,

    the parallelograms indicate the orientation of their wedge product.

Evidently, the absolute value of the wedge product is concentrated near the 5-branes and particularly between them...
&
\adjustbox{raise=-7.6cm}{
\includegraphics[width=9cm]{FluxOfM2M5Intersection}
}
\end{tabular}

\noindent
\hspace{-.1cm}
\begin{tabular}{p{3.6cm}l}
...but the orientation of the wedge product changes sign across the axis connecting the branes, as shown. This means that the flux sourced by this wedge product, according to (15), is, if vanishing at infinity, concentrated between the branes.
&
\adjustbox{raise=-5cm}{
\includegraphics[width=12.5cm]{SourcingM2FluxFromM5Fluxes}
}
\end{tabular}
This sourced flux concentration (indicated in red) witnesses an M2-brane stretching between the two M5-branes. The intersection is known as the {\it M-string}.\end{example}

Notice how in these examples we {\it chose} integral values for the total source brane fluxes. Next in \cref{FluxQuantizationLaws} we discuss how such flux quantization is systematically enforced in the higher gauge theory.

\fi

\section{Flux \& Charge Quantization Laws}
\label{FluxQuantizationLaws}

With the solution space (Prop. \ref{SolutionSpaceViaGaussLaw}) of higher Maxwell-type equations of motion (Def. \ref{HigherMaxwellEquations}) in hand, the question of flux quantization is to further constrain the flux densities such that the total fluxes and their total source charges take values in some discrete space.
The technical issue to be resolved here is that:
\begin{itemize}[leftmargin=.5cm]
\item this is a global condition on the flux densities: The local flux densities may take any value (compatible with the equations of motion) and yet the total accumulation of all these local contributions needs to be constrained;

\item the evident idea of constraining the ordinary integrals of the flux densities (their “periods”) makes sense only for closed differential forms and hence does not work for non-linear Bianchi identities (such as those of the C-field, Ex. \ref{MotionOfCFieldFlux}, and the B\&RR-field, Ex. \ref{MotionOfBRRFieldFluxes}).
\end{itemize}

\smallskip
\noindent
\hspace{-.45cm}
\adjustbox{
  raise=-1.5cm
}{
\def\tabcolsep{5pt}
\begin{tabular}{p{9.5cm}l}
To resolve this, one may first observe that:

\begin{itemize}[leftmargin=.5cm]
\item[$\bullet$] the integrals/periods of ordinary closed differential $n$-forms $f_n$ over $n$-manifolds are in natural correspondence with their de Rham-classes, $[F_n] \in H^n_{\mathrm{dR}}(-)$, which in turn are equivalently their “deformation classes”,
namely their {\it concordance} classes: $H^n_{\mathrm{dR}}(-) \,\simeq\,\Omega^n_{\mathrm{dR}}(-)_{\mathrm{clsd}} \big/_{\!\mathrm{cncrdnc}}$;

\vspace{1mm}
\item[$\bullet$] so that integrality of closed flux density $F_n$ is witnessed by an integral cohomology class $[\rchi] \in H^n(X;\mathbb{Z})$ whose``de Rham character'' image $\mathrm{ch}[\rchi] \in H^n_{\mathrm{dR}}(X)$ coincides with the deformation class $[F_n]$;
\end{itemize}
& \qquad
\adjustbox{
  raise=-2.1cm,
  fbox
}{
  \hspace{.1cm}
  \begin{tikzcd}[sep=0pt]
    &[-5pt]&[-5pt]
    H^2(X; \mathbb{Z})
    \ar[
      dd,
      "{
        \mathrm{ch}
      }"{pos=.4}
    ]
    &
    {[\rchi]}
    \mathrlap{
      \;
      \scalebox{.7}{
        \color{darkblue}
        \bf
        \def\arraystretch{.9}
        \begin{tabular}{c}
          integral
          \\
          charge
        \end{tabular}
      }
    }
    \\
    &&&
    \rotatebox[origin=c]{-90}{$\mapsto$}
    \mathrlap{
      \;
      \scalebox{.7}{
        \color{greenii}
        \bf
        character
      }
    }
    \\[-2pt]
    \Omega^2_{\mathrm{dR}}(X)_\closed
    \ar[
     rr,
     ->>
    ]
    &&
    H^2_{\mathrm{dR}}(X)
    &
    \mathrm{ch}[\rchi]
    \\
    \mathllap{
      \scalebox{.7}{
        \color{darkblue}
        \bf
        \def\arraystretch{.9}
        \begin{tabular}{c}
          flux
          \\
          density
        \end{tabular}
      }
    }
    \hspace{-4pt}
    F_2
      &\mapsto&
    {[F_2]}
    \ar[
      from=ur,
      equals,
      bend left=30,
      "{
        \hspace{-12pt}
        \scalebox{.7}{
        \color{purple}
        \def\arraystretch{.85}
        \begin{tabular}{c}
          integral
          \\
          total flux
        \end{tabular}
        }
      }"{pos=.1}
    ]
  \end{tikzcd}
  \hspace{.55cm}
}
\end{tabular}
}

\vspace{-4mm}
\noindent
and, second, one may observe that this perspective generalizes \cite{FSS23Char}\cite{SS23FQ}:

\medskip
{
\noindent
\hspace{-.6cm}
\def\arraystretch{1.5}
\def\tabcolsep{10pt}
\begin{tabular}{p{8.1cm}|p{8.4cm}|}
\cline{2-2}
\ourrowcolor{lightgray}
Higher Maxwell-type equations have a {\bf characteristic $L_\infty$-algebra} $\mathfrak{a}$:
The flux densities are equivalently $\mathfrak{a}$-valued differential forms, and the Gau{\ss} law \eqref{GaussLawFromBianchiIdentities} is equivalently the condition that these be {\it closed} (i.e.: flat, aka ``Maurer-Cartan element''; in Italian SuGra literature: ``satisfying an FDA'').
&
\adjustbox{raise=-.9cm}{
$
  \hspace{-8pt}
  \def\arraystretch{1.2}
  \def\arraycolsep{5pt}
  \begin{array}{l}
    \\[-9pt]
    \mathrm{SolSpace}(X^d)
    \;\simeq\;
    \\
    \bigg\{
    \overset{
      \mathclap{
        \raisebox{3pt}{
          \scalebox{.7}{
            \color{gray}
            \rm
            flux densities on Cauchy surface
          }
        }
      }
    }{
      \vec B \,\defneq\,
      \big(
        B^{(i)}
        \,\in\,
        \Omega^{\mathrm{deg}_i}_{\mathrm{dR}}(
          \,X^d\,
        )
       \big)_{i \in I}
       }
       \,\bigg\vert\,
       \begin{array}{l}
    \overset{
      \mathclap{
        \raisebox{3pt}{
          \scalebox{.7}{
            \color{gray}
            \rm
            satisfying
            Gau{\ss}'s law
          }
        }
      }
    }{
         \differential
         \, \vec B \,=\,
         \vec P\big( \vec B \big)
    }
    \\
    {}
       \end{array}
    \bigg\}
    \\[10pt]
  \qquad   \;\simeq\;
    \Omega_{\mathrm{dR}}\big(
      X^d
      ;\,
      \mathfrak{a}
    \big)_\closed
    \!\!
    \mathrlap{
      \scalebox{.7}{
        {
          \color{darkblue}
          \def\arraystretch{.9}
          \begin{tabular}{l}
            flat differential forms valued
            \\
            in characteristic $L_\infty$-algebra
          \end{tabular}
        }.
      }
    }
    \\[-9pt]
    {}
    \end{array}
  \hspace{-10pt}
$
}
\\
\cline{2-2}
Also every topological space $\mathcal{A}$ (under mild conditions) has a characteristic $L_\infty$-algebra: Its $\mathbb{R}$-rational  {\bf Whitehead bracket $L_\infty$-algebra} $\mathfrak{l}\mathcal{A}$.
&
\adjustbox{raise=-10pt}{
$
  \scalebox{.8}{
    \color{darkblue}
    \def\arraystretch{.9}
    \begin{tabular}{c}
      (homotopy type of)
      \\
      a topological space
    \end{tabular}
  }
  \mathcal{A}
  \;\;\;
  \underset{
    \mathclap{
    \raisebox{-4pt}{
    \scalebox{.7}{
      \color{greenii}
      \bf
      $\mathbb{R}$-rationalization
    }
    }
    }
  }{
    \rightsquigarrow
  }
  \;\;\;
  \mathfrak{l}\mathcal{A}
  \scalebox{.8}{
    \color{darkblue}
    \def\arraystretch{.9}
    \begin{tabular}{l}
      Whitehead
      \\
      $L_\infty$-algebra
    \end{tabular}
  }
$
}
\\
\cline{2-2}
\ourrowcolor{lightgray}
The {\bf nonabelian} Chern-Dold {\bf character map} turns $\mathcal{A}$-valued maps into closed $\mathfrak{l}\mathcal{A}$-valued differential forms, generalizing the Chern character for $\mathcal{A} = \mathrm{KU}_0$.
&
\hspace{-11pt}
\adjustbox{raise=-8pt}{
$
  \scalebox{.7}{
    \color{darkblue}
    \bf
    charge
  }
  \big(
    \rchi : X^d \to \mathcal{A}
  \big)
  \;\;\;
  \underset{
    \mathclap{
      \raisebox{-5pt}{
        \scalebox{.7}{
          \color{greenii}
          \bf
          character map in
          $\mathcal{A}$-cohomology
        }
      }
    }
  }{
    \;\longmapsto\;
  }
  \;\;\;
  \mathrm{ch}(\rchi)
  \in\,
  \Omega_{\mathrm{dR}}\big(
    X^d; \mathfrak{l}\mathcal{A}
  \big)_\closed
$
}
\\
\cline{2-2}
The {\bf possible flux quantization laws} for a given higher gauge field are those spaces $\mathcal{A}$ whose Whitehead $L_\infty$-algebra is the characteristic one.
&
\adjustbox{raise=-10pt}{\hspace{-2mm}
$
  \mathrm{FluxQuantLaws}
  \;=\;
  \left\{
    \hspace{-4pt}
    \adjustbox{raise=5pt}{$
    \underset{
      {
       \hspace{-9pt}
       \raisebox{-6pt}{
        \scalebox{.7}{
          \color{darkblue}
          \bf
          \def\arraystretch{.85}
          \begin{tabular}{c}
            classifying
            \\
            spaces
          \end{tabular}
        }
        }
      }
      \hspace{-9pt}
    }{
      \mathcal{A}
    }
    $}
    \middle\vert
    \adjustbox{raise=5pt}{$
    \underset{
      {
       \hspace{-10pt}
        \raisebox{-6pt}{
        \scalebox{.65}{
          \color{darkblue}
          \bf
          \def\arraystretch{.9}
          \begin{tabular}{c}
            whose rational homotopy    \\
            encodes the Gau{\ss} law
          \end{tabular}
        }
        }
      }
      \hspace{-10pt}
    }{
    \mathfrak{l}\mathcal{A}
    \,\simeq\,
    \mathfrak{a}
    }
    $}
    \hspace{-3pt}
  \right\}
$
}
\\
\cline{2-2}
\ourrowcolor{lightgray}
Given a flux quantization law $\mathcal{A}$, the corresponding {\bf higher gauge potentials} are deformations of the flux densities into characters of a $\mathcal{A}$-valued map, witnessing the flux densities as reflecting discrete charges quantized in $\mathcal{A}$-cohomology.

(It is not obvious that this reduces to the usual notion of gauge potentials, but it does.)

&
\hspace{-5pt}
\adjustbox{raise=-1.3cm}{
$
  \begin{tikzcd}[row sep=15pt]
    &[+10pt]
    &
    \rchi
    \mathrlap{
        \;\;\;
        \raisebox{0pt}{
          \scalebox{.7}{
            \color{darkblue}
            \bf
            charge
          }
        }
    }
    \ar[
      d,
      shorten=3pt,
      |->,
      "{
        \raisebox{0pt}{
          \scalebox{.7}{
            \color{greenii}
            \bf
            character
          }
        }
      }"
    ]
    \\
    &&
    \mathrm{ch}(\rchi)
    \\
    {
      \scalebox{.7}{
        \color{darkblue}
        \bf
        \def\arraystretch{.85}
        \begin{tabular}{c}
          flux
          \\
          density
        \end{tabular}
      }
    }
    \vec F
    \ar[
      r,
      shorten=7pt,
      |->,
      "{
        \scalebox{.7}{
          \color{greenii}
          \bf
          shape
        }
      }"{swap, yshift=-2pt}
    ]
    &
    \vec F
    \ar[
      from=ur,
      Rightarrow,
      bend left=40pt,
      "{
        \widehat{A}
        \mathrlap{
          \scalebox{.7}{
            \color{orangeii}
            \bf
            \;
            gauge potential
          }
        }
      }"{pos=.35}
    ]
  \end{tikzcd}
$
}
\\
\cline{2-2}
These non-perturbatively completed higher gauge fields form a {\it smooth higher groupoid}: the ``canonical {\bf differential $\mathcal{A}$-cohomology} moduli stack''. Since these are now the flux-quantized on-shell fields, this is the {\bf phase space} of the flux-quantized higher gauge theory (p. \pageref{PhaseSpaceIdea}).
&
\adjustbox{raise=-24pt}{
$
  \begin{array}{l}
  \\[-12pt]
  \overset{
    \mathclap{
      \scalebox{.7}{
        \color{darkblue}
        \bf
        \def\arraystretch{.9}
        \begin{tabular}{c}
          flux-quantized
          \\
          phase space
          \\
          stack is
        \end{tabular}
      }
    }
  }{
    \underset{
    \mathclap{
      \scalebox{.7}{
        \color{darkblue}
        \bf
        \def\arraystretch{.9}
        \begin{tabular}{c}
          differential
          \\
          $\mathcal{A}$-cohomology
          \\
          moduli stack
        \end{tabular}
      }
    }
    }{
    \widehat{\mathcal{A}}(X^d)
    }
  }
  \;:=\;
  \left\{\!
    \left(\!\!
    \def\arraystretch{1.1}
    \begin{array}{rcll}
      \vec F
      &\in&
      \Omega_{\mathrm{dR}}(X^d;\mathfrak{l}\mathcal{A})_{\closed}
      &
      \scalebox{.7}{
        \color{gray}
        flux
      }
      \\
      \rchi
      &\in&
      \mathrm{Map}(X; \mathcal{A})
      &
      \scalebox{.7}{
        \color{gray}
        charge
      }
      \\
      \widehat{A}
      &:&
      \mathrm{ch}(\rchi)
        \Rightarrow
      \vec F
      &
      \scalebox{.7}{
        \color{gray}
        gauge
      }
    \end{array}
    \right)
  \right\}
  \\[-12pt]
  {}
  \end{array}
$
}
\\
\cline{2-2}
\end{tabular}
}

\newpage

This flux-quantized phase space hence subsumes the “solitonic” fields with non-trivial charge sectors $\rchi$, and as such is
a non-perturbative completion of the traditional phase spaces (which correspond to a fixed charge sector only, typically to $\rchi = 0$).

\medskip

Incidentally, it follows that the choice of flux quantization law $\mathcal{A}$ not only defines the
solitonic content of the theory but completely characterizes it:

{
\vspace{-.0cm}
\hspace{-1.73cm}
\def\tabcolsep{25pt}
\def\arraystretch{1.3}
\begin{tabular}{p{8.5cm}|p{5.4cm}|}
\cline{2-2}
\ourrowcolor{lightgray} The shape (topological realization) of this phase space stack is the {\bf space of topological fields},
&
\adjustbox{raise=-5pt}{
$
 \shape \,
 \widehat{\mathcal{A}}\big(X^d\big)
 \;\simeq\;
 \mathcal{A}(X^d)
 \;=\;
 \mathrm{Map}\big(
   X^d
   ,\,
   \mathcal{A}
 \big)
 $
}
\\
\cline{2-1}
which implies that the ordinary homology of the phase space stack constitutes the {\bf topological observables} on the higher gauge theory.
&
\adjustbox{raise=-12pt}{
$
  H_\bullet\big(
    \widehat{\mathcal{A}}(X^d)
    ;\,
    \mathbb{C}
  \big)
  \;\simeq
  H_\bullet\big(
    \mathcal{A}(X^d)
    ;\,
    \mathbb{C}
  \big)
$
}
\\
\cline{2-1}
\ourrowcolor{lightgray}
Hence if we focus only on the solitonic or {\it topological field}-content of the phase space, then we see plain $\mathcal{A}$-cohomology moduli of the Cauchy surface and the full phase space stack only serves to justify this object.
&
\adjustbox{raise=-20pt}{
$
  \begin{array}{l}
  \\[-12pt]
  \overset{
    \mathclap{
      \scalebox{.7}{
        \color{darkblue}
        \bf
        \def\arraystretch{.9}
        \begin{tabular}{c}
          flux-quantized
          \\
          topological
          \\
          phase space
        \end{tabular}
      }
    }
  }{
    \underset{
    \mathclap{
      \scalebox{.7}{
        \color{darkblue}
        \bf
        \def\arraystretch{.9}
        \begin{tabular}{c}
          non-abelian
          \\
          $\mathcal{A}$-cohomology
          \\
          moduli space
        \end{tabular}
      }
    }
    }{
      \mathcal{A}(X^d)
    }
  }
  \;:=\;
  \big\{
      \rchi
      \; \in \;
      \mathrm{Map}(X, \mathcal{A})
  \big\}
  \\[-12pt]
  {}
  \end{array}
  $
}
\\
\cline{2-2}
\multicolumn{1}{p{8cm}}{}
&
\multicolumn{1}{c}{}
\\
\end{tabular}
}

\vspace{-.7cm}
\noindent
We now explain all this in more detail.

\subsection{Total flux as Nonabelian de Rham cohomology}
\label{TotalFluxAsNonabelianDeRhamCohomology}

We explain how higher Bianchi identities \eqref{DualitySymmetricHigherMaxwellEquations} and their corresponding higher Gauss laws \eqref{GaussLawFromBianchiIdentities} are equivalently the closure (flatness) condition on differential forms valued in a characteristic $L_\infty$-algebra (Prop. \ref{FluxSolutionsAsLInfinityValuedForms} below), so that total flux is a class in $\mathfrak{l}\mathcal{A}$-valued nonabelian de Rham cohomology (Def. \ref{NonabelianDeRhamCohomology} below).

\smallskip

The notion of {\it $L_\infty$-} or {\it strong homotopy Lie algebra}
is finally becoming more widely appreciated in physics, where they appear in various guises (cf. \cite{Stasheff16}). Here we are concerned with $L_\infty$-algebras which are (i) nilpotent, (ii) connective (iii) of finite type, in their joint incarnation as higher flux density coefficients and as higher Whitehead brackets (all to be explained in a moment), which one might refer to as the

\medskip
\noindent
{\bf Flux Homotopy Lie algebra triality}:

\vspace{-.45cm}
\noindent
\hspace{-.1cm}
\def\tabcolsep{0pt}
\begin{tabular}{p{9.7cm}l}

\vspace{-1mm}
\begin{itemize}[leftmargin=.4cm]
\item[$\bullet$] With {\it rational homotopy}, we are referring here specifically the fundamental theorem of dg-algebraic rational homotopy theory, mainly due to Quillen, Sullivan and Bousfield \& Gugenheim, as reviewed in \cite[\S 5]{FSS23Char}.

\item[$\bullet$] The {\it FDA method} in supergravity refers to the observations of
\cite{vanNieuwenhuizen83}\cite{DAuriFre82} \cite{CastellaniDAuriaFre91},
as explained and contextualized in \cite{FSS13}\cite{FSS18}\cite{HSS19}, reviewed in \cite{FSS19}.

\item[$\bullet$] The {\it nonabelian character} is the generalization of the Chern-Dold character map from topological K-theory and Whitehead-generalized cohomology to generalized non-abelian cohomology, constructed in \cite{FSS23Char}.
\end{itemize}

& \hspace{1.5mm}
\adjustbox{raise=-5cm}{
\includegraphics[width=7.7cm]{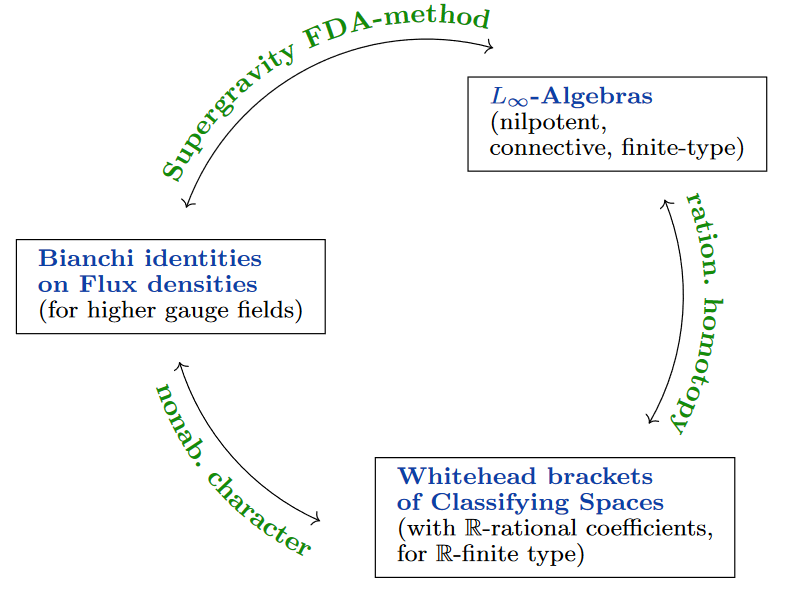}
}
\end{tabular}

\vspace{-2mm}
In particular, this means that $L_\infty$-algebras as used here are {\it not} directly to be understood as generalizations of the gauge Lie algebras familiar from Yang-Mills theory, which are coefficients of the gauge potentials, but instead as the coefficients of their flux densities (cf. Rem. \ref{LInfinityOfGaugePotentialsVsFLuxDensities}).

\vspace{-1mm}
\paragraph{$L_\infty$-Algebras.} Since we are assuming $L_\infty$-algebras to be connective and of finite type (meaning that they are degreewise finite-dimensional and concentrated in non-negative degrees) we may {\it define} them through their Chevalley-Eilenberg (CE) algebras in the following manner, which is not only convenient for dealing with the otherwise intricate sign rules, but also essential to their alternative perspectives in the above triality:

\vspace{-2mm}
\subparagraph{Chevalley-Eilenberg algebras of Lie algebras.} Namely, for $\mathfrak{g}$ a finite-dimensional Lie algebra (our ground field is the real numbers, throughout) with Lie bracket a skew-symmetric linear map $[-,-] \,:\, \mathfrak{g} \otimes \mathfrak{g} \to \mathfrak{g}$, its linear dual vector space $\mathfrak{g}^\ast$ is equipped with the dual bracket $[-,-]^\ast \,:\,\mathfrak{g}^\ast \to \mathfrak{g}^\ast \wedge \mathfrak{g}^\ast$ which extends uniquely to a degree=1 derivation on the graded Grassmann algebra $\wedge^\bullet \mathfrak{g} \,:=\, \underset{n \in \mathbb{N}}{\bigoplus}
\,\underbrace{\mathfrak{g}^\ast \wedge \cdots \wedge \mathfrak{g}^\ast}_{\scalebox{.7}{$n$ factors}}$:

\begin{center}
  \includegraphics[width=\textwidth]{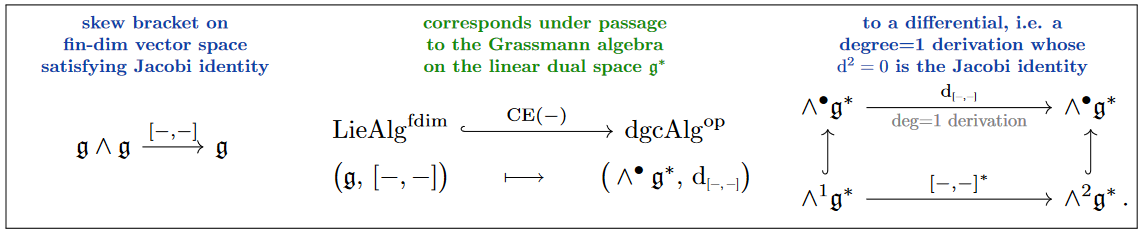}
\end{center}

\vspace{-1mm}
\noindent
One readily checks that this derivation squares to zero iff the bracket satisfies its Jacobi identity(!):
\vspace{1mm}
$$
  \mbox{
    Jacobi identity for $[-,-]$
  }
  \;\;\;\;\;\;
  \Leftrightarrow
  \;\;\;\;\;\;
  \differential_{[-,-]}
  \circ
  \differential_{[-,-]}
  \,=\,
  0
  \,.
$$
The resulting differential graded-commutative (dgc) algebra $(\wedge^\bullet \mathfrak{g}^\ast, \differential_{[-,-]})$ is known as the {\it Chevalley-Eilenberg complex} $\mathrm{CE}(\mathfrak{g})$ whose cochain cohomology computes the Lie algebra cohomology of $\mathfrak{g}$ (with trivial coefficients) --- but the key point at the moment is that its construction is a {\it fully faithful} embedding the category of finite-dimensional Lie algebras into the opposite of that of dgc-algebras.

\subparagraph{$L_\infty$-Algebras of finite type.} With ordinary Lie algebras viewed as special dgc-algebras this way, it is immediate to generalize them to the case where $\mathfrak{g}$ may be a graded vector space of degreewise finite dimension (“of finite type”): Namely, writing
\vspace{1mm}
$$
  (\mathfrak{g}^\vee)_n
  \;\coloneqq\;
  (\mathfrak{g}_n)^\ast,
  \;\;\;\;\;\;\;\;\;\;\;\;
  \wedge^\bullet
  \mathfrak{g}^\vee
  \;\coloneqq\;
  \mathrm{Sym}(\mathfrak{g}^\vee[1])
$$
we can use {\it verbatim} the same construction.
A degree=1 derivation on
$\wedge^\bullet \mathfrak{g}^\vee$
is determined by its restriction to $\wedge^1 \mathfrak{g}^\vee$, where it is a sum of co-$n$-ary linear maps, whose linear duals are identified with $n$-ary degree=(-1) brackets on $\mathfrak{g}[1]$:
\begin{center}
  \includegraphics[width=17cm]{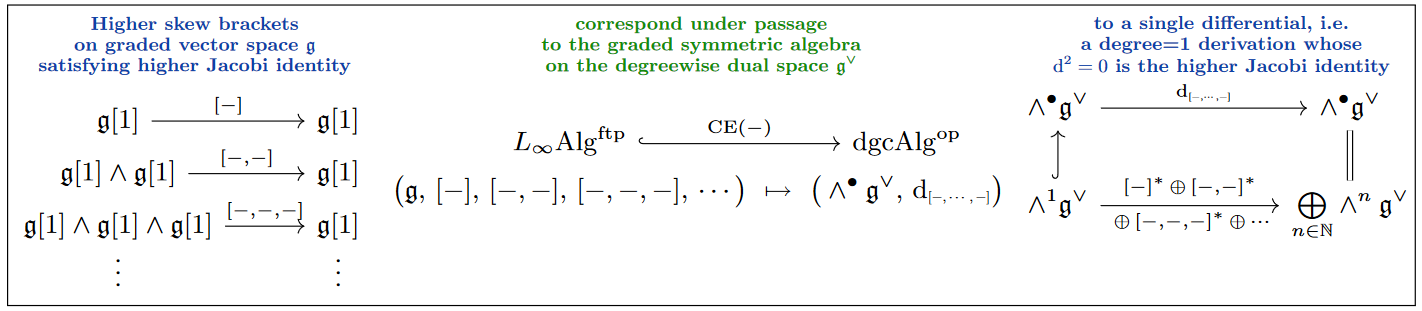}
\end{center}
Here the simple condition that $\differential_{{}_{[\cdots]}}$ be a differential implies a tower of conditions on these brackets, generalizing the Jacobi identity on an ordinary Lie algebra and known as the conditions that make $\big(\mathfrak{g},\, [-],\, [-,-],\, [-,-,-],\, \cdots\big)$ an {\it $L_\infty$-algebra}:
\vspace{2mm}
\begin{equation}
  \label{HigherJacobiIdentity}
  \mbox{
    \begin{tabular}{c}
      Higher Jacobi identity for
      \\
      $[-],\, [-,-],\, [-,-,-],\, \cdots$
    \end{tabular}
  }
  \hspace{.5cm}
  \Leftrightarrow
  \hspace{.5cm}
  \differential_{{}_{[\cdots]}}
  \circ
  \differential_{{}_{[\cdots]}}
  \;=\;
  0
  \,.
\end{equation}

\vspace{1mm}
\noindent In other words, we may identify $L_\infty$-algebras of finite type as the formal dual to dgc-algebras whose underlying graded-commutative algebra is free on a graded vector space.
Several examples are indicated in \eqref{ExamplesOfLInfinityAlgebras}.

\vspace{-1mm}
\subparagraph{Flat $L_\infty$-algebra valued differential forms} now have an immediate definition from this perspective: They are the dg-algebra homomorphism from their CE-algebras into de Rham algebras (aka “Maurer-Cartan elements”):
\begin{equation}
  \label{FlatLInfinityAlgebraValuedDifferentialForms}
  \left.
  \begin{array}{l}
    \mathfrak{a}
    \in
    L_\infty \mathrm{Alg}^{\mathrm{ftp}}
    \\
    X \in \mathrm{SmthMfd}
  \end{array}
  \right\}
  \;\;\;\;\;\;\;
  \vdash
  \;\;\;\;\;\;\;
  \Omega^1_{\mathrm{dR}}(X;\mathfrak{a})_{\closed}
  \;\coloneqq\;
  \mathrm{Hom}_{\mathrm{dgAlg}}\big(
    \mathrm{CE}(\mathfrak{a})
    ,\,
    \Omega^\bullet_{\mathrm{dR}}(X)
  \big)
  \,.
\end{equation}

\noindent
\hspace{-.2cm}
\def\tabcolsep{0pt}
\begin{tabular}{p{6cm}l}
Namely, a graded algebra homomorphism from a CE-algebra sends the algebra generators $\vec b$ to differential forms $\vec B$, and its respect for the differentials imposes on these differential forms exactly the closure/flatness condition.

Examples are shown in \eqref{ExamplesOfFlatForms}.
&
\;\;
\adjustbox{raise=-1cm}{
$
  \begin{tikzcd}[
   row sep=0pt, column sep=12pt
  ]
    \Omega^\bullet_{\mathrm{dR}}(X^d)
    \ar[
      from=rrr,
      "{
        \scalebox{.7}{
          \color{greenii}
          \bf
          \def\arraystretch{.9}
          \begin{tabular}{c}
            dg-algebra
            \\
            homomorphism
          \end{tabular}
        }
      }"{swap}
    ]
    &
    &[+10pt]
    &
    \mathrm{CE}(\mathfrak{g})
    \mathrlap{
      \;
      =
      \mathbb{R}[
        \vec b
        \,
      ]
      \big/
      \big(
        \differential
        \vec b
        \,=\,
        \vec P
        (
          \vec b
        )
      \big)
    }
    \\[+4pt]
    B^{(i)}
    \ar[
      from=rrr,
      |->,
      "{
        \scalebox{.7}{
          \color{greenii}
          \bf
          \def\arraystretch{.9}
          \begin{tabular}{c}
            generator of \scalebox{1.3}{$\mathrm{deg}_i$}
            \\
            sent to
            \scalebox{1.3}{$\mathrm{deg}_i$}-form
          \end{tabular}
        }
      }"{swap}
    ]
    \ar[
      dd,
      |->,
      "{
        \scalebox{.7}{
          \color{greenii}
          \bf
          \def\arraystretch{.9}
          \begin{tabular}{c}
            de Rham
            \\
            differential
          \end{tabular}
        }
      }"{swap}
    ]
    &&&
    b^{(i)}
    \ar[
      ddd,
      |->,
      "{
        \scalebox{.7}{
          \color{greenii}
          \bf
          CE-differential
        }
      }"
    ]
    \\
    \phantom{A}
    \\
    \differential
    \,
    B^{(i)}
    \ar[
      dr,
      equals,
      shorten=-3pt,
      "{
        \scalebox{.7}{
          \color{orangeii}
          \bf
          \def\arraystretch{.9}
          \begin{tabular}{c}
            respect for differentials
            \\
            is the flatness condition
            \\
            hence the Gau{\ss} law
          \end{tabular}
        }
      }"{swap, xshift=3pt}
    ]
    \\
    &
    P^{(i)}\big(
      \vec B
    \big)
    \ar[
      from=rr,
      |->,
      "{
        \scalebox{.7}{
          \color{greenii}
          \bf
          \def\arraystretch{.9}
          \begin{tabular}{c}
            algebra homomorphism
            \\
            preserves
            polynomials
          \end{tabular}
        }
      }"{yshift=-6pt}
    ]
    &&
    P^{(i)}\big(
      \vec b
    \big)
  \end{tikzcd}
$
}
\end{tabular}

\begin{equation}
  \label{ExamplesOfLInfinityAlgebras}
  \def\arraystretch{1.7}
  \begin{array}{|c|c||c|l|l}
    \hline
    \scalebox{.9}{
      $L_\infty$-algebra
    }
    &
    \mathfrak{g}
    &
    \mathfrak{g}^\vee[1]
    &
    \differential_{{}_{[-,\cdots,-]}}
    \\
    \hline
    \hline
     \rowcolor{lightgray}
     \scalebox{.7}{
        \def\arraystretch{.9}
        \begin{tabular}{c}
          line
          \\
          Lie algebra
          \vspacebelow
        \end{tabular}
      }
    &
    \mathfrak{u}(1)
    &
    \mathbb{R}\langle \omega_1 \rangle
    &
    \differential\,
    \omega_1 = 0
    \\
    \hline
    \scalebox{.7}{
      \def\arraystretch{.9}
      \begin{tabular}{c}
        special unitary
        \\
        Lie algebra
        \vspacebelow
      \end{tabular}
    }
    &
    \mathfrak{su}(2)
    &
    \mathbb{R}\big\langle
      \omega_1^{(1)}
      ,\,
      \omega_1^{(2)}
      ,\,
      \omega_1^{(3)}
    \big\rangle
    &
    \differential
    \,
    \omega_1^{(i)}
    =
    \epsilon_{i j k}
    \,
    \omega_1^{(j)}
    \wedge
    \omega_1^{(k)}
    \\
    \hline
    \rowcolor{lightgray}
    \scalebox{.7}{
      \def\arraystretch{.9}
      \begin{tabular}{c}
        line
        \\
        Lie 2-algebra
        \vspacebelow
      \end{tabular}
    }
    &
    b \,
    \mathfrak{u}(1)
    &
    \mathbb{R}\langle
      \omega_2
    \rangle
    &
    \differential\,
    \omega_2 = 0
    \\
    \hline
    \scalebox{.7}{
      \def\arraystretch{.9}
      \begin{tabular}{c}
      string
      \\
      Lie 2-algebra
      \\
      cf. \cite[\S A]{FSS14M57d}
      \end{tabular}
    }
    &
    \mathfrak{string}(3)
    &
    \mathbb{R}\langle
      \omega_1^{(1)}
      ,\,
      \omega_1^{(2)}
      ,\,
      \omega_1^{(3)}
      ,\,
      \omega_2
    \rangle
    &
    \def\arraystretch{1.2}
    \def\arraycolsep{0pt}
    \begin{array}{l}
      {}
      \\[-10pt]
      \differential
      \,\omega_1^{(i)}
      =
      -
      \tfrac{1}{2}
      \epsilon_{i j k}
      \,
      \omega_1^{(j)}
      \wedge
      \omega_1^{(k)}
      \\
      \differential
      \,
      \omega_2
      \;
      =
      \epsilon_{i j k}
      \,
      \omega_1^{(i)}
      \wedge
      \omega_1^{(j)}
      \wedge
      \omega_1^{(k)}
    \end{array}
    \\
    \hline
     \rowcolor{lightgray}
     \scalebox{.7}{
      \def\arraystretch{.9}
      \begin{tabular}{c}
        line
        \\
        Lie 3-algebra
        \vspacebelow
      \end{tabular}
    }
    &
    b^2 \,
    \mathfrak{u}(1)
    &
    \mathbb{R}\langle
      \omega_3
    \rangle
    &
    \differential
    \,
    \omega_3 = 0
    \\
    \hline
    \scalebox{.7}{
      \def\arraystretch{.9}
      \begin{tabular}{c}
        T-duality
        \\
        Lie 3-algebra
        \\
        \cite[\S 7]{FSS18}
      \end{tabular}
    }
    &
    b\mathcal{T}_1
    &
    \mathbb{R}\langle
      \omega^{(i)}_2,
      \omega^{(B)}_2,
      h_3
    \rangle
    &
    \def\arraystretch{1.1}
    \def\arraycolsel{0pt}
    \begin{array}{l}
      {}
      \\[-10pt]
      \differential
      \,
      \omega^{(i)}_2 = 0
      \\
      \differential
      \,
      \omega^{(B)}_2 = 0
      \\
      \differential\,
      h_3 =
      \omega^{(i)}_2
      \wedge
      \omega^{(B)}_2
      \\[-12pt]
      {}
    \end{array}
    \\
    \hline
    \rowcolor{lightgray}
    \scalebox{.7}{
      \def\arraystretch{.9}
      \begin{tabular}{c}
        line
        \\
        Lie 4-algebra
        \vspacebelow
      \end{tabular}
    }
    &
    b^3 \,
    \mathfrak{u}(1)
    &
    \mathbb{R}\langle
      \omega_4
    \rangle
    &
    \differential
    \,
    \omega_4 = 0
    \\
    \hline
    \scalebox{.7}{
      \def\arraystretch{.9}
      \begin{tabular}{c}
        \vspaceabove
        M-theory gauge
        \\
        Lie 7-algebra
        \\
        \cite[\S 4]{Sati10}
        \\
        \cite[\S 2.2]{SatiVoronov22}
        \vspacebelow
      \end{tabular}
    }
    &
    \mathfrak{l}
    \,
    S^4
    &
    \mathbb{R}\langle
      \omega_4
      ,
      \omega_7
    \rangle
    &
    \def\arraystretch{1}
    \def\arraycolsep{0pt}
    \begin{array}{l}
      {}
      \\[-8pt]
      \differential
      \,
      \omega_4 = 0
      \\
      \differential
      \,
      \omega_7 =
      - \omega_4 \wedge \omega_4
      \\[-8pt]
      {}
    \end{array}
    \\
    \hline
    \rowcolor{lightgray}
    \scalebox{.7}{
      \def\arraystretch{.9}
      \begin{tabular}{c}
        cyclified
        \\
        M-theory gauge
        \\
        Lie 7-algebra
        \\
        \cite[Ex. 3.3]{FSS17-Sph}
        \\
        \cite[Ex. 2.47]{BMSS19}
      \end{tabular}
    }
    &
    \mathfrak{l}
    \big(
    \mathcal{L}S^4
    \!\!\sslash\!\! S^1
    \big)
    &
    \mathbb{R}\left\langle
      \def\arraystretch{.9}
      \def\arraycolsep{0pt}
      \begin{array}{l}
      \omega_2
      ,
      \omega_4
      ,
      \omega_6
      \\
      h_3, h_7
      \end{array}
    \right\rangle
    &
    \def\arraystretch{1}
    \begin{array}{l}
      {}
      \\[-8pt]
      \differential
      \,
      h_3 = 0
      \\
      \differential
      \, \omega_2 = 0
      \\
      \differential
      \,
      \omega_4 = h_3 \wedge \omega_2
      \\
      \differential
      \,
      \omega_6 = h_3 \wedge \omega_4
      \\
      \differential
      \,
      h_7
      =
      -\tfrac{1}{2}
      \omega_4 \wedge \omega_4
      +
      \omega_2 \wedge \omega_6
      \mbox{\vspacebelow}
    \end{array}
    \\
    \hline
  \end{array}
\end{equation}
\vspace{.4cm}
\begin{equation}
\label{ExamplesOfFlatForms}
\adjustbox{}{
\footnotesize
\def\tabcolsep{6pt}
\begin{tabular}{ll}
\begin{tabular}{|c|}
\hline
$
  \begin{tikzcd}[
    row sep=-3pt,
    column sep=15pt
  ]
    \mathrm{CE}\big(
      {\color{darkblue}
      b \mathfrak{u}(1)}
    \big)
    \ar[
      rrr
    ]
    &&&[-5pt]
    \Omega^\bullet_{{}_{\mathrm{dR}}}(X)
    \\
    \omega_2
    \ar[
      rrr,
      phantom,
      "{ \longmapsto }"
    ]
    \ar[
      dd,
      phantom,
      "{ \longmapsto }"{sloped},
      "{
        \differential_{{}_{
          b \mathfrak{u}(1)
        }}
      }"{xshift=14pt}
    ]
    &&&
    F
    \ar[
      d,
      phantom,
      "{
        \mapsto
      }"{sloped},
      "{
        \differential_{{}_{\mathrm{dR}}}
      }"{xshift=12pt}
    ]
    \\[10pt]
    &&&
    {\color{darkblue}
    \differential_{{}_{\mathrm{dR}}}
    F}
    \\[5pt]
    0
    \ar[
      rr,
      phantom,
      "{ \longmapsto }"
    ]
    &&
    {\color{darkblue} 0 }
    \ar[
      ur,
      equals,
      darkblue
    ]
  \end{tikzcd}
$
\\
\hline
$
  \begin{tikzcd}[
    row sep=-3pt,
    column sep=5pt
  ]
    \mathrm{CE}\big(
      {
        \color{darkblue}
        \mathfrak{string}(3)
      }
    \big)
    \ar[
      rrr
    ]
    &[+3pt]&[+3pt]&[-35pt]
    \Omega^\bullet_{{}_{\mathrm{dR}}}(X)
    \\
    \omega_1^{(i)}
    \ar[
      phantom,
      "{ \longmapsto }"
    ]
    \ar[
      dd,
      phantom,
      "{ \longmapsto }"{sloped},
      "{
        \differential_{{}_{
          \mathfrak{string}(3)
        }}
      }"{xshift=22pt}
    ]
    &&&
    A^i
    \ar[
      d,
      phantom,
      "{
        \mapsto
      }"{sloped},
      "{
        \differential_{{}_{\mathrm{dR}}}
      }"{xshift=12pt}
    ]
    \\[10pt]
    &&&
    {\color{darkblue}
    \differential_{{}_{\mathrm{dR}}}
    A^i}
    \\[5pt]
    -\tfrac{1}{2}
    \epsilon_{i j k}
    \,
    \omega_1^{(j)}
    \wedge
    \omega_1^{(k)}
    \ar[
      rr,
      phantom,
      "{ \longmapsto }"
    ]
    &&
    {\color{darkblue}
    -\tfrac{1}{2}
    A^i \wedge A^j}
    \ar[
      ur,
      equals,
      darkblue
    ]
    \\[+10pt]
    \omega_2
    \ar[
      rrr,
      phantom,
      "{ \longmapsto }"
    ]
    \ar[
      dd,
      phantom,
      "{ \longmapsto }"{sloped},
      "{
        \differential_{{}_{
          \mathfrak{string}(3)
        }}
      }"{xshift=22pt}
    ]
    &&&
    B
    \ar[
      d,
      phantom,
      "{
        \mapsto
      }"{sloped},
      "{
        \differential_{{}_{\mathrm{dR}}}
      }"{xshift=12pt}
    ]
    \\[13pt]
    &&&
    {\color{darkblue}
    \differential_{{}_{\mathrm{dR}}}
    B}
    \\[5pt]
    \epsilon_{i j k}
    \,
    \omega_1^{(i)}
    \wedge
    \omega_1^{(j)}
    \wedge
    \omega_1^{(k)}
    \ar[
      rr,
      phantom,
      "{ \longmapsto }"
    ]
    &&
    {\color{darkblue}
    \epsilon_{i j k}
    \,
    A^i \wedge A^j \wedge A^k
    }
    \ar[
      ur,
      equals,
      darkblue
    ]
  \end{tikzcd}
$
\\
\hline
$
  \begin{tikzcd}[
    row sep=4pt,
    column sep=5pt
  ]
    \mathrm{CE}\big(
      {\color{darkblue}
      \mathfrak{l}
      \,
      S^4
      }
    \big)
    \ar[
      rrr
    ]
    &[+5pt]&[+5pt]&[-18pt]
    \Omega_{\mathrm{dR}}^\bullet(X)
    \\
    \omega_4
    \ar[
      rrr,
      phantom,
      "{ \longmapsto }"
    ]
    \ar[
      ddd,
      phantom,
      "{ \longmapsto }"{sloped},
      "{
        \differential_{{}_{
          \mathfrak{l}S^4
        }}
      }"{xshift=12pt}
    ]
    &&&
    G_4
    \\
    \\
    &&&
    {\color{darkblue}
    \differential_{{}_{\mathrm{dR}}}
    G_4}
    \\
    0
    \ar[
      rr,
      phantom,
      "{ \longmapsto }"
    ]
    &&
    {\color{darkblue} 0}
    \ar[
      ur,
      equal,
      darkblue
    ]
    \\[10pt]
    \omega_7
    \ar[
      rrr,
      phantom,
      "{ \longmapsto }"
    ]
    \ar[
      ddd,
      phantom,
      "{ \longmapsto }"{sloped},
      "{
        \differential_{{}_{
          \mathfrak{l}
          \,
          S^4
        }}
      }"{xshift=13pt}
    ]
    &&&
    2G_7
    \ar[
      dd,
      phantom,
      "{ \longmapsto }"{sloped},
      "{
        \differential_{{}_{\mathrm{dR}}}
      }"{xshift=12pt}
    ]
    \\
    \\
    &&&
    {\color{darkblue}
    2
    \differential_{{}_{\mathrm{dR}}}
    G_7}
    \\
    -
    \omega_4
    \wedge
    \omega_4
    \ar[
      rr,
      phantom,
      "{ \longmapsto }"
    ]
    &&
    {\color{darkblue}
    - G_4 \wedge G_4}
    \ar[
      ur,
      equals,
      darkblue
    ]
  \end{tikzcd}
$
\\
\hline
\end{tabular}
&
\adjustbox{
  fbox,
  raise=2pt
}{
\begin{tikzcd}[
  row sep=+3.2pt,
  column sep=0pt
]
  \mathrm{CE}\big(
    {\color{darkblue}
    \mathfrak{l}
    (\mathcal{L}S^4 \!\!\sslash\!\! S^1)
    }
  \big)
  \ar[
    rrr
  ]
  &&&[-25pt]
  \Omega^\bullet_{\mathrm{dR}}(X)
  \\
  h_3
  \ar[
    rrr,
    phantom,
    "{ \longmapsto }"
  ]
  \ar[
    ddd,
    phantom,
    "{ \longmapsto }"{sloped},
    "{
      \differential_{{}_{
        \mathfrak{l}(
          \mathcal{L}S^4 \!\sslash\! S^1
        )
      }}
    }"{xshift=23pt}
  ]
  &&&
  H_3
  \ar[
    dd,
    phantom,
    "{ \longmapsto }"{sloped}
  ]
  \\
  \\
  &&&
  {\color{darkblue}
  \differential_{{}_{\mathrm{dR}}}
  H_3}
  \\
  0
  \ar[
    rr,
    phantom,
    "{ \longmapsto }"
  ]
  &&
  {\color{darkblue}
    0
  }
  \ar[
    ur,
    equals,
    darkblue
  ]
  \\[+10pt]
  \omega_2
  \ar[
    rrr,
    phantom,
    "{ \longmapsto }"
  ]
  \ar[
    ddd,
    phantom,
    "{ \longmapsto }"{sloped},
    "{
      \differential_{{}_{
        \mathfrak{l}(
          \mathcal{L}S^4 \!\sslash\! S^1
        )
      }}
    }"{xshift=23pt}
  ]
  &&&
  F_2
  \ar[
    dd,
    phantom,
    "{ \longmapsto }"{sloped}
  ]
  \\
  \\
  &&&
  {\color{darkblue}
  \differential_{{}_{\mathrm{dR}}}
  F_2}
  \\
  0
  \ar[
    rr,
    phantom,
    "{ \longmapsto }"
  ]
  &&
  {\color{darkblue} 0 }
  \ar[
    ur,
    equals,
    darkblue
  ]
  \\[+10pt]
  \omega_4
  \ar[
    rrr,
    phantom,
    "{ \longmapsto }"
  ]
  \ar[
    ddd,
    phantom,
    "{ \longmapsto }"{sloped},
    "{
      \differential_{{}_{
        \mathfrak{l}(
          \mathcal{L}S^4 \!\sslash\! S^1
        )
      }}
    }"{xshift=23pt}
  ]
  &&&
  F_4
  \ar[
    dd,
    phantom,
    "{ \longmapsto }"{sloped}
  ]
  \\
  \\
  &&&
  {\color{darkblue}
  \differential_{{}_{\mathrm{dR}}}
  F_4}
  \\
  h_3 \wedge \omega_2
  \ar[
    rr,
    phantom,
    "{ \longmapsto }"
  ]
  &&
  {\color{darkblue}
    H_3 \wedge F_2
  }
  \ar[
    ur,
    equals,
    darkblue
  ]
  \\[+10pt]
  \omega_6
  \ar[
    rrr,
    phantom,
    "{ \longmapsto }"
  ]
  \ar[
    ddd,
    phantom,
    "{ \longmapsto }"{sloped},
    "{
      \differential_{{}_{
        \mathfrak{l}(
          \mathcal{L}S^4 \!\sslash\! S^1
        )
      }}
    }"{xshift=23pt}
  ]
  &&&
  F_6
  \ar[
    dd,
    phantom,
    "{ \longmapsto }"{sloped}
  ]
  \\
  \\
  &&&
  {\color{darkblue}
  \differential_{{}_{\mathrm{dR}}}
  F_6}
  \\
  h_3 \wedge \omega_4
  \ar[
    rr,
    phantom,
    "{ \longmapsto }"
  ]
  &&
  {\color{darkblue}
    H_3 \wedge F_4
  }
  \ar[
    ur,
    equals,
    darkblue
  ]
  \\[10pt]
  h_7
  \ar[
    rrr,
    phantom,
    "{ \longmapsto }"
  ]
  \ar[
    ddd,
    phantom,
    "{ \longmapsto }"{sloped},
    "{
      \differential_{{}_{
        \mathfrak{l}(
          \mathcal{L}S^4 \!\sslash\! S^1
        )
      }}
    }"{xshift=23pt}
  ]
  &&&
  H_7
  \ar[
    dd,
    phantom,
    "{ \longmapsto }"{sloped}
  ]
  \\
  \\
  &&&
  {\color{darkblue}
  \differential_{{}_{\mathrm{dR}}}
  H_7}
  \\
  \def\arraystretch{.9}
  \begin{array}{c}
    -\tfrac{1}{2}
    \omega_4 \wedge \omega_4
    \\
    + \omega_2 \wedge \omega_6
  \end{array}
  \ar[
    rr,
    phantom,
    "{ \longmapsto }"
  ]
  &&
  {\color{darkblue}
  \def\arraystretch{.9}
  \begin{array}{c}
    -\tfrac{1}{2}
    F_4 \wedge F_4
    \\
    + F_2 \wedge F_6
  \end{array}
  }
  \ar[
    ur,
    equals,
    darkblue
  ]
\end{tikzcd}
}
\end{tabular}
}
\end{equation}

\subparagraph{Flux densities satisfying Gau{\ss} law are flat $L_\infty$-valued differential forms.}
Remarkably, it follows that polynomials $\vec P$ defining Bianchi identities \eqref{DualitySymmetricHigherMaxwellEquations} and Gauss laws \eqref{GaussLawFromBianchiIdentities} are equivalently structure constants of $L_\infty$-algebras $\mathfrak{a}$, such that the Bianchi/Gau{\ss} law is the closure/flatness condition on $\mathfrak{a}$-valued forms:
\begin{equation}
\label{BianchiIdentitiesAsClosedLInfinityValuedForms}
\small
\adjustbox{
  margin=5pt,
  fbox
}{
$
  \def\arraystretch{.2}
  \begin{array}{r}
  \overset{
    \mathclap{
      \raisebox{2pt}{
        \scalebox{.7}{
          \color{purple}
          \bf
          Sheaf of
          closed $L_\infty$-algebra-valued
          differential forms
        }
      }
    }
  }{
  \Omega^1_{\mathrm{dR}}
  \big(
    \underset{
      \mathclap{
        \rotatebox{-30}{
          \rlap{
            \hspace{-15pt}
            \scalebox{.6}{
              \color{gray}
              \def\arraystretch{.9}
              \begin{tabular}{c}
                insert spacetime
                \\
                manifold here
              \end{tabular}
            }
          }
        }
      }
    }{
      -
    }
    ;
    \,\mathfrak{a}
  \big)_{\!\closed}
  \;\;
  =
  \;\;
  \mathrm{Hom}_{{}_{\mathrm{dgAlg}}}
  \big(
    \mathrm{CE}(\mathfrak{a})
    ,\,
    \Omega^\bullet_{\mathrm{dR}}(-)
  \big)
  }
  \;\;
  =
  \;\;
  \Big\{
    \overset{
      \mathclap{
        \raisebox{4pt}{
          \color{darkblue}
          \bf
          \scalebox{.6}{
            systems of flux densities
          }
        }
      }
    }{
    \vec B
    \defneq
    \big(
    B^{\scalebox{.55}{$(i)$}}
    \,\in\,
    \Omega^{\mathrm{deg}_i}_{\mathrm{dR}}(-)
    \big)
    }
    \hspace{.1cm}
    \big\vert
  \hspace{.1cm}
  \overset{
    \mathclap{
      \raisebox{5pt}{
        \scalebox{.6}{
          \color{orangeii}
          \bf
          satisfying this
          Gau{\ss} law
        }
      }
    }
  }{
  \Differential
  \,
  \vec B
  \;
  =
  \;
  \vec P
  \big(
    \vec B
  \big)
  }
  \Big\}
  \\[-2pt]
  \rotatebox{90}{$\xleftrightarrow{\phantom{--}}$}
  \phantom{-------------}
  \\
  \overset{
    \mathclap{
      \rotatebox{30}{
        \rlap{
          \hspace{-24pt}
          \scalebox{.6}{
            \color{purple}
            \bf
            \def\arraystretch{.9}
            \begin{tabular}{c}
              Chevalley-Eilenberg
              \\
              algebra of
            \end{tabular}
          }
        }
      }
    }
  }{
    \mathrm{CE}
  }
  (
    \overset{
      \mathclap{
        \rotatebox{30}{
          \rlap{
            \hspace{-8pt}
            \scalebox{.6}{
              \color{purple}
              \bf
              $L_\infty$-algebra
            }
          }
        }
      }
    }{
      \,\mathfrak{a}\,
    }
  )
  \;\;\;
    =
  \;\;\;
  \overset{
    \mathclap{
      \hspace{15pt}
      \rotatebox{30}{
        \scalebox{.6}{
          \rlap{
            \hspace{-38pt}
            \color{darkblue}
            \bf
            \def\arraystretch{.9}
            \color{darkblue}
            \bf
            \begin{tabular}{c}
              free differential graded-
              \\
              commutative algebra
            \end{tabular}
          }
        }
      }
    }
  }{
  \mathbb{R}
  \Big[
  }
    \big\{
      \overset{
        \mathclap{
          \hspace{11pt}
          \rotatebox{30}{
            \rlap{
              \hspace{-23pt}
              \scalebox{.6}{
                \color{darkblue}
                \bf
                \def\arraystretch{.8}
                \begin{tabular}{r}
                  \\
                  on these graded
                  \\
                  generators
                \end{tabular}
              }
            }
          }
        }
      }{
        b^{(i)}_{\mathrm{deg}_i}
      }
    \big\}_{i \in I}
  \Big]
  \Big/
  \Big(
  \overset{
    \mathclap{
      \raisebox{3pt}{
        \scalebox{.6}{
          \color{orangeii}
          \bf
          \def\arraystretch{.9}
          \begin{tabular}{c}
            satisfying these
            differential relations
          \end{tabular}
        }
      }
    }
  }{
  \Differential
  \,
  \vec b
  \,=\,
  \vec P\big(\vec b\big)
  }
  \! \Big)
  \\
  \\
  \rotatebox{90}{$\xleftrightarrow{\phantom{--}}$}
  \phantom{-------------}
  \\[-10pt]
  \hspace{-12pt}
  \overset{
    \mathclap{
      \hspace{8pt}
      \rotatebox{30}{
        \scalebox{.6}{
          \rlap{
            \hspace{-18pt}
            \color{darkblue}
            \bf
            \def\arraystretch{.9}
            \color{purple}
            \bf
            \begin{tabular}{c}
              $L_\infty$-algebra
            \end{tabular}
          }
        }
      }
    }
  }{
  \mathfrak{a}
  }
  \;\;\;
  =
  \;\;\;
  \overset{
    \mathclap{
      \hspace{5pt}
      \rotatebox{30}{
        \scalebox{.6}{
          \rlap{
            \hspace{-24pt}
            \color{darkblue}
            \bf
            \def\arraystretch{.9}
            \color{darkblue}
            \bf
            \begin{tabular}{c}
              graded
              \\
              vector space
              spanned
            \end{tabular}
          }
        }
      }
    }
  }{
  \mathbb{R}
  }
  \Big\langle
    \big\{
      \overset{
        \mathclap{
          \hspace{11pt}
          \rotatebox{30}{
            \rlap{
              \hspace{-23pt}
              \scalebox{.6}{
                \color{darkblue}
                \bf
                \def\arraystretch{.8}
                \begin{tabular}{r}
                  \\
                  by these graded
                  \\
                  generators
                \end{tabular}
              }
            }
          }
        }
      }{
      v^{\scalebox{.55}{$(i)$}}_{
        \mathrm{deg}_i-1
      }
      }
    \big\}_{i \in I}
  \Big\rangle
  \hspace{.4cm}
  \overset{
    \mathclap{
      \raisebox{3pt}{
        \scalebox{.6}{
          \color{orangeii}
          \bf
          equipped with these higher Lie brackets
        }
      }
    }
  }{
  \big[
    v^{\scalebox{.55}{$(i_1)$}}, \cdots, v^{\scalebox{.55}{$(i_n)$}}
  \big]
  \;\;
  =
  \;\;
  \sum_{i \in I}
  P^{\scalebox{.55}{$(i)$}}
    _{\!\!\scalebox{.55}{$i_1 \cdots i_n$}}
  \;
  v^{\scalebox{.55}{$(i)$}}
  }
  \end{array}
  \hspace{1cm}
$
}
\end{equation}

With Prop. \ref{SolutionSpaceViaGaussLaw}, this means:

\begin{proposition}[Flux solutions as flat $L_\infty$-valued forms]
\label{FluxSolutionsAsLInfinityValuedForms}
Given a higher gauge theory of Maxwell-type (Def. \ref{HigherMaxwellEquations}) with Bianchi identities given by graded-symmetric polynomials $\vec P$ \eqref{DualitySymmetricHigherMaxwellEquations} its space of flux densities solving the higher Maxwell equations is identified with the space of closed differential forms with coefficients in the $L_\infty$-algebra $\mathfrak{a}$ on $I$ $\vec {\mathrm{deg}}$-graded generators with structure constants $\vec P$:
  \begin{equation}
    \label{SolutionSpaceViaLInfinity}
    \def\arraystretch{1}
    \begin{array}{rcl}
    \mathllap{
      \scalebox{.7}{
        \color{darkblue}
        \bf
        \begin{tabular}{c}
          Space of flux densities
          \\
          on spacetime, solving
          \\
          the equations of motion
        \end{tabular}
      }
    }
    \mathrm{SolSpace}(X^D)
    &\defneq&
    \left\{
      \overset{
        \mathclap{
          \raisebox{3pt}{
            \scalebox{.7}{
              \color{gray}
              \rm
              electromagnetic flux densities on spacetime
            }
          }
        }
      }{
      \vec F \,\defneq\,
      \Big(
        F^{(i)}
        \,\in\,
        \Omega^{\mathrm{deg}_i}_{\mathrm{dR}}\big(
          X^D
        \big)
       \Big)_{i \in I}
       }
       \,\middle\vert\,
  \def\arraystretch{1.6}
  \begin{array}{l}
    \overset{
      \mathclap{
        \raisebox{3pt}{
          \scalebox{.7}{
            \color{gray}
            \rm
            Bianchi identities
          }
        }
      }
    }{
    \differential
    \, \vec F \,=\,
    \vec P\big( \vec F \big)
    }
    \\
    \underset{
      \mathclap{
        \raisebox{-1pt}{
          \scalebox{.7}{
            \color{gray}
            \rm
            self-duality
          }
        }
      }
    }{
    \star \, \vec F
    \,=\,
    \vec \mu\big( \vec F \big)
    }
  \end{array}
    \right\}
  \mathrlap{
    \scalebox{.8}{
      \color{darkblue}
      \bf
      covariant form
    }
  }
    \\[30pt]
   &
     \underset{
       \iota^\ast
     }{
       \simeq
     }
    &
    \left\{
    \overset{
      \mathclap{
        \raisebox{3pt}{
          \scalebox{.7}{
            \color{gray}
            \rm
            magnetic flux densities on Cauchy surface
          }
        }
      }
    }{
      \vec B \,\defneq\,
      \Big(
        B^{(i)}
        \,\in\,
        \Omega^{\mathrm{deg}_i}_{\mathrm{dR}}\big(
          \,X^d\,
        \big)
       \Big)_{i \in I}
       }
       \,\middle\vert\,
       \begin{array}{l}
    \overset{
      \mathclap{
        \raisebox{3pt}{
          \scalebox{.7}{
            \color{gray}
            \rm
            Gau{\ss} law
          }
        }
      }
    }{
         \differential
         \, \vec B \,=\,
         \vec P\big( \vec B \big)
    }
    \\
    {}
       \end{array}
    \right\}
  \mathrlap{
    \scalebox{.8}{
      \color{darkblue}
      \bf
      canonical form
    }
  }
    \\[30pt]
   &
     \simeq
    &
    \Omega^1_{\mathrm{dR}}\big(
      X^d
      ;\,
      \mathfrak{a}
    \big)_\closed
    \;\;
    \mathrlap{
      \scalebox{.7}{
        \bf
        \color{purple}
        \begin{tabular}{l}
        space of
        closed (flat)
        \\
        $\mathfrak{a}$-valued
        differential forms
        \end{tabular}
      }
    }
    \end{array}
  \end{equation}
\end{proposition}

\begin{example}
The characteristic $L_\infty$-algebra of ordinary vacuum electromagnetism is the direct sum $b\mathfrak{u}(1) \oplus b \mathfrak{u}(1)$ of two copies of the line Lie 2-algebra, which by the previous example and Prop. \ref{FluxSolutionsAsLInfinityValuedForms} corresponds to:

$$
  \mathrm{SolSpace}_{\mathrm{EM}}(X^3)
  \;\simeq\;
  \Omega^1_{\mathrm{dR}}\big(
    X^3
    ;\,
    b\mathfrak{u}(1)
    \times
    b\mathfrak{u}(1)
  \big)_{\closed}
  \;\simeq\;
  \Omega^2_{\mathrm{dR}}(X^3)
  \times
  \Omega^2_{\mathrm{dR}}(X^3)
  \,.
$$

\vspace{1mm}
\noindent An element here corresponds to
a pair $(E,\, B))$, where
\begin{itemize}
  \item the magnetic flux density is the curvature $B = \mathrm{curv}\big(\widehat A\big)$ of the gauge potential which plays the
  role of the ``canonical coordinate'' on the field space;
  \item the electric flux density $E$ serves as the corresponding ``canonical momentum''.
\end{itemize}
\end{example}

\subparagraph{Total flux in non-abelian de Rham cohomology.} While, with Prop. \ref{FluxSolutionsAsLInfinityValuedForms}, the Gau{\ss} law of the given higher gauge theory of Maxwell-type constrains the flux densities on any Cauchy surface $X^d \hookrightarrow X^D$ to constitute a flat $L_\infty$-algebra valued differential form, the actual value of these differential forms depends on the Cauchy surface, which is an arbitrary choice.
We should, therefore, regard as the {\it total flux} that aspect of the flux densities which is invariant under choice of Cauchy surfaces.

But since the Gau{\ss} law is (by Prop. \ref{SolutionSpaceViaGaussLaw}) nothing but the restriction to the Cauchy surface of the Bianchi identities (on the duality-symmetric flux densities of Def. \ref{HigherMaxwellEquations}), the argument of Prop. \ref{FluxSolutionsAsLInfinityValuedForms} shows that this invariant aspect is the equivalence classes of flux densities under {\it concordance}:
\begin{equation}
  \label{ConcordanceBetweenClosedLInfinityValuedForms}
  \overset{
    \mathclap{
      \raisebox{5pt}{
        \scalebox{.7}{
          \color{darkblue}
          \bf
          deformation of flux densities
        }
      }
    }
  }{
  \vec B_{\color{purple}0}
  \;\sim\;
  \vec B_{\color{purple}1}
  }
  \hspace{.5cm}
  :\Leftrightarrow
  \hspace{.5cm}
  \adjustbox{raise=1pt}{
    $\exists$
  }
  \;
  \vec F \,\in\,
  \Omega^1_{\mathrm{dR}}\big(
    X^d \times [0,1]
  \big)_\closed
  \;\;
  \mbox{with}
  \;\;
  \left\{
  \begin{array}{l}
    \iota^\ast_{\color{purple}0} \vec F
    \,=\,
    \vec B_{\color{purple}0}
    \\
    \iota^\ast_{\color{purple}1} \vec F
    \,=\,
    \vec B_{\color{purple}1}
  \end{array}
  \right.
\end{equation}

\begin{center}
 \includegraphics[width=10cm]{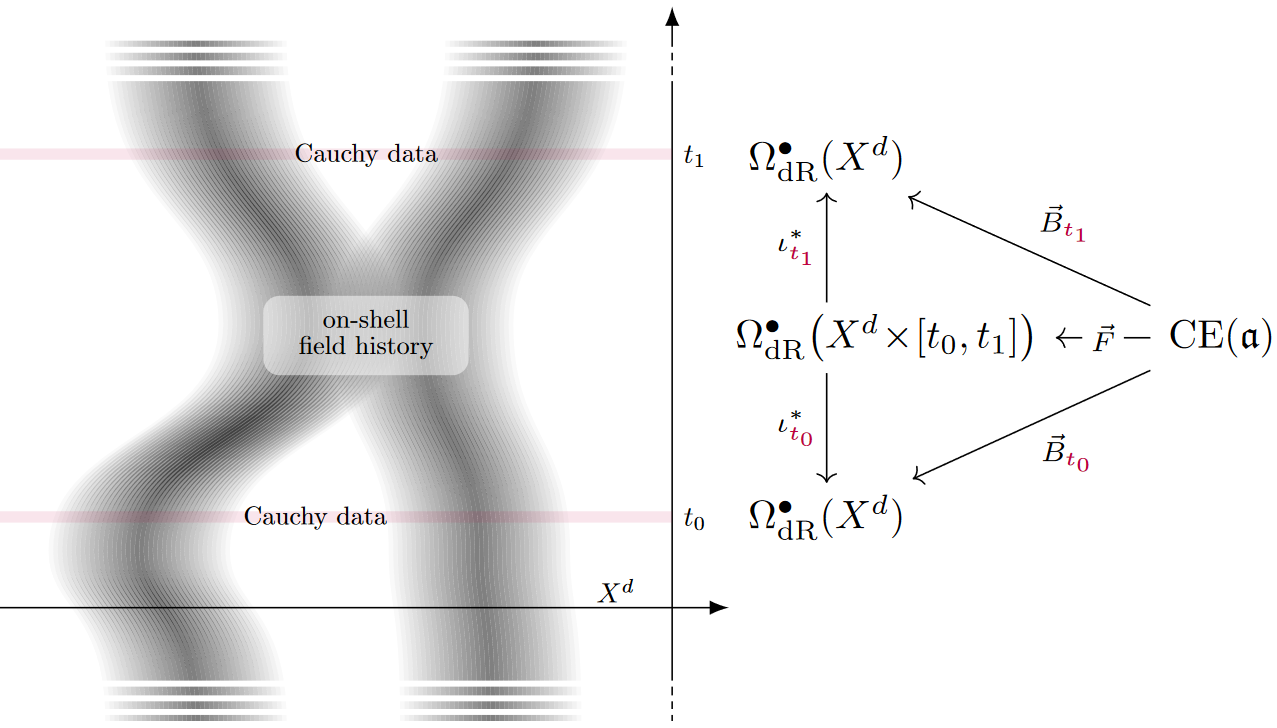}
\end{center}

\begin{definition}[{\bf Non-abelian de Rham cohomology} {\cite[Def. 6.3]{FSS23Char}}]
\label{NonabelianDeRhamCohomology}
Given an $L_\infty$-algebra $\mathfrak{a}$ and a smooth manifold $X^d$, we say that a pair of flat {\it closed $\mathfrak{a}$-valued differential forms} $\vec B_0, \vec B_1 \,\in\, \Omega^1_{\mathrm{dR}}\big(X^d;\mathfrak{a}\big)_{\closed}$ \eqref{FlatLInfinityAlgebraValuedDifferentialForms} are {\it cohomologous} iff they are concordant: iff there exists a closed $\mathfrak{a}$-valued differential form $\vec F$ on the cylinder over $X^d$ whose pullback to the $k$th boundary component equals $\vec B_k$:
\begin{equation}
  \label{ConcordanceBetweenFlatForms}
  \vec B_0
  \;\sim\;
  \vec B_1
  \;\;\;\;\;\;
  \Leftrightarrow
  \;\;\;\;\;\;
  \exists
  \;\;
  \vec F
  \,\in\,
  \Omega^1_{\mathrm{dR}}\big(
    X^d \times [0,1]
    ;\,
    \mathfrak{a}
  \big)_{\closed}
  \;\;\;
  \mbox{with}
  \;\;\;
  \left\{\!\!
  \begin{array}{l}
    \vec B_1 \,=\, \iota_1^\ast \vec F \,,
    \\
    \vec B_0 \,=\, \iota_0^\ast \vec F \,.
  \end{array}
  \right.
\end{equation}
The quotient set by this equivalence relation is $\mathfrak{a}$-valued {\it nonabelian de Rham cohomology} of $X^d$:
\begin{equation}
  \label{NonabelianDeRhamCohomologyExpression}
  H^1_{\mathrm{dR}}\big(
    X^d;\mathfrak{a}
  \big)
  \;\coloneqq\;
  \Omega^1_{\mathrm{dR}}\big(
    X^d
    ;\,
    \mathfrak{a}
  \big)_{\closed}
  \big/
  \!
  \sim
  \,.
\end{equation}
\end{definition}

\begin{remark}[{\bf Conservation of total flux}] $\,$
$\,$

\vspace{-1.5mm}
\noindent
\hspace{-.2cm}
\def\tabcolsep{0pt}
\begin{tabular}{p{8cm}l}
Regarding the image of flux densities in non-abelian de Rham cohomology as expressing their {\it total flux}
it follows immediately that:
\begin{center}
  \it
  Total flux is conserved under time evolution.
\end{center}
&
\hspace{1cm}
\adjustbox{raise=-.8cm}{
$
  \begin{tikzcd}[column sep=small, row sep=0pt]
    \overset{
      \mathclap{
        \raisebox{3pt}{
          \scalebox{.7}{
            \color{darkblue}
            \bf
            \def\arraystretch{.9}
            \begin{tabular}{c}
              closed $\mathfrak{a}$-valued
              \\
              differential forms
            \end{tabular}
          }
        }
      }
    }{
    \Omega^1_{\mathrm{dR}}\big(
      X^d
      ;\,
      \mathfrak{a}
    \big)_{\closed}
    }
    \ar[
      rr,
      ->>
    ]
    && \quad
    \overset{
      \mathclap{
        \raisebox{3pt}{
          \scalebox{.7}{
            \color{darkblue}
            \bf
            \def\arraystretch{.9}
            \begin{tabular}{c}
              nonabelian $\mathfrak{a}$-valued
              \\
              de Rham cohomology
            \end{tabular}
          }
        }
      }
    }{
    H^1_{\mathrm{dR}}\big(
      X^d
      ;\,
      \mathfrak{a}
    \big)
    }
    \\
    \underset{
      \mathclap{
        \raisebox{-3pt}{
          \scalebox{.7}{
            \color{darkblue}
            \bf
            \def\arraystretch{.9}
            \begin{tabular}{c}
              flux densities
            \end{tabular}
          }
        }
      }
    }{
      \vec B
    }
    &\longmapsto&
    \underset{
      \mathclap{
        \raisebox{-3pt}{
          \scalebox{.7}{
            \color{darkblue}
            \bf
            \def\arraystretch{.9}
            \begin{tabular}{c}
              total flux
            \end{tabular}
          }
        }
      }
    }{
    \big[
      \vec B
      \,
    \big]
    }
  \end{tikzcd}
$
}
\end{tabular}
\end{remark}

\vspace{-5mm}
\begin{remark}[{\bf Nonabelian de Rham cohomology as dg-homotopy classes}] $\,$

\noindent
\hspace{-1.5mm}
\def\tabcolsep{0pt}
\begin{tabular}{p{3.5cm}l}
For comparison to the flux quantization rules discussed below in \cref{FluxQuantizationLawsAsNonabelianCohomology} it is useful
to understand this equivalently \cite[Thm. 6.5]{FSS23Char} as the set of dg-homotopy classes of the corresponding dgc-homomorphisms:
&
\hspace{.9cm}
\adjustbox{raise=-2cm}{
\begin{minipage}{13cm}
\begin{equation}
  \label{NonabelianDeRhamCohomologyAsHomotopyClasses}
  \overset{
    \mathclap{
      \raisebox{4pt}{
        \scalebox{.7}{
          \color{darkblue}
          \bf
          \def\arraystretch{1}
          \begin{tabular}{c}
          deformation class
          \\
          of flux densities
          \end{tabular}
        }
      }
    }
  }{
    \big[\vec B\,\big]
  }
  \;\;\;\;\;
  \in
  \;\;\;\;\;
  \overset{
    \mathclap{
      \raisebox{6pt}{
        \scalebox{.7}{
          \color{darkblue}
          \bf
          \def\arraystretch{.9}
          \begin{tabular}{c}
            $\mathfrak{a}$-valued
            \\
            de Rham cohomology
          \end{tabular}
        }
      }
    }
  }{
  H^1_{\mathrm{dR}}
  \big(
    X^d
    ;\,
    \mathfrak{a}
  \big)
  }
  \;\;\;\;
  :=
  \;\;\;\;
  \pi_0
  \left\{
  \begin{tikzcd}[
    column sep=30pt
  ]
    \Omega^\bullet_{\mathrm{dR}}(X^d)
    \ar[
      rr,
      <-,
      gray,
      bend left=60,
      "{
        \color{black}
        \vec B
      }"{description},
      "{\ }"{swap, name=s},
      "{
        \scalebox{.7}{
          {
          \color{greenii}
          \bf
          cocycle
          }
          \rlap{\!(dga-hom)}
        }
      }"{yshift=+7pt}
    ]
    \ar[
      rr,
      <-,
      gray,
      bend right=60,
      "{
        \color{black}
        \vec B'
      }"{description},
      "{\ }"{name=t},
      "{
        \scalebox{.7}{
          {
          \llap{
            another
          }
          \hspace{-1pt}
          \color{greenii}
          \bf
          cocycle
          }
        }
      }"{swap, yshift=-6pt}
    ]
    &&
    \mathrm{CE}(\mathfrak{a})
    \ar[
      from=s,
      to=t,
      shorten=3pt,
      Rightarrow,
      "{
        \scalebox{.7}{
          \hspace{-9pt}
          \def\arraystretch{.9}
          \begin{tabular}{c}
            \bf
            \color{orangeii}
            coboundary
            \\
            (concordance)
          \end{tabular}
          \hspace{-9pt}
        }
      }"{description}
    ]
  \end{tikzcd}
  \right\}
  \,.
\end{equation}
\end{minipage}
}
\end{tabular}
\end{remark}

\begin{example}[{\cite[Prop. 6.4]{FSS23Char}}]

In the case of ordinary electromagnetism and abelian higher gauge fields, hence for $\mathfrak{a} = b^n \mathfrak{u}(1)$ the line Lie $n+1$-algebra, Def. \ref{NonabelianDeRhamCohomology} reduces to the ordinary notions:

\begin{itemize}[leftmargin=.5cm]
\item $\Omega^1_{\mathrm{dR}}\big(X^d; b^n\mathbb{R}\big)_{\closed} \,\simeq\, \Omega^{n+1}(X^d)_{\closed}$
are ordinary closed differential forms;

\item concordance between these coincides with the coboundary relation in the ordinary de Rham complex;

\item $H^1_{\mathrm{dR}}\big(X^d;\, b^n \mathbb{R}\big) \,\simeq\, H^{n+1}_{\mathrm{dR}}\big(X^d\big)$ is ordinary de Rham cohomology;
\end{itemize}
and since the latter also gives the periods of closed differential forms, this recovers indeed the usual notion of total (integrated) flux. Concretely, with $F_2$
the flux density around a magnetic monopole of charge $q$ \eqref{DiracMonopole}, the total flux is as shown:
\vspace{1mm}
$$
  F_2
  \;\coloneqq\;
  q
  \mathrm{dvol}_{S^2}
  \;\in\;
  \Omega^2_{\mathrm{dR}}(S^2)
  \;\xrightarrow{\quad p^\ast_{S^2} \quad}\;
  \Omega^2_{\mathrm{dR}}\big(\mathbb{R}^3 \setminus \{0\}\big)
  \;\simeq\;
  \Omega^2_{\mathrm{dR}}\big(\mathbb{R}^3 \setminus \{0\}
    ;\,
    b\mathbb{R}
  \big)
$$

\noindent
\hspace{-.2cm}
\def\tabcolsep{0pt}
\begin{tabular}{p{7cm}l}

&
\hspace{5pt}
\adjustbox{raise=-.4cm}{
$
  \def\arraystretch{1.3}
  \begin{array}{ccccc}
    H^2_{\mathrm{dR}}\big(
     \mathbb{R}^3 \setminus \{0\}
    \big)
    &\simeq&
    H^2_{\mathrm{dR}}\big(
     S^2
    \big)
    &\simeq&
    \mathbb{R}
    \\
    F_2
    &\mapsto&
    \big[ F_2\,  \big]
    &\mapsto&
    \int_{S^2} F_2
    \mathrlap{
      \,=\, q
      \,.
    }
  \end{array}
$
}
\end{tabular}
\end{example}

With on-shell flux densities thus understood as cocycles in nonabelian de Rham cohomology, we find their flux quantization laws among the corresponding torsion-ful nonabelian cohomology theories:

\subsection{Flux quantization laws as Nonabelian cohomology}
\label{FluxQuantizationLawsAsNonabelianCohomology}

We explain how the $\mathfrak{a}$-valued nonabelian de Rham cohomology of the previous subsection receives character maps from generalized nonabelian cohomology theories whose classifying spaces $\mathcal{A}$ have compatible rational Whitehead $L_\infty$-algebra $\mathfrak{l}\mathcal{A} \,\simeq\, \mathfrak{a}$ -- whence $\mathcal{A}$ encodes a flux quantization law for Bianchi identities characterized by $\mathfrak{a}$, and lifting through the $\mathcal{A}$-character map corresponds to choices of charge quanta which source given total flux.

\paragraph{Classifying spaces for generalized cohomology.} It is a classical fact of algebraic topology --- which may have remained somewhat underappreciated in mathematical physics --- that reasonable generalized cohomology theories have classifying spaces $\mathcal{A}$, in that the sets of cohomology classes assigned to a given domain space (which we take to be a smooth manifold $X^d$) are in natural bijection with the homotopy classes $\pi_0 \mathrm{Map}\big(X,\, \mathcal{A}\big)$ of continuous maps from $X$ into $\mathcal{A}$. (Throughout, it is only the homotopy type of $\mathcal{A}$ that matters.)

\vspace{0cm}
\noindent
\hspace{-.1cm}
\begin{tabular}{p{9cm}l}
The archetypical examples are Eilenberg-MacLane spaces like $K(\mathbb{Z},n)$ which classify ordinary cohomology such as integral cohomology, in any degree $n$. As $n$ ranges, these EM-spaces happen to be loop spaces of each other, up to weak homotopy equivalence: $K(\mathbb{Z},n) \,\simeq\, \Omega K(\mathbb{Z}, n+1)$.

Generalizing from this classical example, one considers Whitehead-generalized cohomology theories which are classified by any sequences of pointed topological spaces $\{E_n\}_{n \in \mathbb{N}}$ equipped with weak homotopy equivalences  $E_n \,\simeq\, \Omega E_{n+1}$, called a {\it spectrum of spaces}.

This implies that each $E_n$ is an infinite-loop space, which makes them be “abelian $\infty$-groups”, reflecting the fact that the homotopy classes of maps into these spaces indeed have the structure of abelian groups.

Perhaps the most familiar example of such {\it abelian} generalized cohomology is topological K-theory, whose classifying space $\mathrm{KU}_0$ may be identified with the space of Fredholm operators on an infinite-dimensional separable complex Hilbert space.

\vspace{1mm}
While Whitehead-generalized cohomology theory has received so much attention that it is now widely understood as the default or even the exclusive meaning of “generalized cohomology”, historically long preceding it is the {\it nonabelian cohomology} of Chern-Weil theory, classified by the original classifying spaces $B G$ of compact Lie groups $G$.

Unless $G$ happens to be abelian itself, this nonabelian cohomology does not assign abelian cohomology groups, nor even any groups at all, but just pointed cohomology sets. Nevertheless, as the historical name “nonabelian cohomology” clearly indicates, these systems of cohomology sets may usefully be regarded as constituting a kind of cohomology theory, too.

& \quad \;
\adjustbox{raise=-13.2cm}{
\includegraphics[width=8cm]{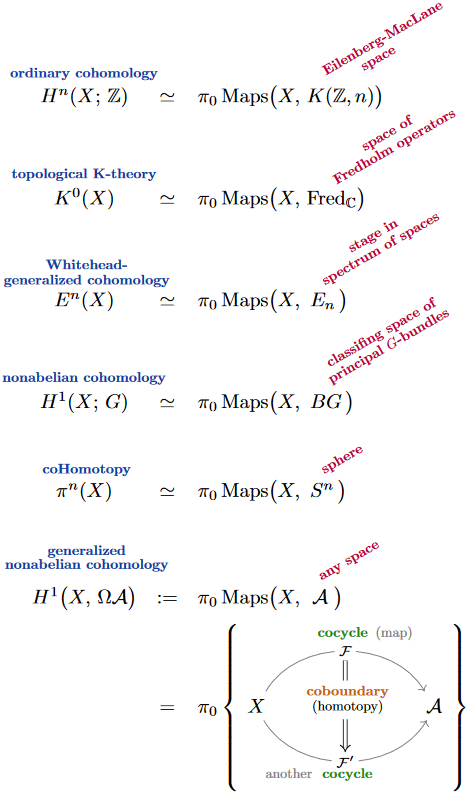}
}
\end{tabular}

In this vein one may observe \cite[\S 2]{FSS23Char} that (the homotopy type of) every connected space $\mathcal{A}$ is equivalently the classifying space of an infinity-group $\Omega \mathcal{A}$, namely of its own loop space regarded as an $A_\infty$-space under concatenation of loops), so that homotopy classes of maps into any connected space are examples of an evident generalization of Chern-Weil-style nonabelian cohomology.

\vspace{1mm}
A fundamental and historical example of such “truly-generalized” nonabelian cohomology is CoHomotopy, whose classifying spaces are the (homotopy types) of spheres.
Notice that “generalized nonabelian cohomology” is really “not necessarily abelian”. It subsumes all the other cases: For $E_\bullet$ a spectrum we have: $E^n(X) \,\simeq\, H^1(X;\Omega E_n)$

\paragraph{Character maps on generalized cohomology.} Moreover, it is classical that, over smooth manifolds, reasonable cohomology theories have their non-torsion content reflected in de Rham cohomology via {\it character maps}:

\begin{equation}
\label{ExamplesOfCharacterMaps}
\hspace{-.5cm}
\adjustbox{scale=.9}{$
  \begin{tikzcd}[
    column sep=20pt,
    row sep=6pt,
    /tikz/column 2/.append style={anchor=base west},
    /tikz/column 4/.append style={anchor=base east},
  ]
    \scalebox{.7}{
      \color{darkblue}
      \bf
      \def\arraystretch{.9}
      \begin{tabular}{c}
        Ordinary
        \\
        integral cohomology
      \end{tabular}
    }
    &[-30pt]
    H^n(
      X
      ;\,
      \mathbb{Z}
    )
    \ar[
      rr,
      "{
        \scalebox{.7}{
          \color{greenii}
          \bf
          de Rham map
        }
      }"
    ]
    &&
    H^n_{\mathrm{dR}}(
      X
    )
    \,\simeq\,
    \mathrm{Hom}_{\mathrm{dgAlg}_{\mathbb{R}}}\big(
      \mathbb{R}[\omega_n]
      ,\,
      H^\bullet_{\mathrm{dR}}(X)
    \big)
    &[-31pt]
    \scalebox{.7}{
      \color{darkblue}
      \bf
      \def\arraystretch{.9}
      \begin{tabular}{c}
        differential forms
        \\
        in degree $n$
      \end{tabular}
    }
    \\
    \scalebox{.7}{
      \color{darkblue}
      \bf
      \def\arraystretch{.9}
      \begin{tabular}{c}
        Traditional
        \\
        nonabelian cohomology
      \end{tabular}
    }
    &
    H^1(
      X
      ;\,
      G
    )
    \ar[
      rr,
      "{
        \scalebox{.7}{
          \color{greenii}
          \bf
        \hspace{-2mm}  Chern-Weil homomorphism
        }
      }"
    ]
    &&
    \mathrm{Hom}_{\mathrm{dgAlg}_{\mathbb{R}}}\big(
      \mathrm{inv}^\bullet(\mathfrak{g})
      ,\,
      H^\bullet_{\mathrm{dR}}(
        X)
    \big)
    &
    \scalebox{.7}{
      \color{darkblue}
      \bf
      \def\arraystretch{.9}
      \begin{tabular}{c}
        differential forms for
        \\
        $\mathfrak{g}$-invariant polynomials
      \end{tabular}
    }
    \\
    \scalebox{.7}{
      \color{darkblue}
      \bf
      \def\arraystretch{.9}
      \begin{tabular}{c}
        Topological
        \\
        K-theory
      \end{tabular}
    }
    &
    K^0(X)
    \ar[
      rr,
      "{
        \scalebox{.7}{
          \color{greenii}
          \bf
          Chern character
        }
      }"
    ]
    &&
    \mathrm{Hom}_{\mathrm{dgAlg}_{\mathbb{R}}}\big(
      \mathbb{R}[
        \omega_0,
        \omega_2,
        \omega_4,
        \cdots
      ]
      ,\,
      H^\bullet_{\mathrm{dR}}(
        X
        )
    \big)
    &
    \scalebox{.7}{
      \color{darkblue}
      \bf
      \def\arraystretch{.9}
      \begin{tabular}{c}
        differential forms
        \\
        in every even degree
      \end{tabular}
    }
    \\
    \scalebox{.7}{
      \color{darkblue}
      \bf
      \def\arraystretch{.9}
      \begin{tabular}{c}
        abelian Whitehead-
        \\
        generalized cohomology
      \end{tabular}
    }
    &
    E^n(X)
    \ar[
     rr,
     "{
       \scalebox{.7}{
         \color{greenii}
         \bf
         Chern-Dold character
       }
     }"
    ]
    &&
    \mathrm{Hom}_{\mathrm{dgAlg}_{\mathbb{R}}}
    \big(
      \wedge^\bullet
      \!
      (
        \pi_\bullet(E)
        \otimes_{{}_\mathbb{Z}}
        \mathbb{R}
      )^\vee
      ,\,
      H^{\bullet+n}_{\mathrm{dR}}(X)
    \big)
    &
    \scalebox{.7}{
      \color{darkblue}
      \bf
      \def\arraystretch{.9}
      \begin{tabular}{c}
        differential forms for
        \\
        rational homotopy groups
        \\
        of the classifying space
      \end{tabular}
    }
    \\[+5pt]
    \scalebox{.7}{
      \color{darkblue}
      \bf
      \def\arraystretch{.9}
      \begin{tabular}{c}
        Generalized
        \\
        non-abelian cohomology
      \end{tabular}
    }
    &
    H^1\big(
      X
      ;\,
      \Omega \mathcal{A}
    \big)
    \ar[
      rr,
      "{
         \scalebox{.7}{
           \color{greenii}
           \bf
           \def\arraystretch{.9}
           \begin{tabular}{c}
             nonabelian
             \\
             character
           \end{tabular}
         }
      }"
    ]
    &&
    H^1_{\mathrm{dR}}(
      X
      ;\,
      \mathfrak{l}A
    )
    \,:=\,
    \mathrm{Hom}_{\mathrm{dgAlg}_{\mathbb{R}}}
    \big(
      \mathrm{CE}(\mathfrak{l}\mathcal{A})
      ,\,
      \Omega^\bullet_{\mathrm{dR}}(X)
    \big)\big/_{\sim}
    &
    \scalebox{.7}{
      \color{darkblue}
      \bf
      \def\arraystretch{.9}
      \begin{tabular}{c}
        differential forms with
        \\
        coefficients in
        \\
        Whitehead $L_\infty$-algebra
      \end{tabular}
    }
  \end{tikzcd}
  $}
\end{equation}

The nonabelian character in the generality of generalized non-abelian cohomology, such as CoHomotopy, is due to
\cite[Def. IV.2]{FSS23Char}, constructed via the fundamental theorem of dg-algebraic rational homotopy theory. We next survey how this works.

\medskip

The key point is that rational homotopy theory characterizes the non-torsion content of (the homotopy type of) a (classifying) space $\mathcal{A}$ by an $L_\infty$-algebra-approximation $\mathfrak{l}\mathcal{A}$ to its loop space $\infty$-group $\Omega \mathcal{A}$.

\begin{proposition}{{\bf Quillen-Sullivan-Whitehead $L_\infty$-algebra} {\rm cf. \cite[Prop. 4.23, 5.6 \& 5.13]{FSS23Char}}}
\label{QuillenSullivanWhiteheadLInfinityAlgebra}
For a topological space $\mathcal{A}$ which is
\begin{itemize}[leftmargin=.8cm]
  \item
  simply connected: $\pi_0 \mathcal{A} = \ast$, $\pi_1 \mathcal{A} = 1$;
  \item of rational finite type:
  $\mathrm{dim}_{\mathbb{Q}}\big(H^n(\mathcal{A}; \mathbb{Q})\big) < \infty$
\end{itemize}
there is a polynomial dgc-algebra over $\mathbb{R}$, unique up to dga-isomorphism, whose
\begin{itemize}[leftmargin=.8cm]
  \item[$\circ$]
   generators are the $\mathbb{R}$-rational homotopy groups of $\mathcal{A}$,
  $$
     \mathrm{CE}(\mathfrak{l}\mathcal{A})
     \;=\;
     \Big(
     \wedge^\bullet
     \big(
       \,{
         \color{darkblue}
         \pi_\bullet
         \big(
           \Omega \mathcal{A}
         \big)
         \!\otimes_{{}_\mathbb{Z}}\!
         \mathbb{R}
       }\,
     \big)^\vee
     ,\,
     \mathrm{d}_{CE(\mathfrak{l}\mathcal{A})}
     \Big)
   $$
   \item[$\circ$]
   cochain cohomology is the ordinary real cohomology of $\mathcal{A}$
  $$
    H^\bullet\big(
      \mathrm{CE}(\mathfrak{l}\mathcal{A})
    \big)
    =
    H^\bullet(\mathcal{A};\, \mathbb{R})
    \,.
  $$
\end{itemize}
\end{proposition}

This dgc-algebra is known as the {\it minimal Sullivan model} of $\mathcal{A}$. By \eqref{HigherJacobiIdentity} it is the Chevalley-Eilenberg algebra of an $L_\infty$-algebra which we denote $\mathfrak{l}\mathcal{A}$: The Whitehead bracket algebra structure on the $\mathbb{R}$-rational homotopy groups of the loop space (think of "$\mathfrak{l}(-)$" as standing for "Lie" or for "loops"):
\begin{equation}
 \label{WhiteheadLInfinityAlgebra}
  \mathfrak{l}\mathcal{A}
  \;=\;
  {
  \color{darkblue}
  \pi_\bullet\big(
    \Omega \mathcal{A}
  \big) \otimes_{\mathbb{Z}} \mathbb{R}
  }
  \,.
\end{equation}
Some {\bf examples} for how to use Prop. \ref{QuillenSullivanWhiteheadLInfinityAlgebra} to compute Sullivan models and hence $\mathbb{R}$-Whitehead $L_\infty$-algebras $\mathfrak{l}\mathcal{A}$ of spaces are spelled out on p. \pageref{ExamplesOfSullivanModels}.

\medskip

Many of the Whitehead $L_\infty$-algebras of familar spaces do not have established names as $L_\infty$-algebras. An interesting exception is the Whitehead $L_\infty$-algebra of the 4-sphere, which happens to coincide
\cite[(13)]{SatiVoronov22}
with what in $D=11$ supergravity-theory is known (quite independently) as the gauge algebra of the C-field \cite[(2.6)]{CJLP98}\cite[\S 4]{Sati10}:

\vspace{1mm}
\begin{equation}
\label{CFieldGaugeAlgebra}
\hspace{-.6cm}
\adjustbox{}{ \small
\def\arraystretch{1.2}
\def\tabcolsep{5pt}
\begin{tabular}{|c|c|c|c|}
  \hline
  \begin{tabular}{c}
    {\bf Homotopy type}
    \\
    (topological space)
  \end{tabular}
  &
  \begin{tabular}{c}
    {\bf Sullivan model}
    \\
    (``FDA'')
  \end{tabular}
  &
  \begin{tabular}{c}
    \bf
    Whitehead
    $L_\infty$-algebra
    \\
    (strong homotopy Lie algebra)
    \\
  \end{tabular}
  \\
  $\mathcal{A}$
  &
  $\mathrm{CE}(\mathfrak{l}\mathcal{A})$
  &
  $\mathfrak{l}\mathcal{A}$
  \\
  \hline
  \hline
  &&
  \\[-7pt]
 $S^4$
   &
  $
  \mathbb{R}
  \left[
  \def\arraystretch{1}
  \def\arraycolsep{1pt}
   \begin{array}{cc}
     \omega_7,
     \\
     \omega_4
   \end{array}
   \right]
   \Big/
   \left(
   \def\arraystretch{1}
   \def\arraycolsep{3pt}
   \begin{array}{cl}
     \differential
     \,\omega_7
     &
     =
     -
     \tfrac{1}{2}\omega_4 \wedge \omega_4
     \\
     \differential
     \,\omega_4 & = 0
   \end{array}
   \right)
  $
  &
  $
  \mathbb{R}
  \left\langle
  \def\arraystretch{1}
  \def\arraycolsep{1pt}
   \begin{array}{cc}
     v_6,
     \\
     v_3
   \end{array}
   \right\rangle
   ,
   \begin{array}{c}
     [v_3 , v_3] = v_6
   \end{array}
   $
  \\[+14pt]
  \scalebox{.8}{
    \color{darkblue}
    \bf
    4-sphere
  }
  &
  \scalebox{.8}{
    \color{darkblue}
    \bf
    \def\arraystretch{.9}
    \begin{tabular}{c}
      abstract Bianchi identity of
      \\
      duality-symmetric C-field fluxes
    \end{tabular}
  }
  &
  \scalebox{.8}{
    \color{darkblue}
    \bf
    C-field gauge algebra
  }
   \\[-7pt]
   &&
   \\
   \hline
\end{tabular}
}
\end{equation}

\begin{remark}[{\bf Prefactors in Sullivan algebras}]

As stated so far, the ubiquitous prefactor $-1/2$ is pure convention, due to the freedom of rescaling generators by rational (or even real) numbers while retaining dga-isomorphy. However, this factor is fixed by requiring certain integrality properties of the generators, see \cite[Prop. 4.6]{FSS19HopfWZ}.
This becomes relevant when regarding the lift back from $\mathfrak{l}S^4$ to $\mathcal{A} \defneq S^4$ as a flux quantization law, because then it implies that the C-field flux densities $G_4$ and $G_7$ in the image of the normalized generators $\omega_4$ and $\omega_7$ satisfy expected integrality conditions \cite[Thm. 4.8]{FSS19HopfWZ}.
We discuss this further below in \cref{CFieldFluxQuantization}.
\end{remark}

{\small
\noindent\label{ExamplesOfSullivanModels}
\hspace{-.2cm}
\def\arraystretch{1}
\def\tabcolsep{5pt}
\begin{tabular}{|l|l|}
\hline
\ourrowcolor{lightgray}
&
\\[-5pt]
\ourrowcolor{lightgray}
\begin{minipage}{9cm}

  {\bf Circle}: $\mathcal{A} \defneq S^1 \simeq B \mathbb{Z}$.

  $
  \big(\pi_\bullet(S^1) \!\otimes_{{}_\mathbb{Z}}\! \mathbb{R}
  \big)^\vee
  \simeq
  \mathbb{R}\langle \omega_1 \rangle
  $,
  \;\;
  $
    H^\bullet(S^1;\mathbb{R})
    \simeq
    \mathbb{R}[
      \omega_1
    ]
  $

  Since
  $\mathbb{R}[\omega_1]$
  is already the correct cohomology ring,

  it must be that $\differential_{S^1} = 0$ and hence
  \begin{equation*}
    \mathrm{CE}\big(
      \mathfrak{l}S^1
    \big)
    \;\simeq\;
    \mathbb{R}[\omega_1]
    \big/
    \big(
      \differential
      \,
      \omega_1 = 0
    \big)
  \end{equation*}

\end{minipage}
&
\begin{minipage}{7.6cm}
  While the circle is not simply connected,
  it is a ``nilpotent space'', and Sullivan's theorem actually applies in this generality.

  Nilpotent spaces have nilpotent fundamental group (e.g.: abelian) such that  all higher homotopy groups are nilpotent modules (e.g.: trivial modules).
\end{minipage}
\\[-5pt]
&
\\
\hline
&
\\[-5pt]
\begin{minipage}{9cm}

  {\bf 2-Sphere}: $\mathcal{A} \defneq S^2$.

  $\big(\pi_\bullet(S^2) \!\otimes_{{}_\mathbb{Z}}\! \mathbb{R}\big)^\vee \,\simeq\,
  \mathbb{R}\langle \omega_2, \,\omega_3 \rangle$,\;
  $H^\bullet(S^2;\mathbb{R})
  \simeq
  \mathbb{R}[\omega_2]/\big(
   \omega_2^2
  \big)
  $

    The differential on $\mathbb{R}[\omega_2, \omega_3]$ needs to remove $\omega_2^2$ and $\omega_3$

    from cohomology, hence it must be that:
  \begin{equation*}
    \mathrm{CE}\big(
      \mathfrak{l}S^2
    \big)
    \;\simeq\;
    \mathbb{R}
    \left[
    \def\arraystretch{1}
    \def\arraycolsep{0pt}
    \begin{array}{l}
      \omega_3,
      \\
      \omega_2
    \end{array}
    \right]
    \Big/
    \left(
    \def\arraystretch{1}
    \def\arraycolsep{1pt}
    \begin{array}{rcl}
      \differential
      \,
      \omega_3 &=&
      -
      \tfrac{1}{2}
      \omega_2 \wedge \omega_2
      \\
      \differential
      \,
      \omega_2 &=& 0
    \end{array}
    \right)
  \end{equation*}

\end{minipage}
&
\begin{minipage}{7.6cm}
  The homotopy group coresponding to the generator $\omega_3$ is that represented by the {\it complex Hopf fibration}
  $$
    S^3
    \xrightarrow{ h_{\mathbb{C}} }
    S^2
    \,.
  $$
\end{minipage}
\\[-5pt]
&
\\
\hline
&
\\[-5pt]
 \ourrowcolor{lightgray}
\begin{minipage}{9cm}

  {\bf 3-Sphere}: $\mathcal{A} \defneq S^3$.

  $
  \big(\pi_\bullet(S^3) \!\otimes_{{}_\mathbb{Z}}\! \mathbb{R}
  \big)^\vee
  \simeq
  \mathbb{R}\langle \omega_3 \rangle
  $,
  \;\;
  $
    H^\bullet(S^3;\mathbb{R})
    \simeq
    \mathbb{R}[
      \omega_3
    ]
  $

  Since
  $\mathbb{R}[\omega_3]$
  is already the correct cohomology ring,

  it must be that $\differential_{S^3} = 0$ and hence
  \begin{equation*}
    \mathrm{CE}\big(
      \mathfrak{l}S^3
    \big)
    \;\simeq\;
    \mathbb{R}[\omega_3]
    \big/
    \big(
      \differential
      \,
      \omega_3 = 0
    \big)
  \end{equation*}

\end{minipage}
&
\begin{minipage}{7.6cm}
  While $S^3 \simeq \mathrm{SU}(2)$, we see that $\mathfrak{l} \, \mathrm{SU}(2)$ is different from $\mathfrak{su}(2)$. But the former captures the cocycles of the latter:
  \vspace{-.3cm}
  $$
    \begin{tikzcd}[sep=0pt]
      \mathfrak{su}(2)
      \ar[rr]
      &&
      \mathfrak{l} \, \mathrm{SU}(2)
      \\
      \mathrm{CE}\big(\mathfrak{su}(2)\big)
      \ar[from=rr]
      &&
      \mathrm{CE}\big(
        \mathfrak{l}
        \,
        \mathrm{SU}(2)
      \big)
      \\
      \mathrm{tr}\big(
        -
        ,\,
        [-,-]
      \big)
      &\mapsfrom&
      \omega_3
    \end{tikzcd}
  $$
\end{minipage}
\\[-5pt]
&
\\
\hline
&
\\[-5pt]
&
\\[-5pt]
\begin{minipage}{9cm}
  {\bf 4-Sphere}: $\mathcal{A} \defneq S^4$.

  $\big(\pi_\bullet(S^4) \!\otimes_{{}_\mathbb{Z}}\! \mathbb{R}\big)^\vee \,\simeq\,
  \mathbb{R}\langle \omega_4, \,\omega_7 \rangle$,\;
  $H^\bullet(S^4;\mathbb{R})
  \simeq
  \mathbb{R}[\omega_4]/\big(
   \omega_4^2
  \big)
  $

    The differential on $\mathbb{R}[\omega_4, \omega_7]$ needs to remove $\omega_4^2$ and $\omega_7$

    from cohomology, hence it must be that:
  \begin{equation*}
    \mathrm{CE}\big(
      \mathfrak{l}S^4
    \big)
    \;\simeq\;
    \mathbb{R}\left[
    \def\arraystretch{1}
    \def\arraycolsep{0pt}
    \begin{array}{l}
      \omega_7,
      \\
      \omega_4
    \end{array}
    \right]
    \Big/
    \left(
    \def\arraystretch{1}
    \def\arraycolsep{1pt}
    \begin{array}{rcl}
      \differential
      \,
      \omega_7 &=&
      -
      \tfrac{1}{2}
      \omega_4 \wedge \omega_4
      \\
      \differential
      \,
      \omega_4 &=& 0
    \end{array}
    \right)
  \end{equation*}
\end{minipage}
&
\begin{minipage}{7.6cm}
  The homotopy group corresponding to the generator $\omega_7$ is that represented by the {\it quaternionic Hopf fibration}
  $$
   S^7
   \xrightarrow{
     h_{\mathbb{H}}
   }
   S^4
  $$
\end{minipage}
\\[-5pt]
&
\\
\hline
&
\\[-4pt]
 \ourrowcolor{lightgray}
\begin{minipage}{9cm}
  {\bf Complex Projective space}: $\mathcal{A} \defneq \mathbb{C}P^n$.

  \vspace{1pt}
  $\big(\pi_\bullet(\mathbb{C}P^n) \!\otimes_{{}_\mathbb{Z}}\! \mathbb{R}\big)^\vee \simeq\,
  \mathbb{R}\langle \omega_2, \,\omega_{2n+1} \rangle$,

  \vspace{1pt}
  $H^\bullet(\mathbb{C}P^n;\mathbb{R})
  \simeq
  \mathbb{R}[\omega_2]/\big(
   \omega_2^{n+1}
  \big)$

    The differential on $\mathbb{R}[\omega_2, \omega_{2n+1}]$ needs to remove $\omega_2^{n+1}$

    from cohomology, hence it must be that:
  \begin{equation*}
    \mathrm{CE}\big(
      \mathfrak{l}
      \,
      \mathbb{C}P^n
    \big)
    \;\simeq\;
    \mathbb{R}
    \left[
    \def\arraystretch{1}
    \def\arraycolsep{0pt}
    \begin{array}{l}
      \omega_{2n+1},
      \\
      \omega_2
    \end{array}
    \right]
    \Big/
    \left(
    \def\arraystretch{1}
    \def\arraycolsep{1pt}
    \begin{array}{lcl}
      \differential
      \,
      \omega_{2n+1} &=&
      -
      \omega_2^{n+1}
      \\
      \differential
      \,
      \omega_2 &=& 0
    \end{array}
    \right)
  \end{equation*}
\end{minipage}
&
\begin{minipage}{7.6cm}
  This is related to the above sequence of examples by the fact that $\mathbb{C}P^n$ is an $S^1$-quotient of $S^{2n+1}$:
  $$
    \begin{tikzcd}
      S^1
      \ar[r, hook]
      &
      S^{2n+1}
      \ar[
        d,
        ->>
      ]
      \\
      &
      \mathbb{C}P^n
    \end{tikzcd}
  $$
\end{minipage}
\\[-5pt]
&
\\
\hline
&
\\[-5pt]
\begin{minipage}{9cm}
  {\bf Infinite Projective space}: $\mathcal{A} \defneq \mathbb{C}P^\infty \simeq B \mathrm{U}(1) \simeq B^2 \mathbb{Z}$.

  $\big(\pi_\bullet(\mathbb{C}P^\infty) \!\otimes_{{}_\mathbb{Z}}\! \mathbb{R}\big)^\vee \,\simeq\,
  \mathbb{R}\langle \omega_2 \rangle$,
  $H^\bullet(\mathbb{C}P^n;\mathbb{R})
  \simeq
  \mathbb{R}[\omega_2]$

  Since $\mathbb{R}[\omega_2]$ is already the correct cohomology ring,

  it must be that $\differential_{\mathbb{C}P^\infty} = 0$:

  \begin{equation*}
    \mathrm{CE}\big(
      \mathfrak{l}
      \,
      \mathbb{C}P^\infty
    \big)
    \;\simeq\;
    \mathbb{R}
    \left[
    \def\arraystretch{1}
    \def\arraycolsep{0pt}
    \begin{array}{l}
      \omega_2
    \end{array}
    \right]
    \big/
    \left(
    \def\arraystretch{1}
    \def\arraycolsep{1pt}
    \begin{array}{lcl}
      \differential
      \,
      \omega_2 &=& 0
    \end{array}
    \right)
  \end{equation*}
\end{minipage}
&
\begin{minipage}{7.6cm}
 This is the Lie 2-algebra of the shifted circle group:
 $$
   \mathfrak{l}
   B \mathrm{U}(1)
   \;\simeq\;
   b \, \mathfrak{u}(1)
 $$
\end{minipage}
\\[-5pt]
&
\\
\hline
&
\\[-5pt]
 \ourrowcolor{lightgray}
\begin{minipage}{9cm}
  {\bf Eilenberg-MacLane space}: $\mathcal{A} \defneq B^n \mathrm{U}(1) \simeq B^{n+1} \mathbb{Z}$.

  $\big(\pi_\bullet(B^{n+1}\mathbb{Z}) \!\otimes_{{}_\mathbb{Z}}\! \mathbb{R}\big)^\vee \,\simeq\,
  \mathbb{R}\langle \omega_{n+1} \rangle$,
  $H^\bullet(B^{n+1}\mathbb{Z})
  \simeq
  \mathbb{R}[\omega_{n+1}]$

  Since $\mathbb{R}[\omega_{n+1}]$ is already the correct cohomology ring,

  it must be that $\differential_{B^{n+1}\mathbb{Z}} = 0$:

  \begin{equation*}
    \mathrm{CE}\big(
      \mathfrak{l}
      \,
      B^{n+1} \mathbb{Z}
    \big)
    \;\simeq\;
    \mathbb{R}
    \left[
    \def\arraystretch{1}
    \def\arraycolsep{0pt}
    \begin{array}{l}
      \omega_{n+1}
    \end{array}
    \right]
    \big/
    \left(
    \def\arraystretch{1}
    \def\arraycolsep{1pt}
    \begin{array}{lcl}
      \differential
      \,
      \omega_{n+1} &=& 0
    \end{array}
    \right)
  \end{equation*}
\end{minipage}
&
\begin{minipage}{7.6cm}
 This is the Lie $(n+1)$-algebra of the circle $(n+1)$-group:
 $$
   \mathfrak{l}
   B^n \mathrm{U}(1)
   \;\simeq\;
   b^n \, \mathfrak{u}(1)
 $$
\end{minipage}
\\[-5pt]
&
\\
\hline
&
\\[-5pt]
\begin{minipage}{9cm}
  {\bf Classifying space}: $\mathcal{A} \defneq B G$ of cpt. 1-conn. Lie group.

  $H^\bullet(B G; \mathbb{R})
  \,\simeq\,
  \mathrm{inv}^\bullet(\mathfrak{g})$
  the invar. polynomials on Lie alg.

  (Chern-Weil theory)

  Since $H^\bullet(B G; \mathbb{R})$ is already a free graded-symmetric ring

  it must be that $\differential_{{}_{B G}} = 0$ (cf. \cite[Lem. 8.2]{FSS23Char}):

  \begin{equation*}
    \mathrm{CE}\big(
      \mathfrak{l}
      \,
      B G
    \big)
    \;\simeq\;
    \mathrm{inv}^\bullet(\mathfrak{g})
    \big/
    \left(
      \differential_{{}_{
        B G
      }}
      =
      0
    \right)
  \end{equation*}
\end{minipage}
&
\begin{minipage}{7.6cm}
 $\mathfrak{l} B G$ captures all the curvature invariants

 hence all the invariant flux densities

 of $\mathfrak{g}$-connections
 $A \,\in\, \Omega^1_{\mathrm{dR}}(X) \otimes \mathfrak{g}$,

 e.g.
 $
   \begin{tikzcd}[sep=0pt]
     \mathrm{CE}\big(
       \mathfrak{l} B
       \mathrm{SU}(2)
     \big)
     \ar[
       rr,
     ]
     &&
     \Omega^\bullet_{\mathrm{dR}}(X)
     \\
     \mathrm{tr}(-,-)
     &\mapsto&
     \delta_{i j}
     F_A^{(i)} \wedge F_A^{(j)}
   \end{tikzcd}
 $
\end{minipage}
\\[-5pt]
&
\\
\hline
\end{tabular}
}

\paragraph{Rational homotopy theory: Discarding torsion in nonabelian cohomology.} From the perspective (above) that any topological space $\mathcal{A}$ serves as the classifying space of a generalized nonabelian cohomology theory, the idea of rational homotopy theory (survey in \cite{Hess06}; \cite[\S 4]{FSS23Char}) becomes that of extracting the {\it non-torsion} content of such a cohomology theory, which we will see is, over smooth manifolds, that shadow of it that is reflected in the non-abelian de Rham cohomology (Def. \ref{NonabelianDeRhamCohomology}) of $\mathfrak{l}\mathcal{A}$-valued differential forms.

\begin{equation}
\label{RationalHomotopyAsNontorsionAspect}
\adjustbox{raise=-.8cm}{
\small
\begin{tikzpicture}
\node
  at (0,0)
  {
$
  \def\arraystretch{2}
  \def\tabcolsep{5pt}
  \begin{tabular}{|c||c|c|}
    \hline
    \bf Homotopy theory
    &
    Rational
    &
    Sullivan model
    \\
    \hline
    \bf
    Nonabelian cohomology
    &
    Non-torsion
    &
    \def\arraystretch{.9}
    de Rham cohomology
    \\
    \hline
  \end{tabular}
$
  };
\draw[
  bend right=70,
  -Latex
]
 (-5.1,.4)
 to
 node[left, xshift=3pt]{
   \scalebox{.7}{
     \color{greenii}
     \bf
     \def\arraystretch{.9}
     \begin{tabular}{c}
       regard spaces as
       \\
       classifying spaces
     \end{tabular}
   }
 }
 (-5.1,-.4);

\end{tikzpicture}
}
\end{equation}
\smallskip

Now, in a sense, the signature of any $\mathcal{A}$-cohomology theory is its (reduced) cohomology groups on spheres,
equal to the homotopy groups of the classifying space:
$$
  \scalebox{.7}{
    \color{darkblue}
    \bf
    \def\arraystretch{.9}
    \begin{tabular}{c}
      reduced $\mathcal{A}$-cohomology
      \\
      of the $n$-sphere
    \end{tabular}
  }
  \widetilde{H}^1\big(
    S^n
    ;\,
    \Omega\mathcal{A}
  \big)
  \;\defneq\;
  \pi_0
  \,
  \mathrm{Map}^{\ast/}
  \big(
    S^n
    ,\,
    \mathcal{A}
  \big)
  \;\;
  \defneq
  \;\;
  \pi_n(\mathcal{A})
  \scalebox{.7}{
    \color{darkblue}
    \bf
    \def\arraystretch{.9}
    \begin{tabular}{c}
      $n$th homotopy group
      \\
      of classifying space
    \end{tabular}
  }
$$
Assuming throughout (for ease of exposition) that $\mathcal{A}$ is simply-connected, the remaining non-trivial homotopy groups are abelian $\pi_{n \geq 2}(i) \in \mathrm{AbGrp}$.
Discarding torsion elements (nilpotent group elements) from these groups is achieved by tensoring with the abelian group of rational numbers:
$$
  \hspace{2cm}
  \begin{tikzcd}[
    row sep=-1pt
  ]
  \hspace{1cm}
  \mathllap{
  \scalebox{.7}{
    \color{darkblue}
    \bf
    \def\arraystretch{.9}
    \begin{tabular}{c}
      reduced $\mathcal{A}$-cohomology
      \\
      of the $n$-sphere
    \end{tabular}
  }
  }
    \widetilde{H}^1(S^n;\Omega\mathcal{A})
    \,\simeq\,
    \pi_n(i)
    \ar[
      rr,
      "{
        \scalebox{.7}{
          {\color{greenii}
          \bf
          rationalization}
        }
      }"
    ]
    &&
    \pi_n(i)
    \otimes_{{}_\mathbb{Z}}
    \mathbb{Q}
  \scalebox{.7}{
    \color{darkblue}
    \bf
    \def\arraystretch{.9}
    \begin{tabular}{c}
      rationalized
      \\
      reduced $A$-cohomology
      \\
      of the $n$-sphere
    \end{tabular}
  }
    \\
    \mbox{
      $[c]$ with
      $k \cdot [c] = 0$
    }
    &\longmapsto&
    {[c]} \otimes 1
    \,=\,
    [c] \otimes k \cdot \tfrac{1}{k}
    \,=\,
    k \cdot [c] \otimes \tfrac{1}{k}
    \,=\,
    0
  \end{tikzcd}
$$
This is a ``projection operation'' (jargon: ``localization''), in that doing it twice has no further effect:
$$
  \begin{tikzcd}[row sep=0pt, column sep=large]
    \mathllap{
      \scalebox{.7}{
        \color{darkblue}
        \bf
        double
        rationalizaton
      }
    }
    \pi_n(i)
    \otimes_{{}_\mathbb{Z}}
    \mathbb{Q}
    \otimes_{{}_\mathbb{Z}}
    \mathbb{Q}
    \ar[
      <->,
      rr,
      "{
        \scalebox{.7}{
          \color{greenii}
          \bf
          isomorphic
        }
      }",
      "{ \sim }"{swap}
    ]
    &&
    \pi_n(i)
    \otimes_{{}_\mathbb{Z}}
    \mathbb{Q}
    \mathrlap{
      \scalebox{.7}{
        \color{darkblue}
        \bf
        single
        rationalizaton
      }
    }
    \\
    {[c]}
    \otimes
    \tfrac{p_1}{q_1}
    \otimes
    \tfrac{p_2}{q_2}
    =
    {[c]}
    \otimes
    \tfrac{p_1}{q_1}
    \otimes
    q_1
    \tfrac{p_2}{q_1 q_2}
    &\longleftrightarrow&
    {[c]}
    \otimes
    \tfrac{p_1 p_2}{q_1 q_2}
  \end{tikzcd}
$$

Hence to have a classifying space for the non-torsion part of $\mathcal{A}$-cohomology means to ask for:
\begin{equation}
\label{RationalizationOfASpace}
\adjustbox{}{
\begin{tabular}{l}
{\bf The rationalization} of $\mathcal{A}$:
\\
\def\arraystretch{2}
\def\tabcolsep{5pt}
\begin{tabular}{|ll|}
  \hline
  \def\arraystretch{.9}
  \begin{tabular}{l}
  A topological space
  \end{tabular}
  &
  $L^{\mathbb{Q}} \mathcal{A}$
  \\
  \def\arraystretch{.9}
  \begin{tabular}{l}
  all whose homotopy groups
  have
  \\
  the structure of $\mathbb{Q}$-vector spaces
  \end{tabular}
  &
  $\pi_n\big(
    L^{\mathbb{Q}} \mathcal{A}
  \big)
  \,\in\, \mathrm{Mod}_{\mathbb{Q}}$
  \\
  \def\arraystretch{.9}
  \begin{tabular}{l}
    equipped with a map from $\mathcal{A}$
  \end{tabular}
  &
  \hspace{-3pt}
  $
    \hspace{10pt}
    \begin{tikzcd}[
      column sep=60pt
    ]
      \mathcal{A}
      \ar[
        r,
        "{
          \eta_{\mathcal{A}}^{\mathbb{Q}}
        }"
      ]
      &
      L^{\mathbb{Q}} \mathcal{A}
    \end{tikzcd}
  $
  \\
  \def\arraystretch{.9}
  \begin{tabular}{l}
    which induces isomorphisms on
    \\
    rationalized
    homotopy groups
  \end{tabular}
  &
  $
    \hspace{-5pt}
    \begin{tikzcd}[
      column sep=30pt
    ]
      \pi_n(i)
      \!\otimes_{{}_{\mathbb{Z}}}\!
      \mathbb{Q}
      \ar[
        r,
        shorten=-2pt,
        "{
          \eta_{\mathcal{A}}^{\mathbb{Q}}
          \otimes_{{}_{\mathbb{Z}}}
          \mathbb{Q}
        }",
        "{ \sim }"{swap}
      ]
      &
      \pi_n\big(
        L^{\mathbb{Q}}\mathcal{A}
      \big)
      \otimes_{{}_{\mathbb{Z}}}
      \mathbb{Q}
    \end{tikzcd}
  $
  \\
  \def\arraystretch{.9}
  \begin{tabular}{l}
    and is universal
    \\
    with this property
  \end{tabular}
  &
  \\[-18pt]
  &
  \\
  \hline
\end{tabular}
\end{tabular}
}
\end{equation}

\noindent
Notice that infinitely many spaces $\mathcal{A}$ share the same rationalization, whence the choice of such an $\mathcal{A}$ as a flux quantization law below is genuine further information.

\smallskip

For example, the rationalization of an integral Eilenberg-MacLane space $B^ \mathbb{Z} \,\defneq\, K(\mathbb{Z}, n)$ classifies ordinary rational cohomology, mapping to ordinary de Rham cohomology:
\begin{equation}
\hspace{-4mm}
  \begin{tikzcd}[
    row sep=2pt,
    column sep=37pt
  ]
    \overset{
      \mathclap{
      \raisebox{7pt}{
      \scalebox{.7}{
        \color{darkblue}
        \bf
        \def\arraystretch{.9}
        \begin{tabular}{c}
          integral
          \\
          EM-space
        \end{tabular}
      }
      }
      }
    }{
      B^n \mathbb{Z}
    }
    \ar[
      rr,
      "{
        \eta^{\mathbb{Q}}_{
          B^n \mathbb{Z}
        }
      }"{swap},
      "{
       \scalebox{.7}{
         \color{greenii}
         \bf
         rationalization
       }
      }"
    ]
    &&
    L^{\mathbb{Q}}
    B^n \mathbb{Z}
    \simeq
    \overset{
      \mathclap{
      \raisebox{7pt}{
      \scalebox{.7}{
        \color{darkblue}
        \bf
        \def\arraystretch{.9}
        \begin{tabular}{c}
          rational
          \\
          EM-space
        \end{tabular}
      }
      }
      }
    }{
      B^n \mathbb{Q}
    }
    \ar[
      rr,
      "{
        \scalebox{.7}{
          \color{greenii}
          \bf
          extension of scalars
        }
      }",
      "{
        B^n
        \left(
          (-)
          \otimes_{\mathbb{Q}}
          \mathbb{R}
        \right)
      }"{swap}
    ]
    &&
    \overset{
      \mathclap{
      \raisebox{7pt}{
      \scalebox{.7}{
        \color{darkblue}
        \bf
        \def\arraystretch{.9}
        \begin{tabular}{c}
          real
          \\
          EM space
        \end{tabular}
      }
      }
      }
    }{
      B^n \mathbb{R}
    }
    \\[+5pt]
    \pi_0
    \mathrm{Map}\big(
      X
      ,\,
      B^n \mathbb{Z}
    \big)
    \ar[
      d,
      phantom,
      "{ \simeq }"{sloped}
    ]
    \ar[
      rr,
      "{
        \pi_0
        \mathrm{Map}\big(
          X
          ,\,
          \eta_{B^n \mathbb{Z}}^{\mathbb{Q}}
        \big)
      }"{swap}
    ]
    &&
    \pi_0
    \mathrm{Map}\big(
      X
      ,\,
      B^n \mathbb{Q}
    \big)
    \ar[
      d,
      phantom,
      "{ \simeq }"{sloped}
    ]
    \ar[
      rr,
      "{
        \pi_0
        \mathrm{Map}\big(\!
          X
          ,
          B^n\left(
            (-)
            \otimes_{\mathbb{Q}}
            \mathbb{R}
          \right)
       \! \big)
      }"{swap}
    ]
    &&
   \;
   \pi_0
    \mathrm{Map}\big(
      X
      ,\,
      B^n \mathbb{R}
    \big)
    \ar[
      d,
      phantom,
      "{ \simeq }"{sloped}
    ]
    \\[+10pt]
    \underset{
      \mathclap{
      \raisebox{-7pt}{
      \scalebox{.7}{
        \color{darkblue}
        \bf
        \def\arraystretch{.9}
        \begin{tabular}{c}
          integral
          \\
          ordinary cohomology
        \end{tabular}
      }
      }
      }
    }{
    H^n(
      X
      ;\,
      \mathbb{Z}
    )
    }
    \ar[
      rrrrr,
      rounded corners,
      to path={
           ([yshift=-00pt]\tikztostart.south)
        -- ([yshift=-10pt]\tikztostart.south)
        -- node[yshift=5pt]{
          \scalebox{.7}{
            \color{greenii}
            \bf
            ordinary character map
          }
        }
           ([yshift=-10pt]\tikztotarget.south)
        -- ([yshift=-00pt]\tikztotarget.south)
      }
    ]
    \ar[
      rr,
      "{
        \scalebox{.7}{
          \color{greenii}
          \bf
          cohomology operation
        }
      }"
    ]
    &&
    \underset{
      \mathclap{
      \raisebox{-7pt}{
      \scalebox{.7}{
        \color{darkblue}
        \bf
        \def\arraystretch{.9}
        \begin{tabular}{c}
          rational
          \\
          ordinary cohomology
        \end{tabular}
      }
      }
      }
    }{
    H^n(
      X
      ;\,
      \mathbb{Q}
    )
    }
    \ar[
      rr,
      "{
        \scalebox{.7}{
          \color{greenii}
          \bf
          cohomology operation
        }
      }"
    ]
    &&
    \underset{
      \mathclap{
      \raisebox{-7pt}{
      \scalebox{.7}{
        \color{darkblue}
        \bf
        \def\arraystretch{.9}
        \begin{tabular}{c}
          real
          \\
          ordinary cohomology
        \end{tabular}
      }
      }
      }
    }{
    H^n(
      X
      ;\,
      \mathbb{R}
    )
    }
    \ar[
      r,
      "{
        \scalebox{.7}{
          \color{greenii}
          \bf
          de Rham
        }
      }",
      "{
        \scalebox{.7}{
          \color{greenii}
          \bf
          isomorphism
        }
      }"{swap}
    ]
    &
    \underset{
      \mathclap{
      \raisebox{-7pt}{
      \scalebox{.7}{
        \color{darkblue}
        \bf
        \def\arraystretch{.9}
        \begin{tabular}{c}
          de Rham
          \\
          cohomology
        \end{tabular}
      }
      }
      }
    }{
      H^n_{\mathrm{dR}}(X)
    }
  \end{tikzcd}
\end{equation}

\medskip
\noindent We may regard this as the archetype of a {\it character map} and ask for its generalization to any $\mathcal{A}$-cohomology theory.
The pivotal observation of \cite{FSS23Char} is that for this purpose one may invoke the fundamental theorem of dg-algebraic rational homotopy theory:

\medskip

\paragraph{The Fundamental Theorem of dg-Algebraic Rational Homotopy Theory} (review in \cite[Prop. 5.6]{FSS23Char}) says that the homotopy theory
of rational spaces (simply-connected with fin-dim rational cohomology) is all encoded by their Whitehead $L_\infty$-algebras \eqref{WhiteheadLInfinityAlgebra}
over the rational numbers.
In particular, for $X$ a CW-complex, the homotopy classes of maps into the rationalization $L^{\mathbb{Q}}\mathcal{A}$ \eqref{RationalizationOfASpace}
of a space $\mathcal{A}$ is identified with dg-homotopy classes of homomorphisms from the rational Sullivan model of $\mathcal{A}$
to the “piecewise $\mathbb{Q}$-polynomial de Rham complex” of the topological space $X$:

\vspace{-.2cm}
\begin{equation}
  \label{FundamentalTheoremOfRHT}
  \begin{tikzcd}
  \mathrm{Map}\big(
    X
    ,\,
    L^{\mathbb{Q}} \mathcal{A}
  \big)_{/ \mathrm{homotopy}}
  \;\;
  \simeq
  \;\;
  \mathrm{Hom}_{\mathrm{dgAlg}}\Big(
    \mathrm{CE}\big(
      \mathfrak{l}^{\mathbb{Q}}
      \mathcal{A}
    \big)
    ,\,
    \Omega^\bullet_{\mathrm{P\mathbb{Q}LdR}}(X)
  \Big)_{
    \!/\mathrm{concordance},
  }
  \end{tikzcd}
\end{equation}
\vspace{-.3cm}

Observing that the right-hand side looks close to the definition of $\mathfrak{l}\mathcal{A}$-valued de Rham cohomology (Def. \ref{NonabelianDeRhamCohomology}), in order to actually connect to such smooth differential forms one needs to extend the ground field scalars from the rational numbers to the real numbers:

\subparagraph{Rational homotopy theory over the Reals} \!\!\!(\cite{BousfieldGugenheim76}, reviewed in \cite[Def. 5.7, Rem. 5.2, Prop. 5.8]{FSS23Char}). The construction \eqref{RationalizationOfASpace} also works over $\mathbb{R}$ (but is then not a “localization”) to give

\smallskip
\begin{equation}
\label{RealRationalization}
\adjustbox{}{$      \small
\def\tabcolsep{2pt}
\begin{tabular}{l}
{\bf The $\mathbb{R}$-rationalization} of $\mathcal{A}$:
\\
\begin{tabular}{|ll|}
  \hline
  &
  \\[-5pt]
  \begin{tabular}{l}
    A topological space
  \end{tabular}
  &
  \hspace{15pt}
  $L^{\mathbb{R}} \mathcal{A}$
  \\
  \begin{tabular}{l}
    equipped with a map
  \end{tabular}
  &
  $
    \hspace{15pt}
    \begin{tikzcd}[column sep=34pt]
      L^{\mathbb{Q}}\mathcal{A}
      \ar[
        rr,
        "{
          \eta^{\mathrm{ext}}_{L^{\mathbb{Q}}\mathcal{A}}
        }"
      ]
      &&
      L^{\mathbb{R}}\mathcal{A}
    \end{tikzcd}
  $
  \\
  \hspace{-.6pt}
  \def\arraystretch{.9}
  \begin{tabular}{l}
    which on homotopy groups
    \\
    is extension of scalars
  \end{tabular}
  &
  \begin{tikzcd}
    \pi_n\big(
      L^{\mathbb{Q}}\mathcal{A}
    \big)
    \ar[
      rr,
      "{
        \pi_n\big(
          \eta^{\mathrm{ext}}_{L^{\mathbb{Q}} \mathcal{A}}
        \big)
      }",
      "{
        =
        \,
        (-)
        \otimes_{\mathbb{Q}}
        \mathbb{R}
      }"{swap}
    ]
    &&
    \pi_n\big(
      L^{\mathbb{R}}\mathcal{A}
    \big)
  \end{tikzcd}
  \\
  \begin{tabular}{l}
  suitably universal as such.
  \end{tabular}
  &
  \\[-5pt]
  &
  \\
  \hline
\end{tabular}
\end{tabular}
$}
\end{equation}

With this “derived extension of scalars” \cite[Lem. 5.3]{FSS23Char} and for $X$ a smooth manifold, the fundamental theorem \eqref{FundamentalTheoremOfRHT} does relate to smooth differential forms via a non-abelian de Rham theorem \cite[Lem. 6.4, Thm. 6.5]{FSS23Char}:

\vspace{-.2cm}
\begin{equation}
  \label{NonabelianDeRhamTheorem}
  \hspace{-5mm}
  \begin{tikzcd}[
    row sep=25pt,
    column sep=10pt
  ]
    \overset{
      \mathclap{
        \raisebox{7pt}{
        \scalebox{.7}{
          \color{darkblue}
          \bf
          \def\arraystretch{.9}
          \begin{tabular}{c}
            non-abelian
            \\
            rational cohomology
          \end{tabular}
        }
        }
      }
    }{
    H^1\big(
      X
      ;\,
      L^{\mathbb{Q}}
      \Omega\mathcal{A}
    \big)
    }
    \ar[
      d,
      equals
    ]
    \ar[
      rr,
      "{
        \scalebox{.7}{
          \color{greenii}
          \bf
          derived extension of scalars
        }
      }"
    ]
    &&
    \overset{
      \mathclap{
        \raisebox{7pt}{
        \scalebox{.7}{
          \color{darkblue}
          \bf
          \def\arraystretch{.9}
          \begin{tabular}{c}
            non-abelian
            \\
            real cohomology
          \end{tabular}
        }
        }
      }
    }{
    H^1\big(
      X
      ;\,
      L^{\mathbb{R}}
      \Omega\mathcal{A}
    \big)
    }
    \ar[
      d,
      equals
    ]
    \ar[
      <->,
      ddr,
      "{
        \scalebox{.7}{
          \color{greenii}
          \bf
          \def\arraystretch{.9}
          \begin{tabular}{c}
            non-abelian
            \\
            de Rham theorem
          \end{tabular}
        }
      }"{sloped},
      "{ \sim }"{sloped, swap}
    ]
    \\
    \pi_0
    \mathrm{Map}\big(
      X
      ,\,
      L^{\mathbb{Q}}\mathcal{A}
    \big)
    \ar[
      rr,
      "{
        \pi_0
        \mathrm{Map}\big(
          X
          ,\,
          \eta_{L^{\mathbb{Q}}\mathcal{A}}^{\mathrm{ext}}
        \big)
      }"
    ]
    \ar[
      <->,
      d,
      "{
        \scalebox{.7}{
          \color{greenii}
          \bf
          \def\arraystretch{.9}
          \begin{tabular}{c}
            fundamental theorem
            \\
            of dg-algebraic
            \\
            rational homotopy
          \end{tabular}
        }
      }"{swap},
      "{ \sim }"{sloped}
    ]
    &&
    \pi_0
    \mathrm{Map}\big(
      X
      ,\,
      L^{\mathbb{R}}\mathcal{A}
    \big)
    \ar[
      <->,
      d,
      "{ \sim }"{sloped}
    ]
    \\
    \mathrm{Hom}_{\mathrm{dgAl}}
    \big(
      \mathrm{CE}(\mathfrak{l}^{\mathbb{Q}}\mathcal{A})
      ,\,
      \Omega^\bullet_{\mathrm{PLdR}}(X)
    \big)_{\!\!/\mathrm{cncd}}
    \ar[
      rr,
      "{
        \scalebox{.7}{
          \color{greenii}
          \bf
          extension of
        }
      }",
      "{
        \scalebox{.7}{
          \color{greenii}
          \bf
          scalars
        }
      }"{swap}
    ]
    &
    \phantom{-----}
    &
    \mathrm{Hom}_{\mathrm{dgAl}}
    \big(
      \mathrm{CE}(\mathfrak{l}\mathcal{A})
      ,\,
      \Omega^\bullet_{\mathrm{dR}}(X)
    \big)_{\!\!/\mathrm{cncd}}
    \ar[
      r,
      phantom,
      "{ \defneq }"
    ]
    &
    \underset{
      \mathclap{
        \raisebox{-5pt}{
          \scalebox{.7}{
            \color{darkblue}
            \bf
            \def\arraystretch{.9}
            \begin{tabular}{c}
              non-abelian
              \\
              de Rham cohomology
            \end{tabular}
          }
        }
      }
    }{
    H^1_{\mathrm{dR}}\big(
      X
      ;\,
      \mathfrak{l}\mathcal{A}
    \big)
    }
  \end{tikzcd}
\end{equation}
\vspace{-.3cm}

\subparagraph{In abelian (i.e., Whitehead-generalized) cohomology theories} both the rationalization step and the subsequent extension of scalars to $\mathbb{R}$ can be more easily described as forming the smash product of the coefficient spectrum with the rational Eilenberg-MacLane spectrum $H\mathbb{R}$ \cite[Ex. 5.7]{FSS23Char}. This is how the Chern-Dold character map over $\mathbb{R}$ is tacitly used in all the literature on abelian (Whitehead-generalized) differential cohomology theory (e.g. \cite[Def. 4.2]{BunkeNikolaus19}):
\vspace{0cm}
\begin{equation}
  \label{RealRationalizationOnSpectra}
  \begin{tikzcd}[column sep=40pt]
    \mathrm{Spectra}
    \ar[
      rr,
      "{
        (-)
          \wedge
        H\mathbb{Q}
      }",
      "{
        \scalebox{.7}{
          \color{greenii}
          \bf
          \def\arraystretch{.9}
          \begin{tabular}{c}
            rationalization
            \\
            localization
          \end{tabular}
        }
      }"{swap}
    ]
    \ar[
      rrrr,
      rounded corners,
      to path={
           ([yshift=00pt]\tikztostart.north)
        -- ([yshift=10pt]\tikztostart.north)
        -- node{
          \colorbox{white}{
          \scalebox{.8}{
            $(-) \wedge H \mathbb{R}$
          }
          }
        }
        node[yshift=9pt]{
          \scalebox{.7}{
            \color{greenii}
            \bf
            rationalization over $\mathbb{R}$
          }
        }
          ([yshift=10pt]\tikztotarget.north)
        -- ([yshift=00pt]\tikztotarget.north)
      }
    ]
    &&
    \mathrm{Spectra}
    \ar[
      rr,
      "{
        (-)
          \wedge_{{}_{H \mathbb{Q}}}
        H\mathbb{R}
      }",
      "{
        \scalebox{.7}{
          \color{greenii}
          \bf
          \def\arraystretch{.9}
          \begin{tabular}{c}
            extension
            \\
            of scalars
          \end{tabular}
        }
      }"{swap}
    ]
    &&
    \mathrm{Spectra}
  \end{tikzcd}
\end{equation}

The point of the non-abelian de Rham theorem \eqref{NonabelianDeRhamTheorem} is to generalize the realifification \eqref{RealRationalizationOnSpectra} of Whitehead-generalized cohomology to generalized non-abelian cohomology, such as to Cohomotopy; and the key result that makes this work is the fundamental theorem of dg-algebraic homotopy theory \eqref{FundamentalTheoremOfRHT}. This, ultimately, is the “reason” why $L_\infty$-valued differential forms relate fluxes to their flux-quantization laws.

\newpage
\paragraph{The general non-abelian character map} is now immediate \cite[Def. IV.2]{FSS23Char}: It is the cohomology operation induced by $\mathbb{R}$-rationalization of classifying spaces \eqref{RealRationalization}, seen under the non-abelian de Rham theorem \eqref{NonabelianDeRhamTheorem}:
\begin{equation}
\label{NonAbDeRhamCohomologyIsTargteOfCharacter}
\hspace{-2cm}
\adjustbox{raise=-17pt}{
  \begin{tikzcd}[
    column sep=17pt
  ]
    H^1(X;\Omega\mathcal{A})
    \ar[
      rr,
      "{
        \scalebox{.7}{
          \color{greenii}
          \bf
          rationalization
        }
      }"
    ]
    \ar[
      d,
      equals
    ]
    \ar[
      rrrrrr,
      rounded corners,
      to path={
           ([yshift=00pt]\tikztostart.north)
        -- ([yshift=8pt]\tikztostart.north)
        -- node{
          \scalebox{.7}{
            \colorbox{white}{
              \color{orangeii}
              \bf
              character map on $\mathcal{A}$-cohomology
            }
          }
        }
           ([yshift=8pt]\tikztotarget.north)
        -- ([yshift=00pt]\tikztotarget.north)
      }
    ]
    &&
    H^1\big(
      X
      ;\,
      L^{\mathbb{Q}}
      \Omega
      \mathcal{A}
    \big)
    \ar[
      rr,
      "{
        \scalebox{.7}{
          \color{greenii}
          \bf
          extension
        }
      }",
      "{
        \scalebox{.7}{
          \color{greenii}
          \bf
          of scalars
        }
      }"{swap}
    ]
    \ar[
      d,
      equals
    ]
    &&
    H^1\big(
      X
      ;\,
      L^{\mathbb{R}}
      \Omega
      \mathcal{A}
    \big)
    \ar[
      d,
      equals
    ]
    \ar[
      rr,
      "{
        \scalebox{.7}{
          \color{greenii}
          \bf
          nonabelian
        }
      }",
      "{
        \scalebox{.7}{
          \color{greenii}
          \bf
          de Rham theorem
        }
      }"{swap}
    ]
    &&
    H^1_{\mathrm{dR}}\big(
      X
      ;\,
      \mathfrak{l}\mathcal{A}
    \big)
    \ar[
      d,
      equals
    ]
    \\
    \pi_0
    \mathrm{Map}\big(
      X
      ,\,
      \mathcal{A}
    \big)
    \ar[
      rr,
      "{
        (\eta_{\mathcal{A}}^{\mathbb{Q}})_\ast
      }"
    ]
    &&
    \pi_0
    \mathrm{Map}\big(
      X
      ,\,
      L^{\mathbb{Q}}\mathcal{A}
    \big)
    \ar[
      rr,
      "{
        (\eta_{L^{\mathbb{Q}}\mathcal{A}}^{\mathrm{ext}})_\ast
      }"
    ]
    &&
    \pi_0
    \mathrm{Map}\big(
      X
      ,\,
      L^{\mathbb{R}}\mathcal{A}
    \big)
    \ar[
      rr,
      "\sim",
      "{
        \scalebox{.7}{
          \color{gray}
          \def\arraystretch{.9}
          \begin{tabular}{c}
            fundamental theorem
            \\
            of dg-algebraic RHT
          \end{tabular}
        }
      }"{swap, yshift=-5pt}
    ]
    &&
    \mathrm{Hom}_{\mathrm{dgAlg}}\big(
      \mathrm{CE}(\mathfrak{l}\mathcal{A})
      ,\,
      \Omega^\bullet_{\mathrm{dR}}(X)
    \big)_{\!/\mathrm{cncrd}}
  \end{tikzcd}
  }
  \hspace{-2cm}
\end{equation}

All the classical abelian character maps \eqref{ExamplesOfCharacterMaps} are special cases of this generalized nonabelian character [FSS23-Char, §7], but now examples in generalized nonabelian cohomology are also included; for instance, there is a character map on Cohomotopy-theory \cite[Ex. 6.11]{FSS23Char}.

\paragraph{Flux quantization in generalized nonabelian cohomology.}  With the generalized nonabelian character map \eqref{NonAbDeRhamCohomologyIsTargteOfCharacter} in hand, we may finally state the general concept of global flux quantization.
Recalling from \cref{TotalFluxAsNonabelianDeRhamCohomology} that the total flux of the higher gauge fields characterized by the $L_\infty$-algebra $\mathfrak{a}$ in encoded in the $\mathfrak{l}\mathcal{A}$-valued nonabelian de Rham cohomology of a Cauchy  surface, it follows

\noindent
\hspace{-.2cm}
\def\tabcolsep{0pt}
\begin{tabular}{p{7.7cm}l}
 that for every choice of classifying space
$\mathcal{A}$ with $\mathfrak{l}\mathcal{A} \,\simeq\, \mathfrak{a}$ the nonabelian character map \eqref{NonAbDeRhamCohomologyIsTargteOfCharacter} may be understood as assigning to discrete charges embodied by $\mathcal{A}$-cohomology classes the corresponding total flux (thereby losing torsion-information encoded in the charges but not in the fluxes).
&
\hspace{-.2cm}
\adjustbox{raise=-1.1cm}{
\begin{minipage}{9.4cm}
\begin{equation}
  \begin{tikzcd}[
    row sep=0pt,
    column sep=large
  ]
    \overset{
      \mathclap{
        \raisebox{4pt}{
          \scalebox{.7}{
          \color{darkblue}
          \bf
          \def\arraystretch{.9}
          \begin{tabular}{c}
            non-abelian
            \\
            cohomology
          \end{tabular}
          }
        }
      }
    }{
      H^1\big(
        X^d
        ;\,
        \Omega\mathcal{A}
      \big)
    }
    \ar[
      rr,
      "{
        \mathrm{ch}_{\mathcal{A}}
      }"{below},
      "{
        \scalebox{.7}{
          \color{greenii}
          \bf
          non-abelian character
        }
      }"
    ]
    &&
    \overset{
      \mathclap{
        \raisebox{6pt}{
          \scalebox{.7}{
            \color{darkblue}
            \bf
            \def\arraystretch{.8}
            \begin{tabular}{c}
              non-abelian
              \\
              de Rham cohomology
            \end{tabular}
          }
        }
      }
    }{
      H^1_{\mathrm{dR}}
     \big(
       X^d
       ;\,
       \mathfrak{l}\mathcal{A}
     \big)
    }
    \\
    \underset{
      \mathclap{
      \raisebox{-4pt}{
        \scalebox{.7}{
        \color{purple}
        \bf
        \def\arraystretch{.9}
        \begin{tabular}{c}
          total charge
        \end{tabular}
        }
      }
      }
    }
    c
    &\mapsto&
      \underset{
        \raisebox{-3pt}{
          \scalebox{.7}{
          \color{darkblue}
          \bf
          \def\arraystretch{.9}
          \begin{tabular}{c}
            total flux
          \end{tabular}
        }
        }
      }{
        \big[
          \vec B
          \,
        \big]
    }
\end{tikzcd}
\end{equation}
\end{minipage}
}
\end{tabular}

\smallskip

Since the total charges in $H^1\big(X^d; \Omega \mathcal{A})\big)$ on the left form a discrete set, we may think of global flux quantization in $\mathcal{A}$-cohomology as {\it lifting} of total fluxes through this character map:

\begin{equation}
  \label{FirstIdeaOfFluxQuantization}
  \hspace{-5mm}
\adjustbox{}{
\def\tabcolsep{3pt}
\begin{tabular}{|l|l|}
\hline
&
\\[-8pt]
\begin{tabular}{p{7.3cm}}
{\bf \small Global flux quantization.}
Higher gauge fields on a spatial Cauchy surface satisfying their Gau{\ss} law constraint
are equivalently closed $L_\infty$-valued forms  for some characteristic $L_\infty$-algebra $\mathfrak{a}$; the global {\it total flux} is their class in nonabelian de Rham cohomology.

A compatible {\it flux quantization law} is a choice of classifying space $\mathcal{A}$
with Whitehead $L_\infty$-algebra $\mathfrak{l}\mathcal{A} \simeq \mathfrak{a}$; and to quantize total flux is to lift it through the {\it character map} to nonabelian $\mathcal{A}$-cohomology.
\end{tabular}
&
\hspace{-4mm}
\adjustbox{raise=0cm}{
$
  \begin{tikzcd}[
    column sep=24pt,
    row sep=28pt
  ]
    &[-10pt]
    &&
    \overset{
      \mathclap{
      \raisebox{3pt}{
      \scalebox{.7}{
       \color{darkblue}
       \bf
        \def\arraystretch{.9}
        \begin{tabular}{c}
          choice of
          \\
          \scalebox{1}{$\mathcal{A}$}-cohomology
          \\
          \color{purple}
          with
          $\mathfrak{l}\mathcal{A} \,\simeq\,
          \mathfrak{a}$
        \end{tabular}
      }
      }
      }
    }{
      H^1\big(
        X^d
        ;\,
        \mathcal{A}
      \big)
    }
    \ar[
      dd,
      "{
        \underset{
          \mathclap{
            \scalebox{.7}{
              \color{greenii}
              \bf
              sourced flux
            }
          }
        }{
        \mathrm{ch}_{\mathcal{A}}(X^d)
        }
      }"{description},
    ]
    \\
    \\
    \ast
    \ar[
      rr,
      "{
        \vec B
      }"{description},
      "{
        \scalebox{.7}{
          \color{darkblue}
          \bf
          \def\arraystretch{.9}
          \begin{tabular}{c}
            flux densities on Cauchy surface
            \\
            satisfying their higher Gau{\ss} law
          \end{tabular}
        }
      }"{swap, yshift=-6pt},
    ]
    \ar[
      uurrr,
      dashed,
      "{
        \scalebox{.7}{
          \color{orangeii}
          \bf
          charge quantum in
          \scalebox{1.2}{$\mathcal{A}$}-cohomology
        }
      }"{sloped},
      "{ c }"{description}
    ]
    &
    &
    \Omega^1_{\mathrm{dR}}(
      X^d
      ;\,
      \mathfrak{a}
    )_{\closed}
    \ar[
      r,
      "{
        \scalebox{.7}{
          \color{greenii}
          \bf
          \def\arraystretch{.9}
          \begin{tabular}{c}
            total
            \\
            flux
          \end{tabular}
        }
      }"{pos=.4}
    ]
    &
    \underset{
      \mathclap{
        \raisebox{-4pt}{
          \hspace{-20pt}
          \scalebox{.7}{
            \color{gray}
            \def\arraystretch{.9}
            \begin{tabular}{c}
              $\mathfrak{a}$-valued
              \\
              de Rham cohomology
            \end{tabular}
          }
        }
      }
    }{
    H^1_{\mathrm{dR}}(
      X^d
      ;\,
      \mathfrak{a}
    )
    }
  \end{tikzcd}
$
}
\\
\hline
\end{tabular}
}
\end{equation}

\smallskip

Notice that such a lift is not {\it just} a (quantization/discretization-){\it condition} on the total fluxes, but also {\it extra structure}, namely a choice of torsion-component of the total charge reflected in total fluxes, as see in $\mathcal{A}$-cohomology:

\noindent
\hspace{-.2cm}
{
\begin{tabular}{ll}
\def\arraystretch{1.2}
\begin{tabular}{l}
Since the character map generally...
\\[+5pt]
...fails to be surjective, i.e., has a cokernel:
\\
  \hspace{5pt}
  $\Rightarrow$
  {\bf flux quantization is a condition} on fluxes
\\[+5pt]
...fails to be injective, i.e., has a kernel:
\\
  \hspace{5pt}
  $\Rightarrow$
  {\bf flux quantization
  is a choice} of ``torsion''
\end{tabular}
&
\hspace{-15pt}
$
  \begin{tikzcd}[
    column sep=20pt
  ]
    \overset{
     \mathclap{
       \raisebox{3pt}{
      \scalebox{.7}{
        \color{darkblue}
        \bf
        \begin{tabular}{c}
          kernel consisting of
          \\
          all compatible charges
        \end{tabular}
        }
      }
      }
    }{
      H^1(X^d;\Omega\mathcal{A})_{[\vec B]}
    }
    \ar[
      rr,
      hook
    ]
    \ar[dd]
    \ar[
      ddrr,
      phantom,
      "{
        \scalebox{.7}{(pb)}
      }"
    ]
    &[-9pt]&[-9pt]
    H^1(X^d;\Omega\mathcal{A})
    \ar[
      dd,
      "{
        \underset{
          \mathclap{
          \scalebox{.7}{
            \color{greenii}
            \bf
            sourced flux
          }
          }
        }{
          \mathrm{ch}_A(X^d)
        }
      }"{description}
    ]
    \ar[
      rr
    ]
    \ar[
      ddrr,
      phantom,
      "{
        \scalebox{.7}{
          (po)
        }
      }"
    ]
    &[-10pt]&[-10pt]
    \ast
    \ar[dd]
    \\
    \\
    \ast
    \ar[
      rr,
      "{
        [\vec B ]
      }"{description},
      "{
        \scalebox{.7}{
          \color{greenii}
          \bf
          \begin{tabular}{c}
            given
            total flux
          \end{tabular}
        }
      }"{swap, yshift=-3pt}
    ]
    &&
    H^1_{\mathrm{dR}}\big(
      X
      ;\,
      \mathfrak{l}A
    \big)
    \ar[
      rr,
      ->>
    ]
    &&
    \underset{
      \mathclap{
        \raisebox{-10pt}{
          \scalebox{.7}{
            \color{darkblue}
            \bf
            \def\arraystretch{.9}
            \begin{tabular}{c}
              cokernel consisting of
              \\
              total fluxes violating
              \\
              the flux quantization law
            \end{tabular}
          }
        }
      }
    }{
    H^1_{\mathrm{dR}}\big(
      X
      ;\,
      \mathfrak{l}A
    \big)
    \big/
    H^1(X^d;\Omega\mathcal{A})
    }
  \end{tikzcd}
$
\end{tabular}
}

However, in a higher gauge theory it is unnatural to have extra structure given by an equality of gauge equivalence classes, instead one should consider a gauge transformation between actual fields. Doing so leads to emergence of the gauge potentials and of the higher phase space stack of the theory, in the next subsection \cref{PhaseSpacesAsDifferentialNonabelianCohomology}.

\newpage
\begin{example}[{\bf Flux quantization laws for ordinary electromagnetism}]
\label{FluxQuantizationLawsForOrdinaryelectromagnetism}
By Ex. \ref{SolutionSpaceOfOrdinaryElectromagnetism}, the characteristic $L_\infty$-algebra of vacuum electromagnetism is two copies of the line Lie 2-algebra $b\mathfrak{u}(1)$. This is the Whitehead $L_\infty$-algebra of the classifying space $B \mathrm{U}(1) \,\simeq\, B^2 \mathbb{Z}$ and hence of its rationalization $B^2 \mathbb{Q}$. Therefore --- among many further variants --- there are the following choices of flux quantization laws for ordinary electromagnetism:

\begin{center}
\def\tabcolsep{3pt}
\begin{tabular}{|l|p{12cm}|}
  \hline
  \rowcolor{lightgray}
  \adjustbox{raise=-5pt}{
  $\;\;\underbrace{B^2 \mathbb{Q}}_{\mathrm{mag}} \times \underbrace{B^2 \mathbb{Q}}_{\mathrm{el}}$
  }
  &
    \vspaceabove
    This choice imposes essentially
    {\it no} flux quantization (it does rule out irrational total fluxes) and as such was the
    tacit choice  since \cite{Maxwell1865}
    until \cite{Dirac1931}.
    \vspacebelow
  \\
  \hline
  \adjustbox{raise=-5pt}{
  $\;\;\underbrace{B^2 \mathbb{Z}}_{\mathrm{mag}} \times \underbrace{B^2 \mathbb{Q}}_{\mathrm{el}}$
  }
  &
  \vspaceabove
  This choice imposes integrality of magnetic charge but no further condition on electric flux --- common choice since \cite{Dirac1931}, for instance in \cite[p. 299]{Alvarez85} \cite[\S 7.1]{Brylinski93}\cite[Ex. 2.1.2]{Freed00}.
  \vspacebelow
  \\
  \hline
  \rowcolor{lightgray}
  \adjustbox{raise=-5pt}{
  $\;\;\underbrace{B^2 \mathbb{Z}}_{\mathrm{mag}} \times \underbrace{B^2 \mathbb{Z}}_{\mathrm{el}}$
  }
  &
  \vspaceabove
  This choice imposes integrality of both magnetic and electric charge --- considered in
  \cite{FreedMooreSegal07a}\cite{FreedMooreSegal07b}\cite[Rem. 2.3]{BeckerBeniniSchenkelSzabo17}
  \cite{LazaroiuShahbazi22}\cite{LazaroiuShahbazi23}
  \vspacebelow
  \\
  \hline
  \adjustbox{raise=-4pt}{
  $
    \;\;
    \underbrace{
      B^2 \mathbb{Z}
    }_{\mathrm{mag}}
    \rtimes
    \underbrace{
    B K
    \ltimes
    B^2 \mathbb{Z}
    }_{\mathrm{el}}
    \;\;
  $
  }
  &
  \vspaceabove
  For a finite group $K \to \mathrm{Aut}(\mathbb{Z})$ --- this choice induces non-commutativity between EL/EL- and EL/M-fluxes, an example of a “non-evident” flux quantization condition considered in \cite{SS23QObs}.
  \vspacebelow
  \\
  \hline
\end{tabular}
\end{center}

\end{example}

\subsection{Phase spaces as Differential nonabelian cohomology}
\label{PhaseSpacesAsDifferentialNonabelianCohomology}

With higher Maxwell-type equations of flux given \cref{EquationsOfMotionOfHigherFlux} and with a compatible flux/charge quantization law $\mathcal{A}$ chosen \cref{FluxQuantizationLawsAsNonabelianCohomology}, the full on-shell field content of the higher gauge theory (including the gauge potentials) and hence its phase space (p. \pageref{PhaseSpaceIdea}) appears as the corresponding “moduli stack” of nonabelian differential cohomology $\widehat {\mathcal{A}}$ evaluated on any Cauchy surface.

\smallskip

\paragraph{Smooth $\infty$-groupids.}
In order to describe this, we need to make free use of the notions of {\it smooth $\infty$-groupoids} presented by simplicial presheaves on Cartesian spaces. Exposition and pointers may be found in this collection \cite{Schreiber24}, a concise compilation of the technical details is given in \cite[\S 1]{FSS23Char}, for more see \cite[\S 3]{SatiSchreiber21}:

In the present context, the point of smooth $\infty$-groupoids is that they provide the {\it joint} home for both {\it fluxes and charges}, namely for

\begin{itemize}[leftmargin=.6cm]
\item[(i)] the sheaves of closed $\mathfrak{a}$-valued differential forms $\Omega^1_{\mathrm{dR}}\big(-;\mathfrak{a}\big)_\closed$, which one may regard as the 0-truncated smooth moduli-stacks (namely: {\it smooth sets}, see \cite{GiotopoulosSati23}) of flux densities

\item[(ii)] the homotopy types of classifying spaces $\mathcal{A}$, which one may regard as the geometrically discrete moduli stacks of charges.
\end{itemize}

\medskip
Once regarded in this joint context, these two moduli-stacks become comparable, as previewed in the diagram on the right, via an object to be denoted $\shape \, \Omega^1_{\mathrm{dR}}\big(-;\mathfrak{a}\big)_\closed$ (described in a moment)
and it is thereby that flux quantization in $\mathcal{A}$-cohomology may be imposed as a {\it local} structure that is equivalent to that of the non-perturbative higher gauge fields themselves.

\begin{equation}
\begin{tikzcd}[
  column sep=40pt,
  row sep=-4pt
]
  \scalebox{.7}{
    \color{darkblue}
    \bf
    smooth sets
  }
  &&&&
  \scalebox{.7}{
    \color{darkblue}
    \bf
    $\infty$-groupoids
  }
  \\
  &&
  \scalebox{.7}{
    \color{darkblue}
    \bf
    smooth $\infty$-groupoids
  }
  \\
  \colorbox{lightgray}{
    $\mathrm{SmthSet}$
  }
  \ar[
    drr,
    hook,
    "{
      \scalebox{.7}{
        \bf
        smooth structure
      }
    }"{yshift=2pt, sloped},
    "{
      \scalebox{.7}{
        (no gauge transf.)
      }
    }"{yshift=-1pt, swap, sloped},
  ]
  \ar[
    dd,
    equals,
    gray
  ]
  &&&&
  \colorbox{lightgray}{
     $\mathrm{Grpd}_\infty$
  }
  \ar[
    dll,
    hook',
    "{
      \scalebox{.7}{
        \bf
        higher gauge transf.
      }
    }"{yshift=2pt, sloped},
    "{
      \scalebox{.7}{
        (discrete smooth struc.)
      }
    }"{swap, yshift=-1pt, sloped}
  ]
  \ar[
    dd,
    equals,
    gray
  ]
  \\
  &&
  \colorbox{lightgray}
  {
    $\mathrm{SmthGrpd}_\infty$
  }
  \ar[
    dd,
    equals,
    gray
  ]
  \\
  \color{gray}
  L^{\mathrm{liso}}\, \mathrm{PSh}(\mathrm{CartSp})
  &&&&
  \color{gray}
  \mathrm{PSh}(\Delta)_{\mathrm{Kan}}
  \\
  &&
  \color{gray}
  L^{\mathrm{lheq}}
  \,
  \mathrm{PSh}\big(
    \mathrm{CartSp}
    ,\,
    \mathrm{PSh}(\Delta)_{\mathrm{Kan}}
  \big)
  \\
  \Omega^1_{\mathrm{dR}}\big(
    -;\mathfrak{a}
  \big)_{\closed}
  \ar[
    uu,
    phantom,
    "{ \in }"{rotate=90}
  ]
  \ar[
    drr,
    "{
      \eta^{\,\scalebox{.7}{$\shape$}}
    }"{description},
    "{
      \scalebox{.7}{
        \color{greenii}
        \bf
        shape unit
      }
    }"{yshift=-3pt, sloped, swap}
  ]
  &&&&
  \mathcal{A}
  \ar[
    uu,
    phantom,
    "{ \in }"{rotate=90}
  ]
  \ar[
    dll,
    "{
      \mathbf{ch}_{\mathcal{A}}
    }"{description},
    "{
      \scalebox{.7}{
        \color{greenii}
        \bf
        differential character
      }
    }"{yshift=-3pt, sloped, swap}
  ]
  \\
  &&
  \shape
  \,
  \Omega^1_{\mathrm{dR}}\big(
    -;\mathfrak{a}
  \big)_{\closed}
  \ar[
    uu,
    phantom,
    "{ \in }"{rotate=90}
  ]
  \\[-10pt]
  \scalebox{.7}{
    \color{darkblue}
    \bf
    \def\arraystretch{.9}
    \begin{tabular}{c}
      moduli of
      \\
      flux densities
    \end{tabular}
  }
  &&&&
  \scalebox{.7}{
    \color{darkblue}
    \bf
    \def\arraystretch{.9}
    \begin{tabular}{c}
      moduli of
      \\
      charges
    \end{tabular}
  }
  \\[-5pt]
  &&
  \scalebox{.7}{
    \color{darkblue}
    \bf
    \def\arraystretch{.9}
    \begin{tabular}{c}
      deformations of
      \\
      flux densities
    \end{tabular}
  }
\end{tikzcd}
\end{equation}

\vspace{-.1cm}

\noindent
\hspace{-.2cm}
\begin{tabular}{p{6.9cm}l}
Notice that when {\it presenting} smooth $\infty$-groupoids by the projective model structure on simplicial presheaves over the site of Cartesian spaces \cite[Ex. 1.20]{FSS23Char}, these moduli objects appear canonically as fibrant objects, so that the only further step for computing the required derived mapping spaces into them is to cofibrantly resolve the domain manifold $X^d$. This is achieved by passage to the {\v C}ech nerve $\widehat{X}{}^d$ of any {\it good} open cover $\big\{U_j \xhookrightarrow{\iota_j} X^d\big\}_{j \in J}$ \cite[Ex. 1.24]{FSS23Char}, as indicated on the right.
&
\qquad
\adjustbox{raise=-2.4cm}{
$
  \begin{array}{c}
  \begin{tikzcd}[
    row sep=-1pt
  ]
    &[-40pt]
    \overset{
      \mathclap{
        \raisebox{3pt}{
          \scalebox{.7}{
            \color{darkblue}
            \bf
            \def\arraystretch{.85}
            \begin{tabular}{c}
              {\v C}ech
              \\
              groupoid
            \end{tabular}
          }
        }
      }
    }{
    \widehat{X}{}^d
    }
    \ar[
      rrrr,
      ->>,
      "{
        \scalebox{.7}{
          \color{greenii}
          \bf
          local homotopy equivalence
        }
      }"{yshift=6pt},
      "{
        \mathrm{lheq}
      }"{description, swap}
    ]
    &[-40pt]
    &&
    &[-40pt]
    \overset{
      \mathclap{
        \raisebox{3pt}{
          \scalebox{.7}{
            \color{darkblue}
            \bf
            \def\arraystretch{.9}
            \begin{tabular}{c}
              smooth
              \\
              manifold
            \end{tabular}
          }
        }
      }
    }{
      X^d
    }
    &[-40pt]
    \\
    &
    \mathrm{Plot}\big(
      \mathbb{R}^n \!\times\! \Delta^2
      ,\,
      \widehat{X}{}^d
    \big)
    \ar[
      rrrr
    ]
    &
    &&
    &
    \mathrm{Plot}\big(
      \mathbb{R}^n \times \Delta^2
      ,\,
      X^d
    \big)
    &
    \\
    &
    (\mathbf{x}, j)
    \ar[
      dr,
      "{
        (\mathbf{x}, j,k)
      }"{sloped}
    ]
    \ar[
      d,
      Rightarrow,
      shorten=5pt,
      "{
        (\mathbf{x}, i, j, k)
      }"{description}
    ]
    \ar[
      rrrr,
      phantom,
      shift right=.7cm,
      "{ \longmapsto }"
    ]
    &&&&
    \mathbf{x}
    \ar[
      dr,
      equals
    ]
    \ar[
      d,
      Rightarrow,
      shorten=5pt,
      equals
    ]
    \\[+25pt]
    (\mathbf{x}, i)
    \ar[
      rr,
      "{
        (\mathbf{x}, i, k)
      }"{swap}
    ]
    \ar[
      ur,
      "{
        (\mathbf{x}, i,j)
      }"{sloped}
    ]
    &{}&
    (\mathbf{x}, k)
    &&
    \mathbf{x}
    \ar[
      rr,
      equals
    ]
    \ar[
      ur,
      equals
    ]
    &{}&
    \mathbf{x}
  \end{tikzcd}
  \\
  \mbox{where}
  \;\;\;\;
  \begin{tikzcd}
    \mathbb{R}^n
    \ar[
      r,
      "{
        (\mathbf{x}, i,j)
      }"
    ]
    \ar[
      rr,
      rounded corners,
      to path={
           ([yshift=-0pt]\tikztostart.south)
        -- ([yshift=-8pt]\tikztostart.south)
        -- node {
          \scalebox{.7}{
            \colorbox{white}{
              $(\mathbf{x},i)$
            }
          }
        }
           ([yshift=-6pt]\tikztotarget.south)
        -- ([yshift=-00pt]\tikztotarget.south)
      }
    ]
    \ar[
      rrr,
      rounded corners,
      to path={
           ([yshift=-0pt]\tikztostart.south)
        -- ([yshift=-18pt]\tikztostart.south)
        -- node[xshift=8pt] {
          \scalebox{.7}{
            \colorbox{white}{
              $\mathbf{x}$
            }
          }
        }
        node[yshift=-4pt, xshift=50pt]{
         \scalebox{.7}{
           smooth
         }
        }
           ([yshift=-18pt]\tikztotarget.south)
        -- ([yshift=-00pt]\tikztotarget.south)
      }
    ]
    &
    U_i \cap U_j
    \ar[r, hook]
    &
    U_i
    \ar[r, hook]
    &
    X
  \end{tikzcd}
  \;\;\;
  \mbox{etc.}
  \end{array}
$
}
\end{tabular}

\vspace{-.6cm}

\noindent
\hspace{-.2cm}
\begin{tabular}{p{6.9cm}l}
\paragraph{Higher deformations of flux densities.}
Recall \eqref{ConcordanceBetweenFlatForms} that a coboundary in $\mathfrak{a}$-valued de Rham cohomology is a “concordance” of flux densities, to be thought of as a path of smooth variations of the flux densities, subject to their Bianchi identities.

But in higher gauge theories there are also non-trivial deformations-of-deformations varying over the higher dimensional $n$-simplices $\Delta^n_{\mathrm{geo}}$, forming the following simplicial object:
&
\qquad
\adjustbox{raise=-2.7cm}{
$
  \begin{tikzcd}
    \overset{
      \mathclap{
        \raisebox{10pt}{
          \scalebox{.7}{
            \color{darkblue}
            \bf
            \def\arraystretch{.9}
            \begin{tabular}{c}
              deformation paths
              \\
              of flux densities
            \end{tabular}
          }
        }
      }
    }{
    \Omega^1_{\mathrm{dR}}
    \big(
      -
      \times
      [0,1]
      ;\,
      \mathfrak{a}
    \big)_{\closed}
    }
    \ar[
      dd,
      shift left=18pt,
      "{
        (-)_0
      }"{description},
      "{
        \scalebox{.7}{
          \color{greenii}
          \bf
          \def\arraystretch{.9}
          \begin{tabular}{c}
            take starting point
            \\
            of deformation path
          \end{tabular}
        }
      }"
    ]
    \ar[
      dd,
      shift right=18pt,
      "{
        (-)_1
      }"{description},
      "{
        \scalebox{.7}{
          \color{greenii}
          \bf
          \def\arraystretch{.9}
          \begin{tabular}{c}
            take endpoint of
            \\
            deformation path
          \end{tabular}
        }
      }"{swap, xshift=-4pt}
    ]
    \ar[
      from=dd
    ]
    \ar[
      r,
      phantom,
      "{ \defneq }"
    ]
    &
    \Big\{
      \vec B_0
      \xrightarrow{\; \vec B_{[0,1]} \;}
      \vec B_1
    \Big\}
    \\
    \\
    \underset{
      \mathclap{
        \raisebox{-10pt}{
          \scalebox{.7}{
            \color{darkblue}
            \bf
            \def\arraystretch{.9}
            \begin{tabular}{c}
              flux densitites satisfying
              \\
              their Bianchi identities
            \end{tabular}
          }
        }
      }
    }{
    \Omega^1_{\mathrm{dR}}
      (
        -
        ;\,
        \mathfrak{a}
      )_\closed
    }
    \ar[
      r,
      phantom,
      "{ \defneq }"
    ]
    &
    \big\{
      \vec B
    \big\}
  \end{tikzcd}
$
}
\end{tabular}

$$
  \shape
  \,
  \Omega^1_{\mathrm{dR}}\big(
    -;
    \mathfrak{a}
  \big)_{\closed}
  \;
  =
  \;
  \left(
  \hspace{-8pt}
  \adjustbox{}{
  \begin{tikzcd}
    &[50pt]
    {}
    \ar[
      dd,
      -,
      dotted,
      shift right=24
    ]
    \ar[
      dd,
      -,
      dotted,
      shift right=16
    ]
    \ar[
      dd,
      -,
      dotted,
      shift right=8
    ]
    \ar[
      dd,
      -,
      dotted,
      shift right=0
    ]
    \ar[
      dd,
      -,
      dotted,
      shift left=8
    ]
    \ar[
      dd,
      -,
      dotted,
      shift left=16
    ]
    \ar[
      dd,
      -,
      dotted,
      shift left=24
    ]
    \\
    \\
    \adjustbox{}{
       \begin{tikzpicture}
         \draw
           (.02, .03) node {
             \scalebox{.5}{$3$}
           };
         \draw[gray]
           (90:.7) to (.02, .2);
         \draw[gray]
           (120+90:.7) to (-.06, .03);
         \draw[gray]
           (240+90:.7) to (+.1, .04);
         \draw[
          fill=gray,
          draw opacity=.6,
          fill opacity=.6
         ]
           (90:.7) -- (120+90:.7) -- (240+90:.7) -- cycle;
          \draw (90:.81) node {\scalebox{.7}{$1$}};
          \draw (240+90-4:.9) node {\scalebox{.7}{$2$}};
          \draw (120+90+4:.9) node {\scalebox{.7}{$0$}};
       \end{tikzpicture}
    }
    &[70pt]
    \mathllap{
          \scalebox{.7}{
            \color{darkblue}
            \bf
            \def\arraystretch{.9}
            \begin{tabular}{c}
              deformation paths
              \\
              of deformation paths
              \\
              of deformation paths
              \\
              of flux densities
            \end{tabular}
      }
    }
    \Omega^1_{\mathrm{dR}}\big(
      -
      \!\times \Delta^3_{\mathrm{geo}}
      ;\,
      \mathfrak{a}
    \big)_\closed
    \ar[
      dd,
      shift right=51pt,
      "{
        (-)_{[1,2,3]}
      }"{description}
    ]
    \ar[
      dd,
      shift right=17.5pt,
      "{
        (-)_{[0,2,3]}
      }"{description}
    ]
    \ar[
      dd,
      shift left=17.5pt,
      "{
        (-)_{[0,1,3]}
      }"{description}
    ]
    \ar[
      dd,
      shift left=51pt,
      "{
        (-)_{[0,1,2]}
      }"{description}
    ]
    \\
    \\
    \adjustbox{}{
       \begin{tikzpicture}
         \draw[
          fill=gray,
          draw opacity=.6,
          fill opacity=.6
         ]
           (90:.7) -- (120+90:.7) -- (240+90:.7) -- cycle;
          \draw (90:.81) node {\scalebox{.7}{$1$}};
          \draw (240+90-4:.9) node {\scalebox{.7}{$2$}};
          \draw (120+90+4:.9) node {\scalebox{.7}{$0$}};
       \end{tikzpicture}
    }
    &[70pt]
    \mathllap{
          \scalebox{.7}{
            \color{darkblue}
            \bf
            \def\arraystretch{.9}
            \begin{tabular}{c}
              deformation paths
              \\
              of deformation paths
              \\
              of flux densities
            \end{tabular}
      }
    }
    \Omega^1_{\mathrm{dR}}\big(
      -
      \!\times \Delta^2_{\mathrm{geo}}
      ;\,
      \mathfrak{a}
    \big)_\closed
    \ar[
      dd,
      shift left=15,
      "{
        (-)_{[1,2]}
      }"{description}
    ]
    \ar[
      from=dd,
      shift left=7.5
    ]
    \ar[
      dd,
      shift left=0,
      "{
        (-)_{[0,2]}
      }"{description}
    ]
    \ar[
      from=dd,
      shift right=7.5
    ]
    \ar[
      dd,
      shift right=15,
      "{
        (-)_{[0,1]}
      }"{description}
    ]
    \ar[
      r,
      phantom,
      "{ \defneq }"
    ]
    &
    \adjustbox{raise=-30pt}{
     \includegraphics[width=4cm]{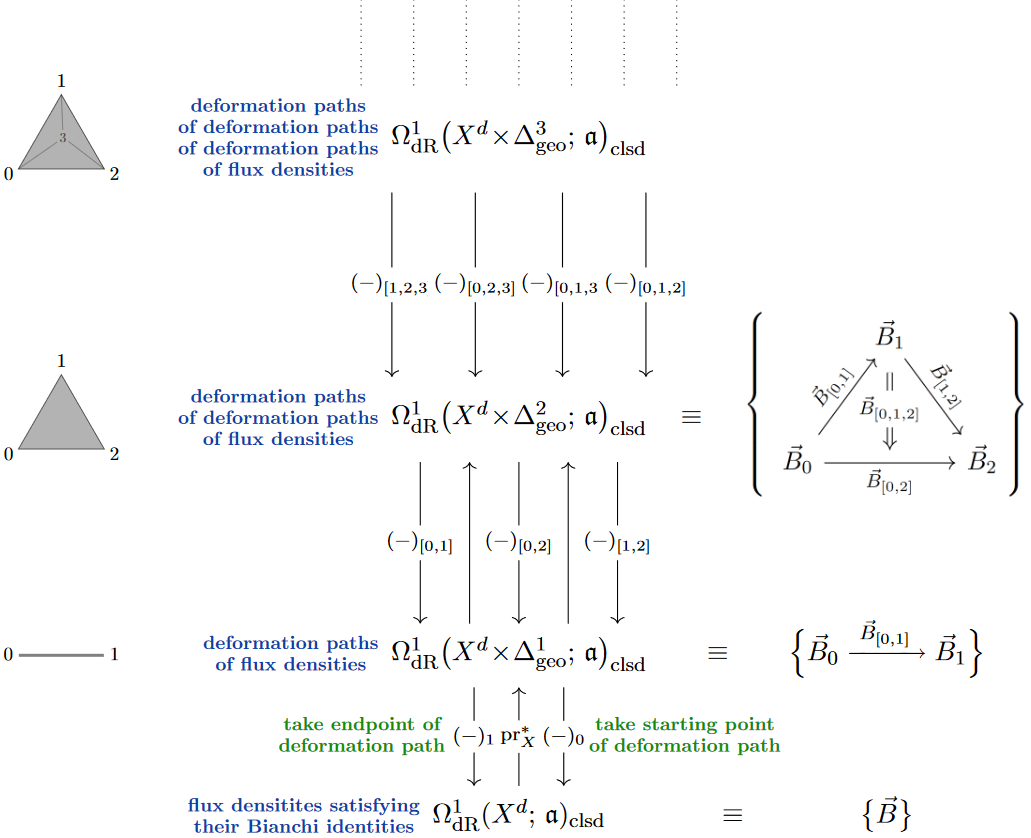}
    }
    \\
    \\
    \adjustbox{}{
       \begin{tikzpicture}
         \draw[line width=1.2, gray]
           (-.6, .07) -- (+.6,.07);
         \draw
           (-.75,0) node {
            \scalebox{.7}{$0$}
           };
         \draw
           (+.75,0) node {
            \scalebox{.7}{$1$}
           };
       \end{tikzpicture}
    }
    &
    \mathllap{
          \scalebox{.7}{
            \color{darkblue}
            \bf
            \def\arraystretch{.9}
            \begin{tabular}{c}
              deformation paths
              \\
              of flux densities
            \end{tabular}
      }
    }
    \Omega^1_{\mathrm{dR}}
    \big(
      -
      \!\times
      \Delta^1_{\mathrm{geo}}
      ;\,
      \mathfrak{a}
    \big)_{\closed}
    \ar[
      dd,
      shift left=18pt,
      "{
        (-)_0
      }"{description},
      "{
        \scalebox{.7}{
          \color{greenii}
          \bf
          \def\arraystretch{.9}
          \begin{tabular}{c}
            take starting point
            \\
            of deformation path
          \end{tabular}
        }
      }"{xshift=2pt}
    ]
    \ar[
      dd,
      shift right=18pt,
      "{
        (-)_1
      }"{description},
      "{
        \scalebox{.7}{
          \color{greenii}
          \bf
          \def\arraystretch{.9}
          \begin{tabular}{c}
            take endpoint of
            \\
            deformation path
          \end{tabular}
        }
      }"{swap, xshift=-5pt}
    ]
    \ar[
      from=dd
    ]
    \ar[
      r,
      phantom,
      "{ \defneq }"
    ]
    &
    \Big\{
      \vec B_0
      \xrightarrow{\; \vec B_{[0,1]} \;}
      \vec B_1
    \Big\}
    \\
    \\
    &
    \mathllap{
          \scalebox{.7}{
            \color{darkblue}
            \bf
            \def\arraystretch{.9}
            \begin{tabular}{c}
              flux densitites satisfying
              \\
              their Bianchi identities
            \end{tabular}
          }
    }
    \Omega^1_{\mathrm{dR}}
      (
        -
        ;\,
        \mathfrak{a}
      )_\closed
      \ar[
        r,
        phantom,
        "{ \defneq }"
      ]
      &
      \big\{
        \vec B
      \big\}
  \end{tikzcd}
  }
  \hspace{-8pt}
  \right)
$$
This is a Kan-simplicial presheaf \cite[Def. 9.1, Prop. 5.10]{FSS23Char}  that we may think of as the {\it shape} or {\it smooth path $\infty$-groupoid}
\cite[p. 144]{SatiSchreiber21}
of the 0-truncated moduli stack of flux densities.
It is in this object that flux densities become comparable to their charges:

\begin{itemize}
\item[(i)] There is an evident inclusion
$
  \begin{tikzcd}
  \Omega^1_{\mathrm{dR}}\big(
    -;\mathfrak{a}
  \big)_{\closed}
  \ar[
    rr,
    "{
      \scalebox{.7}{
        \color{greenii}
        \bf
        shape unit
      }
    }"
  ]
  &&
  \shape
  \,
  \Omega^1_{\mathrm{dR}}\big(
    -;\mathfrak{a}
  \big)_{\closed}
  \end{tikzcd}
$ \cite[(9.3)]{FSS23Char}, which we may identify as the {\it shape unit} of the moduli of flux densities;

\item[(ii)] given an identification $\mathfrak{a} \simeq \mathfrak{l}\mathcal{A}$ with a Whitehead $L_\infty$-algebra \eqref{FirstIdeaOfFluxQuantization}, then the fundamental theorem of dg-algebraic rational homotopy theory \eqref{FundamentalTheoremOfRHT} furthermore says \cite[Lem. 9.1]{FSS23Char} that we have a \mbox{(homotopy-)} equivalence to the $\mathbb{R}$-rationalization $L^{\mathbb{R}} \mathcal{A}$ of $\mathcal{A}$ \eqref{RealRationalization}, so that rationalization gives a {\it differential character map} \cite[Def. 9.2]{FSS23Char}:

$$
  \begin{tikzcd}[
    column sep=60pt
  ]
    \mathcal{A}
    \ar[
      r,
      "{
        \scalebox{.7}{
          \color{greenii}
          \bf
          \begin{tabular}{c}
            rationalization
          \end{tabular}
        }
      }"
    ]
    \ar[
      rrr,
      rounded corners,
      to path={
           ([yshift=-00pt]\tikztostart.south)
        -- ([yshift=-12pt]\tikztostart.south)
        -- node[yshift=-5pt]
           {
             \scalebox{.7}{
               \color{greenii}
               \bf
               differential character map
             }
           }
           node[yshift=4pt]{
             \scalebox{.7}{
               $\mathbf{ch}$
             }
           }
           ([yshift=-8pt]\tikztotarget.south)
        -- ([yshift=-00pt]\tikztotarget.south)
      }
    ]
    &[+10pt]
    L^{\mathbb{Q}}
    \mathcal{A}
    \ar[
      r,
      "{
        \scalebox{.7}{
          \color{greenii}
          \bf
          \def\arraystretch{.9}
          \begin{tabular}{c}
            extension
            \\
            of scalars
          \end{tabular}
        }
      }"
    ]
    &
    L^{\mathbb{R}}
    \mathcal{A}
    \ar[
      r,
      "{
        \scalebox{.7}{
          \color{greenii}
          \bf
          \def\arraystretch{.9}
          \begin{tabular}{c}
            fundamental thm. of RHT
            \\
            piecewise smooth version
          \end{tabular}
        }
      }",
      "{ \sim }"{swap}
    ]
    &[+50pt]
    \shape
    \,
    \Omega^1_{\mathrm{dR}}\big(
      -;
      \mathfrak{l}\mathcal{A}
    \big)_{\closed}
  \end{tikzcd}
$$
\end{itemize}

\smallskip

\begin{equation}
\label{LocalFluxQuantization}
\adjustbox{}{
\hspace{-.6cm}
\def\tabcolsep{4pt}
\begin{tabular}{|p{10cm}|l|}
\hline
&
\\[-7pt]
{\bf Local flux quantization: Gauge potentials in differential cohomology.}
This way one may now {\it locally} implement flux quantization, by taking the higher gauge field fields on $X^d$ to be {\it homotopies}
deforming flux densities $\vec B$ into the differential character of local charges $\rchi$.

On equivalence classes, this reproduces the quantization of total fluxes \eqref{FirstIdeaOfFluxQuantization} and thereby lifts it to
a local structure. Indeed, the higher gauge fields defined this way are the cocyles of the nonabelian {\it differential}
$\mathcal{A}$-cohomology \cite[Def. 9.3]{FSS23Char}.
&
\adjustbox{raise=-1.45cm}{
  \hspace{-15pt}
  \begin{tikzcd}[
    column sep=11pt,
    row sep=30pt
  ]
    \widehat{X}{}^d
    \ar[
      rr,
      dashed,
      "{
        \scalebox{.7}{
          \color{greenii}
          \bf
          charges
        }
      }",
      "{ \rchi }"{
        swap,
        name=s
      }
    ]
    \ar[
      dd,
      dashed,
      "{
        \scalebox{.7}{
          \color{greenii}
          \bf
          \def\arraystretch{.9}
          \begin{tabular}{c}
            flux
            \\
            densities
          \end{tabular}
        }
      }"{swap, xshift=3pt},
      "{
        \vec B
      }"{
        name=t
      }
    ]
    &&
    \mathcal{A}
    \ar[
      dd,
      "{
        \mathbf{ch}
      }"{swap},
      "{
        \scalebox{.7}{
          \color{greenii}
          \bf
          \def\arraystretch{.9}
          \begin{tabular}{c}
            differential
            \\
            character
          \end{tabular}
        }
      }"{xshift=-6pt}
    ]
    \\
    \\
    \Omega^1_{\mathrm{dR}}\big(
      -;
      \mathfrak{a}
    \big)_{\closed}
    \ar[
      rr,
      "{
        \eta^{\,\scalebox{.62}{$\shape$}}
      }",
      "{
        \scalebox{.7}{
          \color{greenii}
          \bf
          \def\arraystretch{.9}
          \begin{tabular}{c}
            shape
            \\
            unit
          \end{tabular}
        }
      }"{swap}
    ]
    &&
    \shape
    \;
    \Omega^1_{\mathrm{dR}}\big(
      -;
      \mathfrak{a}
    \big)_{\closed}
    \ar[
      from=s,
      to=t,
      Rightarrow,
      shorten=7pt,
      "{
        \widehat{A}
      }"{swap},
      "{
        \scalebox{.7}{
          \color{orangeii}
          \bf
          \def\arraystretch{.9}
          \begin{tabular}{c}
            gauge
            \\
            potentials
          \end{tabular}
        }
      }"{xshift=-11pt, yshift=+5pt}
    ]
  \end{tikzcd}
  \hspace{-14pt}
  }
\\[+2pt]
\hline
\end{tabular}
}
\end{equation}

\vspace{.1cm}

\noindent
In terms of physics these homotopies turn out to reflect the expected higher {\it gauge potentials} --- which is not entirely obvious from the definition but follows by examination:

\begin{example}[{\bf Higher $\mathrm{U}(1)$-gauge potentials in ordinary differential cohomology}]
\label{HigherCircleGaugePotentials}
The data $\widehat A \,:\, \rchi \Rightarrow \vec B$ in \eqref{LocalFluxQuantization} is equivalent \cite[Prop. 9.5]{FSS23Char}...

\begin{itemize}[leftmargin=.8cm]
\item[(A)] ...for the case $\mathfrak{a} = b\mathfrak{u}(1)$ and $\mathcal{A} \,\defneq\, B\mathrm{U}(1) \,=\, B^2 \mathbb{Z}$ (Ex. \ref{FluxQuantizationLawsForOrdinaryelectromagnetism}):

to that of connections on $\mathrm{U}(1)$-principal bundles -- which of course is the traditional data for the gauge potential of ordinary electromagnetism \cite{WuYang75}, cf. \cite{WuYang06}\cite[Ex. 5.5]{EguchiGilkeyHanson80} \cite[\S 6.1]{RudolphSchmidt17};

\item[(B)] ...for the case of  $\mathfrak{a} = b^2 \mathfrak{u}(1)$ and $\mathcal{A} = B^2 \mathrm{U}(1) = B^3 \mathbb{Z}$:

to that of 3-cocyles in Deligne cohomology (often equivalently regarded as connections on ``bundle gerbes''), this being the traditional understanding of the B-field gauge potential in string theory \cite{Gawedzki88}  \cite[\S 6]{FreedWitten99} \cite{CareyJohnsonMurray04} \cite{BonoraRuffinoSavelli08};

\item[(C)] ...for the case of $\mathfrak{a} = b^3 \mathfrak{u}(1)$ and  $\mathcal{A} = B^3 \mathrm{U}(1) = B^4 \mathbb{Z}$:

to that of 4-cocyles in Deligne cohomology (also regarded as connections on ``bundle 2-gerbes''), which was one of the proposed models
for the C-field gauge potential (in the case where the class $\tfrac{1}{2}p_1[T Y^{11}]$ of spacetime is even, otherwise the expected
half-integral shift has been added ``by hand'')
\cite{AschieriJurco04} \cite{HopkinsSinger05}\cite{DiaconescuFreedMoore07}\cite{FSS15CField}.

\end{itemize}

\end{example}

\begin{remark}[{\bf Shortcoming of higher $\mathrm{U}(1)$-charge quantization}]
\label{ShortcomingOfHigherCircleFluxQuantization}
For a long time, these examples \ref{HigherCircleGaugePotentials} used to be the state of the art in understanding flux quantization of higher gauge fields. But notice that in all three items the flux-quantization of the duality-partner fields (and hence of the canonical momenta) have been ignored. For item (A) this can readily be rectified, since here the partner (electric) field can be flux-quantized in the same way (and later has been, Ex. \ref{FluxQuantizationLawsForOrdinaryelectromagnetism}),  but in items (B) and (C) it is actually impossible to model the dual fields (with flux densities $H_7$ and $G_7$, respectively) as higher $\mathrm{U}(1)$-gauge fields (nor even as generalized higher abelian gauge fields, Ex. \ref{WhiteheadGeneralizedDifferentialCohomology}), since their Bianchi identities are non-linear (by Ex. \ref{MotionOfBRRFieldFluxes} and Ex. \ref{MotionOfCFieldFlux}, respectively), cf. \cref{CFieldFluxQuantization}.
\end{remark}

\begin{example}[{\bf abelian Whitehead-generalized differential cohomology}]
\label{WhiteheadGeneralizedDifferentialCohomology}

For $E_\bullet$ a spectrum of spaces and $\mathcal{A} \defneq E_n$, the data $\hat A \,\colon\,\chi \Rightarrow \vec B$ in \eqref{LocalFluxQuantization} is equivalent \cite[Ex. 9.1]{FSS23Char} to cocycles in the "canonical" version of the differential cohomology theory $\widehat{E}^n(-)$, as originally introduced in \cite{HopkinsSinger05}, cf. \cite[p. 88]{Bunke12}; for exposition see also in this collection the contribution \cite{Debray24}.

For applications to flux quantization, the most prominent example of such abelian generalized differential cohomology remain flavors
of differential K-theory, to which we come in \cref{RRFieldFluxQuantization}.
\end{example}

\medskip

\ifdefined\shortversion
\else

\subsection{Background fluxes as Twisting of cohomology}

(...)

\fi

\medskip

\section{Examples in String-/M-Theory}
\label{ExamplesInStringTheory}

While flux quantization is an issue in any higher gauge theory, the examples where it has received most (essentially all) of the attention are those of evident relevance in string theory — which is what we focus on in the following.

\smallskip

While string theory is an attempt to understand the all-important but elusive non-perturbative behaviour of Yang-Mills theories (notably quantum chromodynamics) by regarding quarks confined by color flux tubes as endpoints of open strings stuck on intersecting branes in an unobserved higher dimensional spacetime (cf. \cite{Polyakov12} \cite{HariDass24}), ironically also string physics itself has really been understood only perturbatively (namely by replacing Feynman diagrams in ordinary worldline perturbative quantum field theory with worldsheet $n$-point functions of a 2d SCFT).

\smallskip
However, since flux quantization laws (as discussed in \cref{FluxQuantizationLaws}) are hypotheses/prescriptions for otherwise missing non-perturbative degrees of freedom of the string’s background fields, their investigation goes towards the heart of the open problem finding a non-perturbative completion of string theory itself, famous under the working title {\it M-theory} \cite{HoravaWitten96} \cite{Duff96} \cite{Duff99}.

\smallskip
For instance, the traditional {\it Hypothesis K} (\cref{RRFieldFluxQuantization}) that RR-field fluxes are quantized in topological K-theory has been motivated/justified \cite[\S 3]{Witten98} as describing – or in fact pre-scribing – the stable end results of the tachyon condensation of open string modes stretching between D-brane/anti D-brane pairs, a process which cannot be followed by string perturbation theory, but which is expected (“Sen's conjecture” \cite{Sen98}) to find the non-perturbative true vacuum state where D-brane/anti D-brane pairs have mutually annihilated as far as possible. Indeed, at least in practice, RR-field flux quantization in topological K-theory has become the widely-accepted definition of stable D-brane vacua, and as such must be understood as a partial proposal for the nature of non-perturbative string theory.

\smallskip
On the other hand, strongly-coupled string theory at large-scale/low-energy is also famously argued to be described by D=11 supergravity, whence it stands to reason that flux quantization of the supergravity C-field in 11d should go further still towards the full non-perturbative definition of string theory (hence of M-theory). While the details are subtle and generally deserve more attention, the systematic understanding of non-linear flux quantization reviewed above provides a systematic mathematical theory that clearly delineates the available choices of non-perturbative completions and allows one to rigorously derive their consequences.

\subsection{RR-field flux quantization in 10d}
\label{RRFieldFluxQuantization}

Recall the Gauss law of the unbounded RR-field flux densities (Ex. \ref{MotionOfUnboundedRRFieldFluxes}, Prop. \ref{SolutionSpaceViaGaussLaw}) as commonly expected in {\it massive} type IIA supergravity and ignoring (as commonly done, but see \cite{BMSS19} for possible justification) the non-linear Bianchi identity (Ex. \ref{MotionOfBRRFieldFluxes}) of the dual B-field flux $H_7$ (whence we now notationally suppress $H_7$ altogether, as usual) its admissible flux quantization laws have classifying spaces $\mathcal{A}$ whose $\mathbb{R}$-Sullivan algebra looks as follows:
\begin{equation}
  \label{SolutionSpaceOfUnboundedRRRecalled}
  \mathrm{SolSpace}
  =
  \left\{
  \!\!\!
  \def\arraystretch{1.2}
    \begin{array}{l}
      H_3 \in \Omega^3_{\mathrm{dR}}(X^9)
      \\
      F_{2 \bullet} \in \Omega^{2\bullet}_{\mathrm{dR}}(X^9)
    \end{array}
  \,\middle\vert\,
  \def\arraystretch{1.2}
  \def\arraycolsep{2pt}
  \begin{array}{l}
    \differential\, H_3 = 0
    \\
    \differential\, F_{2\bullet} =
    H_3 \wedge F_{2 \bullet -2}
  \end{array}
  \!
  \right\}
  \;\;
  \Rightarrow
  \;\;
  \mathrm{CE}\big(
    \mathfrak{l}
    \mathcal{A}
  \big)
  \,=\,
  \mathbb{R}[h_3, f_{2\bullet}]
  \Big/
  \left(
    \def\arraycolsep{2pt}
    \begin{array}{l}
      \differential\, h_3 = 0
      \\
      \differential\, f_{2 \bullet}
      =
      h_3 \wedge f_{2\bullet -2}
    \end{array}
  \right).
\end{equation}
Now it so happens that a space $\mathcal{A}$ with this property is given \cite[p. 6]{FreedHopkinsTeleman07} \cite[Lem. 2.31]{BMSS19} by the classifying space $\mathrm{KU}_0$ for complex topological K-theory in degree=0, homotopy-quotiented by an action of the projective unitary group $\mathrm{PU}(\HilbertSpace{H})$ on an essentially unique separably infinite-dimensional complex Hilbert space $\HilbertSpace{H}$:
\vspace{-.1cm}
\begin{equation}
\label{ClassifyingSpaceForTwistedKTheory}
  \begin{tikzcd}[
    column sep=30pt
  ]
    \mathrm{KU}_0
    \ar[
      r,
      "{
        \mathrm{hofib}_p
      }"
    ]
    &
    \overbrace{
    \mathrm{KU}_0
    \sslash
    \mathrm{PU}(\HilbertSpace{H})
    }^{ \mathcal{A} \,:\defneq\, }
    \ar[
      r,
      "{ p }"
    ]
    &
    B \, \mathrm{PU}(\HilbertSpace{H})
    \,\simeq\,
    B^3 \mathbb{Z}
    \,,
  \end{tikzcd}
\end{equation}

\noindent
\hspace{-.2cm}
\def\tabcolsep{0pt}
\begin{tabular}{p{9.3cm}l}
Accordingly, the generalized nonabelian cohomology theory classified by $\mathcal{A} \defneq \mathrm{KU}_0 \sslash \mathrm{PU}(\HilbertSpace{H})$ decomposes over the ordinary integral cohomology in degree=3, with fibers being the abelian Whitehead-generalized cohomology of topological K-theory.
As such it may and traditionally is understood as an abelian but {\it twisted} cohomology theory:  {\it Twisted topological K-theory}
\cite[Def. 3.3]{AtiyahSegal04} \cite[(2.6)]{FreedHopkinsTeleman07}\cite[Ex. 4.5.4]{SatiSchreiber21} \cite[Ex. 3.4]{FSS23Char}.
For a review of twisted K-theory see also the contribution \cite{Rosenberg24} in this collection.
&
\hspace{12pt}
\adjustbox{raise=-1.9cm}{
\begin{minipage}{7.4cm}
\begin{equation}
  \label{TwistedKtheory}
  \begin{tikzcd}[
    column sep=15pt
  ]
    &[30pt]&[-25pt]
    \mathrm{KU}_0 \!\sslash\!
    \mathrm{PU}(H)
    \ar[
      dd,
      "{
        \scalebox{.7}{
          \color{darkblue}
          \bf
          \;\;\;\;
          \def\arraystretch{.9}
          \begin{tabular}{c}
            classifying fibration
            \\
            for twisted
            K-theory
          \end{tabular}
        }
      }"{description}
    ]
    \\
    \\
    X^9
    \ar[
      rr,
      dashed
    ]
    \ar[
      uurr,
      dashed,
      "{
        \scalebox{.7}{
          \color{greenii}
          \bf
          RR-field charges
        }
      }"{sloped}
    ]
    \ar[
      dr,
      shorten >=-5pt,
      dashed,
      "{
        \scalebox{.7}{
          \color{greenii}
          \bf
          \def\arraystretch{.9}
          \begin{tabular}{c}
            background
            \\
            B-field
            charges
          \end{tabular}
        }
      }"{sloped, swap}
    ]
    &&
    \scalebox{.85}{
      $B \mathrm{PU}(H)$
    }
    \ar[
      dl,
      "{ \sim }"{sloped}
    ]
    \\
    &
    \scalebox{.85}{
      $B^2 \mathrm{U}(1)$
    }
  \end{tikzcd}
\end{equation}
\end{minipage}
}
\end{tabular}

\newpage
Moreover, under this decomposition, the nonabelian character map on \eqref{ClassifyingSpaceForTwistedKTheory} is \cite[Prop. 10.1]{FSS23Char} the {\it twisted Chern character} (the archetypical example which gives its name to the more general “character map”).
Therefore, choosing \eqref{ClassifyingSpaceForTwistedKTheory} as the flux quantization law for the unbounded RR-fields means to hypothesize/declare that RR-field flux and hence D-brane charge is quantized in twisted topological K-theory, with the twisted de Rham cohomology-classes of the RR-field flux densities just being the image of these K-theory classes under the twisted Chern character. This is the {\it Hypothesis K} (our terminology) originally due to \cite{MinasianMoore97} \cite{Witten98}, with further perspectives added by \cite{FreedHopkins00}\cite{BouwknegtMathai01} and others.

\begin{remark}[Comparison to the literature]
The original motivation close to the above logic via the character map may be found in \cite{Witten98}, in the paragraph wrapping p. 9-10.
When comparing to \cite{MinasianMoore97} beware that these authors, and the literature following them, take the $\mathbb{R}$-rational D-brane charge to be expressed by the Chern character {\it multiplied} with the square root of the A-hat genus of the tangent bundle of spacetime. However, since this term is multiplicatively invertible, this is not intrinsic to the notion of D-brane charge and may be disregarded for the purpose of charge quantization (cf. \cite[ftn. 12]{FreedHopkins00}); its role is rather as a technical convenience making the Chern character natural under push-forward \cite[\S 2]{BrodzkiMathaiRosenbergSzabo08}.
\end{remark}

\vspace{-.1cm}
\noindent
\hspace{-.2cm}
\def\tabcolsep{0pt}
\begin{tabular}{p{10.5cm}l}
An influential argument why {\it Hypothesis K} \eqref{ClassifyingSpaceForTwistedKTheory} should be singled out among (the infinitude of) other compatible choices of RR-field flux quantization was the observation \cite[\S 3]{Witten98}
that the equivalence relation on virtual vector bundles which characterizes topological K-theory $\mathrm{KU}_0(X)$ on compact Hausdorff spaces plausibly mimics the expected mechanism (“Sen's conjecture” \cite{Sen98}) of D-brane/anti D-brane-annihilation via tachyon condensation, by which pairs $(\HilbertSpace{W},  \overline{\HilbertSpace{W}}$) of isomorphic but opposite Chan-Paton bundles on the worldvolume of coincident D-branes should mutually annihilate.
&
\hspace{10pt}
\adjustbox{raise=-3.4cm}{
\label{ChanPatonFactorAnnihilation}
\includegraphics[width=6cm]{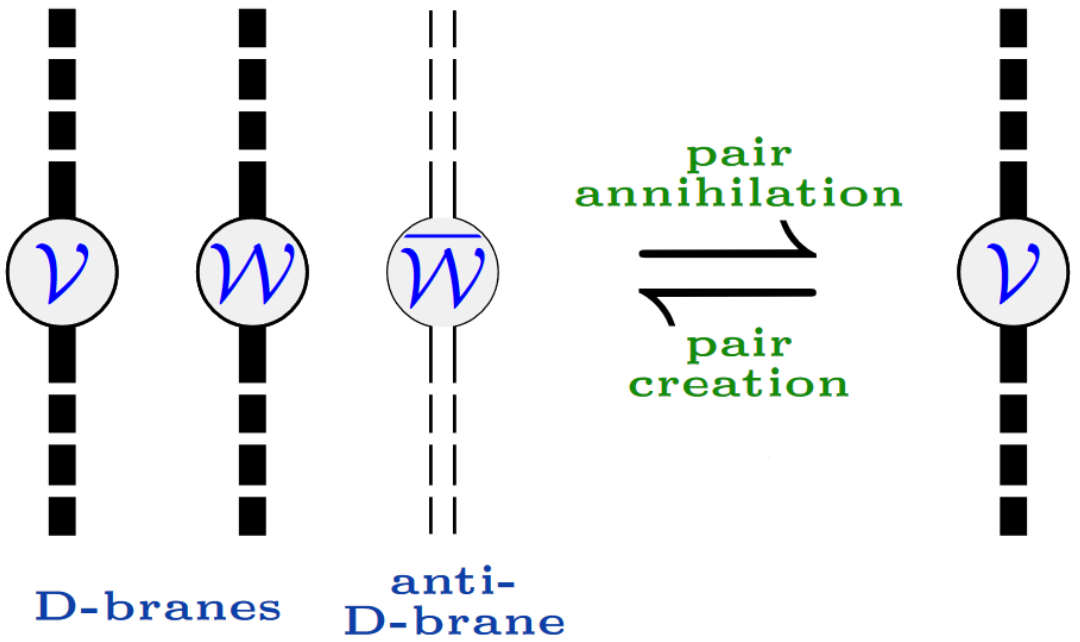}
}
\end{tabular}

\smallskip

We highlight that this is a heuristic argument: There is no string-theoretic computation that actually verifies this intuition (cf. commentary by \cite[p. 32]{Erler13}). In fact, \cite[ftn. 2]{Witten98} already points out that, on closer inspection, it is less clear how the picture should work.
This is noteworthy in view of the fact that the starting point \eqref{SolutionSpaceOfUnboundedRRRecalled} of {\it Hypothesis K} is a little shaky (as discussed after Ex. \ref{MotionOfCFieldFlux}): There is room to doubt that {\it Hypothesis K} is quite the correct flux-quantization law for the RR-field, after all; see also the concerns raised by \cite[\S 4.5.2 \& \S 4.6.5]{BDHKMMS02}\cite[p. 1]{FredenhagenQuella05}\footnote{\cite[p. 1]{FredenhagenQuella05}: ``{It might surprise that despite all the progress that has been made in understanding branes on group manifolds, there are usually not enough D-branes known to explain the whole charge group predicted by (twisted) K-theory. [...] it is fair to say that a satisfactory answer is still missing.}''}\cite[\S 8]{Evslin06}\cite[p. 32]{Erler13}.\footnote{\cite[p. 32]{Erler13}: ``{ It would also be interesting to see if these developments can shed light on the long-speculated relation between string field theory and the K-theoretic description of D-brane charge. We leave these questions for future work.}''}

\medskip

This motivates having a closer look at the flux quantization of the M-theoretic avatar/origin of the RR-fields: The C-field.

\subsection{C-Field flux quantization in 11d}
\label{CFieldFluxQuantization}

Given the C-field‘s Gauss law (Ex. \ref{MotionOfCFieldFlux}, Prop. \ref{SolutionSpaceViaGaussLaw})
its admissible flux quantization laws have a classifying space $\mathcal{A}$ whose $\mathbb{R}$-Sullivan algebra is as shown on the right here:
$$
  \mathrm{SolSpace}
  \;=\;
  \left\{
  \!
  \def\arraystretch{1.1}
  \begin{array}{l}
    B_4 \,\in\, \Omega^4_{\mathrm{dR}}(X^{10})
    \\
    B_7 \,\in\, \Omega^7_{\mathrm{dR}}(X^{10})
  \end{array}
  \;\middle\vert\;
  \def\arraystretch{1.1}
  \begin{array}{rcl}
    \mathrm{d}
    \,
    B_4 &=& 0
    \\
    \mathrm{d}
    \,
    B_7 &=&
    -\tfrac{1}{2}
    B_4 \wedge B_4
  \end{array}
  \!
  \right\}
  \;\;
  \Rightarrow
  \;\;
  \mathrm{CE}(\mathfrak{l}\mathcal{A})
  \;
  =
  \;
  \mathbb{R}\big[
    b_4, b_7
  \big]
  \Big/\!
  \left(
  \def\arraycolsep{2pt}
  \begin{array}{rcl}
    \mathrm{d}
    \,
    b_4 &=& 0
    \\
    \mathrm{d}
    \,
    b_7 &=&
    -\tfrac{1}{2}
    b_4 \wedge b_4
  \end{array}
 \! \right)
  .
$$
We review now two possible choices of such flux quantization laws $\mathcal{A}$ for the C-field that have been considered in the literature (here we denote them by $\mathcal{A}_{{}_{\mathrm{DFM}}}$ and $\mathcal{A}_{{}_{\mathrm{FSS}}}$, respectively), both of which, while quite distinct from each other, being an "evident" choice from their respective natural perspective.
\footnote{The are other proposals that advocate seeking a generalized cohomology underlying the fields in M-theory, see
\cite{Sati05}\cite{Sati06}. }

Recall again that, besides these "evident" choices, there is an infinitude of admissible variant flux quantization laws which differ in their torsion content. In the present case, any such choice is a hypothesis/definition concerning aspects of the elusive M-theory. Careful investigation of the implications of the "evident" flux quantization laws of the C-field may not only serve to decide if either is "correct" (which is not always straightforward to decide, as long as a plausibly complete definition of M-theory remains missing), but also to understand how variant flux quantization laws would have to be chosen if the "evident" ones are deemed to have undesired implications.

\medskip

\paragraph{DFM-like flux-quantization.}
If one takes the point of view that a higher $\mathrm{U}(1)$-flux quantization law as in Ex. \ref{HigherCircleGaugePotentials} is the most natural starting point, which naively demands $G_4$ to be quantized in integral 4-cohomology with classifying space $B^4 \mathbb{Z} \,\defneq\, K(\mathbb{Z}, 4)$, then one is naturally led to consider the deformation of this situation which just adds-on the condition that half the cup-square of this 4-class be trivialized in rational cohomology.

\noindent
\hspace{-.2cm}
\def\tabcolsep{0pt}
\begin{tabular}{p{8.5cm}l}
In terms of classifying spaces, this means to pass to the homotopy fiber, here to be denoted $\mathcal{A}_{{}_{\mathrm{DFM}}}$, of the map that classifies (minus) half the cup-square cohomology operation on integral 4-cohomology. This has the required Sullivan model, as shown (these kinds of computations are reviewed in \cite[§1, §5]{FSS23Char}):
&
\;\;
\adjustbox{raise=-.9cm}{
$
  \begin{tikzcd}[
    row sep=0pt
  ]
    \scalebox{.7}{
      \color{greenii}
      \bf
      classifying map
    }
    &[-30pt]
    B^4 \mathbb{Z}
    \ar[
      rr,
      "{
        -\tfrac{1}{2}\mathrm{sq}
      }"
    ]
    &\phantom{A}&
    B^8 \mathbb{Q}
    \\
    \scalebox{.7}{
      \color{greenii}
      \bf
      Sullivan model
    }
    &
    \frac{
     \mathbb{R}[
        g_4
      ]
    }
    {
    (
      \differential
      \,
      g_4 = 0
    )
    }
    \ar[
      from=rr,
      "{
        -\tfrac{1}{2}
        g_4 \wedge g_4
        \;\,\mapsfrom\;\,
        q_8
      }"{swap}
    ]
    &&
    \frac{
     \mathbb{R}[
        q_8
      ]
    }
    {
    (
      \differential
      \,
      q_8 = 0
    )
    }
    \\
    \scalebox{.7}{
      \color{greenii}
      \bf
      cohomology operation
    }
    &
    H^4(-;\mathbb{Z})
    \ar[
      rr,
      "{
        -
        \tfrac{1}{2}
        \mathrm{sq}_\ast
        =
        -\tfrac{1}{2}(-)^{\cup^2}
      }"
    ]
    &&
    H^8(-;\mathbb{Q})
  \end{tikzcd}
$
}
\end{tabular}

\vspace{.2cm}

\begin{equation}
\label{HomotopyFiberOfFractionalCupSquare}
\adjustbox{}{
\def\arraystretch{.9}
\begin{tabular}{ccccc}
\scalebox{.7}{
\color{darkblue}
\bf
\def\arraystretch{.9}
\begin{tabular}{c}
  homotopy fiber
  \\
  of classifying map
  \\
  of {\color{orangeii} fractional}
  \\
  cup-square operation
\end{tabular}
}
&&
\scalebox{.7}{
\color{darkblue}
\bf
\def\arraystretch{.9}
\begin{tabular}{c}
  presented as
  \\
  principal 8-bundle
  \\
  over classifying space
\end{tabular}
}
&&
\scalebox{.7}{
\color{darkblue}
\bf
\def\arraystretch{.9}
\begin{tabular}{c}
  image in
  \\
  Sullivan models
\end{tabular}
}
\\[-5pt]
$
  \begin{tikzcd}[sep=15pt]
    \mathcal{A}_{{}_{\mathrm{DFM}}}
    \ar[
      rr,
      "{\ }"{pos=.9, name=s}
    ]
    \ar[
      dd,
      "{\ }"{pos=.9, swap, name=t}
    ]
    &&
    \ast
    \ar[
      dd,
      "{ 0 }"
    ]
    \\
    \\
    B^4 \mathbb{Z}
    \ar[
      rr,
      "{
        \scalebox{.7}{$-
        \color{orangeii}\tfrac{1}{2}$}
        \mathrm{sq}
      }"
    ]
    &&
    B^8 \mathbb{Q}
    \ar[
      from=s,
      to=t,
      Rightarrow,
      shorten=15pt,
      "{
        \scalebox{.7}{(hpb)}
      }"
    ]
  \end{tikzcd}
$
&
$
\;\;
\underset{\mathrm{hmtp}}{\simeq}
\;\;
$
&
$
  \begin{tikzcd}[sep=15pt]
    \mathcal{A}_{{}_{\mathrm{DFM}}}
    \ar[
      rr,
      "{\ }"{pos=.9, name=s}
    ]
    \ar[
      dd,
      ->>,
      "{\ }"{pos=.9, swap, name=t}
    ]
    \ar[
      ddrr,
      phantom,
      "{
        \scalebox{.7}{(pb)}
      }"
    ]
    &&
    E B^7 \mathbb{Q}
    \ar[
      dd,
      ->>,
      "{ p }"
    ]
    \\
    \\
    B^4 \mathbb{Z}
    \ar[
      rr,
      "{
        -
        \scalebox{.7}{\color{orangeii}$\tfrac{1}{2}$}
        \mathrm{sq}
      }"
    ]
    &&
    B^8 \mathbb{Q}
  \end{tikzcd}
$
& \;\;
$
    \xmapsto{
      \mathrm{CE}(
        \mathfrak{l}(-)
      )
    }
$
\;
&
\;\;
$
  \begin{tikzcd}[row sep=10, column sep=30pt]
    \frac{
      \mathbb{R}[
        g_4,\, g_7
      ]
    }{
    \left(
      \scalebox{.6}{$
      \def\arraycolsep{0pt}
      \begin{array}{l}
        \differential\, g_4 = 0
        \\
        \differential\,  g_7
        =
        -
        \tfrac{1}{2}
        g_4 \wedge g_4
      \end{array}
      $}
    \right)
    }
    \ar[
      from=rr,
      "{
        \scalebox{.7}{$
        \def\arraystretch{.85}
        \begin{array}{r}
          g_7 \,\mapsfrom\, g_7
          \\
          -
          \tfrac{1}{2}
          g_4 \wedge g_4 \,\mapsfrom\, q_8
        \end{array}
        $}
      }"{swap}
    ]
    \ar[
      from=dd,
      hook'
    ]
    \ar[
      rrdd,
      phantom,
      "{
        \scalebox{.7}{(po)}
      }"
    ]
    &&
    \frac{
      \mathbb{R}[g_7, q_8]
    }{
      \left(
      \scalebox{.6}{$
      \def\arraycolsep{0pt}
      \begin{array}{l}
        \differential\, q_8 = 0
        \\
        \differential\, g_7 = q_8
      \end{array}
      $}
      \right)
    }
    \ar[
      from=dd,
      hook',
      "{
        \scalebox{.7}{$
        \def\arraystretch{.9}
        \begin{array}{c}
          q_8
          \\
          \mapsup
          \\
          q_8
        \end{array}
        $}
      }"{swap}
    ]
    \\
    \\
    \frac{
      \mathbb{R}[g_4]
    }{
      (
      \differential
      \,
      g_4 = 0
      )
    }
    \ar[
      from=rr,
      "{
        -
        \scalebox{.7}{$\tfrac{1}{2}$}
        g_4 \wedge g_4
        \,\mapsfrom\,
        q_8
      }"{swap}
    ]
    &&
    \frac{
      \mathbb{R}[q_8]
    }{
    (
      \differential
      \,
      q_8 = 0
    )
    }
  \end{tikzcd}
$
\end{tabular}
}
\end{equation}

This flux quantization law corresponds essentially to  the model of the C-field considered
in \cite{DiaconescuFreedMoore07}\footnote{
  Beware that much of the content of \cite{DiaconescuFreedMoore07} is concerned with first adjoining an $E_8$ Yang-Mills field to the C-field and then imposing gauge equivalences which make this field disappear again up to gauge equivalence; see (3.11) there. \label{RoleOfE8InDFM}
} \cite{Moore05} following \cite{HopkinsSinger05} --  when specialized to the case where the Pontrjagin classes of spacetime vanish (such as for near horizon geometries of flat singular branes, \cref{SingularVersusSolitonicBranes}, by \cite[Prop. 22]{SS21-M5}), namely it enforces integer $G_4$-flux quantization much as in Ex. \ref{HigherCircleGaugePotentials} while implementing $J_8 \,:=\, \tfrac{1}{2}G_4 \wedge G_4$ as an electric source term, but  essentially no $G_7$-flux quantization is enforced.

By the long exact sequence of homotopy groups associated with the fiber sequence \eqref{HomotopyFiberOfFractionalCupSquare} it follows that the homotopy groups of $\mathcal{A}_{{}_{\mathrm{DFM}}}$ are concentrated in degrees 4 and 7, which implies that the flat singular branes (cf. \cref{SingularVersusSolitonicBranes}) it predicts are exactly integer numbers of M5-branes and any (rational) number of M2-branes (cf. \cref{HigherFluxesAndTheirBraneSources})
$$
  \pi_n\big(
    \mathcal{A}_{{}_{\mathrm{DFM}}}
  \big)
  \;=\;
  \left\{
  \begin{array}{ccl}
    \mathbb{Z} &\vert& n = 4
    \\
    \mathbb{Q} &\vert& n = 7
    \\
    0 &\vert& \mbox{otherwise}
  \end{array}
  \right.
  \;
  \Rightarrow
  \;
  \def\arraystretch{1.6}
  \begin{array}{c}
    \scalebox{.7}{
      \def\arraystretch{.9}
      \color{darkblue}
      \bf
      \begin{tabular}{c}
        charges of flat
        \\
        singular M5-branes
      \end{tabular}
    }
    =
    H^1\big(
      \mathbb{R}^{10,1}
      \setminus
      \mathbb{R}^{5,1}
      ;\,
      \Omega \mathcal{A}_{{}_{\mathrm{DFM}'}}
    \big)
    \;\simeq\;
    \pi_4\big(
      \mathcal{A}_{{}_{\mathrm{DFM}'}}
    \big)
    \;\simeq\;
    \mathbb{Z}
    \,,
    \\
    \scalebox{.7}{
      \def\arraystretch{.9}
      \color{darkblue}
      \bf
      \begin{tabular}{c}
        charges of flat
        \\
        singular M2-branes
      \end{tabular}
    }
    =
    H^1\big(
      \mathbb{R}^{10,1}
      \setminus
      \mathbb{R}^{2,1}
      ;\,
      \Omega \mathcal{A}_{{}_{\mathrm{DFM}'}}
    \big)
    \;\simeq\;
    \pi_7\big(
      \mathcal{A}_{{}_{\mathrm{DFM}'}}
    \big)
    \;\simeq\;
    \mathbb{Q}
    \,.
  \end{array}
$$
Of course, with charge-quantization in generalized non-abelian cohomology (\cref{FluxQuantizationLawsAsNonabelianCohomology})
it is straightforward to fix this, by forming instead the homotopy fiber of the integral
$-\mathrm{sq} : B^4 \mathbb{Z} \to B^8 \mathbb{Z}$ and using the freedom in isomorphy of Sullivan models to rescale the generator $g_7$ by 2:
\begin{equation}
\label{HomotopyFiberOfCupSquare}
\hspace{-5mm}
\adjustbox{}{
\def\arraystretch{.9}
\begin{tabular}{ccccc}
\scalebox{.7}{
\color{darkblue}
\bf
\def\arraystretch{.9}
\begin{tabular}{c}
  homotopy fiber
  \\
  of classifying map
  \\
  of {\color{orangeii} integral}
  \\
  cup-square operation
\end{tabular}
}
&&
\scalebox{.7}{
\color{darkblue}
\bf
\def\arraystretch{.9}
\begin{tabular}{c}
  presented as
  \\
  bundle 6-gerbe
  \\
  over classifying space
\end{tabular}
}
&&
\scalebox{.7}{
\color{darkblue}
\bf
\def\arraystretch{.9}
\begin{tabular}{c}
  image in
  \\
  Sullivan models
\end{tabular}
}
\\[-5pt]
$
  \begin{tikzcd}[sep=15pt]
    \mathcal{A}_{{}_{\mathrm{DFM}'}}
    \ar[
      rr,
      "{\ }"{pos=.9, name=s}
    ]
    \ar[
      dd,
      "{\ }"{pos=.9, swap, name=t}
    ]
    &&
    \ast
    \ar[
      dd,
      "{ 0 }"
    ]
    \\
    \\
    B^4 \mathbb{Z}
    \ar[
      rr,
      "{
        -
        \mathrm{sq}
      }"
    ]
    &&
    B^8 \mathbb{Z}
    \ar[
      from=s,
      to=t,
      Rightarrow,
      shorten=15pt,
      "{
        \scalebox{.7}{(hpb)}
      }"
    ]
  \end{tikzcd}
$
&
$
\;\;
\underset{\mathrm{hmtp}}{\simeq}
\;\;
$
&
$
  \begin{tikzcd}[sep=15pt]
    \mathcal{A}_{{}_{\mathrm{DFM}'}}
    \ar[
      rr,
      "{\ }"{pos=.9, name=s}
    ]
    \ar[
      dd,
      ->>,
      "{\ }"{pos=.9, swap, name=t}
    ]
    \ar[
      ddrr,
      phantom,
      "{
        \scalebox{.7}{(pb)}
      }"
    ]
    &&
    E B^6 \mathrm{U}(1)
    \ar[
      dd,
      ->>,
      "{ p }"
    ]
    \\
    \\
    B^4 \mathbb{Z}
    \ar[
      rr,
      "{
        -
        \mathrm{sq}
      }"
    ]
    &&
    B^7 \mathrm{U}(1)
  \end{tikzcd}
$
& \;\;
$
    \xmapsto{
      \mathrm{CE}(
        \mathfrak{l}(-)
      )
    }
$
\;\;
&
$
  \begin{tikzcd}[row sep=10, column sep=30pt]
    \frac{
      \mathbb{R}[
        g_4,\,
        {\color{orangeii}2}g_7
      ]
    }{
    \left(
      \scalebox{.6}{$
      \def\arraycolsep{0pt}
      \begin{array}{l}
        \differential\, g_4 = 0
        \\
        \differential\,
        {\color{orangeii}2}g_7
        =
        -
        g_4 \wedge g_4
      \end{array}
      $}
    \right)
    }
    \ar[
      from=rr,
      "{
        \scalebox{.7}{$
        \def\arraystretch{.85}
        \begin{array}{r}
          {\color{orangeii}2}g_7
          \,\mapsfrom\,
          {\color{orangeii}2}g_7
          \\
          -
          g_4 \wedge g_4 \,\mapsfrom\, q_8
        \end{array}
        $}
      }"{swap}
    ]
    \ar[
      from=dd,
      hook'
    ]
    \ar[
      rrdd,
      phantom,
      "{
        \scalebox{.7}{(po)}
      }"
    ]
    &&
    \frac{
      \mathbb{R}[
      {\color{orangeii}2}g_7
      ,
      q_8
      ]
    }{
      \left(
      \scalebox{.6}{$
      \def\arraycolsep{0pt}
      \begin{array}{l}
        \differential\, q_8 = 0
        \\
        \differential\,
        {\color{orangeii}2}g_7
        =
        q_8
      \end{array}
      $}
      \right)
    }
    \ar[
      from=dd,
      hook',
      "{
        \scalebox{.7}{$
        \def\arraystretch{.9}
        \begin{array}{c}
          q_8
          \\
          \mapsup
          \\
          q_8
        \end{array}
        $}
      }"{swap}
    ]
    \\
    \\
    \frac{
      \mathbb{R}[g_4]
    }{
      (
      \differential
      \,
      g_4 = 0
      )
    }
    \ar[
      from=rr,
      "{
        -
        g_4 \wedge g_4
        \,\mapsfrom\,
        q_8
      }"{swap}
    ]
    &&
    \frac{
      \mathbb{R}[q_8]
    }{
    (
      \differential
      \,
      q_8 = 0
    )
    }
  \end{tikzcd}
$
\end{tabular}
}
\end{equation}
This adjusted flux-quantization law ``$\mathcal{A}_{{}_{\mathrm{DFM}'}}$'' now enforces the desired M2-charge quantization:
$$
  \pi_n\big(
    \mathcal{A}_{{}_{\mathrm{DFM}'}}
  \big)
  \;=\;
  \left\{
  \begin{array}{ccl}
    \mathbb{Z} &\vert& n = 4
    \\
    \mathbb{Z} &\vert& n = 7
    \\
    0 &\vert& \mbox{otherwise}
  \end{array}
  \right.
  \;
  \Rightarrow
  \;
  \def\arraystretch{1.6}
  \begin{array}{c}
    \scalebox{.7}{
      \def\arraystretch{.9}
      \color{darkblue}
      \bf
      \begin{tabular}{c}
        charges of flat
        \\
        singular M5-branes
      \end{tabular}
    }
    =
    H^1\big(
      \mathbb{R}^{10,1}
      \setminus
      \mathbb{R}^{5,1}
      ;\,
      \Omega \mathcal{A}_{{}_{\mathrm{DFM}'}}
    \big)
    \;\simeq\;
    \pi_4\big(
      \mathcal{A}_{{}_{\mathrm{DFM}'}}
    \big)
    \;\simeq\;
    \mathbb{Z}
    \,,
    \\
    \scalebox{.7}{
      \def\arraystretch{.9}
      \color{darkblue}
      \bf
      \begin{tabular}{c}
        charges of flat
        \\
        singular M2-branes
      \end{tabular}
    }
    =
    H^1\big(
      \mathbb{R}^{10,1}
      \setminus
      \mathbb{R}^{2,1}
      ;\,
      \Omega \mathcal{A}_{{}_{\mathrm{DFM}'}}
    \big)
    \;\simeq\;
    \pi_7\big(
      \mathcal{A}_{{}_{\mathrm{DFM}'}}
    \big)
    \;\simeq\;
    \mathbb{Z}
    \,.
  \end{array}
$$
(cf. the claim in \cite[\S 5]{Moore05}).

But it still does not {\it predict} the half-integral shift of the M5-brane charge, nor the correction of the electric source by the $I_8$-term in the case that the Pontrjagin classes of spacetime do not vanish, even though these effects can be added ``by hand''.
In order to see these effects arise automatically we turn to yet another possible flux-quantization law of the C-field:

\medskip

\paragraph{FSS flux-quantization.}

Another perspective is to regard the baseline of all flux quantization to be that classified by the point $\mathcal{A}_0 \defneq \ast$ (for the trivial theory) and to obtain non-trivial classifying spaces from this maximally unbiased starting point by iterated attachment of cells in the sense of CW-complexes (e.g. \cite[p. 5]{Hatcher02}).
The {\it minimal} choice of C-field flux quantization in this sense, requiring the minimum number 1 of cell attachments, is to take $\mathcal{A}_{{}_{\mathrm{FSS}}} \defneq S^4$ to be the (homotopy type of) the 4-sphere \cite[\S 2.5]{Sati13} (which is a valid choice of C-field flux quantization, by the examples on p. \pageref{ExamplesOfSullivanModels}).

\noindent
\hspace{-.2cm}
\def\tabcolsep{0pt}
\begin{tabular}{p{8.2cm}l}
The  generalized nonabelian cohomology theory classified by the $n$-spheres is known as {\it Cohomotopy}
\cite{Borsuk1936}\cite{Pontrjagin38}\cite{Spanier49}
\cite{Peterson56},
being dual to the unstable homology theory constituted by the homotopy groups of spaces.
& \qquad
\adjustbox{raise=-.8cm}{\small
\begin{tabular}{ll}
\def\tabcolsep{5pt}
\def\arraystretch{1.4}
\begin{tabular}{|c|p{3.75cm}|c|}
  \hline
  \scalebox{1}{
    \def\arraystretch{.9}
    \begin{tabular}{c}
      \vspaceabove
      Unstable
      \\
      \bf
      homology theory
      \vspacebelow
    \end{tabular}
  }
  &
  \scalebox{1}{
    \def\arraystretch{.9}
    \begin{tabular}{c}
      Unstable (nonabelian)
      \\
      \bf
      cohomology theory
    \end{tabular}
  }
  \\
  \hline
  \hline
  \scalebox{.9}{
    \color{darkblue}
    \bf
    homotopy
  }
  &
  \scalebox{.9}{
    \color{darkblue}
    \bf
    cohomotopy
  }
  \\
  $
  \pi_n(X)
  \defneq
  \pi_0
    \mathrm{Map}^{\ast/}\big(
    S^n
    ,
    X
  \big)
  $
  &
  $
  \pi^n(X)
  \defneq
  \pi_0
    \mathrm{Map}^{\ast/}\big(
    X
    ,
    S^n
  \big)
  $
  \\
  \hline
\end{tabular}
\end{tabular}
}
\end{tabular}

\vspace{1mm}
\noindent
Therefore, the hypothesis that the proper classifying space for C-field flux quantization is the 4-sphere may be called {\it Hypothesis H}, for ``{\it H}omotopy cohomology theory'' \cite{FSS20-H}\cite{SatiSchreiber20}\cite{GradySati2021}
\cite{FSS19HopfWZ}\cite{FSS20TwistedString}\cite{SS21-M5}, review in \cite[\S 12]{FSS23Char}.

\medskip

\noindent
\hspace{-.2cm}
\def\tabcolsep{0pt}
\begin{tabular}{p{9.9cm}l}
{\bf Hypothesis H on flat spacetimes.} The non-torsion homotopy groups of $S^4$ are exactly in degrees 4 and in degree 7 (whose generator is the quaternionic Hopf fibration, cf. \cite[p. 4]{FSS20-H}). This implies that Hypothesis H gives the expected integral charge quantization for flat singular M5-branes (cf. \cref{SingularVersusSolitonicBranes} \& \cref{HigherFluxesAndTheirBraneSources}) {\it and} for  flat singular M2-branes -- the ``Page charge'' \eqref{PageChargeFlux}.
&
\;\;\;
\adjustbox{raise=-1cm}{
\def\tabcolsep{5pt}
\begin{tabular}{|l|}
    \hline
    \!\!\!\!\!\!
    $
    \def\arraystretch{1.4}
    \begin{array}{l}
    \mbox{
      \color{darkblue}
      $
    \pi^4\big(
      \mathbb{R}^{10,1}
      \setminus
      \mathbb{R}^{5,1}
    \big)
    $}
    \;=\;
    \pi^4\big(
      \mathbb{R}^{5,1}
      \times
      \mathbb{R}_+
      \times
      S^4
    \big)
    \\
    \;=\;
    \pi^4(S^4)
    \,=\,
    \pi_4(S^4)
    \,=\,
    \mbox{\color{darkblue}
    $\mathbb{Z}$
    }
    \end{array}
  $
\\
\hline
\!\!\!    $
    \def\arraystretch{1.4}
    \begin{array}{l}
    {
    \mbox{
    \color{darkblue}
    $
    \pi^4\big(
      \mathbb{R}^{10,1}
      \setminus
      \mathbb{R}^{2,1}
    \big)
    $}
    }
    \;=\;
    \pi^4\big(
      \mathbb{R}^{2,1}
      \times
      \mathbb{R}_+
      \times
      S^7
    \big)
    \\
    \;=\;
    \pi^4(S^7)
    \,=\,
    \pi_7(S^4)
    \,=\,
    \mbox{
      \color{darkblue}
      $\mathbb{Z}$
    }
    \color{gray}
    \oplus
    \mathbb{Z}_{12}
    \end{array}
  $
\\
\hline
\end{tabular}
}
\end{tabular}
\begin{equation}
  \label{PureFlatM2BraneCharge}
  \begin{tikzcd}
    S^3 \ar[r, hook]
    &[-10pt]
    \overset{
     \mathclap{
       \scalebox{.7}{
         \color{darkblue}
         \bf
         \def\arraystretch{.9}
         \begin{tabular}{c}
           quaternionic
           \\
           Hopf fibration
         \end{tabular}
       }
     }
    }{
      S^7
    }
    \ar[d, "{ h_{\mathbb{H}} }"]
    \\
    & S^4
  \end{tikzcd}
  \hspace{1cm}
  \big[
    S^7
      \xrightarrow{h_{\mathbb{H}}}
    S^4
  \big]
  \;=\;
  1
  \in
  \mathbb{Z}
  \hookrightarrow
  \pi_7(S^4)
  \;\;\;
  \begin{tikzcd}[
    row sep=2pt
  ]
    &&
    S^7
    \ar[
      dd,
      "{ h_{\mathbb{H}} }"
    ]
    &[-10pt]
    \in \pi^7(X)
    \ar[
      dd,
      shift left=8pt,
      "{ (h_{\mathbb{H}})_\ast }"
    ]
    &[-20pt]
    \scalebox{.7}{
      \color{darkblue}
      \bf
      \def\arraystretch{.8}
     \begin{tabular}{c}
       pure
       \\
       $\mathrm{M}_2$-brane charges
     \end{tabular}
    }
    \\
    && & &
    \scalebox{.7}{
      mapping into
    }
    \\
    X^{10}
    \ar[
      uurr,
      dashed,
      "{ c_{6} }"
    ]
    \ar[
      rr,
      dashed,
      "{
        c_3
        \,=\,
        (h_{\mathbb{H}})_\ast c_6
      }"{swap}
    ]
    &&
    S^4
    &
    \in
    \pi^4(X)
    &
    \scalebox{.7}{
      \color{darkblue}
      \bf
      \def\arraystretch{.8}
     \begin{tabular}{c}
       full
       \\
       $\mathrm{M}$-brane charges
     \end{tabular}
    }
  \end{tikzcd}
\end{equation}

\vspace{-1mm}
\noindent On the other hand, $S^4$ also has plenty of torsion homotopy groups \cite[(22)]{SS23MF}: Under Hypothesis H, each of them is a prediction of novel ``fractional M-brane'' species (such as of fractional M2-branes, cf. \cite[\S 2.2]{AharonyBergmanJafferis08}, of order 12, even in flat space) which do not manifest as BPS-states of supergravity.

Notice that the prediction of stable non-BPS branes carrying torsion charges is a generic property of flux quantization laws (in fact: their defining property, cf. \eqref{RationalHomotopyAsNontorsionAspect}), and is well-familiar in the context of Hypothesis K (\cref{RRFieldFluxQuantization}): cf.
\cite{Braun2000}\cite{BrunnerDistler02}\cite{BrunnerDistlerMahajan02}.

The specific torsion content in the C-field that is implied by Hypothesis H has the following consequences:

\vspace{1mm}
\noindent
{\bf Divided powers of M5-brane charge.} Noticing that the generator $S^4 \to B^4 \mathbb{Z}$ of $\pi_4\big(B^4 \mathbb{Z}\big) \simeq \mathbb{Z}$ induces a cohomology operation $\pi^4(-) \to H^4(-;\mathbb{Z})$, there is an integer-cohomology class $\gamma_4$ underlyig the Cohomotopical C-field charge. This has the following properties:

\begin{itemize}[leftmargin=.7cm]
\item[{\bf (i)}] $\gamma_4 \cup \gamma_4$ is divisible by 2, thus making the C-field's electric source $-\tfrac{1}{2} G_4 \wedge G_4$ be integral

\cite[Prop. 2.7 (iii)]{GradySati2021}, cf. \cite[p. 29]{DiaconescuFreedMoore07};

\item[{\bf (ii)}] $\gamma_4 \cup \gamma_4 \cup \gamma_4$ is divisbible by 6, thus making the 11d CS-term, locally $\tfrac{1}{6}C_3 \wedge G_4 \wedge G_4$, be globally well-defined

\cite[p. 12]{GradySati2021}, cf. \cite[p. 10]{Witten97}.
\end{itemize}

\noindent

\vspace{-3pt}
\noindent
\hspace{-.2cm}
\def\tabcolsep{0pt}
\begin{tabular}{p{9.6cm}l}
{\bf Worldvolumes in Cobordism cohomology.}
The Pontrjagin theorem \cite{Pontrjagin38} \cite[\S IX]{Kosinski93} identifies unstable $n$-Cohomotopy with (unstable, framed) Cobordism,
&
\hspace{15pt}
\adjustbox{raise=-.5cm}{
$
  \begin{tikzcd}[column sep=20pt]
    \overset{
      \mathclap{
       \scalebox{.7}{
         \color{darkblue}
         \bf
          $n$-Cohomotopy
        }
      }
    }{
     \pi^n
     \big(
       X^d
     \big)
    }
     \ar[
       rr,
       "{
         \scalebox{.7}{
           \color{greenii}
           \bf
           \def\arraystretch{.9}
           \begin{tabular}{c}
             Pontrjagin
             \\
             theorem
           \end{tabular}
         }
       }"{pos=.6},
       "{ \sim }"{swap, pos=.6}
     ]
     &&
     \mathrm{Cob}^n_{\mathrm{Fr}}\big(
       X^d
     \big)
     \hspace{-11pt}
     \adjustbox{raise=4pt}{
         \raisebox{3pt}{
           \scalebox{.7}{
             \color{darkblue}
             \bf
             \def\arraystretch{.9}
             \begin{tabular}{c}
               cobordism classes
               \\
               of normally framed
               \\
               submanifolds of
               \\
               codimension=$n$.
             \end{tabular}
           }
         }
    }
  \end{tikzcd}
$
}
\end{tabular}
suggestive of the
 {\it worldvolumes of  solitonic M-branes} (\cref{SingularVersusSolitonicBranes}) carrying Cohomotopy charge, cf. \cite[\S2.1]{SatiSchreiber20}\cite[p. 13]{SS23MF}\cite[\S 2.4]{SS22Config}.
Here the cobordism relation reflects \mbox{(anti-)}brane

\noindent
\hspace{-1.5mm}
\def\tabcolsep{0pt}
\begin{tabular}{p{3.4cm}l}
 annihilation (and creation) much as expected in K-theory (cf. p. \pageref{ChanPatonFactorAnnihilation}).
{\it Stabilized}  Cohomotopy is equivalent to algebraic K-theory over the ``absolute base field $\mathbb{F}_1$'' \cite[Thm. 5.9]{ChuLorscheidSanthanam12}.
&
\hspace{12pt}
\adjustbox{
  raise=-1.9cm
}{
\adjustbox{
  scale=1.1,
  raise=-1.8cm
}{
\begin{tikzpicture}[
  rotate=90
]

  \draw[fill=lightgray, draw opacity=0, fill opacity=.4]
    (-3,-3) rectangle (0,3);

  \draw[line width=5, lightgray]
    (0,3) to (0,-3);

  \begin{scope}

  \clip (-2.2,-2.2) rectangle (0,2.2);

  \draw[line width=2]
    (0,0) circle (2);

  \end{scope}

  \draw[->, orangeii]
    (90:2) to  (90:2.5);

  \draw[->, orangeii]
    (90+30:2) to  (90+30:2.5);

  \draw[->, orangeii]
    (90+60:2) to  (90+60:2.5);

  \draw[->, orangeii]
    (90+90:2) to  (90+90:2.5);

  \draw[->, orangeii]
    (-90:2) to  (-90:2.5);

  \draw[->, orangeii]
    (-90-30:2) to  (-90-30:2.5);

  \draw[->, orangeii]
    (-90-60:2) to  (-90-60:2.5);

  \draw[fill=black]
    (0,2) circle (.1);

  \draw[fill=white]
    (0,-2) circle (.1);

  \draw
    (-.4, 3.1)
    node
    {
      \rlap{
      \scalebox{.5}{
      \color{orangeii}
      \bf
      \def\arraystretch{.9}
      \begin{tabular}{c}
        normal
        \\
        framing
        \\
        in space
      \end{tabular}
      }
      }
    };

  \draw
    (+.3, 2.5)
    node
    {
      \rlap{
      \scalebox{.5}{
      \color{darkblue}
      \bf
      \begin{tabular}{c}
        brane
      \end{tabular}
      }
      }
    };

  \draw
    (-.4, -1.85)
    node
    {
      \rlap{
      \scalebox{.5}{
      \color{orangeii}
      \bf
      \def\arraystretch{.9}
      \begin{tabular}{c}
        opposite
        \\
        normal
        \\
        framing
      \end{tabular}
      }
      }
    };

  \draw
    (.3, -1.3)
    node
    {
      \rlap{
      \scalebox{.5}{
      \color{darkblue}
      \bf
      \begin{tabular}{c}
        anti-brane
      \end{tabular}
      }
      }
    };

  \draw
    (-1.9, -1.1)
    node
    {
      \rlap{
      \scalebox{.5}{
      \color{orangeii}
      \bf
      \begin{tabular}{c}
        normal framing
        \\
        in spacetime
      \end{tabular}
      }
      }
    };

  \draw
    (-2.7, +1.5)
    node
    {
      \rlap{
      \scalebox{.5}{
      \color{darkblue}
      \bf
      \begin{tabular}{c}
        spacetime
      \end{tabular}
      }
      }
    };

  \node[
    rotate=90,
    xscale=-1
  ]
  at
    (-1,0)
    {
      $
        {
          \xleftrightharpoons{
          \adjustbox{scale={-1}{1}}{
            \tiny
            \color{greenii}
            pair creation
          }
          }
        }
      $
    };

  \node[
    rotate=90
  ]
    at (-1,-.38)
    {
      \raisebox{10pt}{
        \tiny
        \color{greenii}
        annihilation
      }
    };

  \node
    at (0,+.9)
    {
      \scalebox{.5}{
      \color{darkblue}
      \bf
        space
      }
    };

\end{tikzpicture}
}
\hspace{-.5cm}
\adjustbox{
  scale=1.1,
  raise=-1.6cm
}{
\begin{tikzpicture}
\def\reduce{.4}
\begin{scope}[shift={(-.8,0)}]
  \draw[fill = black]
    (-.05,1.5-\reduce) rectangle (+.05,-1.5+\reduce);
  \draw[fill=white]
    (0,0) circle (.23);
  \draw[fill=lightgray, fill opacity=.6]
    (0,0) circle (.23);
  \begin{scope}
    \clip
      (0,0) circle (.22);
    \draw (0,0)
      node
      {
        \color{blue}
        $f$
      };
  \end{scope}
  \node
    at
    (-.8,0)
    {
      \scalebox{.6}{
        \color{orangeii}
        \bf
        \def\arraystretch{.8}
        \begin{tabular}{c}
          framing
          \\
          charge
        \end{tabular}
      }
    };
\end{scope}

\begin{scope}[shift={(-0,0)}]
  \draw[fill = black]
    (-.05,1.5-\reduce) rectangle (+.05,-1.5+\reduce);
  \draw[fill=white]
    (0,0) circle (.23);
  \draw[fill=lightgray, fill opacity=.6]
    (0,0) circle (.23);
  \begin{scope}
    \clip
      (0,0) circle (.22);
    \draw (0,0)
      node
      {
        \color{blue}
        $w$
      };
  \end{scope}
\end{scope}

\begin{scope}[shift={(+.8,0)}]
  \draw[fill=lightgray, fill opacity=.6]
    (0,0) circle (.23);
  \draw[fill=white, draw opacity=0]
    (-.05,1.5-\reduce) rectangle (+.05,-1.5+\reduce);
  \draw
    (-.05,1.5-\reduce) to (-.05,-1.5+\reduce);
  \draw
    (+.05,1.5-\reduce) to (+.05,-1.5+\reduce);
  \draw[fill=white, draw opacity=0]
    (0,0) circle (.23);
  \draw[fill=lightgray, draw opacity=0, fill opacity=.6]
    (0,0) circle (.23);
  \begin{scope}
    \clip
      (0,0) circle (.22);
    \draw (0,0)
      node
      {
        \color{blue}
        \raisebox{1pt}{
          $\overline{w}$
        }
      };
  \end{scope}
\end{scope}

  \draw
    (2,0)
    node
    {
      \scalebox{2.4}{
        $\rightleftharpoons$
      }
    };

\begin{scope}[shift={(+3.2,0)}]
  \draw[fill = black]
    (-.05,1.5-\reduce) rectangle (+.05,-1.5+\reduce);
  \draw[fill=white]
    (0,0) circle (.23);
  \draw[fill=lightgray, fill opacity=.6]
    (0,0) circle (.23);
  \begin{scope}
    \clip
      (0,0) circle (.22);
    \draw (0,0)
      node
      {
        \color{blue}
        $f$
      };
  \end{scope}
\end{scope}

 \draw[white,line width=1.4]
   (-1, 1.32-\reduce) to (3.4, 1.32-\reduce);
 \draw[white,line width=1.4]
   (-1, 1.1-\reduce) to (3.4, 1.1-\reduce);
 \draw[white,line width=1.4]
   (-1, .89-\reduce) to (3.4, .89-\reduce);

 \draw[white,line width=1.4]
   (-1, -1.1+\reduce) to (3.4, -1.1+\reduce);
 \draw[white,line width=1.4]
   (-1, -.89+\reduce) to (3.4, -.89+\reduce);

  \draw
    (2,-.6)
    node
    {
      \tiny
      \color{greenii}
      \bf
      \def\arraystretch{.75}
      \begin{tabular}{c}
        creation /
        \\
        annihilation
      \end{tabular}
    };

\draw
  (-.4,-1.8+\reduce)
  node
  {
    \tiny
    \color{darkblue}
    \bf
    M-branes
  };

\draw
  (+.8,-1.8+\reduce)
  node
  {
    \tiny
    \color{darkblue}
    \bf
    \def\arraystretch{.75}
    \begin{tabular}{c}
      anti-
      \\
      brane
    \end{tabular}
  };

\end{tikzpicture}
}
}
\end{tabular}

\medskip

\noindent
{\bf Hypothesis H on gravitational backgrounds.}
The equations of motion of $D=11$ supergravity are in fact subject to ``higher curvature corrections'' which shift the $G_4$-flux by $\tfrac{1}{2}\big(\tfrac{1}{2}p_1\big)$
(cf. \cite[p. 8]{Tsimpis04})
and shift the Bianchi identity for $G_7$ by a term proportional to $I_8 \defneq \tfrac{1}{48}\big(p_2 - (\tfrac{1}{2}p_1)^2\big)$ (cf. \cite[\S 4]{SoueresTsimpis17}), where $p_n$ denotes the $n$th Pontrjagin form of the Levi-Civita connection (the gravitational field) on spacetime (cf. \cite[Ex. 8.1 (ii)]{FSS23Char}). In order to reflect such extra coupling of the C-field flux to gravitational background charges, the flux-quantizing cohomology theory must be twisted, somehow, by the tangent bundle of spacetime, induced by an action of a Spin-group on the classiying space $\mathcal{A}$.

While Spin-groups do not seem to act usefully on $\mathcal{A}_{{}_{\mathrm{DFM}}}$, preventing a {\it systematic} coupling of this model to gravitational charges, it is noteworthy that
$\mathrm{Spin}(5)$ acts, of course, canonically on $\mathcal{A}_{{}_{\mathrm{FSS}}} \,\defneq\,  S^4 \,\simeq\, S(\mathbb{R}^5)$.

\noindent
\hspace{-.2cm}
\def\tabcolsep{0pt}
\begin{tabular}{p{6cm}l}
This means in particular  that for $D=11$ supergravity on 8-manifolds $X^{10} \simeq T^2 \times X^8$ equipped with tangential $\mathrm{Spin}(5)$-structure $\tau$, there is a canonical notion of tangentially {\it twisted 4-Cohomotopy} \cite[\S 2]{FSS20-H} \cite[Ex. 3.8]{FSS23Char}, given by homotopy classes of sections of the 4-spherical fibration associated to the tangential structure
-- cf. the analogous case of twisted K-theory \eqref{TwistedKtheory}.
&
\hspace{0cm}
\adjustbox{raise=-2.1cm}{
\begin{minipage}{11.5cm}
\begin{equation}
  \label{TwistedFourCohomotopy}
  \overset{
    \mathclap{
      \;\;\;\;\;
      \scalebox{.7}{
        \color{darkblue}
        \bf
        \def\arraystretch{.9}
        \begin{tabular}{c}
          tangentially
          \\
          twisted
          \\
          4-Cohomotopy
        \end{tabular}
      }
    }
  }{
  \pi^{\tau}\big(
    X^{10}
  \big)
  }
  \;
  \defneq
  \;
  \left\{
  \hspace{-5pt}
  \adjustbox{raise=.2cm}{
  \begin{tikzcd}[
    column sep=30pt
  ]
    &[+20pt]
    &[-50pt]
    S^4 \sslash \mathrm{Spin}(5)
    \ar[
      dd,
      "{
        \scalebox{.7}{
          \color{darkblue}
          \bf
          \def\arraystretch{.9}
          \begin{tabular}{c}
            classifying fibration
            for
            \\
            twisted 4-Cohomotopy
          \end{tabular}
        }
      }"{description}
    ]
    \\
    \\
    T^2 \times X^8
    \ar[
      uurr,
      dashed,
      "{
        \scalebox{.7}{
          \color{greenii}
          \bf
          C-field charges
        }
      }"{sloped},
      "{ c_3 }"{swap}
    ]
    \ar[
      rr,
      "{ \tau }",
      "{
        \scalebox{.7}{
          \color{greenii}
          \bf
          tnagential twist
        }
      }"{swap}
    ]
    \ar[
      dr,
      "{
        \scalebox{.7}{
          \color{greenii}
          \bf
          \def\arraystretch{.9}
          \begin{tabular}{c}
            tangent bundle /
            \\
            gravitational charges
          \end{tabular}
        }
      }"{swap, sloped}
    ]
    &&
    B \mathrm{Spin}(5)
    \ar[dl]
    \\
    &
    B\mathrm{Spin}(8)
  \end{tikzcd}
  }
  \hspace{-5pt}
  \right\}_{
  \mathrlap{
    \hspace{-4pt}
    \Big/
    \hspace{-4pt}
    \scalebox{.7}{
      \def\arraystretch{.9}
      \begin{tabular}{l}
       rel
       \\
       hmtp
     \end{tabular}
    }
  }
  }
\end{equation}
\end{minipage}
}
\end{tabular}

Now C-field flux quantization in twisted Cohomotopy does imply the expected half-integral shifted quantization of the $G_4$-flux, as follows \cite[\S 3.4, Prop. 3.13]{FSS20-H}:

\vspace{-.1cm}
\begin{equation}
\label{ProofOfShiftedFluxQuantizationOfCField}
\begin{tikzcd}[
 row sep=6pt,
 column sep=3pt
]
  \scalebox{.8}{
    \color{darkblue}
    \bf
    \def\arraystretch{.9}
    \begin{tabular}{c}
      cocycle in
      \\
      tangentially twisted
      \\
      4-cohomotopy
    \end{tabular}
  }
  &
  X
  \ar[
    rr,
    dashed,
    "{ c_3 }"
  ]
  \ar[
    dr,
    "{
      \vdash \, \mathrm{Fr}(X)
    }"{swap, sloped}
  ]
  &&
  S^4\!\sslash\!\mathrm{Spin}(5)
  \;\simeq\;
  B \mathrm{Spin}(4)
  \ar[dl]
  \\[-14pt]
  &
  &
  B \mathrm{Spin}(d)
  \\[-2pt]
  \scalebox{.8}{
    \color{darkblue}
    \bf
    \def\arraystretch{.9}
    \begin{tabular}{c}
      induced charge in
      \\
      real cohomology
    \end{tabular}
  }
  &
  H^\bullet\big(
    X;
    \,
    \mathbb{R}
  \big)
  \ar[
    from=rr,
    "{ c_3^\ast }"{description}
  ]
  &&
  H^\bullet\big(
    B \mathrm{Spin}(4)
    ;\,
    \mathbb{R}
  \big)
  \,=\,
  \mathbb{R}
  \big[
    p_1
    ,\;
    \rchi_4
  \big]
  \\[-10pt]
  &
  \big[p_1(\nabla)\big]
  &\longmapsfrom&
  p_1
  \;
  \scalebox{.7}{
    \color{darkblue}
    \bf
    first Pontrjagin class
  }
  \\[-10pt]
  &
  \big[G_4\big]
  &\longmapsfrom&
  \tfrac{1}{2}\rchi_4
  \scalebox{.7}{
    \color{darkblue}
    \bf
    fractional Euler class
  }
  \\
  \scalebox{.8}{
    \color{darkblue}
    \bf
    \def\arraystretch{.9}
    \begin{tabular}{c}
      induced charge in
      \\
      integral cohomology
    \end{tabular}
  }
  &
  H^\bullet(
    X
    ;\,
    \mathbb{Z}
  )
  \ar[
    from=rr,
    "{
      c_3^\ast
    }"{description}
  ]
  &&
  H^\bullet\big(
    B \mathrm{Spin}(4)
    ;\,
    \mathbb{Z}
  \big)
  \,=\,
  \mathbb{Z}
  \big[
    \tfrac{1}{2}p_1
    ,\;
    \tfrac{1}{2}\rchi_4
    +
    \tfrac{1}{4}p_1
  \big]
  \\[-10pt]
    \scalebox{.7}{
      \color{orangeii}
      \bf
      \def\arraystretch{.9}
      \begin{tabular}{c}
        integral class of
        \\
        shifted C-field flux
      \end{tabular}
    }
  &
  \underbrace{
  \big[G_4 + \tfrac{1}{4}p_1(\nabla)\big]
  }_{ [\widetilde G] }
  &\longmapsfrom&
  \tfrac{1}{2}\rchi_4
  +
  \tfrac{1}{4}p_1
  \;\;
  \scalebox{.7}{
    \bf
    \color{darkblue}
    \def\arraystretch{.9}
    \begin{tabular}{c}
      universal integral
      \\
      characteristic class
    \end{tabular}
  }
\end{tikzcd}
\end{equation}
\vspace{-.3cm}

However, for this twisting to preserve the distinction between M2- and M5-brane charges and hence the quaternionic Hopf fibration \eqref{PureFlatM2BraneCharge},
one actually has \cite[Prop. 2.20]{FSS20-H}\cite[Prop. 2.2]{FSS22Twistorial}
to regard
$\mathrm{Spin}(5)$ as the quaternionic unitary group $\mathrm{Sp}(2)$
which acts canonically also on $S^7 = S(\mathbb{H}^2)$.

\noindent
\hspace{-.2cm}
\def\tabcolsep{0pt}
\begin{tabular}{p{10cm}l}
While both groups are abstractly isomorphic, $\mathrm{Spin}(5) \,\simeq\, \mathrm{Sp}(2)$, they are {\it not} isomorphic {\it as subgroups} of $\mathrm{Spin}(8)$, {\it but} they are mapped into each other under the triality automorphism $\mathrm{tri} : \mathrm{Spin}(8) \xrightarrow{\sim} B \mathrm{Spin}(8)$.
\cite[Prop. 2.17]{FSS20-H}.

This means (i) that one actually needs $\mathrm{Sp}(2)$-structure on spacetime to couple both M5- as well as M2-brane charges to gravitational charges,
(ii) which differs from $\mathrm{Spin}(5)$-structure by triality
and (iii) the gravitational charges in degree=4 are invariant under this transformation,  but the degree=8 charges pick up the $I_8$-term \cite[Lem. 2.19]{FSS20-H}.
& \qquad \quad
\adjustbox{
  raise=-1.5cm
}{
$
\begin{tikzcd}[
    column sep=0pt,
    row sep=14pt
  ]
    B \mathrm{Sp}(2)
    \ar[rr, "{ \sim }"]
    \ar[d, hook]
    &&
    B \mathrm{Spin}(5)
    \ar[d, hook]
    \\[-5pt]
    B \mathrm{Spin}(8)
    \ar[
      rr,
      "{ B \mathrm{tri} }"{
        yshift=2pt
      },
      "{\sim}"{swap}
    ]
    &&
    B
    \mathrm{Spin}(8)
    \\[-10pt]
    H^\bullet\big(
      B \mathrm{Sp}(2)
      ;\,
      \mathbb{R}
    \big)
    \ar[
      from = rr,
      "{ (B \mathrm{tri})^\ast }"{swap}
    ]
    &&
    H^\bullet\big(
      B \mathrm{Spin}(5)
      ;\,
      \mathbb{R}
    \big)
    \\[-14pt]
    \tfrac{1}{2}p_1
    &\mapsfrom&
    \tfrac{1}{2}p_1
    \\[-14pt]
    (\tfrac{1}{4}p_1)^2
    -
    24 \cdot I_8
    &\mapsfrom&
    \tfrac{1}{4}p_2
\end{tikzcd}
$
}
\end{tabular}

\noindent
\hspace{-.2cm}
\def\tabcolsep{0pt}
\begin{tabular}{p{9.4cm}l}
For $\tau$ an $\mathrm{Sp}(2)$-structure on spacetime, the flux densities in the image of the twisted character on $\tau$-twisted 4-Cohomotopy are shown on the right\footnotemark\;
\cite[Prop. 3.20]{FSS20-H}\cite[Thm. 2.14]{FSS22Twistorial}.
&
\hspace{-1.4cm}
\adjustbox{raise=-.6cm}{
\begin{minipage}{9.2cm}
\begin{equation}
  \label{FluxesQuantizedInSp2TwistedCohomotopy}
  \adjustbox{raise=10pt}{$
  \def\arraystretch{1.7}
  \begin{array}{l}
    \differential G_4
    =
    0
    \,,
    \;
    \big[\widetilde G_4\big]
    =
    \mathrlap{
    \big[
      G_4 + \tfrac{1}{4}p_1
    \big]
    \in
    H^4\big(
      X^d; \mathbb{Z}
    \big)
    }
    \\
    \differential G_7
    =
    -\tfrac{1}{2}
    \widetilde G_4
    \wedge
    \big(
      \widetilde G_4 - \tfrac{1}{2}p_1
    \big)
    -
    12 \cdot I_8
  \end{array}
  $}
\end{equation}
\end{minipage}
}
\end{tabular}

\footnotetext{
A gravitational shift in the Bianchi identity for $G_7$ is expected but has remained undetermined \cite[(4.16)]{Tsimpis04} \cite[(4.11)]{SoueresTsimpis17}.
On the factor 12 in \eqref{FluxesQuantizedInSp2TwistedCohomotopy} cf.
\cite[pp. 12-13 \& \S3.8]{FSS20-H}\cite[Rem. 7 \& 8]{SS21-M5}\cite[Rem. 4.1]{SS23MF}; and notice that this term disappears in \eqref{HypothesisHConclusion} below.}

\medskip

\subsection{B-Field flux quantization in 6d}
\label{BFieldFluxQuantizationIn6d}

The (5+1)-dimensional worldvolume $\Sigma^6$ of an M5-brane sigma-model is to carry a 3-form flux $H_3$ which in simple (decoupled) sitations satisfies the equations of motion of a self-dual higher gauge field (Ex. \ref{MotionOfSelfDualHigherGaugeField}) \cite{ClausKalloshVanProeyen98} \cite{Witten10}.
The effect of the self-duality on the phase space is, with
Prop. \ref{SolutionSpaceViaGaussLaw}, that the evident Gau{\ss} law $\differential \, H_3 \,=\, 0$ is imposed on a {\it single} flux density, in contrast to the non-self-dual case (Ex. \ref{SolutionSpaceOfOrdinaryElectromagnetism}). Therefore a traditional flux quantization of the self-dual field is in a {\it single} copy
\cite[p. 32]{FreedMooreSegal07b}
of ordinary differential cohomology in degree=3 (cf. Ex. \ref{HigherCircleGaugePotentials}, \cite[\S 3.2]{SS23FQ}), in contrast to the two copies seen for non-self dual abelian gauge fields Ex. \ref{FluxQuantizationLawsForOrdinaryelectromagnetism}.

\smallskip

\noindent
\hspace{-.2cm}
\def\tabcolsep{0pt}
\begin{tabular}{p{7cm}l}
\hspace{11pt} But in general the $H_3$-flux on the Fivebrane is not actually closed, rather it is sourced by the pullback of $G_4$, and its self-duality is subtle \cite[(36, 40)]{HoweSezgin97}
\cite{HoweSezginWest97}\cite[(5.57, 5.82)]{Sorokin00}\cite{GSS-6dSuperFlux}.
&
\hspace{-.4cm}
\adjustbox{raise=-1.1cm}{
\begin{minipage}{11cm}
\begin{equation}
  \label{MotionOfBFieldOnFivebrane}
  \begin{tikzcd}[row sep=-10pt]
    &&
    \adjustbox{scale=.9, fbox}{$
      \def\arraystretch{.9}
      \begin{array}{c}
        \mathrm{d} \, H_3
      \end{array}
      \begin{array}{c}
        =
      \end{array}
      \begin{array}{c}
        \phi^\ast \widetilde G_4
      \end{array}
    $}
    \\[-8pt]
    \adjustbox{scale=.9, fbox}{
      \def\arraystretch{.9}
      \begin{tabular}{c}
        B-field flux on
        \\
        5-brane worldvolume
        \\
        $\phi : \Sigma^{6}
          \xrightarrow{\;\;}
         Y^{11}$
      \end{tabular}
    }
    \ar[
      urr,
      <->,
      rounded corners,
      to path={
        ([xshift=-0pt]\tikztostart.east)
        --
        ([xshift=30pt]\tikztostart.east)
        --
        ([xshift=-14pt,]\tikztotarget.west)
        --
        ([xshift=-0pt]\tikztotarget.west)
      }
    ]
    \ar[
      drr,
      <->,
      rounded corners,
      to path={
        ([xshift=-0pt]\tikztostart.east)
        --
        ([xshift=30pt]\tikztostart.east)
        --
        ([xshift=-13pt]\tikztotarget.west)
        --
        ([xshift=-0pt]\tikztotarget.west)
      }
    ]
    \\[-8pt]
    &&
    \adjustbox{scale=.9, fbox}{
        subtle self-duality
    }
  \end{tikzcd}
\end{equation}
\end{minipage}
}
\end{tabular}

\smallskip

\noindent
\hspace{-.2cm}
\def\tabcolsep{0pt}
\begin{tabular}{p{7cm}l}
\hspace{12pt}Hence the quantization of $H_3$ depends on the quantization of $G_4$. Assuming Hypothesis H for the latter and since the homotopy fiber of the $\mathbb{H}$-Hopf fibration is $S^3$, it is natural to quantize $H_3$ in 3-Cohomotopy twisted by the C-field's Cohomotopy-charge, via the $\mathbb{H}$-Hopf fibration
\cite[Prop. 3.20]{FSS20-H}\cite[(10)]{FSS20TwistedString}. In fact this extends from Cauchy surfaces $\Sigma^5$ to the worldvolume $\Sigma^6$  \cite{GSS-6dSuperFlux}.
&
\hspace{12pt}
\adjustbox{raise=-2cm}{
\begin{minipage}{10cm}
\label{TwistedThreeCohomotopy}
\begin{equation}
  \overset{
    \mathclap{
      \raisebox{3pt}{
        \scalebox{.7}{
          \color{darkblue}
          \bf
          \def\arraystretch{.9}
          \begin{tabular}{c}
            C-field-twisted
            \\
            3-Cohomotopy
          \end{tabular}
        }
      }
    }
  }{
  \pi^{ \phi^\ast \! c_3 }\big(
    \Sigma^5
  \big)
  }
  \;\;
  \defneq
  \left\{
  \hspace{-5pt}
  \adjustbox{raise=.2cm}{
  \begin{tikzcd}[
    column sep=37pt,
    row sep=20pt
  ]
    \Sigma^5
    \ar[
      dd,
      "{ \phi }"{swap},
      "{
        \scalebox{.7}{
          \color{greenii}
          \bf
        }
      }",
      "{
        \scalebox{.7}{
          \color{darkblue}
          \bf
          \def\arraystretch{.9}
          \begin{tabular}{c}
            fivebrane
            \\
            worldvolume
            \\
            embedding
          \end{tabular}
        }
      }"{xshift=-3pt}
    ]
    \ar[
      rr,
      dashed,
      "{
        \scalebox{.7}{
          \def\arraystretch{.9}
          \begin{tabular}{c}
            {\color{greenii}
            \bf
            B-field charges} on
            \\
            5-brane worldvolume
          \end{tabular}
        }
      }",
      "{ b_2 }"{swap}
    ]
    &&
    S^7 \!\sslash\! \mathrm{Sp}(2)
    \ar[
      dd,
      "{
        \scalebox{.7}{
          \color{darkblue}
          \bf
          \def\arraystretch{.9}
          \begin{tabular}{c}
            \scalebox{1.3}{$\mathbb{H}$}-Hopf fibration as
            \\
            classifying fibration for
            \\
            twisted 3-Cohomotopy
          \end{tabular}
        }
      }"{description}
    ]
    \\
    \\
    X^{10}
    \ar[
      rr,
      "{ c_3 }",
      "{
        \scalebox{.7}{
          \begin{tabular}{c}
            {\color{greenii}
            \bf
            C-field charges} on
            \\
            spacetime
            \eqref{TwistedFourCohomotopy}
          \end{tabular}
        }
      }"{swap}
    ]
    &&
    S^4 \!\sslash\! \mathrm{Sp}(2)
  \end{tikzcd}
  }
  \hspace{-5pt}
  \right\}_{
  \mathrlap{
    \hspace{-3pt}
    \Big/
    \hspace{-5pt}
    \scalebox{.7}{
      \def\arraystretch{.9}
      \begin{tabular}{c}
        rel
        \\
        hmtp
      \end{tabular}
    }
  }
  }
\end{equation}
\end{minipage}
}
\end{tabular}

\noindent
Moreover, for the topological analysis of WZW terms we may consider $\Sigma^7$ an ``extended worldvolume'', namely a closed 7-manifold which is a cobordism from $\varnothing$ to the fivebrane worldvolume $\Sigma^6$ and back to $\varnothing$,
and over which the worldvolume fields $(\phi, H_3)$ extend, cf. \cite[\S 2]{FSS19HopfWZ}. Then the M2-brane background flux exerting
\linebreak
\noindent
\hspace{-.2cm}
\def\tabcolsep{0pt}
\begin{tabular}{ll}
\begin{tabular}{l}
Lorentz force on the 5-brane is
as on the
\\
right \cite[(8)]{Page83}\cite[(43)]{DuffStelle91},
\end{tabular}
&
\hspace{-.7cm}
\adjustbox{raise=.1cm}{
\begin{minipage}{11.15cm}
\begin{equation}
  \label{PageChargeFlux}
  \scalebox{.7}{
    \color{darkblue}
    \bf
    \def\arraystretch{.9}
    \begin{tabular}{c}
      flux density of Page
      \\
      charge of M2-branes
    \end{tabular}
  }
  \;
  2\widetilde{G}_7
  :=
  2\big(
    \phi^\ast G_7
  +
  \tfrac{1}{2}
  H_3 \wedge \phi^\ast \widetilde G_4
    \big)
  \in
  \Omega^7_{\mathrm{dR}}\big(\Sigma^7\big)
\end{equation}
\end{minipage}
}
\end{tabular}

\noindent
which we are showing scaled by 2 in accord with \eqref{HomotopyFiberOfCupSquare} and since this is how it actually appears as the {\it Hopf-WZ term} in the Fivebrane's worldvolume theory \cite[(2.8)]{Intriligator00}.

The combined C-field- \& B-field flux quantization ought to imply (as for any Dirac charge quantization condition) the flux density \eqref{PageChargeFlux} to have integer-valued integral over $\Sigma^7$:

\begin{itemize}[leftmargin=.6cm]
\item[(i)]  its integral is the total magnetic charge of singular M2-branes enclosed by $\Sigma^7$, and these ought to appear in integer numbers

\item[(ii)] its exponentiated integral mod $\mathbb{Z}$ is an anomaly in the Hopf-WZ term in the fivebrane's action functional, whose vanishing is the ``level quantization''-condition for 7d Chern-Simons theory with Lagrange density \eqref{PageChargeFlux}.
\end{itemize}

\noindent

\noindent
\hspace{0cm}
\def\tabcolsep{0pt}
\begin{tabular}{p{6.7cm}l}
The obstruction, under Hypothesis H, to \eqref{PageChargeFlux} being integral,
hence the M5-'s Hopf-WZ-term anomaly, turns out to be $24 I_8 = \rchi_8$ of the given $\mathrm{Sp}(2)$-structure.
\cite[Thm. 4.8, Ex. 3.2]{FSS19HopfWZ}.
& \;\;
\adjustbox{raise=-.9cm}{
$
  \begin{tikzcd}[
   column sep=12pt
  ]
    &
    B \widehat{\mathrm{Sp}(2)}
    \ar[
      d
    ]
    \ar[
      r
    ]
    \ar[
      dr,
      phantom,
      "{
        \scalebox{.7}{
          (hpb)
        }
      }"
    ]
    &[+2pt]
    \ast
    \ar[d]
    \\
    X^{10}
    \ar[
      r,
      "{
        \tau
      }"
    ]
    \ar[
      ur,
      dashed,
      "{
        \widehat{ \tau }
      }"{description},
      "{
        \scalebox{.7}{
          \color{greenii}
          \bf
          \def\arraystretch{.9}
          \begin{tabular}{c}
            M-Fivebrane
            \\
            str.
          \end{tabular}
        }
      }"{yshift=1pt, sloped, pos=.3}
    ]
    &
    B \mathrm{Sp}(2)
    \ar[
      r,
      "{
        24 I_8
      }"
    ]
    &
    B^8 \mathbb{Z}
  \end{tikzcd}
  \!
    \Rightarrow
  \begin{tikzcd}[
    row sep=9pt
  ]
    \Sigma^7
    \ar[
      d,
      "{ \phi }"{swap}
    ]
    \ar[
      r,
      "{ b_2 }"
    ]
    \ar[
      rr,
      rounded corners,
      to path={
           ([yshift=00pt]\tikztostart.north)
        -- ([yshift=8pt]\tikztostart.north)
        -- node{
          \scalebox{.7}{
            \colorbox{white}
            {
              \color{greenii}
              \bf
              quantized Hopf-WZ/Page-charge
            }
          }
        }
           ([yshift=8pt]\tikztotarget.north)
        -- ([yshift=00pt]\tikztotarget.north)
      }
    ]
    &
    S^7
      \sslash
    \widehat{ \mathrm{Sp}(2) }
    \ar[
      d,
      "{
        h_{\mathbb{H}}
        \sslash
        \widehat{ \mathrm{Sp}(2) }
      }"{xshift=1pt}
    ]
    \ar[
      r,
      dashed,
      "{
        \exists
      }"
    ]
    &
    B^7 \mathbb{Z}
    \\
    X^{10}
    \ar[
      r,
      "c_3"
    ]
    &
    S^4
      \!\sslash\!
    \widehat{\mathrm{Sp}(2)}
  \end{tikzcd}
$
}
\end{tabular}

In particular, the Page charge \eqref{PageChargeFlux} of {\it flat} M2-branes (Ex. \ref{FlatBranes}) is integral, being (the Whitehead-integral formula for) the {\it Hopf invariant} of the cohomotopical C-field charge $c_3 : S^7 \to S^4$, cf. also \cite[\S 2.7]{SS23MF}.

\noindent
\hspace{-.2cm}
\def\tabcolsep{0pt}
\begin{tabular}{p{5cm}l}
In conclusion, the Hypothesis $\widehat{\mbox{H}}$ that the C-field is flux-quantized in tangentially $\widehat{\mathrm{Sp}(2)}$-twisted 4-Cohomotopy,
and that the Fivebrane's worldvolume B-field is flux-quantized in the correspondingly twisted 3-Cohomotopy
implies the expected quantization of $G_4$-flux/M5-charge and of $G_7$-flux/M2-charge.
&
\hspace{-.9cm}
\adjustbox{raise=-2.1cm}{
\begin{minipage}{13.2cm}
\begin{equation}
\label{HypothesisHConclusion}
\adjustbox{
  raise=-1cm
}{$
  \begin{tikzcd}[
    column sep=20pt,
    row sep=10pt
  ]
    &[+20pt]&[-50pt]
    S^7
      \sslash
    \widehat{\mathrm{Sp}(2)}
    \ar[
      d,
      shorten=-2pt,
      "{
        h_{\mathbb{H}}
        \sslash
        \widehat{\mathrm{Sp}(2)}
      }"{xshift=1pt}
    ]
    \\
    \Sigma^7
    \ar[
      d,
      shorten=-2pt,
      "{ \phi }"{swap}
    ]
    \ar[
      urr,
      dashed,
      "{
        \scalebox{.7}{
          \color{greenii}
          \bf
          \def\arraystretch{.9}
          \begin{tabular}{c}
            worldvolume
            \\
            B-field charges
          \end{tabular}
        }
      }"{sloped},
      "{
        b_2
      }"{swap}
    ]
    &&
    S^4
      \!\sslash\!
    \widehat{\mathrm{Sp}(2)}
    \ar[
      d,
      shorten=-2pt
    ]
    \\
    T^2 \!\times\! X^8
    \ar[
      dr,
      shorten=-2pt,
      "{
        \scalebox{.7}{
          \color{greenii}
          \bf
          \def\arraystretch{.9}
          \begin{tabular}{c}
            tan. bundle/
            \\
            gravitational charges
          \end{tabular}
        }
      }"{swap, pos=.4, sloped}
    ]
    \ar[
      rr,
      shorten=-1pt,
      "{
        \widehat{\tau}
      }",
      "{
        \scalebox{.7}{
          \color{greenii}
          \bf
          \def\arraystretch{.8}
          \begin{tabular}{c}
            M-Fivebrane
            \\
            str.
          \end{tabular}
        }
      }"{swap, yshift=1pt, pos=.6}
    ]
    \ar[
      urr,
      dashed,
      "{
        c_3
      }"{swap, pos=.7},
      "{
        \scalebox{.7}{
          \color{greenii}
          \bf
          C-field charges
        }
      }"{sloped},
    ]
    &&
    B \, \widehat{\mathrm{Sp}(2)}
    \ar[
      dl,
      shorten=-2pt
    ]
    \\
    &
    B \mathrm{Spin}(8)
  \end{tikzcd}
  \hspace{-8pt}
  \Rightarrow
  \left\{
  \hspace{-10pt}
  \begin{array}{l}
  \def\arraystretch{1.4}
  \begin{array}{l}
    \differential  G_4 = 0
    \\
    \differential  H_3
    =
    \phi^\ast \widetilde G_4
    \\
    \differential G_7 =
    -\tfrac{1}{2}\widetilde G_4
    \wedge
    \big(
      \widetilde G_4 - \tfrac{1}{2}p_1
    \big)
  \end{array}
  \\
  \\
  \begin{array}{r}
    \big[
      \underbrace{
       G_4
       +
       \tfrac{1}{4}p_1
       }_{ \widetilde{G}_4 }
    \big]
    \! \in \! H^4\big(X^{10};\, \mathbb{Z}\big)
    \\
    \big[
      \underbrace{
       \phi^\ast(2 G_7)
       +
       H_3
       \wedge
       \phi^\ast \widetilde{G}_4
       }_{
         2\widetilde{G}_7
       }
    \big]
   \! \in  \! H^7\big(\Sigma^7;\, \mathbb{Z}\big)
  \end{array}
  \end{array}
  \right.
  \hspace{-2cm}
$}
\end{equation}
\end{minipage}
}
\end{tabular}

\medskip

\subsection{Green-Schwarz mechanism in 11d}
\label{GreenSchwarzMechanicsm}

Just as the B-field
\eqref{MotionOfBFieldOnFivebrane}
on the M-Fivebrane worldvolume (and we will see in a moment that this is not a coincidence),
\linebreak
\noindent
\hspace{-.2cm}
\def\tabcolsep{0pt}
\begin{tabular}{p{12.1cm}l}
so the B-field flux density $H_3$ in heterotic supergravity is famously not closed,
but is sourced by
the second Chern form of a $G$-Yang-Mills gauge potential $\widehat{A}$
minus the first fractional Pontrjagin form of the spin-connection $\widehat \omega$ on spacetime.
& \quad
\adjustbox{raise=-.4cm}{
\begin{minipage}{4.8cm}
\begin{equation}
  \label{GreenSchwarzRelationForHeteroticString}
  \differential \, H_3
  \;=\;
  c_2\big(\widehat{A}\big)
  -
  \tfrac{1}{2}p_1
  \big(\widehat{\omega}\big)
\end{equation}
\end{minipage}
}
\\[-.1cm]
This sourcing of $H_3$-flux is (the reflection of) the {\it Green-Schwarz mechanism} \cite{GreenSchwarz84} \cite[p. 49]{CHSW85}, which implies that for $G = E_8 \times E_8$ the heterotic superstring on such a background is anomaly-free --- iff the de Rham coboundary \eqref{GreenSchwarzRelationForHeteroticString} is flux quantized to a coboundary on {\it integral} cohomology classes (\cite[(2.11)]{Witten00}).

There is a higher Lie group, the $\mathrm{String}^{c_2}$ 2-group (see pointers in \cite[App.]{FSS14M57d}) whose classifying space is the {\it universal} space carrying a homotopy of this form
\cite[\S 2]{Sati11StrucII}\cite[\S 2.2]{SSSStringStruc12}.
& \;\;\;
\adjustbox{raise=-1cm}{
$
  \begin{tikzcd}[
    row sep=10pt
  ]
    &
    B G
    \ar[
      dr,
      "{ c_2 }"
    ]
    \\
    X^{10}
    \ar[
      ur,
      dashed,
      "{
        \scalebox{.7}{
          \color{greenii}
          \bf
          \def\arraystretch{.85}
          \begin{tabular}{c}
            gauge field
            \\
            charges
          \end{tabular}
        }
      }"{sloped},
      "{\ }"{swap, name=s}
    ]
    \ar[
      dr,
      dashed,
      "{
        \scalebox{.7}{
          \color{greenii}
          \bf
          \def\arraystrech{.85}
          \begin{tabular}{c}
            gravity
            \\
            charges
          \end{tabular}
        }
      }"{swap, sloped},
      "{\ }"{name=t}
    ]
    \ar[
      from=s,
      to=t,
      Rightarrow,
      dashed,
      end anchor={[xshift=2pt]},
      "{
        \scalebox{.7}{
          \color{orangeii}
          \bf
          \def\arraystretch{.9}
          \begin{tabular}{c}
            B-field
            \\
            charge
          \end{tabular}
        }
      }"{pos=.4}
    ]
    &&
    \mathbf{B}^4 \mathbb{Z}
    \\
    &
    B\mathrm{Spin}
    \ar[
      ur,
      "{
        \scalebox{.7}{$\tfrac{1}{2}p_1$}
      }"{swap}
    ]
  \end{tikzcd}
$
}
\\[-.7cm]
This means that the gauge-, gravitational- and B-field charges of heterotic supergravity may {\it jointly} be understood as a single higher gauge field with higher structure group the $\mathrm{String}^{c_2}$ 2-group, in differential (stacky) refinement (``twisted differential String-structures'') of the resulting diagram on the right.
& \;\;
\adjustbox{raise=-.2cm}{
$
  \begin{tikzcd}[
    column sep=7pt,
    row sep=10pt
  ]
    &[+12pt]
    &[-15pt]
    B G
    \ar[
      dr,
      "{
        c_2
      }"
    ]
    &[-9pt]
    \\
    X^{10}
    \ar[
      r,
      dashed,
      shorten=-1.5pt,
      "{
        \scalebox{.7}{
          \color{greenii}
          \bf
          \def\arraystretch{.9}
          \begin{tabular}{c}
            \scalebox{1.4}{$\mathrm{String}^{c_2}$}
            \\
            2-gauge field
            \\
            charges
          \end{tabular}
        }
      }"
    ]
    &
    B \mathrm{String}^{c_2}
    \ar[
      ur,
      "{\ }"{swap, name=s}
    ]
    \ar[
      dr,
      "{\ }"{name=t, pos=.57}
    ]
    \ar[
      from=s,
      to=t,
      Rightarrow,
      shift left=4pt,
      "{
        \scalebox{.7}{
          (hpb)
        }
      }"{pos=.45}
    ]
    &&
    B^4 \mathbb{Z}
    \\
    &
    &
    B \mathrm{Spin}
    \ar[
      ur,
      "{
        \scalebox{.7}{$\tfrac{1}{2}p_1$}
      }"{swap}
    ]
  \end{tikzcd}
$
}
\end{tabular}

This 2-group gauge theoretic flux quantization of (the Green-Schwarz mechanism in) heterotic supergravity has been discussed in
\cite[p. 13]{SSS09LInfinity}
\cite{SSSStringStruc12}
\cite[\S 3.7, \S 3.8]{FSS14M57d}
\cite{FSS15CField}, a corresponding construction (for the translation see \cite{Capotosti16}) in terms of bundle gerbes (and for the special case $c_2 = 0$) is in \cite{Waldorf13}. The terminology, at least, of 2-group gauge theory for GS-mechanisms has recently become popular in the non-mathematical physics literature, which also speaks of ``1-form symmetries'' (e.g. \cite{BeniniEtAl19}) or ``categorical symmetries'' (e.g. \cite{CDIS22}, cf. \cite{SchreiberSkoda09}).

\medskip

However, as with the RR-fields in type II supergravity  \cref{RRFieldFluxQuantization}, the flux-quantization and hence {\it partial} non-perturbative completion of the fields in heterotic supergravity somewhat begs the question: How does it connect to the expected {\it full} non-perturbative completion via M-theory?

\smallskip

Indeed the GS-mechanism has been argued to lift to M-theory, where the heterotic spacetime $X^{10}$
appears as a {\it pair} of ``$\mathrm{MO}_{9}$-planes'' in $Y^{11}$ each carrying {\it one} of the two copies of $E_8$ gauge fields \cite{HoravaWitten96}, cf. e.g. \cite[\S 1.3]{Dumitru22}, which is thought to, somehow, be the restriction of an $E_8$-gauge field $\widehat{A}_{E_8}$ on all of $Y^{11}$ itself,  modifying the relation \eqref{MotionOfBFieldOnFivebrane} to
\cite[(2.2)]{Witten97}\cite[(3.9)]{DiaconescuFreedMoore07}:
\begin{equation}
  \label{MTheoreticGreenSchwarz}
  \hspace{-4mm}
  \begin{tikzcd}
  \scalebox{.7}{
    \color{darkblue}
    \bf
    \def\arraystretch{.9}
    \begin{tabular}{c}
      M-theoretic avatar
      \\
      of GS-mechanism
      \eqref{GreenSchwarzRelationForHeteroticString}
    \end{tabular}
  }
  \;\;
  \differential\,
  H_3
  \,=\,
  G_4
  -
  \tfrac{1}{4}p_1\big(
    \widehat{\omega}
  \big)
  +
  c_2\big(
    \widehat{A}_{E_8}
  \big) \;
  \ar[
    rr,
    Rightarrow,
    shorten=-3pt,
    "{
      \scalebox{.7}{
        \color{greenii}
        \bf
        heterotic line bundle
      }
    }"{yshift=2pt},
    "{
      \scalebox{.8}{
        $
          S\left(\mathrm{U}(1)^2\right)
          \,\subset\,
          E_8
        $
      }
    }"{swap, yshift=-2pt}
  ]
  &\phantom{AA}&
  \;
  \differential
  \,
  H_3
  \,=\,
  G_4
  -
  \tfrac{1}{4}p_1\big(
    \widehat{\omega}
  \big)
  -
  F_2 \wedge F_2
  \,.
  \end{tikzcd}
\end{equation}
\vspace{-.2cm}

\noindent
While the ontology of $\widehat{A}_{E_8}$ had remained mysterious \cite{EvslinSati03}, for quasi-realistic phenomenology it is  ``heterotic line bundles'' that matter \cite{AGLP12}, reducing the structure group along $S\big( \mathrm{U}(1)^{n} \big) \hookrightarrow \mathrm{SU}(n) \hookrightarrow E_8$ with $2 \leq n \leq 5$. For $n = 2$ the resulting Green-Schwarz type Bianchi identity \eqref{MTheoreticGreenSchwarz}
does have a flux quantization compatible with the C-field flux quantization $\mathcal{A}_{{}_{\mathrm{FSS}}}$ in \cref{CFieldFluxQuantization}:

\noindent
\hspace{-.15cm}
\def\tabcolsep{0pt}
\begin{tabular}{p{3.8cm}l}
The quaternionic Hopf fibration \eqref{PureFlatM2BraneCharge} factors through the {\it twistor fibration} $t_{\mathbb{H}} : \mathbb{C}P^3 \to S^4$, the tangentially twisted flux quantization law $\mathcal{A} \defneq \mathbb{C}P^3$ is admissible for M-theory with heterotic line bundles and implies all the desired total charge quantizations \cite[Thm. 2.14, (6)]{FSS22Twistorial}.
&\;
\adjustbox{raise=-2.5cm}{
$
  \begin{tikzcd}[
    column sep=4pt
  ]
    &[+10pt]&[-30pt]
    \mathbb{C}P^3
      \!\sslash\!
    \widehat{\mathrm{Sp}(2)}
    \ar[
      d,
      shorten=-1pt,
      "{
        t_{\mathbb{H}}
        \sslash
        \mathrm{Sp}(2)
      }"{swap}
    ]
    \\
    &&
    S^4 \!\sslash\!
    \widehat{\mathrm{Sp}(2)}
    \ar[
      d,
      shorten=-1pt
    ]
    \\
    T^2 / \mathbb{Z}_2
    \!\times\!
    X^8
    \ar[
      rr,
      "{
        \widehat{\tau}
      }",
      "{
       \scalebox{.7}{
         \color{greenii}
         \bf
         \def\arraystretch{.9}
         \begin{tabular}{c}
           M-Fivebrane
           \\
           str.
         \end{tabular}
       }
      }"{swap}
    ]
    \ar[
      urr,
      dashed,
      "{
        \scalebox{.7}{
          \color{greenii}
          \bf
          C-field charges
        }
      }"{sloped},
      "{ c_3 }"{swap, pos=.6}
    ]
    \ar[
      uurr,
      rounded corners,
      dashed,
      to path={
           ([yshift=00pt]\tikztostart.north)
        -- ([yshift=40pt]\tikztostart.north)
        -- node[yshift=5pt, sloped] {
          \scalebox{.7}{
            \color{greenii}
            \bf
            heterotic line bundle
          }
        }
           node[yshift=-5pt] {
             \scalebox{.7}{$a_1$}
           }
           ([xshift=00pt, yshift=-2pt]\tikztotarget.west)
      },
      "{
        \scalebox{.7}{
          \color{greenii}
          \bf
          \begin{tabular}{c}
            heterotic
            \\
            line bundle
          \end{tabular}
        }
      }"{sloped}
    ]
    \ar[
      dr,
      "{
        \scalebox{.7}{
          \color{greenii}
          \bf
          \def\arraystretch{.9}
          \begin{tabular}{c}
            tan. bundle/
            \\
            gravitational
            \\
            charges
          \end{tabular}
        }
      }"{swap, sloped, pos=.3}
    ]
    &&
    B\widehat{\mathrm{Sp}(2)}
    \ar[dl]
    \\
    &
    B\mathrm{Spin}(8)
  \end{tikzcd}
  \adjustbox{raise=.8cm}{$
  \Rightarrow
  \left\{\!\!
  \def\arraystretch{1.4}
  \def\arraycolsep{2pt}
  \begin{array}{lr}
    \differential \, F_2 = 0
    &
    {\big[F_2\big]} \! \in \! H^2\big(X^{10}; \mathbb{Z}\big)
    \\
    \differential \, H_3 =
    G_4 - \tfrac{1}{4}p_1 - F_2 \wedge F_2
    \\[+10pt]
    \differential \, G_4 = 0
    &
    {\big[\widetilde G_4\big]}
    \!\in \! H^4\big(X^{10}; \mathbb{Z}\big)
    \\
    \differential \, G_7 =
    - \tfrac{1}{2}\widetilde G_4
    \wedge
    \big(
      \widetilde G_4
      -
      \tfrac{1}{2}p_1
    \big)
    &
    {\big[\widetilde G_7\big]}
    \!\in \! H^7\big(X^{10}; \mathbb{Z}\big)
  \end{array}
  \right.
  $}
$
}
\end{tabular}

For $\mathrm{MO}_9$-s but also for (M-Fivebranes probing) ADE-singularities in heterotic M-theory, this flux quantization restricts to \eqref{GreenSchwarzRelationForHeteroticString}, by \cite[Thm. 1.1]{SS20EquTw}.

Remarkably, the mechanism behind this flux-quantized Green-Schwarz mechanism lifts, at least for flat branes, to C-field fluxes seen not just in rational but in any {\it complex-oriented} Whitehead-generalized cohomology theory \cite[\S 2.9]{SS23MF}. This includes complex K-theory but also elliptic cohomology and complex Cobordism and might finally explain the role these cohomology theories play for flux quantization in M-theory.

\medskip

\noindent
{\bf Keywords:} higher gauge theory, flux quantization, charge quantization, homotopy theory, rational homotopy theory, de Rham cohomology, homotopy Lie algebras, L-infinity algebras, generalized cohomology, twisted cohomology, differential cohomology, K-theory, Chern-Dold character, electromagnetism, supergravity, string theory, M-theory, duality, monopoles, solitons, branes


\newpage


\begin{thebibliography}{10}
\small

\bibitem[Abrikosov 1957]{Abrikosov1957}
Abrikosov A.:
{\it On the Magnetic properties of superconductors of the second group}, Sov. Phys. JETP {\bf 5} (1957), 1174-1182; Zh. Eksp. Teor. Fiz. {\bf 32} (1957),
1442-1452, [\href{https://inspirehep.net/literature/9138}{\tt spire:9138}].

\bibitem[Acharya et al. 1999]{AFHS99}
Acharya B., Figueroa-O'Farrill J., Hull C., Spence B., {\it Branes at conical singularities and holography}, Adv. Theor. Math. Phys. {\bf 2} (1999),
1249-1286,
 [\href{https://dx.doi.org/10.4310/ATMP.1998.v2.n6.a2}{\tt
  doi:10.4310/ATMP.1998.v2.n6.a2}],
[\href{https://arxiv.org/abs/hep-th/9808014}{\tt arXiv:hep-th/9808014}].

\bibitem[Aharony et al. 2008]{AharonyBergmanJafferis08}
Aharony O., Bergman O. and Jafferis D., {\it Fractional M2-branes}, JHEP {\bf 0811} (2008) 043, \newline
[\href{https://doi.org/10.1088/1126-6708/2008/11/043}{\tt doi:10.1088/1126-6708/2008/11/043}],
[\href{https://arxiv.org/abs/0807.4924}{\tt arXiv:0807.4924}].

\bibitem[Alfonsi 2024]{Alfonsi24}
Alfonsi, L., {\it Higher geometry in physics}, in: {\it Enc. Math. Phys. 2nd ed.}, Elsevier (2024)
[\href{https://arxiv.org/abs/2312.07308}{\tt arXiv:2312.07308}].


\bibitem[Alvarez 1985]{Alvarez85}
Alvarez O., {\it Topological quantization and cohomology}, Commun. Math. Phys. {\bf 100} 2 (1985), 279-309, \newline
[\href{https://projecteuclid.org/euclid.cmp/1103943448}{\tt euclid.cmp/1103943448}].

\bibitem[Anderson et al. 2012]{AGLP12}
Anderson L., Gray J., Lukas A. and Palti E., {\it Heterotic Line Bundle Standard Models}, JHEP {\bf 113} (2012) 06,
[\href{https://doi.org/10.1007/JHEP06(2012)113}{\tt doi:10.1007/JHEP06(2012)113}],
[\href{https://arxiv.org/abs/1202.1757}{\tt arXiv:1202.1757}].



\bibitem[Aschieri \& Jur{\v c}o 2004]{AschieriJurco04}
Aschieri P. and Jur{\v c}o B., {\it Gerbes, M5-Brane Anomalies and $E_8$ Gauge Theory}, JHEP {\bf 0410} (2004) 068,
[\href{https://doi.org/10.1088/1126-6708/2004/10/068}{\tt doi:10.1088/1126-6708/2004/10/068}],
[\href{https://arxiv.org/abs/hep-th/0409200}{\tt arXiv:hep-th/0409200}].

\bibitem[Atiyah \& Segal 2004]{AtiyahSegal04}
Atiyah M. and Segal G.,
{\it Twisted K-theory}, Ukrainian Math. Bull. {\bf 1} 3 (2004),
[\href{https://arxiv.org/abs/math/0407054}{\tt arXiv:math/0407054}].


\bibitem[Bandos et al. 1998]{BandosBerkovitsSorokin98}
Bandos I., Berkovits N., and Sorokin D.,
{\it Duality-Symmetric Eleven-Dimensional Supergravity and its Coupling to M-Branes}, Nucl. Phys. B {\bf 522} (1998), 214-233,
[\href{https://doi.org/10.1016/S0550-3213(98)00102-3}{\tt doi:10.1016/S0550-3213(98)00102-3}], \newline
[\href{https://arxiv.org/abs/hep-th/9711055}{\tt arXiv:hep-th/9711055}],

\bibitem[Becker et al. 2017]{BeckerBeniniSchenkelSzabo17}
Becker C, Benini M, Schenkel A and Szabo R., {\it Abelian duality on globally hyperbolic spacetimes}, Commun. Math. Phys. {\bf 349}
(2017), 361-392,
[\href{https://doi.org/10.1007/s00220-016-2669-9}{\tt
doi:10.1007/s00220-016-2669-9}],
[\href{https://arxiv.org/abs/1511.00316}{\tt arXiv:1511.00316}].


\bibitem[Beekman et al. 2011]{BeekmanZaanen11}
Beekman A. J., Zaanen J.,
{\it Electrodynamics of Abrikosov vortices: the field theoretical formulation}, Front. Phys. {\bf 6} (2011), 357–369,
[\href{https://doi.org/10.1007/s11467-011-0205-0}{\tt doi:10.1007/s11467-011-0205-0}].

\bibitem[Benini et al. 2019]{BeniniEtAl19}
Benini F., Cordova C., Hsin P., {\it On 2-Group Global Symmetries and Their Anomalies}, J. High Energ. Phys. {\bf2019} (2019)
118,
[\href{https://doi.org/10.1007/JHEP03(2019)118}{\tt doi:10.1007/JHEP03(2019)118}],
[\href{https://arxiv.org/abs/1803.09336}{\tt arXiv:1803.09336}].


\bibitem[Bergman et al. 1999]{BGH99}
Bergman O., Gimon E. G., Ho{\v r}ava P.,
{\it Brane Transfer Operations and T-Duality of Non-BPS States}, JHEP {\bf 9904} (1999) 010,
[\href{https://doi.org/10.1088/1126-6708/1999/04/010}{\tt doi:10.1088/1126-6708/1999/04/010}],
[\href{https://arxiv.org/abs/hep-th/9902160}{\tt arXiv:hep-th/9902160}].

\bibitem[Bergshoeff et al. 1998]{BergshoeffLozanoOrtin98}
Bergshoeff E., Lozano Y. and Ortin T., {\it Massive Branes}, Nucl. Phys. B {\bf 518} (1998), 363-423,
[\href{https://doi.org/10.1016/S0550-3213(98)00045-5}{\tt doi:10.1016/S0550-3213(98)00045-5}],
[\href{https://arxiv.org/abs/hep-th/9712115}{\tt arXiv:hep-th/9712115}].

\bibitem[Blaschke \& Gieres 2021]{BlaschkeGieres21}
Blaschke D. N. and Gieres F., {\it On the canonical formulation of gauge field theories and Poincar{\'e} transformations},
Nucl. Phys. B {\bf 965} (2021) 115366,
[\href{https://doi.org/10.1016/j.nuclphysb.2021.115366}{\tt doi:10.1016/j.nuclphysb.2021.115366}],
[\href{https://arxiv.org/abs/2004.14406}{\tt arXiv:2004.14406}].

\bibitem[Blumenhagen et al. 2013]{BLT13}
Blumenhagen R., L{\"u}st D, Theisen S., {\it Basic Concepts of String Theory}, Springer (2013),
[\href{https://doi.org/10.1007/978-3-642-29497-6}{\tt doi:10.1007/978-3-642-29497-6}].

\bibitem[Bonora et al. 2008]{BonoraRuffinoSavelli08}
Bonora L., Ruffino  F. F., Savelli R.,
{\it Classifying A-field and B-field configurations in the presence of D-branes}, JHEP {\bf 0812} (2008) 078,
[\href{https://arxiv.org/abs/0810.4291}{\tt arXiv:0810.4291}],
[\href{https://doi.org/10.1088/1126-6708/2008/12/078}{\tt doi:10.1088/1126-6708/2008/12/078}].

\bibitem[Borsten et al. 2024]{BFJKNRSW24}
Borsten, L.,
 Farahani M. J.,  Jur{\v c}o B., Kim H.,  N{\'a}ro{\v z}n{\' y} J., Rist D.,  Saemann C., and Wolf M.,
{\it Higher Gauge Theory},
in: {\it Enc.  Math. Phys. 2nd ed.}, Elsevier (2024), [\href{https://arxiv.org/abs/2401.05275}{\tt arXiv:2401.05275}].

\bibitem[Borsuk 1936]{Borsuk1936}
Borsuk K., {\it Sur les groupes des classes de transformations continues}, CR Acad. Sci. Paris {\bf 202} (1936),
1400-1403, [\href{http://www.numdam.org/articles/10.24033/asens.603/}{\tt  numdam/10.24033/asens.603}].

\bibitem[Bousfield \& Gugenheim 1976]{BousfieldGugenheim76}
Bousfield A. and Gugenheim V, {\it On PL deRham theory and rational homotopy type}, Memoirs of the AMS {\bf 179} (1976),
[\href{https://bookstore.ams.org/memo-8-179}{\tt ams:memo-8-179}].

\bibitem[Bouwknegt \& Mathai 2001]{BouwknegtMathai01}
Bouwknegt P., Mathai V.,
{\it D-branes, B-fields and twisted K-theory}, J. High Energy Phys.  {\bf 003} (2000) 007,
[\href{https://arxiv.org/abs/hep-th/0002023}{\tt arXiv:hep-th/0002023}],
[\href{https://doi.org/10.1088/1126-6708/2000/03/007}{\tt doi:10.1088/1126-6708/2000/03/007}].

\bibitem[Braun 2000]{Braun2000}
Braun V., {\it K-Theory Torsion},
[\href{https://arxiv.org/abs/hep-th/0005103}{\tt arXiv:hep-th/0005103}].

\bibitem[Braunack-Mayer et al. 2019]{BMSS19}
Braunack-Mayer V., Sati H., and Schreiber U.,
{\it Gauge enhancement of Super M-Branes via rational parameterized stable homotopy theory},
Commun. Math. Phys.
{\bf 371} (2019) 197–265, [\href{https://arxiv.org/abs/1806.01115}{\tt arXiv:1806.01115}],
[\href{https://doi.org/10.1007/s00220-019-03441-4}{\tt doi:10.1007/s00220-019-03441-4}].

\bibitem[Brodzki et al. 2008]{BrodzkiMathaiRosenbergSzabo08}
Brodzki J., Mathai V., Rosenberg J. and Szabo R., {\it D-Branes, RR-Fields and Duality on Noncommutative Manifolds}, Commun. Math. Phys. {\bf 277} (2008),
643-706, [\href{https://doi.org/10.1007/s00220-007-0396-y}{\tt doi:10.1007/s00220-007-0396-y}], \newline
[\href{https://arxiv.org/abs/hep-th/0607020}{\tt arXiv:hep-th/0607020}].


\bibitem[Brunner et al. 2002a]{BrunnerDistler02}
Brunner I. and Distler J., {\it Torsion D-Branes in Nongeometrical Phases}, Adv. Theor. Math. Phys. {\bf 5} (2002), 265-309,
[\href{https://doi.org/10.4310/ATMP.2001.v5.n2.a3}{\tt doi:10.4310/ATMP.2001.v5.n2.a3}],
[\href{https://arxiv.org/abs/hep-th/0102018}{\tt arXiv:hep-th/0102018}].

\bibitem[Brunner et al. 2002b]{BrunnerDistlerMahajan02}
Brunner I., Distler J. and Mahajan R., {\it Return of the Torsion D-Branes},  Adv. Theor.  Math. Phys. {\bf 5} (2002), 311-352,
[\href{ https://dx.doi.org/10.4310/ATMP.2001.v5.n2.a4}{\tt
doi:10.4310/ATMP.2001.v5.n2.a4}],
[\href{https://arxiv.org/abs/hep-th/0106262}{\tt
arXiv:hep-th/0106262}].

\bibitem[Brylinski 1993]{Brylinski93}
Brylinski J.-L.,
{\it Loop spaces, characteristic classes and geometric quantization}, Birkh{\"a}user (1993), \newline
[\href{https://www.springer.com/gp/book/9780817647308}{\tt doi:10.1007/978-0-8176-4731-5}].

\bibitem[Bunke 2012]{Bunke12}
Bunke U.,
{\it Differential cohomology},
[\href{https://arxiv.org/abs/1208.3961}{\tt arXiv:1208.3961}].

\bibitem[Bunke \& Nikolaus 2019]{BunkeNikolaus19}
Bunke U., Nikolaus T., {\it Twisted differential cohomology}, Algebr. Geom. Topol. {\bf 19} 4 (2019),
1631-1710,
[\href{https://projecteuclid.org/euclid.agt/1566439272}{\tt euclid:agt/1566439272}],
[\href{https://arxiv.org/abs/1406.3231}{\tt arXiv:1406.3231}].


\bibitem[Candelas et al. 1985]{CHSW85}
Candelas P, Horowitz G., Strominger A. and Witten E., {\it Vacuum configurations for superstrings},
Nucl. Phys. B {\bf 258} (1985), 46-74,
[\href{https://doi.org/10.1016/0550-3213(85)90602-9}{\tt doi:10.1016/0550-3213(85)90602-9}].

\bibitem[Capotosti 2016]{Capotosti16}
Capotosti A., {\it From String structures to Spin structures on loop spaces}, Ph.D. thesis, Rome (2016),
\newline
[\href{https://ncatlab.org/nlab/files/Capotosti-FromStringStructures.pdf}{\tt ncatlab.org/nlab/files/Capotosti-FromStringStructures.pdf}]


\bibitem[Carey et al. 2004]{CareyJohnsonMurray04}
Carey A., Johnson S., and Murray M.,
{\it Holonomy on D-Branes}, J. Geom. Phys. {\bf 52} 2 (2004), 186-216,
[\href{https://doi.org/10.1016/j.geomphys.2004.02.008}{\tt doi:10.1016/j.geomphys.2004.02.008}],
[\href{https://arxiv.org/abs/hep-th/0204199}{\tt arXiv:hep-th/0204199}].



\bibitem[Cartan1924]{Cartan24}
{\'E.} Cartan,
{\it Sur les vari{\'e}t{\'e}s {\`a} connexion affine, et la th{\'e}orie de la relativit{\'e} g{\'e}n{\'e}ralis{\'e}e (premi{\`e}re partie) (Suite)}, Annales scientifiques de l’{\'E}.N.S. 3e s{\'e}rie, tome {\bf 41} (1924), 1-25,
[\href{http://www.numdam.org/item?id=ASENS\_1924\_3\_41\_\_1\_0}{\tt numdam:ASENS\_1924\_3\_41\_\_1\_0}].

\bibitem[Castellani et al. 1991]{CastellaniDAuriaFre91}
Castellani L., D'Auria R. and Fr{\'e} P.,
{\it Supergravity and Superstrings -- A Geometric Perspective}, World Scientific, (1991),
[\href{https://doi.org/10.1142/0224}{\tt doi:10.1142/0224}].

\bibitem[Chu et al. 2012]{ChuLorscheidSanthanam12}
Chu C. , Lorscheid O. and Santhanam R., {\it Sheaves and K-theory for $\mathbb{F}_1$-schemes},
Adv. Math. {\bf 229} 4 (2012), 2239-2286,
[\href{https://doi.org/10.1016/j.aim.2011.12.023}{\tt doi:10.1016/j.aim.2011.12.023}],
[\href{https://arxiv.org/abs/1010.2896}{\tt arXiv:1010.2896}].


\bibitem[Claus et al. 1998]{ClausKalloshVanProeyen98}
Claus P., Kallosh R. and Van Proeyen A.,
{\it M5-brane and superconformal $(0,2)$ tensor multiplet in 6 dimensions}, Nucl. Phys. B {\bf 518} (1998), 117-150,
[\href{https://doi.org/10.1016/S0550-3213(98)00137-0}{\tt doi:10.1016/S0550-3213(98)00137-0}],
[\href{https://arxiv.org/abs/hep-th/9711161}{\tt doi:hep-th/9711161}].


\bibitem[Cordova et al. 2022]{CDIS22}
Cordova C., Dumitrescu T. T., Intriligator K. and Shao S.-H.,
{\it Snowmass White Paper: Generalized Symmetries in Quantum Field Theory and Beyond},
[\href{https://arxiv.org/abs/2205.09545}{\tt arXiv:2205.09545}].

\bibitem[Cremmer et al. 1978]{CremmerJuliaScherk79}
Cremmer E., Julia B., Scherk J., {\it Supergravity in theory in 11 dimensions}, Phys. Lett. B {\bf 76} (1978), 409-412,
[\href{https://doi.org/10.1016/0370-2693(78)90894-8}{\tt doi:10.1016/0370-2693(78)90894-8}].


\bibitem[Cremmer et al. 1998]{CJLP98}
Cremmer E., Julia B., Lu H. and Pope C., {\it Dualisation of Dualities, II: Twisted self-duality of doubled fields and superdualities},
Nucl. Phys. B {\bf 535} (1998), 242-292, [\href{https://doi.org/10.1016/S0550-3213(98)00552-5}{\tt doi:10.1016/S0550-3213(98)00552-5}],
[\href{https://arxiv.org/abs/hep-th/9806106}{\tt arXiv:hep-th/9806106}].

\bibitem[Crnkovi{\'c} \& Witten 1987]{CrnkovicWitten87}
Crnkovi{\'c} {\v C}, Witten E., {\it Covariant Description of Canonical Formalism in Geometrical Theories},
in: {\it Three Hundred Years of Gravitation}, Cambridge University Press (1987), 676-684,
[\href{https://www.cambridge.org/us/universitypress/subjects/physics/cosmology-relativity-and-gravitation/three-hundred-years-gravitation?format=PB&isbn=9780521379762}{\tt ISBN:9780521379762]}].

\bibitem[D'Auria \& Fr{\'e} 1982]{DAuriFre82}
D'Auria R. and Fr{\'e} P.,
{\it Geometric Supergravity in $D=11$ and its hidden supergroup},
Nucl. Phys. B {\bf 201} (1982), 101-140,
[\href{https://doi.org/10.1016/0550-3213(82)90376-5}{\tt doi:10.1016/0550-3213(82)90376-5}].

\bibitem[de Boer et al. 2002]{BDHKMMS02}
de Boer J., Dijkgraaf R., Hori K., Keurentjes A., Morgan J, Morrison D. and Sethi S., {\it Triples, Fluxes, and Strings},
Adv. Theor. Math. Phys. {\bf 4} (2002), 995-1186,
[\href{https://dx.doi.org/10.4310/ATMP.2000.v4.n5.a1}{\tt
doi:10.4310/ATMP.2000.v4.n5.a1}],
[\href{https://arxiv.org/abs/hep-th/0103170}{\tt arXiv:hep-th/0103170}].

\bibitem[Debray 2024]{Debray24}
Debray A.,
{\it Differential Cohomology}, in: {\it Enc. Math. Phys. 2nd ed.}, Elsevier (2024),
[\href{https://arxiv.org/abs/2312.14338}{\tt arXiv:2312.14338}].

\bibitem[de Wit \& Louis 1999]{deWitLouis99}
de Wit B. and Louis J., {\it Supersymmetry and Dualities in various dimensions}, NATO Sci. Ser. C {\bf 520} (1999) 33-101
[\href{https://arxiv.org/abs/hep-th/9801132}{\tt arXiv:hep-th/9801132}]
[\href{https://inspirehep.net/literature/453367}{\tt inspire:453367}]

\bibitem[Diaconescu et al. 2007]{DiaconescuFreedMoore07}
Diaconescu D.-E., Freed D. and Moore G., {\it The M-theory 3-form and $E_8$-gauge theory}, in: {\it Elliptic Cohomology Geometry},
Cambridge University Press (2007),
[\href{https://doi.org/10.1017/CBO9780511721489}{\tt doi:10.1017/CBO9780511721489}],
[\href{https://arxiv.org/abs/hep-th/0312069}{\tt hep-th/0312069}].



\bibitem[Dirac 1931]{Dirac1931}
 Dirac, P.A.M.: {\it Quantized Singularities in the Electromagnetic Field}, Proc. Royal Society A {\bf 133} (1931), 60-72, [\href{http://rspa.royalsocietypublishing.org/content/133/821/60.short}{\tt doi:doi:10.1098/rspa.1931.0130}].


\bibitem[Duff 1996]{Duff96}
Duff M., {\it M-Theory (the Theory Formerly Known as Strings)}, Int. J. Mod. Phys. A {\bf 11} (1996), 5623-5642,
[\href{https://doi.org/10.1142/S0217751X96002583}{\tt doi:10.1142/S0217751X96002583}],
[\href{https://arxiv.org/abs/hep-th/9608117}{\tt arXiv:hep-th/9608117}].


\bibitem[Duff 1999]{Duff99}
Duff M.,
{\it The World in Eleven Dimensions: Supergravity, Supermembranes and M-theory}, IoP (1999), \newline
[\href{https://www.crcpress.com/The-World-in-Eleven-Dimensions-Supergravity-supermembranes-and-M-theory/Duff/9780750306720}{\tt ISBN:9780750306720}].

\bibitem[Duff et al. 1987]{DHIS87}
Duff M., Howe P, Inami T., and Stelle K., {\it Superstrings in $D=10$ from Supermembranes in $D=11$}, Phys. Lett. B {\bf 191} (1987), 70-74,
[\href{https://doi.org/10.1016/0370-2693(87)91323-2}{\tt doi:10.1016/0370-2693(87)91323-2}],
[\href{https://inspirehep.net/literature/245249}{\tt spire:245249}].


\bibitem[Duff, Khuri \& Lu 1992]{DuffKhuriLu92}
Duff M., Khuri R. R., and Lu J. X.,
{\it String and Fivebrane Solitons: Singular or Non-singular?}, Nucl.Phys. B {\bf 377} (1992), 281-294,
[\href{https://doi.org/10.1016/0550-3213(92)90025-7}{\tt doi:10.1016/0550-3213(92)90025-7}],
[\href{https://arxiv.org/abs/hep-th/9112023}{\tt arXiv:hep-th/9112023}].

\bibitem[Duff, Khuri \& Lu 1995]{DuffKhuriLu95}
Duff M., Khuri R. R., and Lu J. X.,
{\it String Solitons}, Phys. Rept. {\bf 259} (1995), 213-326,
[\href{https://doi.org/10.1016/0370-1573(95)00002-X}{\tt doi:10.1016/0370-1573(95)00002-X}],
[\href{https://arxiv.org/abs/hep-th/9412184}{\tt arXiv:hep-th/9412184}].


\bibitem[Duff \& Lu 1994]{DuffLu1994}
Duff M. and Lu J. L,
{\it Black and super p-branes in diverse dimensions}, Nucl. Phys. B {\bf 416} (1994), 301-334,
[\href{https://doi.org/10.1016/0550-3213(94)90586-X}{\tt doi:10.1016/0550-3213(94)90586-X}],
[\href{https://arxiv.org/abs/hep-th/9306052}{\tt arXiv:hep-th/9306052}].


\bibitem[Duff et al. 1991]{DuffStelle91}
Duff M and Stelle K.,
{\it Multi-membrane solutions of $D=11$ supergravity}, Phys. Lett. B {\bf 253} 1-2 (1991), 113-118,
[\href{https://doi.org/10.1016/0370-2693(91)91371-2}{\tt doi:10.1016/0370-2693(91)91371-2}].

\bibitem[Dumitru 2022]{Dumitru22}
Dumitru S., {\it The strongly coupled $E_8 \times E_8$ heterotic string: Geometry \& Phenomenology},
[\href{https://arxiv.org/abs/2206.12310}{\tt arXiv:2206.12310}],
[\href{https://inspirehep.net/literature/2100628}{\tt spire:2100628}].


\bibitem[Eguchi et al. 1980]{EguchiGilkeyHanson80}
Eguchi T., Gilkey P. and Hanson A., {\it Gravitation, gauge theories and differential geometry}, Phys. Rept. {\bf 66} 6 (1980), 213-393, [\href{https://doi.org/10.1016/0370-1573(80)90130-1}{\tt doi:10.1016/0370-1573(80)90130-1}].

\bibitem[Erler 2013]{Erler13}
Erler T.,
{\it Analytic Solution for Tachyon Condensation in Berkovits’ Open Superstring Field Theory}, JHEP {\bf 1311} (2013) 007,
[\href{https://doi.org/10.1007/JHEP11(2013)007}{\tt doi:10.1007/JHEP11(2013)007}],
[\href{https://arxiv.org/abs/1308.4400}{\tt arXiv:1308.4400}].


\bibitem[Evslin 2006]{Evslin06}
Evslin J.,
{\it What Does(n’t) K-theory Classify?}, Second Modave Summer School in Mathematical Physics, \newline
[\href{https://arxiv.org/abs/hep-th/0610328}{\tt arXiv:hep-th/0610328}].

\bibitem[Evslin \& Sati 2003]{EvslinSati03}
Evslin J. and Sati H.,
{\it SUSY vs $E_8$ Gauge Theory in 11 Dimensions}, JHEP {\bf 0305} (2003) 048,
[\href{https://doi.org/10.1088/1126-6708/2003/05/048}{\tt doi:10.1088/1126-6708/2003/05/048}],
[\href{https://arxiv.org/abs/hep-th/0210090}{\tt arXiv:hep-th/0210090}].



\bibitem[Faraday 1852]{Faraday1852}
Faraday, M.: {\it Delienation of Lines of Magnetic Force by iron filings}, \S37 in: {\it Experimental Researches in Electricity}
Twenty-Ninth Series, Philosophical Transactions of the Royal Society of London {\bf 142} (1852), 137-159,
[\href{https://doi.org/10.1098/rstl.1852.0012}{\tt doi:10.1098/rstl.1852.0012}], [\href{https://www.jstor.org/stable/108540}{\tt jstor:108540}].


\bibitem[Fazzi M. 2017]{Fazzi17}
Fazzi M.,
{\it Higher-dimensional field theories from type II supergravity}, PhD thesis,  ULB Brussels, \newline
[\href{https://arxiv.org/abs/1712.04447}{\tt arXiv:1712.04447}].



\bibitem[Fiorenza et al. 2014]{FSS14M57d}
Fiorenza D., Sati H., and Schreiber U.,
{\it Multiple M5-branes, String 2-connections, and 7d nonabelian Chern-Simons theory},
Adv. Theor. Math. Phys.,
{\bf 18} 2 (2014) 229-321,
[\href{https://dx.doi.org/10.4310/ATMP.2014.v18.n2.a1}{\tt doi:10.4310/ATMP.2014.v18.n2.a1}],
[\href{https://arxiv.org/abs/1201.5277}{\tt arXiv:1201.5277}].

\bibitem[Fiorenza et al. 2015a]{FSS15CField}
Fiorenza D., Sati H., and Schreiber U., {\it The moduli 3-stack of the C-field}, Commun. Math. Phys. {\bf 333} 1 (2015), 117-151,
[\href{ https://doi.org/10.1007/s00220-014-2228-1}{\tt doi:10.1007/s00220-014-2228-1}],
[\href{https://arxiv.org/abs/1202.2455}{\tt arXiv:1202.2455}].


\bibitem[Fiorenza et al. 2015b]{FSS13}
Fiorenza D., Sati H., and Schreiber U., {\it Super Lie $n$-algebra extensions, higher WZW models and super $p$-branes with tensor multiplet fields},
Int. J. Geom. Meth. Mod. Phys. {\bf 12} 02 (2015) 1550018, [\href{https://arxiv.org/abs/1308.5264}{\tt arXiv:1308.5264}],
[\href{http://www.worldscientific.com/doi/abs/10.1142/S0219887815500188}{\tt doi:10.1142/S0219887815500188}].

\bibitem[Fiorenza et al. 2017]{FSS17-Sph}
Fiorenza D., Sati H., and Schreiber U., {\it Rational sphere valued supercocycles in M-theory and type IIA string theory}, J. Geom. Phys.
{\bf 114} (2017), 91-108,
[\href{http://dx.doi.org/10.1016/j.geomphys.2016.11.024}{\tt doi:10.1016/j.geomphys.2016.11.024}],
[\href{https://arxiv.org/abs/1606.03206}{\tt arXiv:1606.03206}].



\bibitem[Fiorenza et al. 2018]{FSS18}
Fiorenza D., Sati H., and  Schreiber U.,
{\it T-Duality from super Lie $n$-algebra cocycles for super $p$-branes}, Adv. Theor. Math. Phys. {\bf 22} 5 (2018),
 1209-1270,
[\href{http://dx.doi.org/10.4310/ATMP.2018.v22.n5.a3}{\tt doi:10.4310/ATMP.2018.v22.n5.a3}],
[\href{https://arxiv.org/abs/1611.06536}{\tt arXiv:1611.06536}].



\bibitem[Fiorenza et al. 2019]{FSS19}
 Fiorenza D., Sati H., and Schreiber U., {\it The rational higher structure of M-theory}, Fortschritte der Physik
{\bf 67} 8-9 (2019)
[\href{https://doi.org/10.1002/prop.201910017}{\tt doi:10.1002/prop.201910017}],
[\href{https://arxiv.org/abs/1903.02834}{\tt arXiv:1903.02834}].


\bibitem[Fiorenza et al. 2020]{FSS20-H}
Fiorenza D., Sati H., and  Schreiber U., {\it Twisted Cohomotopy implies M-theory anomaly cancellation on 8-manifolds},
Comm Math. Phys. {\bf 377} (2020), 1961-2025,
[\href{https://doi.org/10.1007/s00220-020-03707-2}{\tt doi:10.1007/s00220-020-03707-2}],
[\href{https://arxiv.org/abs/1904.10207}{\tt arXiv:1904.10207}].


\bibitem[Fiorenza et al. 2021a]{FSS19HopfWZ}
Fiorenza D., Sati H., and Schreiber U.,
{\it Twisted Cohomotopy implies M5 WZ term level quantization},
Commun. Math. Phys. {\bf 384} (2021), 403-432,
[\href{https://doi.org/10.1007/s00220-021-03951-0}{\tt doi:10.1007/s00220-021-03951-0}],
[\href{https://arxiv.org/abs/1906.07417}{\tt arXiv:1906.07417}].

\bibitem[Fiorenza et al. 2021b]{FSS20TwistedString}
Fiorenza D., Sati H., and Schreiber U.:
{\it Twisted cohomotopy implies twisted String structure on M5-branes},
J. Math. Phys. {\bf 62} (2021),  042301 ,
[\href{https://doi.org/10.1063/5.0037786}{\tt doi:10.1063/5.0037786}],
[\href{https://arxiv.org/abs/2002.11093}{\tt arXiv:2002.11093}].

\bibitem[Fiorenza et al. 2022]{FSS22Twistorial}
Fiorenza D., Sati H., and Schreiber U.,
{\it Twistorial Cohomotopy Implies Green-Schwarz anomaly cancellation},
Rev. Math. Phys.
{\bf 34} 5 (2022) 2250013,
[\href{https://doi.org/10.1142/S0129055X22500131}{\tt doi:doi.org/10.1142/S0129055X22500131}],
[\href{https://arxiv.org/abs/2008.08544}{\tt arXiv:2008.08544}].


\bibitem[Fiorenza et al. 2023]{FSS23Char}
Fiorenza D., Sati H., and Schreiber U.,
{\it The Character map in Nonabelian Cohomology --- Twisted, Differential and Generalized},
World Scientific (2023),
[\href{https://doi.org/10.1142/13422}{\tt doi:10.1142/13422}],
[\href{https://arxiv.org/abs/2009.11909}{\tt arXiv:2009.11909}].

\bibitem[Frankel 1997]{Frankel97}
Frankel T., {\it The Geometry of Physics -- An Introduction}, Cambridge University Press (1997, 2004, 2012), \newline
[\href{https://doi.org/10.1017/CBO9781139061377}{\tt doi:10.1017/CBO9781139061377}].



\bibitem[Fredenhagen \& Quella 2005]{FredenhagenQuella05}
Fredenhagen S. and Quella T., {\it Generalised permutation branes}, JHEP {\bf 0511} (2005) 004,
[\href{https://doi.org/10.1088/1126-6708/2005/11/004}{\tt
doi:10.1088/1126-6708/2005/11/004}],
[\href{https://arxiv.org/abs/hep-th/0509153}{\tt arXiv:hep-th/0509153}].



\bibitem[Freed 2000]{Freed00}
Freed D. S.,
{\it Dirac charge quantization and generalized differential cohomology},
Surveys in Differential Geometry,
Int. Press, Somerville, MA, 2000, pp. 129-194,
[\href{https://dx.doi.org/10.4310/SDG.2002.v7.n1.a6}{\tt doi:10.4310/SDG.2002.v7.n1.a6}],
[\href{https://arxiv.org/abs/hep-th/0011220}{\tt arXiv:hep-th/0011220}].

\bibitem[Freed \& Hopkins 2000]{FreedHopkins00}
Freed D. S., Hopkins M.,
{\it On Ramond-Ramond fields and K-theory}, JHEP {\bf 0005} (2000) 044,
[\href{https://doi.org/10.1088/1126-6708/2000/05/044}{\tt doi:10.1088/1126-6708/2000/05/044}],
[\href{https://arxiv.org/abs/hep-th/0002027}{\tt arXiv:hep-th/0002027}].


\bibitem[Freed et al. 2007a]{FreedHopkinsTeleman07}
Freed D. S., Hopkins M. and Teleman C., {\it Twisted equivariant K-theory with complex coefficients}, J. Topology {\bf 1} 1 (2007),
16-44, [\href{https://doi.org/10.1112/jtopol/jtm001}{\tt doi:10.1112/jtopol/jtm001}],
[\href{https://arxiv.org/abs/math/0206257}{\tt arXiv:math/0206257}].



\bibitem[Freed et al. 2007b]{FreedMooreSegal07a}
Freed D. S., Moore G. W., Segal G.,
{\it The Uncertainty of Fluxes},
Commun. Math. Phys. {\bf 271} (2007) 247-274,
[\href{https://doi.org/10.1007/s00220-006-0181-3}{\tt doi:10.1007/s00220-006-0181-3}],
[\href{https://arxiv.org/abs/hep-th/0605198}{\tt arXiv:hep-th/0605198}].


\bibitem[Freed et al. 2007c]{FreedMooreSegal07b}
Freed D. S., Moore G. W., Segal G.,
{\it Heisenberg Groups and Noncommutative Fluxes}, Annals Phys. {\bf 322} (2007), 236-285,
[\href{https://doi.org/10.1016/j.aop.2006.07.014}{\tt doi:10.1016/j.aop.2006.07.014}],
[\href{https://arxiv.org/abs/hep-th/0605200}{\tt arXiv:hep-th/0605200}].

\bibitem[Freed \& Witten 1999]{FreedWitten99}
Freed D. and Witten E., {\it Anomalies in String Theory with D-Branes}, Asian J. Math. {\bf 3} 4 (1999),
819-852,
[\href{https://dx.doi.org/10.4310/AJM.1999.v3.n4.a6}{\tt
doi:10.4310/AJM.1999.v3.n4.a6}],
[\href{https://arxiv.org/abs/hep-th/9907189}{\tt arXiv:hep-th/9907189}].

\bibitem[Gawedzki 1988]{Gawedzki88}
Gawedzki K., {\it Topological Actions in two-dimensional Quantum Field Theories}, in: {\it Nonperturbative Quantum Field Theory},
Nato Science Series B {\bf 185}, Springer (1986),
[\href{https://doi.org/10.1007/978-1-4613-0729-7_5}{\tt doi:10.1007/978-1-4613-0729-7\_5}].


\bibitem[Giotopoulos \& Sati 2023]{GiotopoulosSati23}
Giotopoulos G. and Sati H., {\it Field Theory via Higher Geometry I: Smooth Sets of Fields},
[\href{https://arxiv.org/abs/2312.16301}{\tt arXiv:2312.16301}].

\bibitem[Giotopoulos et al. 2024a]{GSS-11dSuperFlux}
Giotopoulos G., Sati H. and Schreiber S., {\it Flux Quantization on 11d Superspace},
J. of High Energy Physics {\bf 2024} 82 (2024)
[\href{https://arxiv.org/abs/2403.16456}{\tt arXiv:2403.16456}],
[\href{https://doi.org/10.1007/JHEP07(2024)082}{\tt doi:10.1007/JHEP07(2024)082}].

\bibitem[Giotopoulos et al. 2024b]{GSS-6dSuperFlux}
Giotopoulos G., Sati H. and Schreiber S., {\it Flux Quantization on M5-Branes},
J. of High Energy Physics {\bf 2024} 140 (2024)
[\href{https://arxiv.org/abs/2406.11304}{\tt arXiv:2406.11304}], [\href{https://doi.org/10.1007/JHEP10(2024)140}{\tt doi:10.1007/JHEP10(2024)140}]


\bibitem[Grady \& Sati 2021]{GradySati2021}
Grady D. and Sati H., {\it Differential cohomotopy versus differential cohomology for M-theory and differential lifts of Postnikov towers},
J. Geom. Phys. {\bf 165} (2021) 104203,
[\href{https://doi.org/10.1016/j.geomphys.2021.104203}{\tt doi:10.1016/j.geomphys.2021.104203}],
[\href{https://arxiv.org/abs/2001.07640}{\tt arXiv:2001.07640}].

\bibitem[Green \& Schwarz 1984]{GreenSchwarz84}
Green M. and Schwarz J.,
{\it Anomaly Cancellation in Supersymmetric $D=10$ Gauge Theory and Superstring Theory}, Phys. Lett. B {\bf 149} (1984), 117-122,
[\href{https://doi.org/10.1016/0370-2693(84)91565-X}{\tt doi:10.1016/0370-2693(84)91565-X}].

\bibitem[Hanany \& Witten 1997]{HananyWitten97}
Hanany A.  and  Witten E., {\it Type IIB Superstrings, BPS Monopoles, And Three-Dimensional Gauge Dynamics}, Nucl. Phys. B {\bf 492} (1997),
152-190,  [\href{https://doi.org/10.1016/S0550-3213(97)80030-2}{\tt doi:10.1016/S0550-3213(97)80030-2}],
[\href{https://arxiv.org/abs/hep-th/9611230}{\tt arXiv:hep-th/9611230}].

\bibitem[Hanany \& Zaffaroni 1998]{HananyZaffaroni98}
Hanany A.  and Zaffaroni A.,
{\it Chiral Symmetry from Type IIA Branes},
Nucl. Phys. B {\bf 509} (1998), 145-168,
[\href{https://doi.org/10.1016/S0550-3213(98)00045-5}{\tt doi:10.1016/S0550-3213(98)00045-5}],
[\href{https://arxiv.org/abs/hep-th/9706047}{\tt arXiv:hep-th/9706047}].

\bibitem[Hari Dass 2024]{HariDass24}
Hari Dass N. D.,
{\it Strings to Strings -- Yang-Mills Flux Tubes, QCD Strings and Effective String Theories}, Lecture Notes in Physics {\bf 1018}, Springer (2024), [\href{https://doi.org/10.1007/978-3-031-35358-1}{\tt doi:10.1007/978-3-031-35358-1}].



\bibitem[Hatcher 2002]{Hatcher02}
Hatcher A.,
{\it Algebraic Topology}, Cambridge University Press (2002), [\href{https://www.cambridge.org/gb/academic/subjects/mathematics/geometry-and-topology/algebraic-topology-1?format=PB&isbn=9780521795401}{\tt ISBN:9780521795401}].

\bibitem[Hehl et al. 2016]{HIO16}
Hehl F. W.,  Itin Y. and Obukhov Y. N.,
{\it On Kottler's path: origin and evolution of the premetric program in gravity and in electrodynamics}, Int. J. Modern Phys. D
{\bf 25} 11 (2016) 1640016,
[\href{https://doi.org/10.1142/S0218271816400162}{\tt doi:10.1142/S0218271816400162}],
[\href{https://arxiv.org/abs/1607.06159}{\tt arXiv:1607.06159}].


\bibitem[Hehl \& Obukhov2003]{HehlObhukov03}
Hehl F. W. and  Obukhov Y. N.,
{\it Foundations of Classical Electrodynamics -- Charge, Flux, and Metric}, Progress in Mathematical Physics {\bf 33}, Springer (2003),
[\href{https://doi.org/10.1007/978-1-4612-0051-2}{\tt doi:10.1007/978-1-4612-0051-2}].


\bibitem[Henneaux et al. 1992]{HenneauxTeitelboim92}
Henneaux, M. and Teitelboim C.,
{\it Quantization of Gauge Systems},
Princeton University Press (1992),
[\href{https://doi.org/10.2307/j.ctv10crg0r}{\tt doi:10.2307/j.ctv10crg0r}].

\bibitem[Hess 2006]{Hess06}
Hess H.,
{\it Rational homotopy theory: a brief introduction}, in: {\it Interactions between Homotopy Theory and Algebra},
Contemporary Mathematics {\bf 436} AMS 2007,
[\href{https://www.ams.org/books/conm/436/}{\tt
ams.org/books/conm/436}],
[\href{https://arxiv.org/abs/math/0604626}{\tt arXiv:math/0604626}].

\bibitem[Hopkins \& Singer 2005]{HopkinsSinger05}
Hopkins M. and Singer I.,
{\it Quadratic Functions in Geometry, Topology, and M-Theory}, J. Differential Geom. {\bf 70} 3 (2005), 329-452,
[\href{https://doi.org/10.4310/jdg/1143642908}{\tt doi:10.4310/jdg/1143642908}],
[\href{https://arxiv.org/abs/math/0211216}{\tt arXiv:0211216}].


\bibitem[Ho{\v r}ava \& Witten 1996]{HoravaWitten96}
Ho{\v r}ava P. and Witten E.,
{\it Heterotic and Type I string dynamics from eleven dimensions}, Nucl. Phys. B {\bf 460} (1996), 506-524,
[\href{https://doi.org/10.1016/0550-3213(95)00621-4}{\tt doi:10.1016/0550-3213(95)00621-4}],
[\href{https://arxiv.org/abs/hep-th/9510209}{\tt arXiv:hep-th/9510209}].

\bibitem[Howe \& Sezgin 1997]{HoweSezgin97}
Howe P. and Sezgin E.,
{\it $D=11$, $p=5$},
Phys. Lett. B {\bf 394} (1997), 62-66, [\href{https://arxiv.org/abs/hep-th/9611008}{\tt arXiv:hep-th/9611008}],
[\href{https://doi.org/10.1016/S0370-2693(96)01672-3}{\tt doi:10.1016/S0370-2693(96)01672-3}].

\bibitem[Howe, Sezgin \& West 1997]{HoweSezginWest97}
Howe P., Sezgin E., West P., {\it The six-dimensional self-dual tensor}, Physics Letters B {\bf 400} 3–4 (1997) 255-259
[\href{https://doi.org/10.1016/S0370-2693(97)00365-1}{\tt doi:10.1016/S0370-2693(97)00365-1}]



\bibitem[Huerta et al 2019]{HSS19}
Huerta J., Sati H. and  Schreiber U.,
{\it Real ADE-equivariant (co)homotopy and Super M-branes}
Commun. Math. Phys. {\bf 371} (2019) 425-524,
[\href{https://doi.org/10.1007/s00220-019-03442-3}{\tt doi:10.1007/s00220-019-03442-3}],
[\href{https://arxiv.org/abs/1805.05987}{\tt arXiv:1805.05987}].

\bibitem[Hyperphysics]{HyperphysicsMagneticFlux}
{\it Magnetic Flux}, [\href{http://hyperphysics.phy-astr.gsu.edu/hbase/magnetic/fluxmg.html}{\tt hyperphysics.phy-astr.gsu.edu/hbase/magnetic/fluxmg.html}].

\bibitem[Intriligator 2000]{Intriligator00}
Intriligator K., {\it Anomaly Matching and a Hopf-Wess-Zumino Term in 6d, $\mathcal{N} = (2,0)$ Field Theories}, Nucl. Phys. B {\bf 581} (2000), 257-273,
[\href{https://doi.org/10.1016/S0550-3213(00)00148-6}{\tt doi:10.1016/S0550-3213(00)00148-6}],
[\href{https://arxiv.org/abs/hep-th/0001205}{\tt arXiv:hep-th/0001205}].


\bibitem[Kosinski 1993]{Kosinski93}
A. Kosinski,
{\it Differential manifolds}, Academic Press (1993), [\href{https://www.sciencedirect.com/bookseries/pure-and-applied-mathematics/vol/138/suppl/C}{\tt ISBN:978-0-12-421850-5}].


\bibitem[Lazaroiu \& Shahbazi 2022]{LazaroiuShahbazi22}
Lazaroiu C and Shahbazi C. S.,
{\it The duality covariant geometry and DSZ quantization of abelian gauge theory}, Adv. Theor. Math. Phys. {\bf 26} (2022),
2213–2312, [\href{https://dx.doi.org/10.4310/ATMP.2022.v26.n7.a5}{\tt doi:10.4310/ATMP.2022.v26.n7.a5}],
[\href{https://arxiv.org/abs/2101.07236}{\tt arXiv:2101.07236}].

\bibitem[Lazaroiu \& Shahbazi 2023]{LazaroiuShahbazi23}
Lazaroiu C and Shahbazi C. S.,
{\it The geometry and DSZ quantization of four-dimensional supergravity},
Lett. Math. Phys. {\bf 113}  (2023) 4,
[\href{https://doi.org/10.1007/s11005-022-01626-y}{\tt doi:10.1007/s11005-022-01626-y}],
[\href{https://arxiv.org/abs/2101.07778}{\tt arXiv:2101.07778}].

\bibitem[Loudon et al. 2009]{LoudonMidgley09}
Loudon J. C. and Midgley P. A., {\it Imaging Flux Vortices in Type II Superconductors with a Commercial Transmission Electron Microscope},
Ultramicroscopy {\bf 109} 6 (2009), 700-729,
[\href{https://doi.org/10.1016/j.ultramic.2009.01.008}{\tt doi:10.1016/j.ultramic.2009.01.008}],
[\href{https://arxiv.org/abs/0807.2401}{\tt arXiv:0807.2401}].



\bibitem[Marolf 2001]{Marolf01}
Marolf D.,
{\it T-duality and the case of the disappearing brane}, JHEP {\bf 0106} (2001) 036,
[\href{https://arxiv.org/abs/hep-th/0103098}{\tt arXiv:hep-th/0103098}],
[\href{https://doi.org/10.1016/S0550-3213(97)80030-2}{\tt doi:10.1016/S0550-3213(97)80030-2}].


\bibitem[Martin09]{Martin09}
Martin, T. (ed.), {\it Faraday’s diary of experimental investigation 1820-1862}, HR Direct, Riverton, UT (2009),  \newline
[\href{http://faradaysdiary.com}{\tt faradaysdiary.com}].

\bibitem[Mathai \& Sati 2004]{MathaiSati04}
Mathai V. and Sati H.,
{\it Some Relations between Twisted K-theory and $E_8$ Gauge Theory}, J. High Energy Phys. {\bf 2004} 03 (2004) 016,
[\href{https://doi.org/10.1088/1126-6708/2004/03/016}{\tt doi:10.1088/1126-6708/2004/03/016}],
[\href{https://arxiv.org/abs/hep-th/0312033}{\tt arXiv:hep-th/0312033}].

\bibitem[Maxwell 1865]{Maxwell1865}
Maxwell J. C.,
{\it A Dynamical Theory of the Electromagnetic Field}, Phil. Trans. Royal Soc. London {\bf 155} (1865),
459-512, [\href{https://doi.org/10.1098/rstl.1865.0008}{\tt doi:10.1098/rstl.1865.0008}].

\bibitem[Miemiec et al. 2006]{MiemiecSchnakenburg06}
Miemiec A. and Schnakenburg I., {\it Basics of M-Theory}, Fortsch. Phys. {\bf 54} (2006), 5-72, \newline
[\href{https://doi.org/10.1002/prop.200510256}{\tt doi:10.1002/prop.200510256}],
[\href{https://arxiv.org/abs/hep-th/0509137}{\tt arXiv:hep-th/0509137}].


\bibitem[Minasian \& Moore 1997]{MinasianMoore97}
Minasian R. and Moore G., {\it K-theory and Ramond-Ramond charge}, JHEP {\bf 9711} (1997) 002,
[\href{https://doi.org/10.1088/1126-6708/1997/11/002}{\tt doi:10.1088/1126-6708/1997/11/002}],
[\href{https://arxiv.org/abs/hep-th/9710230}{\tt arXiv:hep-th/9710230}].


\bibitem[Moore 2005]{Moore05}
Moore G.,
{\it Anomalies, Gauss laws, and Page charges in M-theory}, Comptes Rendus Physique {\bf 6} (2005), 251-259,
[\href{https://doi.org/10.1016/j.crhy.2004.12.005}{\tt doi:10.1016/j.crhy.2004.12.005}],
[\href{https://arxiv.org/abs/hep-th/0409158}{\tt arXiv:hep-th/0409158}].

\bibitem[Mkrtchyan \& Valach 2023]{MkrtchyanValach23}
Mkrtchyan K. and Valach F.,
{\it Democratic actions for type II supergravities}, Phys. Rev. D {\bf 107} 6 (2023) 066027,
[\href{https://doi.org/10.1103/PhysRevD.107.066027}{\tt doi:10.1103/PhysRevD.107.066027}],
[\href{https://arxiv.org/abs/2207.00626}{\tt arXiv:2207.00626}].


\bibitem[Nielsen et al. 1973]{NielsenOlesen73}
Nielsen, H. and Olesen P,
{\it Vortex-line models for dual strings}, Nucl. Phys. B {\bf 61} (1973),
45-61, [\href{https://doi.org/10.1016/0550-3213(73)90350-7}{\tt doi:10.1016/0550-3213(73)90350-7}].

\bibitem[Page 1983]{Page83}
Page D.,
{\it Classical stability of round and squashed seven-spheres in eleven-dimensional supergravity}, Phys. Rev. D {\bf 28} 12 (1983),
2976-2982, [\href{https://doi.org/10.1103/PhysRevD.28.2976}{\tt doi:10.1103/PhysRevD.28.2976}].

\bibitem[Peterson 1956]{Peterson56}
F. Peterson,
{\it Some Results on Cohomotopy Groups}, Amer. J. Math. {\bf 78} 2 (1956), 243-258,
[\href{https://www.jstor.org/stable/2372514}{\tt jstor:2372514}].


\bibitem[Polyakov 2012]{Polyakov12}
Polyakov A. M.,
{\it From Quarks to Strings}, chapter 44 {\it Quarks, strings and beyond} in {\it The Birth of String Theory},
Cambridge University Press (2012), 544-551,
[\href{https://doi.org/10.1017/CBO9780511977725.048}{\tt doi:10.1017/CBO9780511977725.048}],
[\href{https://arxiv.org/abs/0812.0183}{\tt arXiv:0812.0183}].



\bibitem[Pontrjagin 1938]{Pontrjagin38}
Pontrjagin L.,
{\it Classification of continuous maps of a complex into a sphere}, Communication I, Doklady Akademii Nauk SSSR {\bf 19} 3 (1938),
147-149, [\href{https://www.routledge.com/Selected-Research-Papers/Pontryagn/p/book/9782881241055}{\tt ISBN:9782881241055}].


\bibitem[Rosenberg 2024]{Rosenberg24}
Rosenberg J.,
{\it Twisted cohomology}, Enc. Math. Phys., Elsevier (2024),
[\href{https://browse.arxiv.org/abs/2401.03966}{\tt
arxiv:2401.03966}].

\bibitem[Rudolph \& Schmidt 2017]{RudolphSchmidt17}
Rudolph G. and Schmidt M., {\it Differential Geometry and Mathematical Physics: Part II. Fibre Bundles, Topology and Gauge Fields}, 
Springer (2017),
[\href{https://link.springer.com/book/10.1007/978-94-024-0959-8}{\tt doi:10.1007/978-94-024-0959-8}].

\bibitem[Sati 2005]{Sati05}
H. Sati,
{\it Flux Quantization and the M-Theoretic Characters},
Nucl. Phys. B {\bf 727} (2005), 461-470, \newline
[\href{https://doi.org/10.1016/j.nuclphysb.2005.09.008}{\tt
doi:10.1016/j.nuclphysb.2005.09.008}],
[\href{https://arxiv.org/abs/hep-th/0507106}
{\tt arXiv:hep-th/0507106}].
 	

\bibitem[Sati 2006]{Sati06}
H. Sati,
{\it Duality symmetry and the form fields of M-theory},
J. High Energy Phys. {\bf 0606} (2006) 062,  \newline
[\href{https://doi.org/10.1088/1126-6708/2006/06/062}{\tt
doi:10.1088/1126-6708/2006/06/062}],
 [\href{https://arxiv.org/abs/hep-th/0509046}{\tt
 arXiv:hep-th/0509046}].


\bibitem[Sati 2010]{Sati10}
Sati H.,
{\it Geometric and topological structures related to M-branes},
in:
{\it Superstrings, Geometry, Topology, and $C^\ast$-algebras},
Proc. Symp. Pure Math. {\bf 81}, AMS, Providence (2010), 181-236,
[\href{https://doi.org/10.1090/pspum/081}{\tt doi:10.1090/pspum/081}],
[\href{https://arxiv.org/abs/1001.5020}{\tt arXiv:1001.5020}].

\bibitem[Sati 2011]{Sati11StrucII}
Sati H.
{\it Geometric and topological structures related to M-branes II: Twisted String and $\mathrm{String}^c$ structures},
J. Australian Math. Soc. {\bf 90} (2011), 93-108,
[\href{https://doi.org/10.1017/S1446788711001261}{\tt doi:10.1017/S1446788711001261}],
[\href{https://arxiv.org/abs/1007.5419}{\tt arXiv:1007.5419}].

\bibitem[Sati 2013]{Sati13}
Sati H., {\it Framed M-branes, corners, and topological invariants}, J. Math. Phys. {\bf 59} (2018) 062304,
[\href{https://arxiv.org/abs/1310.1060}{\tt arXiv:1310.1060}]
[\href{https://doi.org/10.1063/1.5007185}{\tt doi:10.1063/1.5007185}].


\bibitem[Sati \& Schreiber 2020]{SatiSchreiber20}
Sati H. and Schreiber U.,
{\it Equivariant Cohomotopy implies orientifold tadpole cancellation}, J. Geom. Phys. {\bf 156} (2020) 103775,
[\href{https://doi.org/10.1016/j.geomphys.2020.103775}{\tt doi:10.1016/j.geomphys.2020.103775}],
[\href{https://arxiv.org/abs/1909.12277}{\tt arXiv:1909.12277}].

\bibitem[Sati \& Schreiber 2020]{SS20EquTw}
Sati H. and Schreiber U.,
{\it The character map in equivariant twistorial Cohomotopy}, \newline
[\href{https://arxiv.org/abs/2011.06533}{\tt arXiv:2011.06533}].


\bibitem[Sati \& Schreiber 2021a]{SS21-M5}
Sati H. and Schreiber U.,
{\it Twisted Cohomotopy implies M5-brane anomaly cancellation}, Lett. Math. Phys. {\bf 111} (2021) 120,
[\href{https://doi.org/10.1007/s11005-021-01452-8}{\tt doi:10.1007/s11005-021-01452-8}],
[\href{https://arxiv.org/abs/2002.07737}{\tt arXiv:2002.07737}].


\bibitem[Sati \& Schreiber 2021b]{SatiSchreiber21}
Sati H. and Schreiber U., {\it Equivariant principal $\infty$-bundles},
[\href{https://arxiv.org/abs/2112.13654}{\tt arXiv:2112.13654}].

\bibitem[Sati \& Schreiber 2022]{SS22Config}
Sati H. and Schreiber U.,
{\it Differential Cohomotopy implies intersecting brane observables
via configuration spaces and chord diagrams},
Adv. Theor. Math. Phys. {\bf 26} 4 (2022),
957-1051, [\href{https://arxiv.org/abs/1912.10425}{\tt arXiv:1912.10425}], \newline
[\href{https://doi.org/10.4310/ATMP.2022.v26.n4.a4}{\tt
doi:10.4310/ATMP.2022.v26.n4.a4}].
.


\bibitem[Sati \& Schreiber 2023a]{SS23MF}
Sati H. and Schreiber U.,
{\it M/F-Theory as $M\!f$-Theory},
Rev. Math. Phys. {\bf 35} 10 (2023) 2350028,
[\href{https://doi.org/10.1142/S0129055X23500289}{\tt doi:10.1142/S0129055X23500289}],
[\href{https://arxiv.org/abs/2103.01877}{\tt arXiv:2103.01877}].

\bibitem[Sati \& Schreiber2023b]{SS23FQ}
Sati H. and Schreiber U.,
{\it Flux Quantization on Phase Space},
Annales Henri Poincar{\'e} (2024)
[\href{https://arxiv.org/abs/2312.12517}{\tt arXiv:2312.12517}], [\href{https://doi.org/10.1007/s00023-024-01438-x}{\tt doi:10.1007/s00023-024-01438-x}].

\bibitem[Sati \& Schreiber 2023c]{SS23QObs}
Sati H. and Schreiber U.,
{\it Quantum Observables on Quantized Fluxes},
Annales Henri Poincar{\'e} (2024)
[\href{https://arxiv.org/abs/2312.13037}{\tt arXiv:2312.13037}].

\bibitem[Sati et al 2009]{SSS09LInfinity}
Sati H., Schreiber U. and Stasheff J, {\it $L_\infty$-algebra connections and applications to String- and Chern-Simons $n$-transport},
in {\it Quantum Field Theory}, Birkh{\"a}user (2009), 303-424,
[\href{https://doi.org/10.1007/978-3-7643-8736-5_17}{\tt doi:10.1007/978-3-7643-8736-5\_17}],
[\href{https://arxiv.org/abs/0801.3480}{\tt arXiv:0801.3480}].


\bibitem[Sati et al. 2012]{SSSStringStruc12}
Sati H., Schreiber U and Stasheff J., {\it Twisted Differential String and Fivebrane Structures}, Commun. Math. Phys. {\bf 315} (2012),
169-213,  [\href{https://link.springer.com/article/10.1007/s00220-012-1510-3}{\tt doi:10.1007/s00220-012-1510-3}],
[\href{https://arxiv.org/abs/0910.4001}{\tt arXiv:0910.4001}].



\bibitem[Sati \& Voronov 2022]{SatiVoronov22}
Sati H. and  Voronov A.,
{\it Mysterious Triality and M-Theory},
[\href{https://arxiv.org/abs/2212.13968}{\tt arXiv:2212.13968}].



\bibitem[Schreiber 2024]{Schreiber24}
Schreiber U.,
{\it Higher Topos Theory in Physics}, Enc. Math. Phys. 2nd ed., Elsevier (2024), \newline [\href{https://arxiv.org/abs/2311.11026}{\tt arXiv:2311.11026}].

\bibitem[Schreiber \& {\v S}koda 2009]{SchreiberSkoda09}
Schreiber U. and {\v S}koda Z., {\it Categorified symmetries}, 5th Summer School of Modern Mathematical Physics, SFIN XXII Series {\bf A1} (2009),
397-424,
[\href{https://arxiv.org/abs/1004.2472}{\tt arXiv:1004.2472}],
[\href{https://inspirehep.net/literature/851901}{\tt spire:851901}].

\bibitem[Sen 1998]{Sen98}
Sen A., {\it Tachyon Condensation on the Brane Antibrane System}, JHEP {\bf 9808}  (1998) 012,
[\href{https://arxiv.org/abs/hep-th/9805170}{\tt arXiv:hep-th/9805170}],
[\href{https://doi.org/10.1088/1126-6708/1998/08/012}{\tt doi:10.1088/1126-6708/1998/08/012}].


\bibitem[Sezgin 2023]{Sezgin21}
Sezgin E.,
{\it Survey of supergravities}, in: {\it Handbook of Quantum Gravity}, Springer (2023),
[\href{https://arxiv.org/abs/2312.06754}{\tt arXiv:2312.06754}],
[\href{https://doi.org/10.1007/978-981-19-3079-9}{\tt doi:10.1007/978-981-19-3079-9}].

\bibitem[Sorokin 2000]{Sorokin00}
D. Sorokin,
{\it Superbranes and Superembeddings},
Phys. Rept. {\bf 329} (2000), 1-101,
[\href{https://arxiv.org/abs/hep-th/9906142}{\tt arXiv:hep-th/9906142}], \newline
[\href{https://doi.org/10.1016/S0370-1573(99)00104-0}{\tt doi:10.1016/S0370-1573(99)00104-0}].


\bibitem[Sou{\`e}res \& Tsimpis 2017]{SoueresTsimpis17}
Sou{\'e}res B. and Tsimpis D.,
{\it The action principle and the supersymmetrisation of Chern-Simons terms in eleven-dimensional supergravity},
Phys. Rev. D {\bf 95}  (2017) 026013,
[\href{https://doi.org/10.1103/PhysRevD.95.026013}{\tt doi:10.1103/PhysRevD.95.026013}], \newline
[\href{https://arxiv.org/abs/1612.02021}{\tt arXiv:1612.02021}].

\bibitem[Spanier 1949]{Spanier49}
E. Spanier,
{\it Borsuk’s Cohomotopy Groups}, Ann. Math. {\bf 50} 1 (1949), 203-245,
[\href{http://www.jstor.org/stable/1969362}{\tt jstor:1969362}].

\bibitem[Stasheff 2016]{Stasheff16}
Stasheff J., {\it Higher homotopy structures, then and now}, talk at  {\it Higher Structures in Geometry and Physics}, MPI Bonn (2016),
[\href{https://arxiv.org/abs/1809.02526}{\tt arXiv:1809.02526}].

\bibitem[Tanii 1996]{Tanii96}
Tanii Y.,
{\it Introduction to Supergravities in Diverse Dimensions}, in {\it YITP Workshop on Supersymmetry}, Kyoto (1996)
[\href{https://arxiv.org/abs/hep-th/9802138}{\tt arXiv:hep-th/9802138}]

\bibitem[Timm 2020]{Timm20}
Timm C., {\it Theory of Superconductivity}, lecture notes (2020-2023),
\newline
[\href{https://tu-dresden.de/mn/physik/itp/cmt/ressourcen/dateien/skripte/Skript_Supra.pdf}{\tt tu-dresden.de/mn/physik/itp/cmt/ressourcen/dateien/skripte/Skript\_Supra.pdf}]

\bibitem[Tsimpis 2004]{Tsimpis04}
Tsimpis D.,
{\it 11D supergravity at $\mathcal{O}(\ell^3)$},
JHEP {\bf 0410} (2004) 046,
[\href{https://doi.org/10.1088/1126-6708/2004/10/046}{\tt doi:10.1088/1126-6708/2004/10/046}],
[\href{https://arxiv.org/abs/hep-th/0407271}{\tt arXiv:hep-th/0407271}].


\bibitem[van Nieuwenhuizen 1983]{vanNieuwenhuizen83}
van Nieuwenhuizen P., {\it Free Graded Differential Superalgebras}, in: {\it Group Theoretical Methods in Physics},
Lecture Notes in Physics {\bf 180}, Springer (1983), 228–247,
[\href{https://doi.org/10.1007/3-540-12291-5_29}{\tt doi:10.1007/3-540-12291-5\_29}].

\bibitem[Waldorf 2013]{Waldorf13}
Waldorf K., {\it String Connections and Chern-Simons Theory}, Trans. Amer. Math. Soc. {\bf 365} (2013), 4393-4432,
[\href{https://doi.org/10.1090/S0002-9947-2013-05816-3}{\tt doi:10.1090/S0002-9947-2013-05816-3}],
[\href{https://arxiv.org/abs/0906.0117}{\tt arXiv:0906.0117}].



\bibitem[Witten 1997]{Witten97}
Witten E.,
{\it On Flux Quantization In M-Theory And The Effective Action}, J. Geom. Phys. {\bf 22} 1 (1997), 1-13,
[\href{https://doi.org/10.1016/S0393-0440(96)00042-3}{\tt doi:10.1016/S0393-0440(96)00042-3}],
[\href{https://arxiv.org/abs/hep-th/9609122}{\tt arXiv:hep-th/9609122}].


\bibitem[Witten 1998]{Witten98}
Witten E., {\it D-Branes And K-Theory}, JHEP {\bf 9812}  (1998) 019,
[\href{https://doi.org/10.1088/1126-6708/1998/12/019}{\tt doi:10.1088/1126-6708/1998/12/019}],
[\href{https://arxiv.org/abs/hep-th/9810188}{\tt arXiv:hep-th/9810188}].

\bibitem[Witten 2000]{Witten00}
Witten E.,
{\it World-Sheet Corrections Via D-Instantons}, JHEP {\bf 2000} (2000) 030,
[\href{https://arxiv.org/abs/hep-th/9907041}{\tt arXiv:hep-th/9907041}],
[\href{https://doi.org/10.1088/1126-6708/2000/02/030}{\tt doi:10.1088/1126-6708/2000/02/030}].


\bibitem[Witten 2010]{Witten10}
Witten E.
{\it Geometric Langlands From Six Dimensions}, in {\it A Celebration of the Mathematical Legacy of Raoul Bott},
CRM Proceedings \& Lecture Notes {\bf 50}, AMS (2010),
[\href{https://bookstore.ams.org/crmp-50}{\tt ams:crmp-50}]
[\href{https://arxiv.org/abs/0905.2720}{\tt arXiv:0905.2720}].

\bibitem[Wu \& Yang 1975]{WuYang75}
Wu T. T. and  Yang C. N., {\it Concept of nonintegrable phase factors and global formulation of gauge fields}, Phys. Rev. D {\bf 12} (1975),
3845-3857,
[\href{https://doi.org/10.1103/PhysRevD.12.3845}{\tt doi:10.1103/PhysRevD.12.3845}].

\bibitem[Wu \& Yang 2006]{WuYang06}
Wu A. C. T. and Yang C. N., {\it Evolution of the concept of vector potential in the description of the fundamental interactions},
Int. J. Modern Physics A {\bf 21} 16 (2006), 3235-3277,
[\href{https://doi.org/10.1142/S0217751X06033143}{\tt doi:10.1142/S0217751X06033143}].


\end{thebibliography}
\end{document}